\documentclass{ws-rv975x65}
\usepackage{subfigure}
\usepackage{ws-rv-thm}
\usepackage{ws-rv-van}
\usepackage{hyperref}
\usepackage{slashed}

\usepackage{tikz}
\usepackage{gnuplot-lua-tikz}

\def\lsim{\raise0.3ex\hbox{$<$\kern-0.75em\raise-1.1ex\hbox{$\sim$}}}
\def\gsim{\raise0.3ex\hbox{$>$\kern-0.75em\raise-1.1ex\hbox{$\sim$}}}

\newcommand{\cO}{{\cal O}}
\newcommand{\cL}{{\cal L}}

\newcommand{\cD}{{\cal D}}

\newcommand{\D}{\mbox{\rm d}}

\newcommand{\beq} {\begin{equation}}
\newcommand{\eeq} {\end{equation}}

\newcommand{\be}{\begin{equation}}
\newcommand{\ee}{\end{equation}}
\newcommand{\bea}{\begin{eqnarray}}
\newcommand{\eea}{\end{eqnarray}}
\newcommand{\bean}{\begin{eqnarray*}}
\newcommand{\eean}{\end{eqnarray*}}
\newcommand{\bit}{\begin{itemize}}
\newcommand{\eit}{\end{itemize}}

\newcommand{\md}{\mathrm{d}}
\newcommand{\vecx}{{\mathbf x}}
\newcommand{\vecnull}{{\mathbf 0}}
\newcommand{\vecp}{{\mathbf p}}

\newcommand{\nc}[1]{\newcommand{#1}}
\nc{\hmu}{\hat{\mu}}
\nc{\beqa}{\begin{eqnarray}}
\nc{\eeqa}{\end{eqnarray}}

\begin{document}

\chapter{THERMODYNAMICS OF STRONG-INTERACTION MATTER FROM LATTICE QCD}

\author{Heng-Tong Ding}
\address{Key Laboratory of Quark \& Lepton Physics (MOE), Institute of Particle
Physics, Central China Normal University, Wuhan, 430079, China}

\author{Frithjof Karsch}
\address{Physics Department, Brookhaven National Laboratory, Upton, NY 11973, USA \\
and \\ Fakult\"at f\"ur Physik, Universit\"at Bielefeld, D-33615 Bielefeld, Germany}

\author{Swagato Mukherjee}
\address{Physics Department, Brookhaven National Laboratory, Upton, NY 11973, USA}

\authormark{H.-T. Ding, F. Karsch and S. Mukherjee}

\begin{abstract}
We review results from lattice QCD calculations on the thermodynamics
of strong-interaction matter with emphasis on input these calculations
can provide to the exploration of the phase diagram and properties
of hot and dense matter created in heavy ion experiments. This review
is organized as follows:
\begin{itemize}
\item[\ref{sec:intro})] Introduction
\item[\ref{sec:basics})] QCD thermodynamics on the lattice
\item[\ref{sec:phase})] QCD phase diagram at high temperature
\item[\ref{sec:bulk})] Bulk thermodynamics
\item[\ref{sec:fluctuation})] Fluctuations of conserved charges
\item[\ref{sec:transport})] Transport properties
\item[\ref{sec:hadrons})] Open heavy flavors and heavy quarkonia
\item[\ref{sec:externalB})] QCD in external magnetic fields
\item[\ref{sec:summary})] Summary
\end{itemize}
\end{abstract}
\body

\section{Introduction}
\label{sec:intro}

It has long been recognized that under extreme conditions of
high temperature or densities matter interacting through the
strong force -- strong-interaction matter -- cannot exist in the
form of nuclear matter formed from hadrons. The copious production
of new resonances \cite{Hagedorn:1965st} and the intrinsic size of
the nucleons put a natural limit on the range of validity of
hadron physics \cite{Baym:1979,Celik:1980td}. In the context of 
thermodynamics of strong-interaction matter this found its expression in 
the formulation of the hadron resonance
gas model \cite{Hagedorn:1965st,Redlich:2015bga} which led to the postulate 
of a limiting temperature -- Hagedorn temperature -- for the thermodynamics of 
ordinary nuclear matter.

With the formulation of Quantum Chromodynamics (QCD) 
\cite{Politzer:1973fx,Gross:1973id}
as the fundamental theory of strong interactions it has soon been 
realized that matter formed from strongly interacting particles may 
be converted into a new form of 
matter~\cite{Cabibbo:1975ig,Collins:1974ky} --the quark-gluon 
plasma \cite{Shuryak:1980tp} --
in which the dominant degrees of freedom are the constituents of hadrons, 
i.e. quarks and gluons. At low temperatures and densities they are confined 
in hadrons. However, at high temperature and densities they can 
move ``freely'' over macroscopic distances in bulk strong-interaction matter.

\begin{figure}[b]
\begin{center}
    \includegraphics[width=0.75\textwidth]{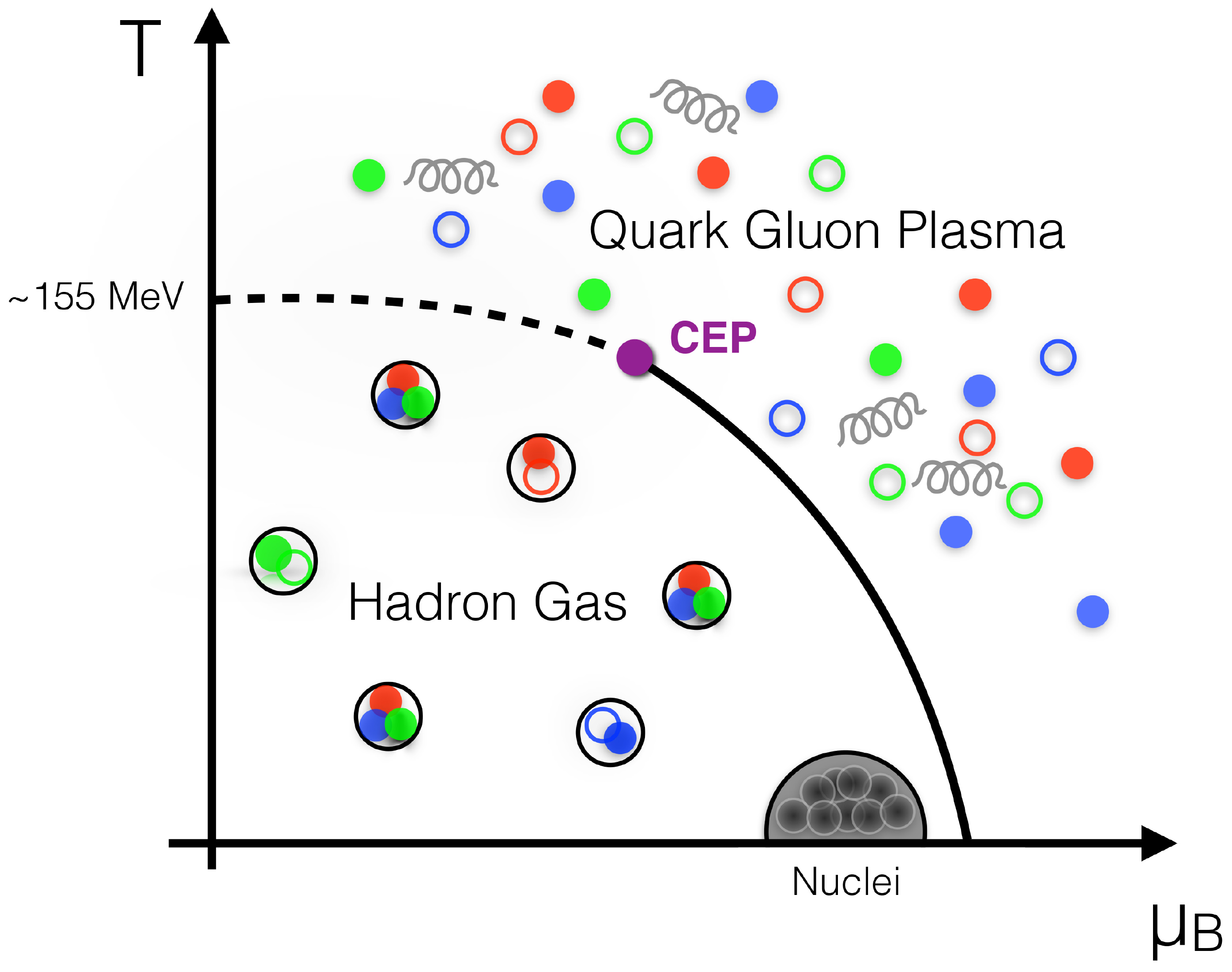}
     \caption{QCD phase diagram in the temperature ($T$) and baryon 
chemical potential ($\mu_B$) plane. At vanishing $\mu_B$ lattice QCD 
calculations show
that the transition is not a phase transition but a continuous crossover,
reflecting pseudo-critical behavior in the vicinity of the true chiral
transition. At $\mu_B>0$ a second order critical end point (CEP) may
exist, which would be followed by a line of first order transitions at 
larger values of $\mu_B$. Constraints on the location of a CEP will come from
lattice QCD calculations.}
\label{fig:phasediagram}
  \end{center}
\end{figure}

It has been speculated that the transition from confined to deconfined
matter, like in condensed matter systems, goes along \cite{Pisarski:1983ms} 
with the restoration
of global symmetries of the underlying microscopic theory describing
the interaction among constituents of strong-interaction matter, 
i.e. QCD. The analysis of these symmetries led to the phenomenology 
of a complex phase structure of strong-interaction matter, leading
to homogeneous and inhomogeneous as well as color-superconducting
phases. This bears many analogies to phase structures known in
condensed matter systems \cite{Rajagopal:2000wf,Fukushima:2010bq}. 
Only little of this complex structure is verified through first 
principle lattice QCD calculations. 
In Fig.~\ref{fig:phasediagram} we show a sketch of the QCD phase diagram
that only highlights those aspects that we believe can be addressed 
at present through lattice QCD calculations. We will discuss the
phase diagram  in detail in Section~\ref{sec:phase}.

Phase transitions as well as properties of matter close to such
transitions arise from complex long range interactions in multi-particle
systems on multiple length scales and at the same time often
develop simple universal behavior. Their analysis requires 
non-perturbative techniques. With the formulation of lattice QCD, i.e. 
the introduction of a discrete space-time lattice as regulator for 
quantum field theories like QCD \cite{Wilson:1974sk} a framework 
for new non-perturbative approaches to the study of strong-interaction 
matter had been established. It became possible to perform first-principle 
numerical calculations -- lattice QCD simulations \cite{Creutz:1980zw} -- 
by which a systematic study of the phase structure and basic bulk properties 
of strong-interaction matter \cite{McLerran:1980pk,Kuti:1980gh,Engels:1980ty}
could be performed for the first time. We will give a brief introduction
to lattice QCD in Section~\ref{sec:basics}.
Today these calculations
are highly advanced and continuum extrapolated results for the
QCD transition temperature and the equation of state exist
at vanishing net baryon number density. We will
review these results and report on the status of the lattice
QCD program that aims at an extension of these calculations to
non-zero net baryon number density or equivalently non-zero baryon
chemical potential in Section~\ref{sec:bulk}.

Properties of strong-interaction matter are studied intensively
in large experimental programs at the
Relativistic Heavy Ion Collider (RHIC) at Brookhaven National
Laboratory (BNL), USA, and the Large Hadron Collider (LHC) as well
as the Super Proton Synchrotron (SPS) at the European Research Center
(CERN). Experiments at these accelerators are devoted to the exploration
of the properties of hot and dense
matter created in collisions of ultra-relativistic heavy ions.
Through the variation of beam energies at RHIC, i.e. the beam energy
scan (BES), it also became possible to explore systematically the phase 
structure of strong-interaction matter at non-zero net baryon number density.
This allows to search systematically for the chiral critical point or
critical end point (CEP) in the QCD phase diagram. The CEP is a postulated 
second order phase transition point which is expected to mark
the endpoint of a line of first order phase transitions that separates
the low temperature, low density hadronic phase from a low temperature,
large baryon number quark-gluon plasma phase (also the existence of this
first order line at present is not confirmed through lattice QCD
calculations). The existence of a CEP in the QCD phase diagram is
imprinted in the properties of fluctuations of conserved charges,
e.g. net baryon number, electric charge or strangeness fluctuations.
These conserved charge fluctuations can be calculated in lattice QCD at 
vanishing baryon
chemical potential and the calculations can be extended to non-zero values 
of the chemical potential using Taylor expansions of thermodynamic 
observables. They can directly
be compared to experimental measurements of conserved charge fluctuations
performed at the LHC and the BES at RHIC and may provide insight into
the existence and location of a CEP in the QCD phase diagram. We 
will report in Section~\ref{sec:fluctuation} on the status of lattice QCD calculations of 
conserved charge fluctuations and their usage to analyze freeze-out
conditions in heavy ion experiments.

Heavy ion experiments at RHIC provided striking evidence for the highly
correlated, non-perturbative structure of strong-interaction matter in the
high temperature phase. In particular, the observation of strong elliptic 
flow and very efficient energy loss of high momentum quarks
traversing through the medium (jet quenching) paved the way for
a picture of a strongly interacting
medium that has been called an `almost perfect liquid'. These 
measurements suggest that strong-interaction matter at temperatures
close to but above the transition temperature has quite unique transport
properties, (i) a small shear viscosity to entropy ratio that may be close
to the conformal limit value $\eta/s =1/4\pi$ calculated in conformal
field theories (AdS/CFT limit), (ii) a large bulk viscosity that may 
diverge in the massless QCD limit at the chiral phase transition temperature,
(iii) small electrical conductivity and diffusion coefficients. These 
transport properties can systematically be analyzed in lattice QCD,
although this requires the reconstruction of continuous spectral functions 
from a finite set of observables, which 
is in general an ill-posed problem.  The calculation of transport
properties in lattice QCD will provide crucial input to the
modeling of the hydrodynamic expansion of hot and dense matter formed
in heavy ion collisions. Spectral functions also carry 
information on in-medium modifications of hadrons. The dissociation of
heavy quark bound states has long been advocated as one of the striking 
signatures for the formation of a deconfined color screened medium 
\cite{Matsui:1986dk} while the melting of light quark bound states
is intimately related with the restoration of chiral symmetry 
\cite{Rapp:1999ej}. 
We will report on the status of calculations of transport coefficients
and in-medium hadron properties in Sections~\ref{sec:transport} and~\ref{sec:hadrons}.

Finally, in Section~\ref{sec:externalB} we will briefly review a relatively new but
very exciting field of QCD thermodynamics that addresses the behavior
of strong-interaction matter in strong external magnetic fields.
Strong, static magnetic fields will certainly influence
the thermodynamics and lead to modifications of the transition
temperature as well as the equation of state. 
Strong fields are generated in the early phase of heavy ion
collisions \cite{Kharzeev:2012ph,Kharzeev:2007jp} but weaken 
rapidly during the subsequent expansion phase. While lattice QCD calculations 
will allow to understand
basic effects that occur in matter exposed to strong fields it
is at present not clear to what extent equilibrium lattice QCD 
calculations can contribute to a quantitative analysis of experimental
data.

\section{QCD thermodynamics on the lattice}
\label{sec:basics}

\subsection{Path integral formulation of QCD thermodynamics}

The equilibrium thermodynamics of elementary particles interacting only through the
strong force is controlled by the QCD partition function which can be expressed in
terms of a Euclidean path integral. The grand canonical partition function, $Z
(T,V,\vec{\mu})$, is given as an integral over the fundamental quark ($\bar{\psi},\;
\psi$) and gluon ($A_\nu$) fields. In addition to its dependence on volume ($V$),
temperature ($T$) and  a set of $N_f$ chemical potentials, $\vec{\mu}\equiv
(\mu_u, \mu_d, \mu_s,...)$ the partition function implicitly depends on 
the masses, $\vec{m}\equiv (m_u, m_d, m_s,...)$, of the $N_f$ different quark 
flavors.
In this review we will most of the time discuss the thermodynamics of QCD
with 2 light quarks ($u, d$) which are assumed to have degenerate masses,
$m_u=m_d$, and a heavier strange quark ($s$), with mass $m_s$. This
often is referred to as (2+1)-flavor QCD, or $N_f=2+1$. 

In Euclidean space-time, which is
obtained from the Minkowski formulation by substituting $t\rightarrow -i\tau$
with $\tau \in \mathbb{R} $, the QCD Lagrangian is given by
\bea
\cL^{E}_{QCD}&=&\mathcal{L}^{E}_{gluon} + \mathcal{L}^{E}_{fermion} \nonumber \\
 &=& - \frac{1}{4}F^{\mu\nu}_{a}(x)F^{a}_{\mu\nu}- 
\sum_{f=u,d,s...}\bar{\psi}^{\alpha}_f(x)\left(\slashed{D}^{E}_{\alpha\beta}+m_f\delta_{\alpha\beta}\right)\psi^{\beta}_f(x) \; ,
\eea
where Greek letters are spinor indices, $a=1,..,N_c^2-1$ is the color index, 
$N_c$ is the number of colors ($N_c=3$ for QCD) and $m_f$ is the mass of quarks
with flavor $f$. The covariant derivative $\slashed{D}^E$ and the field 
strength tensor $F^a_{\mu\nu}$ are given by 
\bea
\slashed{D}^{E}&=&\gamma_{\mu}^{E}D^{E}_{\mu} = \left(\partial_{\mu}+ig\frac{\lambda_a}{2}A^a_{\mu}\right)\gamma^{E}_{\mu}\; , \\
F^{a}_{\mu\nu} &=& \partial_{\mu}A^{a}_{\nu}-\partial_{\nu}A_{\mu}^{a} - g f^{abc}A^{b}_{\mu}A^{c}_{\nu} \; .
\eea
Here $A^{a}_{\mu}$ are the gauge fields, 
$\psi^{\alpha}_f$ ($\bar{\psi}^{\alpha}_f$) are the quark (anti-quark)
fields, $\lambda^a$ are the generators of SU($N_c$), $f_{abc}$ are the 
corresponding structure constants, $\gamma^{E}_{\mu}$ are the Euclidean 
Dirac matrices obeying 
$\{ \gamma^{E}_{\mu},\gamma^{E}_{\nu}\}= 2\delta_{\mu\nu}$,
and $g$ is the bare coupling constant.

In the Euclidean path-integral formalism the partition function of QCD is then 
given by
\be
\mathcal{Z}(T,V,\vec{\mu})=\int\prod_{\mu}\cD A_{\mu}\hspace{-0.1cm}
\prod_{f=u,d,s...} \hspace{-0.1cm}
\cD\psi_f\cD\bar{\psi}_f\ {\rm e}^{-S_E(T,V,\vec{\mu})} \; ,
\label{eq:partition}
\ee
with the Euclidean action 
\begin{eqnarray}
S_E(T,V,\vec{\mu}) &\equiv&  - \int\limits_0^{1/T}
\hspace{-0.1cm} \D x_0 \hspace{-0.1cm} 
\int\limits_V \hspace{-0.1cm} \D^3 {\bf x} \;
\cL^{E} (\vec{\mu}) \; .
\label{lagrangian}
\end{eqnarray}
Here we have suppressed the dependence of the Euclidean Lagrangian and
action on the fields $(A_{\mu},\bar{\psi}_f, \psi_f)$ but have stressed
explicitly their dependence on the various quark chemical 
potentials that couple to the conserved quark number currents
\be
\cL^{E} (\vec{\mu}) = \cL^{E}_{QCD} +\sum_{f=u,d,s..} 
\mu_f \bar{\psi}_f \gamma_0 \psi_f \; .
\ee

The thermal expectation value of physical observables $\cO$ can be obtained 
through
\bea
\left<\cO\right> &=& \frac{1}{Z(T,V,\vec{\mu})} 
\int\prod_{\mu}\cD A_{\mu}\prod_{f}\cD\psi_f\cD\bar{\psi}_f\, \cO\, 
{\rm e}^{-S_E(T,V,\vec{\mu})}  \; .
\label{eq:thermal_average}
\eea

Basic thermodynamic quantities like the pressure ($P$), energy density ($\epsilon$)
or net-quark number density $n_f$ can be obtained from the logarithm of the partition
function using standard thermodynamic relations,
\begin{eqnarray}
\frac{P}{T^4} &=& \frac{1}{VT^3} \ln Z(T,V,\vec{\mu})\; , \\
\frac{\epsilon}{T^4} &=&  - \frac{1}{VT^4} 
\left. \frac{\partial
 \ln Z(T,V,\vec{\mu})}{\partial 1/T} \right|_{\vec{\mu} /T ~{\rm fixed}}
\; , \\
\frac{n_f}{T^3} &=&  \frac{1}{VT^3} \frac{\partial
 \ln Z(T,V,\vec{\mu})}{\partial \hat{\mu}_f} \; ,
\label{e3pmu}
\end{eqnarray}
where we introduced the chemical potentials in units of temperature,
$\hat{\mu}_f = \mu_f/T$. In fact, the QCD partition function
depends on chemical potentials only through these dimensionless combinations, 
which are the logarithms of the fugacities, $z_f\equiv \exp (\mu_f/T)$.

As will become clear in the next subsection, the numerical analysis of
thermodynamic observables at non-vanishing chemical potential,
$\vec{\mu}\ne 0$, is difficult in lattice regularized QCD. A viable
approach that circumvents the so-called sign-problem in lattice QCD at
$\vec{\mu}\ne 0$, suitable for moderate values of the chemical potentials,
is to consider Taylor expansions for thermodynamic observables.
The starting point for such an analysis is the
Taylor expansion of the pressure. In (2+1)-flavor QCD it reads
\begin{equation}
\frac{P}{T^4} = \sum_{i,j,k} \frac{1}{i!j!k!} \chi_{ijk}^{uds}(T) 
\left( \frac{\mu_u}{T}\right)^i
\left( \frac{\mu_d}{T}\right)^j
\left( \frac{\mu_s}{T}\right)^k \; , 
\label{Taylor}
\end{equation}
where the expansion coefficients are dimensionless, generalized susceptibilities that can be 
evaluated at $\vec{\mu} =  0$,
\begin{equation}
\chi^{uds}_{ijk}(T)= \left. 
\frac{\partial^{i+j+k}P/T^4}{\partial \hat\mu_u^i \partial \hat\mu_d^j \partial \hat\mu_s^k}\right|_{\vec\mu=0}\;.
\label{eq:fluct1}
\end{equation}
We will present results from Taylor expansions of QCD thermodynamics in 
several chapters of this review.
Other thermodynamic observables and expressions
more suitable for the calculation of $P/T^4$ and $\epsilon/T^4$ in
lattice regularized QCD will be introduced in later sections where they
appear.

\subsubsection{The high temperature, ideal gas limit}

At very high temperatures, all quark masses are small on the scale of
the temperature and massless QCD becomes a good approximation. 
Furthermore, because of asymptotic freedom also interactions among quarks 
and gluons become small at high temperature. QCD thermodynamics thus
approaches that of a non-interacting, massless quark-gluon gas. In the
temperature range accessible to heavy ion collisions the massless limit
of 3-flavor QCD is most relevant ($N_f=3$).
In this limit the pressure is given by
\begin{equation}
\left( \frac{P}{T^4} \right)_{\rm ideal} =
\frac{8 \pi^2}{45} + \sum_{f=u,d,s} \left[\frac{7 \pi^2}{60} +
\frac{1}{2}  \left(\frac{\mu_f}{T}\right)^2 
+ \frac{1}{4 \pi^2} \left(\frac{\mu_f}{T}\right)^4 
\right] \quad,
\label{eq:free}
\end{equation}
where the first term gives the contribution of gluons and the sum
yields the contribution of quarks with different flavor degrees of 
freedom. 

The thermodynamics of QCD at high temperatures can be systematically
analyzed in perturbation theory. This, however, becomes complicated beyond
${\cal O}(g^2)$ due to the appearance of non-perturbative length scales of 
${\cal O}(gT)$ and ${\cal O}(g^2T)$
reflecting electric and magnetic screening lengths \cite{Linde:1980ts}, which
require the resummation of diagrams leading to the so-called hard thermal
loop perturbation theory \cite{Braaten:1989mz,Haque:2014rua}, or the explicit 
integration over hard scales leading to the dimensional reduction 
scheme \cite{Braaten:1995cm,Hietanen:2008tv}.

\subsubsection{The low temperature hadron resonance gas approximation}

At low temperature quarks and gluons are confined in colorless
hadrons, i.e. baryons and mesons are the relevant degrees 
of freedom. In fact, it turns out that even at temperatures close
to the transition region from hadronic matter to the high temperature 
quark-gluon plasma phase a non-interacting gas constructed from all 
experimentally known resonances 
does provide quite a  good description of QCD thermodynamics.  This had 
been anticipated by R. Hagedorn when he formulated the hadron resonance
gas (HRG) \cite{Hagedorn:1965st,Redlich:2015bga} model prior to QCD.

In an HRG model the pressure and other thermodynamic observables are 
easily obtained from the logarithm of the partition function, 
\begin{eqnarray}
\ln{\cal Z}_{HRG}(T,V,\vec{\mu}) 
=\sum_{i\in\;mesons}\hspace{-3mm} \ln{\cal Z}^{M}_{m_i}(T,V,\vec{\mu})
+\hspace{-3mm} 
\sum_{i\in\;baryons}\hspace{-3mm} \ln{\cal Z}^{B}_{m_i}(T,V,\vec{\mu})\; ,
\label{eq:ZHRG}
\end{eqnarray}
where $\vec{\mu}=(\mu_B,\ \mu_Q,\ \mu_S)$ is the set of baryon number,
electric charge and strangeness chemical potentials, respectively.
The partition functions for mesons ($M$) or baryons ($B$) are given by
\begin{equation}
\ln{\cal Z}^{M/B}_{m_i}(T,V, \vec{\mu})
=\mp{V\over{2\pi^2}}\int_0^\infty dk\, k^2\,
\ln(1\mp z_ie^{-\varepsilon_i/T}) \quad ,
\label{eq:ZMB}
\end{equation}
with energies $\varepsilon_i^2=k^2+m_i^2$ and fugacities
\begin{equation}
z_i=\exp\big((B_i\mu_B+Q_i\mu_Q+S_i\mu_S)/T\big) \quad .
\label{eq:fuga}
\end{equation}
Of course, $B_i=0$ for all mesons and $B_i=\pm 1$ for baryons.
The set of hadron chemical potentials, related to conserved quantum
numbers, and the set of quark flavor chemical potentials are easily
related to each other,
\begin{eqnarray}
\mu_u&=&\frac{1}{3}\mu_B + \frac{2}{3}\mu_Q \; , \nonumber \\
\mu_d&=&\frac{1}{3}\mu_B - \frac{1}{3}\mu_Q \; ,\nonumber \\
\mu_s&=&\frac{1}{3}\mu_B - \frac{1}{3}\mu_Q - \mu_S \; .
\label{potential}
\end{eqnarray} 
With this it is straightforward to rewrite the Taylor series given in 
Eq.~\ref{Taylor} in terms of quark chemical potentials, also in terms of 
baryon number, electric charge and strangeness chemical potentials,
\begin{equation}
\frac{P}{T^4} = \sum_{i,j,k} \frac{1}{i!j!k!} \chi_{ijk}^{BQS}(T) 
\left( \frac{\mu_B}{T}\right)^i
\left( \frac{\mu_Q}{T}\right)^j
\left( \frac{\mu_S}{T}\right)^k \; , 
\label{TaylorBQS}
\end{equation}
where the expansion
coefficients\footnote{We will in general suppress super- and subscripts
of the generalized susceptibilities, if a 
subscript is zero, i.e. $\chi_{102}^{BQS}\equiv \chi_{12}^{BS}$ or 
$\chi_{200}^{BQS}\equiv \chi_{2}^{B}$.} 
$\chi^{BQS}_{ijk}$ can again be evaluated at $\vec{\mu} =  0$,
\begin{equation}
\chi^{BQS}_{ijk}(T)= \left. 
\frac{\partial^{i+j+k}P/T^4}{\partial \hat\mu_B^i \partial \hat\mu_Q^j \partial \hat\mu_S^k}\right|_{\vec\mu=0}\;.
\label{eq:fluct1BQS}
\end{equation}

\subsection{Lattice Regularization and Continuum Limit}

Lattice gauge theory suggested by K. G. Wilson in 1974~\cite{Wilson:1974sk} is 
based on the Euclidean path integral formalism and provides a particular
regularization scheme for QCD by introducing a finite lattice spacing as 
cut-off. In this way space-time is discretized
and the path integral becomes a finite, yet high dimensional integral over the
gauge and fermion field variables. The discretization of space and time 
introduces a systematic cut-off dependence in all observables, which vanishes 
when the lattice spacing is taken to zero, i.e. in the continuum limit.  
Many theoretical and technical details can be found in excellent
textbooks such as Ref.~\citen{Montvay94,Rothe05,Degrand06,Gattringer10}. 
Here we will only give a brief 
introduction that may help to make this review self-contained.

By introducing a hyper-cubic lattice of size $N_{\sigma}^3\times N_{\tau}$ 
with a small but finite lattice spacing $a$ 
the calculation of observables given by Eq.~\ref{eq:thermal_average} 
reduces to the evaluation of finite, but high dimensional integrals.
Rather than calculating the partition function directly one usually 
is concerned with the calculation of expectation values, 
Eq.~\ref{eq:thermal_average}.
This can be done by exploiting well-known Monte Carlo simulation 
techniques \cite{Creutz:1980zw}. 
In order to preserve gauge invariance of the discretized
action one introduces the gauge degrees of freedom as variables living
on links between neighboring sites of the lattice, 
$U_{x,\hat{\mu}} \equiv \exp(i a g A_\mu (x))$, while the fermionic
degrees of freedom, $\bar{\psi}_x$ and $\psi_x$, are defined on the sites of 
the lattice.
The latter are anti-commuting Grassmann variables and the former are
$N_c\times N_c$ matrices that are elements of the $SU(N_c)$ color group.

The volume $V$ and the temperature $T$ are related to the spatial and temporal 
extents of the lattice, respectively,
\bea
V=(aN_{\sigma})^3,~~~~~~T=\frac{1}{aN_{\tau}}  \;.
\eea
Here $N_{\sigma}$ and $N_{\tau}$ are the number of sites in spatial and 
temporal directions. A consequence of introducing a discrete
space-time lattice is that aside from Lorentz symmetry also some of the 
symmetries of QCD get explicitly broken by the finite lattice cut-off. 
They will be recovered 
in the continuum limit, $a\rightarrow 0$ at fixed $V$ and $T$.
Asymptotically, for small values of the bare gauge coupling $g$ this 
is controlled by the QCD $\beta$-function
\be 
a \Lambda_L = \left( \frac{1}{b_0g^2}\right)^{b_1/2b_0^2} {\rm e}^{-1/2b_0g^2} \; , 
\label{betafunction}
\ee
with $b_0=\left(\frac{11}{3}N_c-\frac{2}{3}N_f\right)/16\pi^2$ and
$b_1=\left( \frac{34}{3}N_c^2 -\left( \frac{10}{3}N_c + \frac{N_c^2-1}{N_c}
\right)N_f\right)/(16\pi^2)^2$.

Keeping physical observables constant when approaching the
continuum limit also requires a proper tuning of the bare quark masses. 
In the discretized
version of QCD (lattice QCD) all quark masses are naturally expressed
in units of the lattice spacing, $\hat{m}_f\equiv m_f a$. Taking the 
continuum limit thus requires to take the limit, $g^2 \rightarrow 0 $  
and $\hat{m}_f\rightarrow 0$ while keeping some physical observables, i.e. 
one per flavor degree of freedom, constant. This defines ``lines of constant 
physics'' (LCP) in the space  of quark masses and the gauge coupling.
In the case of (2+1)-flavor QCD with degenerate light quark masses, $m_u=m_d$,
one thus can determine the bare quark mass parameters 
$\hat{m}_l\equiv \hat{m}_u$ and the strange quark mass $\hat{m}_s$ using
two physical observables. An often used approach is to fix the strange
quark mass using the (fictitious) $\eta_{s\bar s}$ meson mass, which
only contains strange quarks and is quite insensitive to light quark 
mass values. This mass is matched to the lowest order chiral perturbation
theory estimate $m_{\eta_{s \bar s}}=\sqrt{2 m_K^2-m_{\pi}^2}$. 
The light quark mass is then fixed using a constant ratio 
$\hat{m}_s/\hat{m}_l$, the value corresponding to the physical pion mass
value is $\hat{m}_s/\hat{m}_l=27.5$ \cite{Agashe:2014kda}.

The lattice regularized partition function may be written as
\bea
\mathcal{Z}(T,V,\vec{\mu})&=&\int\prod_{x,\hat{\mu}}{\rm d} U_{x,\hat{\mu}}
\prod_{x,f}{\rm d} \psi_{x,f} {\rm d}\bar{\psi}_{x,f} \ {\rm e}^{-S_f-S_g} 
\; , \nonumber \\
&=&\int\prod_{x,\hat{\mu}}{\rm d} U_{x,\hat{\mu}}\
\prod_{f} \mathrm{\det} M_{f} (\mu_f) \ {\rm e}^{-S_g} \; ,
\label{Zlat}
\eea
where $S_g$ and $S_f$ are the discretized versions of the gluonic and fermonic 
part of the action $S_E$. As the action is bilinear in the quark fields
these can be integrated out, which gives rise to the determinant
of fermion matrix $M_f$. Quite often exploratory lattice QCD studies
are performed in the so-called ``quenched approximation". This refers
to the calculation of observables that contain fermionic degrees of freedom
as valence quarks. However, back-reaction of fermions on the gauge fields
(virtual quark loops) are not included in the observables. This is achieved
by fixing $\mathrm{\det} M_{f}=const.$ in Eq.~\ref{Zlat}, i.e. the determinant
does not contribute in the generation of gauge field configurations nor does 
it contribute in the calculation of expectation values.

The fermion matrix depends on the quark chemical potential
$\mu_f$ which is introduced on the time-like links of the lattice 
\cite{Hasenfratz:1983ba}. Although other formulations are possible 
\cite{Gavai:1985ie} a common approach is to replace the link variables,
$U_{x,\hat{0}}\rightarrow \exp{(\mu_f a)} U_{x,\hat{0}}$ and
$U^\dagger_{x,\hat{0}}\rightarrow \exp{(-\mu_f a)} U^\dagger_{x,\hat{0}}$.
The fermion determinant is positive definite only for $\mu_f \equiv 0$. 
For non-zero values
of the chemical potential $\mathrm{\det} M_{f} (\mu_f)$ is a complex 
function. Still the partition function is real, i.e. integration over
the gauge fields eliminates the imaginary contributions. However, the 
real part of the fermion determinant still changes sign, which prohibits the 
application of conventional Monte Carlo simulation techniques. This
is known as the 'sign-problem' in finite density QCD. There are various
attempts to circumvent this sign-problem in numerical simulations performed
directly at non-vanishing $\vec{\mu}$. At present complex Langevin
simulation techniques \cite{Aarts:2011ax,Aarts:2013uxa,Sexty:2014dxa} 
and the integration over Lefschetz thimbles 
\cite{Cristoforetti:2012su,Fujii:2013sra,Cristoforetti:2014gsa} are
most actively being explored. In this review we will focus
on the Taylor expansion approach \cite{Gavai:2003mf,Allton:2003vx} for the 
evaluation of thermodynamic 
observables at non-zero chemical potential. This only requires numerical
simulations at vanishing baryon chemical potential. However, it is naturally
limited in applicability through the radius of convergence of the Taylor
series which remains to be determined for QCD.

\subsection{Fermion discretization schemes}
When discretizing the Euclidean action one has quite some freedom. 
Discretization schemes that differ by higher order corrections in the
cut-off are equivalent to the extent that they yield the same
physical answers in the continuum limit.
The basic construction principle of a lattice discretized version of QCD
is that one wants to start with a theory that is manifestly gauge invariant.
In addition one tries to 
choose discretization schemes that preserve as many of the symmetries of 
QCD as possible already at non-zero values of the lattice spacing. 
The discretization of the gluonic part of the action, 
$S_g$, is rather straightforward and well-established improvement
schemes (Symanzik improvement \cite{Symanzik:1983dc,Symanzik:1983gh}) 
are known that allow to eliminate
systematically higher order cut-off errors.
 
Some difficulties already arise in the naive discretization 
of the fermionic part of the Euclidean action, $S_f$. When one simply 
approximates the covariant derivative in the fermion action by  
nearest-neighbor differences of the quark fields additional, unphysical degrees
of freedom arise even in the continuum limit owing to the periodicity
of the fermion dispersion relation on the lattice.
This is known as the fermion doubling problem 
which is closely related to the explicit breaking of chiral symmetry in
most commonly used fermion discretization schemes \cite{Nielsen:1981hk}.
 
\begin{figure}[t]
\begin{center}
\includegraphics[width=0.51\textwidth]{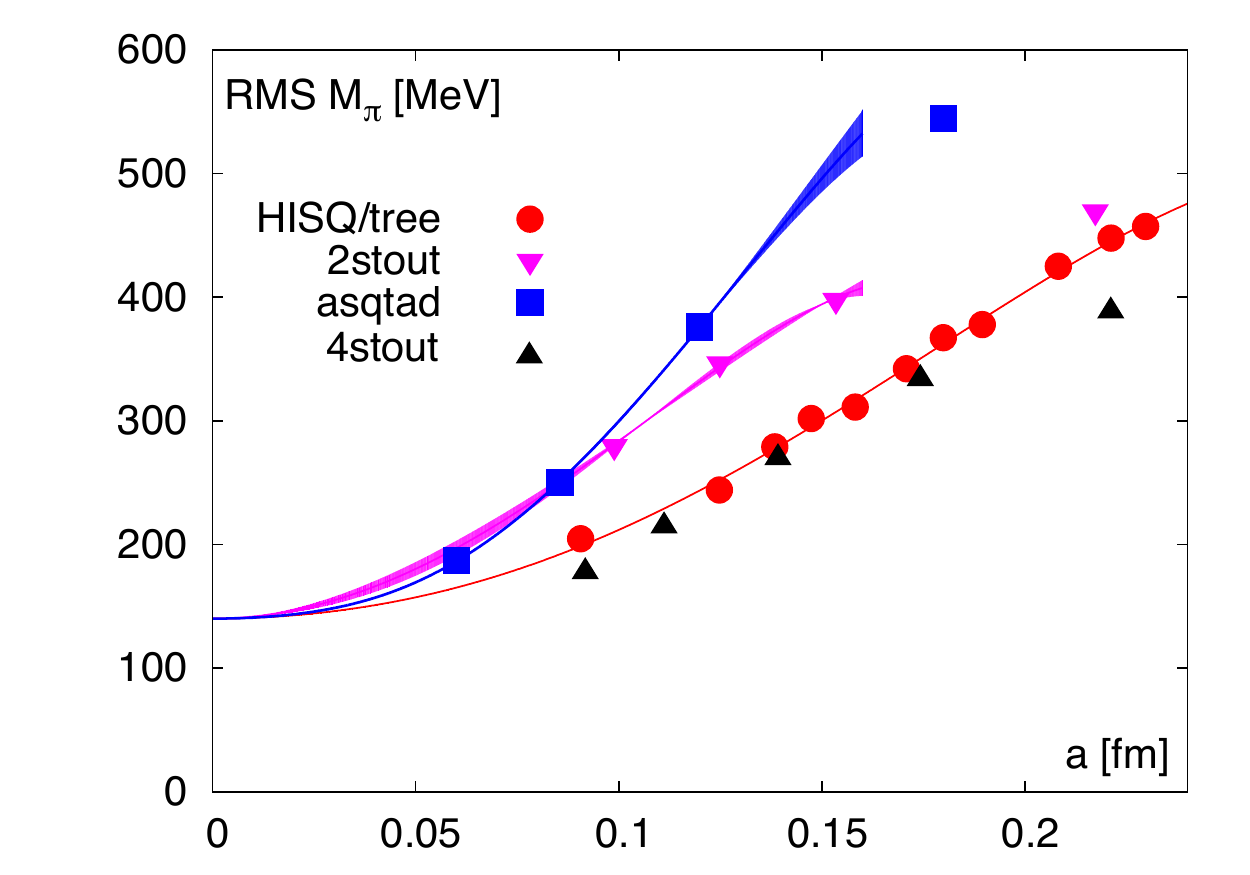}\includegraphics[width=0.51\textwidth]{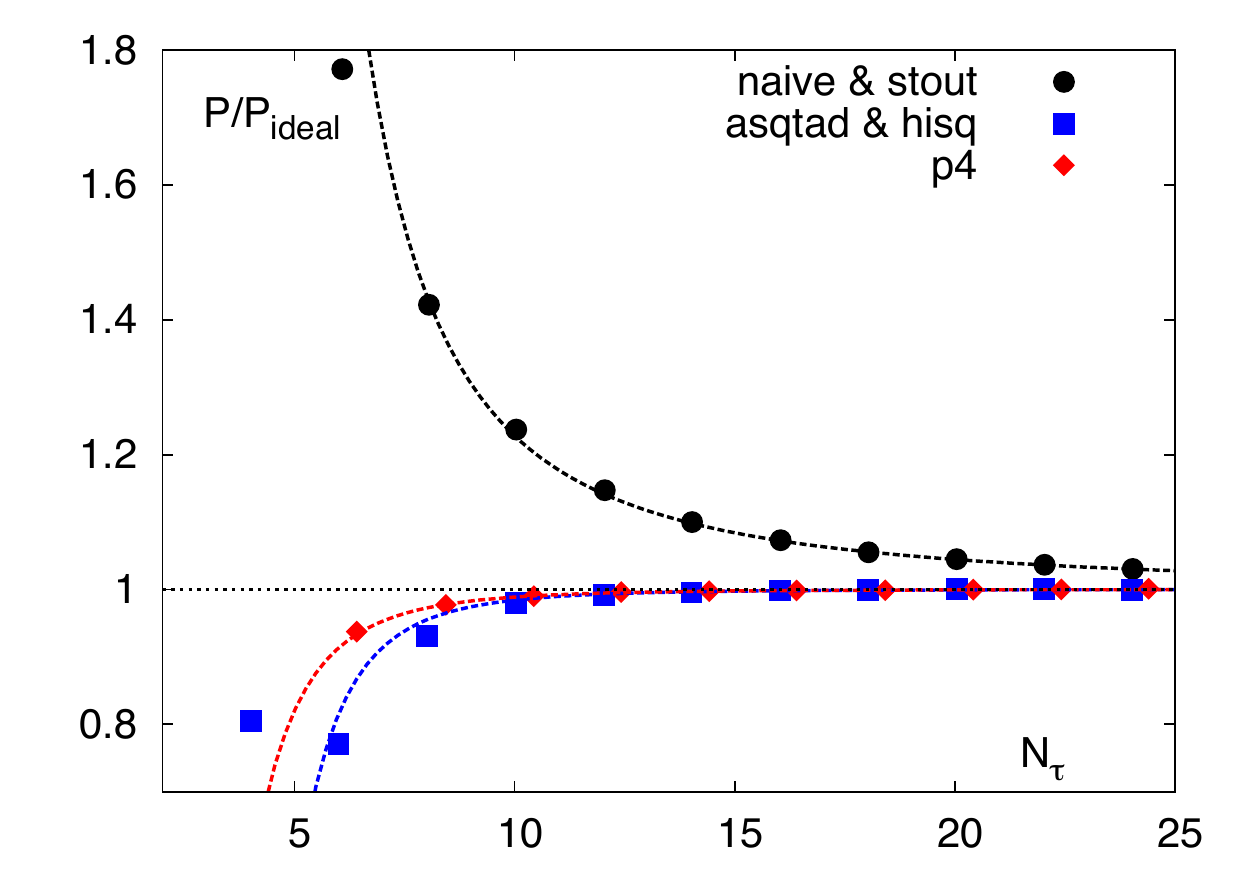}

\end{center}
\caption{(Left) Root mean squared pion mass ($M^{\rm RMS}_{\pi}$) as a function of lattice spacing $a$. RMS $M_{\pi}$ is defined as the rooted sum of the mass squared of 16 pseudo scalar states divided by 4. The lattice cutoff
effects are smaller with smaller $M^{\rm RMS}_{\pi}$. (Right) Ratio of quark
contribution to the pressure, obtained from lattice QCD calculations in the 
infinite temperature, ideal gas limit, to the
pressure of an ideal quark gas in the continuum ($P_{\rm ideal}$) as a 
function of $N_{\tau}$. $P/P_{\rm ideal}$
approaches unity in the continuum limit.
}
\label{fig:staggered}
\end{figure}

There are several discretization schemes for the fermionic part
of the action such as Wilson fermions~\cite{Wilson:1974sk},
 staggered fermions~\cite{Kogut:1974ag}, Domain Wall fermions~\cite{Kaplan:1992bt,Shamir:1993zy,Shamir:1992im,Furman:1994ky} as well as overlap fermions~\cite{Neuberger:1998wv,Neuberger:1997fp}. 
They differ to the extent they preserve chiral symmetry and/or eliminate
the problem of fermion doublers. 
Wilson fermions avoid the doubling problem by
adding a dimension five operator to the naively discretized 
fermion action~\cite{Wilson:1974sk}. A variant of this is the so-called 
clover-improved Wilson fermion action where lattice cutoff effects are 
reduced from $\mathcal{O}(a)$ to 
$\mathcal{O}(a^2)$~\cite{Sheikholeslami:1985ij}. 
Most commonly used in QCD thermodynamics calculations is the staggered fermion
discretization scheme in which the Dirac 4-spinor is spread over several
lattice sites \cite{Kogut:1974ag}. This reduces the doubler problem but does 
not eliminate
it completely. However, it preserves a $U(1)_{even}\times U(1)_{odd}$ 
remnant of chiral symmetry, which allows independent rotations of the
quark fields on even and odd sites of the lattice, respectively. This
is of great advantage in the study of chiral symmetry
breaking at non-zero temperature as it allows to introduce an order parameter,
the chiral condensate, that is sensitive to the spontaneous breaking of this 
continuous symmetry. The reduction of the doubler problem in the staggered 
fermion approach
leads to the introduction of additional heavier states for each fermion
flavor. In the continuum limit these so-called taste symmetry partners will 
become degenerate with the Goldstone mode that corresponds to the broken 
$U(1)_{even}\times U(1)_{odd}$ symmetry.
The Goldstone boson thus has another 15 heavier partners (tastes) which become
degenerate with the Goldstone boson only in the continuum limit. 
To reduce the taste symmetry breaking, i.e. reduce the mass difference between 
physical states and their heavier taste partners two improved staggered 
actions, the Highly Improved Staggered Quarks (HISQ)~\cite{PhysRevD.75.054502} 
and the stout smeared action~\cite{Morningstar:2003gk}, are commonly used in 
lattice studies of QCD thermodynamics. 
In Fig.~\ref{fig:staggered}~(left) we show some results for the cut-off 
dependence of the root-mean-square mass values $M_{\pi}^{\rm RMS}$ of the Goldstone particle and 
its 15 taste partners calculated in different staggered fermion 
discretization schemes. As seen in the left panel of Fig.~\ref{fig:staggered} 
the 4stout and HISQ actions have the smallest $M_{\pi}^{\rm RMS}$.
The right hand panel of this figure shows the 
influence of cut-off effects on the calculation of the fermion contribution
to the pressure in the high temperature ideal gas limit. These cut-off
effects are independent of the taste improvement schemes and solely arise
from the strategy used to discretize the covariant derivatives appearing in 
the fermion action. The stout-smeared actions utilize the naive
1-link discretization scheme that leads to ${\cal O} (a^2)$ errors in the 
ideal gas  limit. In the HISQ, asqtad and p4 \cite{Heller:1999xz} actions 
3-link terms are added to the action that eliminate the leading order cut-off 
effects in the ideal gas limit. In these cases discretization errors only 
start at ${\cal O} (a^4)$.

Domain Wall fermions (DWF) and overlap fermions allow for an exact 
representation of chiral symmetry even at non-zero lattice spacing.
In the Domain Wall discretization scheme the physical chiral Dirac fermions 
are constructed on two 4-dimensional hyper-surfaces at the edges of a  
5-dimensional lattice with extent $L_s$ in that fifth direction. 
These hyper-surfaces represent the usual 4-dimensional space-time and on 
each of
the surfaces fermions with one given chirality exist. There persists a small
explicit breaking of chiral symmetry which vanishes exponentially in the
limit $L_s\rightarrow \infty$. In DWF calculations this gives rise to
a small additive renormalization of the quark masses,
the so-called residual mass \cite{Antonio:2008zz} $m_{res}$.
Controlling and reducing these residual mass effects
is one of the important improvement steps in calculations with DWF 
\cite{Renfrew:2009wu}.
Overlap fermions on the other hand preserve exact chiral symmetry on 
a 4-dimensional lattice by obeying the Ginsparg-Wilson relation at non-zero
lattice spacing~\cite{Ginsparg:1981bj}. 

Numerical calculations with Domain Wall as well as overlap fermions
are quite time consuming. However, with increasing speed of super-computers
calculations with physical light quark masses become feasible and chiral
fermions have been used recently also for QCD thermodynamics 
studies\cite{Buchoff:2013nra,Cossu:2013uua,Bhattacharya:2014ara,Dick:2015twa}. 
In particular in the analysis of subtle aspects of the QCD transition related
to the temperature dependence of the axial anomaly, calculations with
fermions obeying exact chiral symmetry already at non-zero values of
the cut-off are mandatory.

\section{QCD phase diagram at high temperature}
\label{sec:phase}

Our thinking about the phase structure of strong-interaction matter centers around
two very basic concepts in strong interaction physics -- confinement and chiral
symmetry breaking. The former expresses the fact that only colorless states, baryons
and mesons, can exist in the vacuum and are observed experimentally.  This gave rise
to the concept of a linearly rising, confining potential exhibited between quarks and
anti-quarks,
\begin{equation}
V_{q\bar{q}} (r) = -\frac{\alpha (r)}{r} +\sigma r \; ,
\label{Cornell}
\end{equation} 
with $\sigma$ being the string tension and $\alpha (r)$ the running coupling of QCD.

Chiral symmetry breaking, on the other hand is a mandatory feature of strong
interactions needed to explain the appearance of a light, almost massless, particle
in the hadron spectrum -- the pions. In the limit of vanishing quark masses the QCD
Lagrangian has a build-in chiral symmetry. It is invariant under independent global
rotations of quarks with left and right handed chirality in flavor space as 
well as chiral rotations of single flavor quark spinors. 
This gives rise to the $U(N_f)_L\times U(N_f)_R$ chiral
symmetry which is equivalent to $SU(N_f)_L\times SU(N_f)_R\times U(1)_A\times
U_V(1)$.  The latter $U_V(1)$ reflects the conservation of baryon number, and the axial
$U(1)_A$ symmetry, although an exact symmetry of the classical QCD Lagrangian, is
explicitly broken by quantum fluctuations. The global $SU(N_f)_L\times SU(N_f)_R$
flavor symmetry, on the other hand, is broken spontaneously in massless QCD, giving
rise to massless Goldstone modes and a non-vanishing chiral condensate,
\begin{equation}
\langle \bar{\psi}\psi \rangle_f = \frac{T}{V} 
\frac{\partial \ln Z}{\partial m_f} \;\; ,\;\ f=1,...,N_f .
\label{ppbar}
\end{equation}
The masses of the light pseudo-scalar (Goldstone) pions are then understood as 
arising from the small non-zero values of
the light up and down quark masses.  This is reflected in the Gell-Mann-Oakes-Renner
relation between the pion mass ($m_\pi$), the quark masses ($m_u, m_d)$, the pion
decay constant $(f_\pi$) and the non-vanishing chiral condensate $\langle
\bar{\psi}\psi\rangle=\langle\bar{\psi}\psi\rangle_u + \langle\bar{\psi}\psi\rangle_d$ that arises from the spontaneous breaking of chiral symmetry,
\begin{equation}
f_\pi^2 m_\pi^2 =\frac{1}{2}(m_u+m_d) \langle \bar{\psi}\psi\rangle\; .
\end{equation} 
The fact that QCD describes confinement as well as spontaneous breaking of chiral
symmetry is not evident from the QCD Lagrangian or the perturbative treatment of
strong interactions described by it.  It requires a non-perturbative analysis --
lattice QCD calculations -- to firmly establish the confining and symmetry breaking
features of QCD. 
 
Asymptotic freedom of QCD suggests that non-perturbative effects are suppressed at
high temperatures and hot strong-interaction matter approaches ideal gas behavior
asymptotically. It thus is expected that non-perturbative condensates also disappear
at high temperatures. For exact global symmetries of QCD this is expected to
happen through a true phase transition. Basic features of this transition can be
understood by invoking universality arguments \cite{Pisarski:1983ms}. 

\begin{figure}[t]
\begin{center}
\includegraphics[scale=0.7]{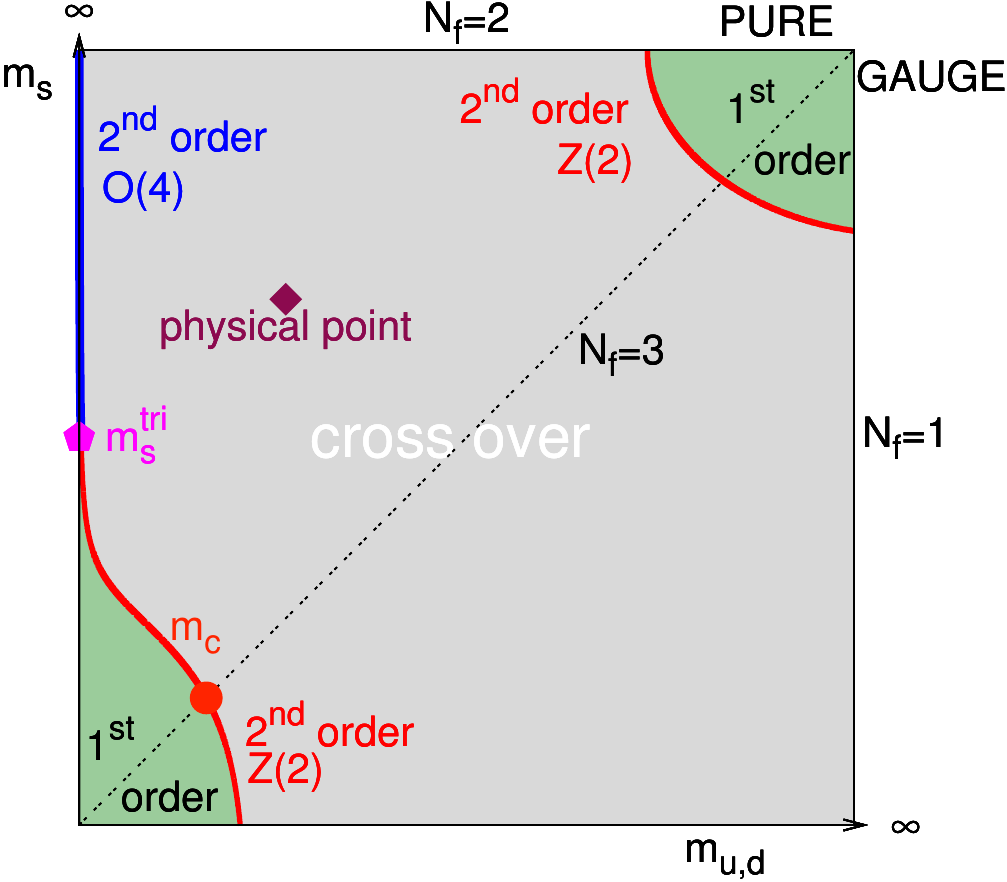}
\end{center}
\caption{A sketch of the nature of the QCD transition as functions of the two degenerate
light (up and down) quarks with masses, $m_{u,d}\equiv m_l$, and a heavier strange
quark with mass, $m_s$, at zero baryon chemical potential.}
\label{fig:Columbia}
\end{figure}

The nature of the QCD transition depends crucially on the values of the quark masses and
the number of flavors ($N_f$). Fig.~\ref{fig:Columbia} shows a sketch of the nature of
the QCD transition as functions of the quark masses for a theory with two degenerate
light (up and down) quarks with masses, $m_{u,d}\equiv m_l=m_u=m_d$, and a heavier
strange quark with mass, $m_s$, at zero baryon chemical potential. 
In the limit $m_l\to\infty$ and $m_s\to\infty$ fermions decouple and the
thermodynamics of a pure $SU(3)$ gauge theory ($N_f=0$) is recovered. The
$SU(3)$ gauge theory has an exact $Z(3)$ symmetry and the deconfimement 
transition is first order. This first
order region extends to lower quark masses and ends at a second order
critical line that belongs to the universality class of the 3-d, Z(2) symmetric  
Ising model 
\cite{Saito:2011fs,Saito:2013vja}. For three degenerate flavors $N_f=3$
with small quark masses $m_l=m_s\to0$ the chiral transition is known to be first
order \cite{Karsch:2001nf,deForcrand:2006pv}. Recent lattice QCD studies
\cite{Endrodi:2007gc,Ding:2011du} with improved actions suggest that the extent of this
first order region is quite small, i.e. 
limited to $m_l=m_s\lesssim m_s^{phys}/270$ where
$m_s^{phys}$ is the physical value of the strange quark mass.

An additional ingredient in the discussion of the order of the transition
in the chiral limit arises from the role of the axial anomaly. The
nature of the chiral transition for the massless $N_f=2$ theory , i.e. 
for $m_l\to0$ and $m_s\to\infty$, depends on the magnitude of the axial 
$U_A(1)$ symmetry breaking. If this 
remains significant close to the transition temperature then the relevant 
symmetry becomes isomorphic to that of the 3-d $O(4)$ spin model and the 
transition is expected to be second order belonging to that universality class
\cite{Pisarski:1983ms,Butti:2003nu}. However, if $U_A(1)$ symmetry breaking
becomes negligible near the chiral transition temperature, the relevant 
symmetry becomes
isomorphic to $O(2)\times O(4)$ and the transition be either first order
\cite{Pisarski:1983ms} or second order \cite{Pelissetto:2013hqa,Grahl:2013pba}. In
the intermediate quark mass region there is no true phase transition, rather a
crossover takes place from the hadronic to the quark-gluon plasma phase. 

All the first order regions
are separated from the crossover region by lines of second order phase transitions
belonging to the 3-d $Z(2)$ universality class. The first order region for
the $N_f=2+1$ case, the second order $Z(2)$ line 
separating the $N_f=2+1$ first order and the crossover regions and the 
second order $O(4)$ line for the $N_f=2$ case are supposed
to meet at a tri-critical point characterized by a certain value of the 
strange quark mass, $m_s^{tric}$.  Although, it is well established that in 
the real world, i.e. for the physical values of the quark masses, the 
transition is a crossover \cite{Aoki:2006we,Bhattacharya:2014ara}, the 
location of the physical point with respect to $m_s^{tric}$ has not been 
established and even $m_s^{tric} \rightarrow \infty$ cannot be ruled out.  
More specifically, it is yet
unclear whether $m_s^{phys}>m_s^{tric}$ or $m_s^{phys}=m_s^{tric}$ or $m_s^{phys}<m_s^{tric}$. If
$m_s^{phys}>m_s^{tric}$ then in the limit of $m_l\to0$ one should find a second order
transition belonging to the $3$-d $O(4)$ universality class, if $m_s^{phys}=m_s^{tric}$ the tri-critical 
point is a Gaussian fixed point of the 3-dimensional $\phi^6$ model and its critical
exponents take the mean field values~\cite{Riedel72} and if
$m_s^{phys}<m_s^{tric}$ then in the $m_l\to0$ limit one may cross through a
second order transition belonging to the 3-d $Z(2)$ universality class and 
then may end up in the first order transition region. 

In the following discussion we assume $m_s^{phys}>m_s^{tric}$.

\subsection{Chiral transition in (2+1)-flavor QCD} 

In the vicinity of the chiral phase transition, the free energy density may be
expressed as a sum of a singular and a regular part
\begin{equation}
f = -\frac{T}{V} \ln Z\equiv f_{sing}(t,h)+ f_{reg}(T,m_l,m_s,\vec{\mu}) \; .
\label{free_energy}
\end{equation}
The parameter $h$ represents the dimensionless explicit chiral symmetry breaking
(magnetic) field and $t$ incorporates all the `thermal' variables that do not
explicitly break the chiral symmetry, e.g. at leading order,    
\begin{equation}
t = \frac{1}{t_0} \left[ \frac{T-T_c^0}{T_c^0} + 
\kappa_q \left( \frac{\mu_q}{T} \right)^2 \right]
\;, \qquad \mathrm{and} \qquad
h = \frac{1}{h_0} \frac{m_l}{m_s}
\;,
\label{reduced}
\end{equation}
where $\mu_q\equiv \mu_u=\mu_d$.
$T_c^0$ denotes the (unknown) phase transition temperature in the chiral limit
$m_l\to0$ and for $\mu_q=0$. The scaling variables $t$, $h$ are normalized by two
unknown parameters $t_0$ and $h_0$. The other unknown parameter $\kappa_q$ controls
how the chiral transition temperature $T_c^0$ changes as a function of $\mu_q$.  All
these 4 unknown quantities $(T_c, \kappa_q, t_0, h_0)$ are unique to QCD, 
similar to the low energy constants in
the chiral Lagrangian. These parameters can be determined by analyzing the scaling
behaviors, arising from the singular part of the free energy, of the chiral order
parameter and susceptibilities using lattice QCD.

\subsubsection{Pseudo-critical behavior at vanishing chemical potential}
Sufficiently close to the QCD chiral transition the renormalization
group invariant dimensionless order parameter $M_b$, constructed out of the 
light quark chiral condensate, defined in Eq.~\ref{ppbar}, and the strange 
quark mass obeys the scaling relation  
\begin{equation}
M_b \equiv \frac{m_s \langle \bar{\psi}\psi \rangle_l}{T^4} = 
h^{1/\delta} f_G(z) + \mathrm{regular\ terms} \;,
\label{Mb}
\end{equation}
in terms of a single scaling variable $z=t/h^{1/\beta\delta}$. The critical exponents
$\beta$ and $\delta$ and the scaling function $f_G(z)$ uniquely characterize the
universality class of the chiral phase transition. As discussed before, for continuum
QCD the relevant universality class is expected to be the same as that of 3-d $O(4)$
spin models. However, the situation is more subtle for QCD on the lattice.
For example, since for staggered fermions away from the
continuum limit there is only one Goldstone boson in the chiral limit the relevant
universality class is that of the 3-d $O(2)$ spin model. Table~\ref{tab:parameter}
summarizes the critical exponents and other relevant quantities for the 3-d $O(2)$
and $O(4)$ universality classes. 

\begin{table}[b]
\tbl{Critical exponents $\alpha$, $\beta$, $\gamma$, $\delta$ for the 3-dimensional $O(N)$
universality classes \cite{Engels:2001bq,Engels:2003nq}.  Only two of the four
critical exponents are independent. They are related to each other through the scaling 
relations
$\gamma=\beta(\delta-1)$ and $\alpha+2\beta+\gamma=2$. The last two columns 
give the
location of the maxima of the scaling functions $f'_G(z)$ and $f_\chi (z)$,
respectively \cite{Engels:2001bq,Engels:2003nq}.
\vspace{0.3cm}
 }
{\begin{tabular}{c c c c c c c}
\hline \hline
$N$ & $\alpha$ & $\beta$ & $\gamma$ & $\delta$ & $z_t$ & $z_p$ \\
\hline
2 & -0.017 & 0.349 & 1.319  & 4.779 & 0.46(20) & 1.56(10) \\
4 & -0.213 & 0.380 & 1.453 & 4.824 & 0.73(10) & 1.35(3) \\
\hline \hline
\end{tabular}}
\label{tab:parameter}
\end{table}

\begin{figure}[t]
\begin{center}
\includegraphics[scale=0.49]{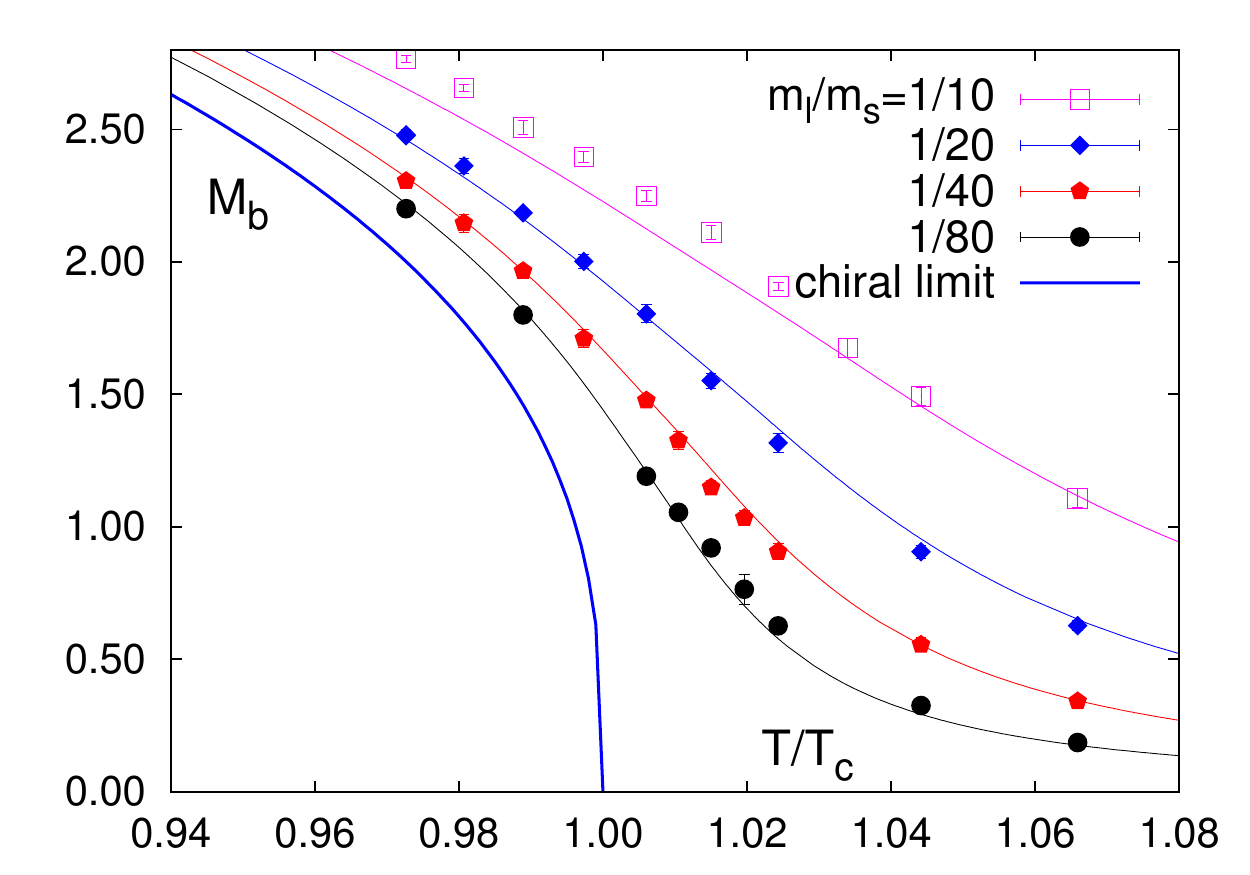}
\includegraphics[scale=0.49]{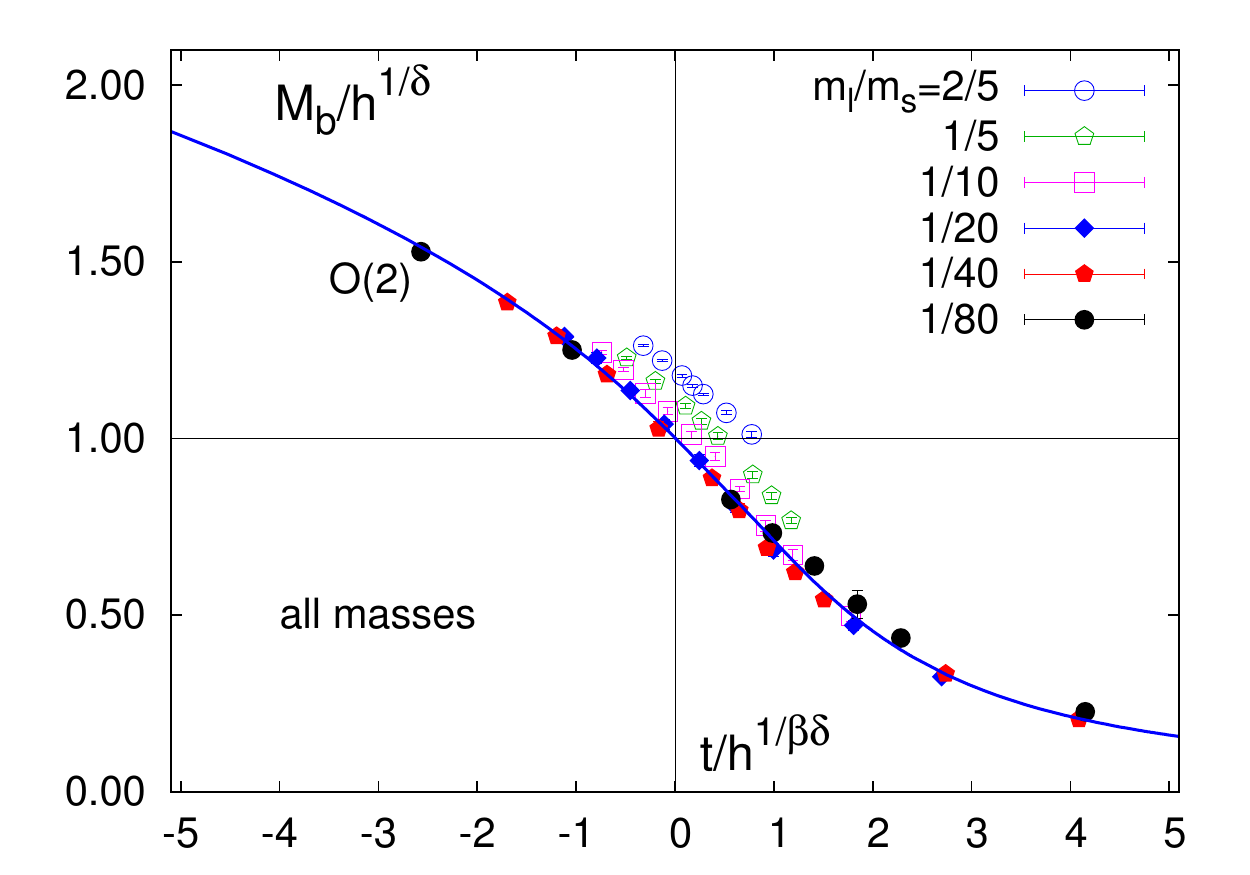}
\end{center}
\caption{$O(N)$ scaling of the chiral order parameter $M_b$ introduced
in Eq.~\ref{Mb} for  (2+1)-flavor QCD \cite{Ejiri:2009ac}.}
\label{fig:MEOS}
\end{figure}

Fig.~\ref{fig:MEOS} shows results for the chiral order parameter $M_b$ calculated in
(2+1)-flavor QCD \cite{Ejiri:2009ac} for several values of the light to strange quark
mass ratio $m_l/m_s$ while keeping $m_s$ fixed to its physical value $m_s^{phys}$. In
the continuum limit the physical value of this ratio is
$m_l^{phys}/m_s^{phys}=1/27.5$ \cite{Agashe:2014kda}. The left hand panel in 
Fig.~\ref{fig:MEOS} shows the variation of $M_b$ with temperature calculated on
rather coarse lattices ($N_\tau =4$) for several values of
$m_l/m_s$. In the continuum limit the smallest of these ratios would correspond to
pion masses of about $80$~MeV. Clearly, the transition sharpens as the light quark
masses become smaller. The right hand panel of Fig.~\ref{fig:MEOS} shows a comparison
of the scaled order parameter $M_b/h^{1/\delta}$ (see Eq.~\ref{Mb}) 
with the $O(2)$ scaling function. As
can be seen, for small enough light quark masses, $m_l\lesssim m_s^{phys}/20$, the
scaled order parameter for different values of the light quark mass collapses
on a unique curve and compares well with the $O(2)$ scaling function. While these lattice
results suggest that the chiral transition for $N_f=2$ theory belongs to the 3-d
$O(N)$ universality class one should bear in mind that these results are not
continuum extrapolated and discretization effects may change the conclusion. 
In fact, lattice studies performed with the unimproved staggered action even 
further away from the continuum limit suggest that in this case the 
transition may even be first order \cite{Bonati:2014kpa}.

The chiral order parameter $M_b$ can be used to define two susceptibilities 
obtained by either taking a derivative with respect to `thermal' variable $t$, 
which give the mixed susceptibility 
$\chi_t \sim \partial M_b/\partial t\sim \partial^2 f_s/\partial t \partial h$, 
or with respect to the symmetry breaking variable,
$\chi_h \sim \partial M_b/\partial h\sim \partial^2 f_s/\partial h^2$. 
These two susceptibilities are the only
two second derivatives of the free energy of an $O(N)$ symmetric theory like QCD that
diverge at the critical point\footnote{In general, there is a third 
susceptibility, the specific heat, $C_V$, which is obtained as a second 
derivative of the free energy with respect to the `thermal' variable $t$,
$C_V \sim \partial^2 f_s/\partial t^2$. In $O(N)$ symmetric theories, 
however, the relevant critical exponent $\alpha$ is negative. The specific 
heat, thus, does not diverge at $T_c$ in the chiral limit.}, i.e. 
at $T=T_c^0$ for $m_l=0$. Their singular
behavior is controlled by two scaling functions, $f'_G(z)={\rm d} f_G(z)/{\rm d}z$,
and $f_\chi(z)=(f_G(z) -(z/\beta)f'_G(z))/\delta$, respectively. Away from the chiral
limit, these two scaling functions have maxima at universal values of $z$, i.e.
$z=z_t$ and $z=z_p$ (see Tab.~\ref{tab:parameter}), which can be used to define
pseudo-critical temperatures.

\begin{figure}[t]
\begin{center}
\includegraphics[scale=0.49]{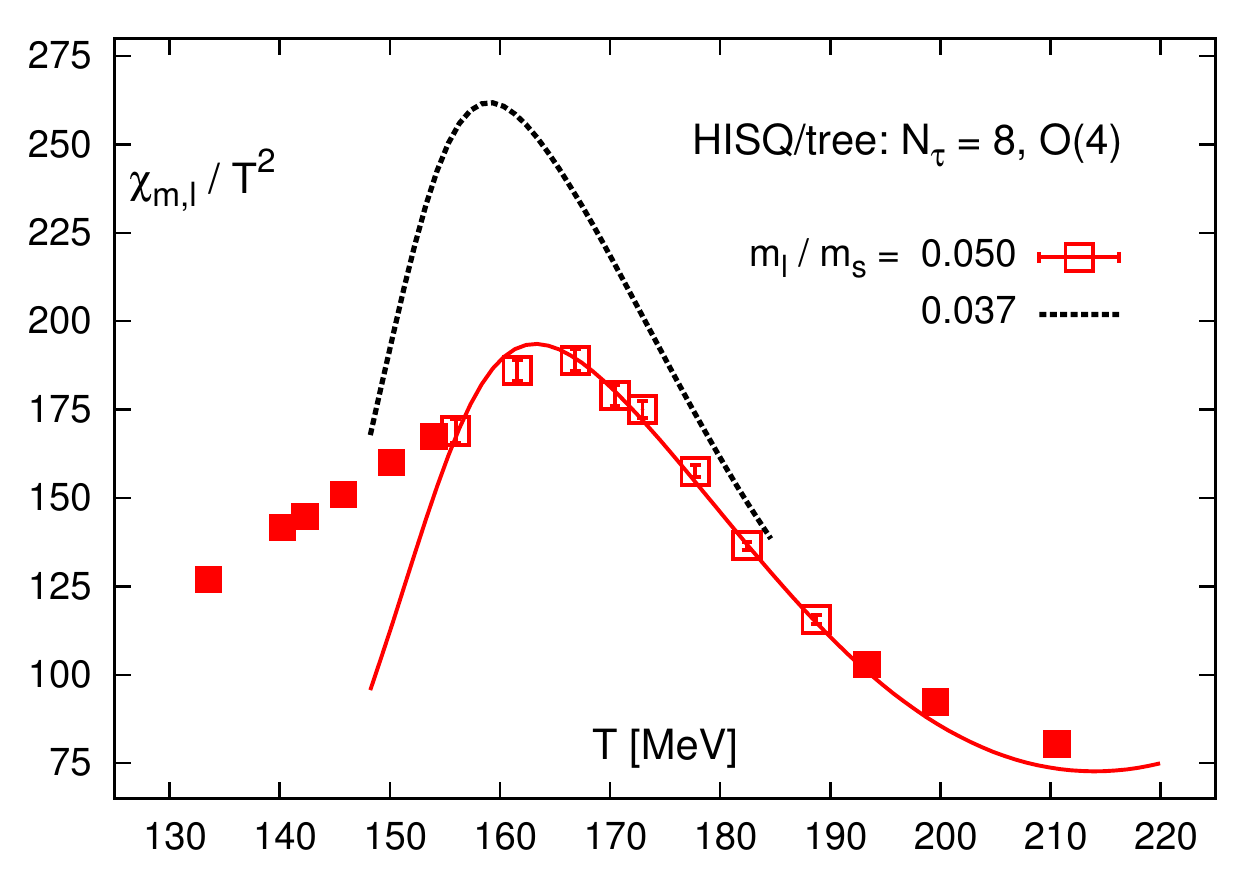}
\includegraphics[scale=0.49]{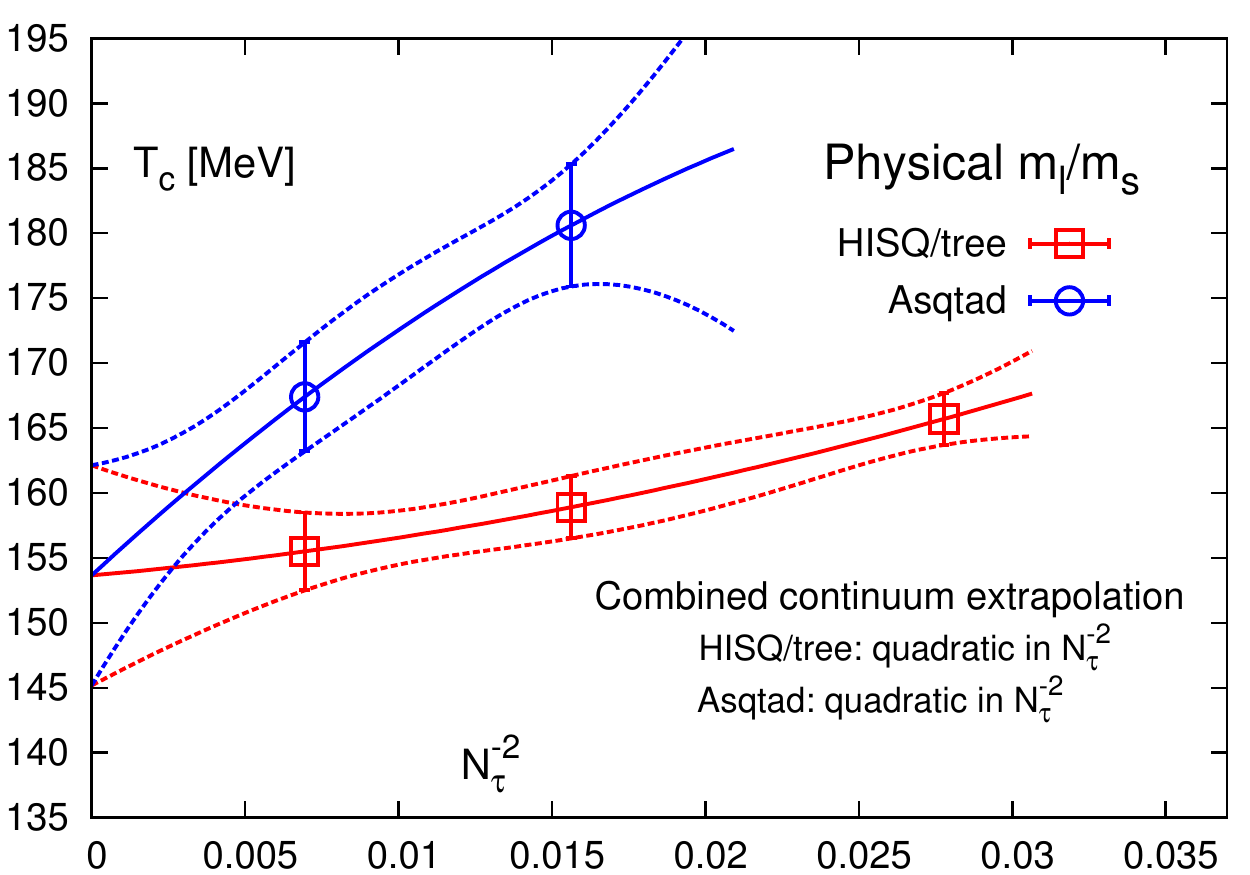}
\end{center}
\caption{(Left) $O(4)$ scaling of the chiral susceptibility and its extrapolation to
the physical light quark mass for (2+1)-flavor QCD \cite{Bazavov:2011nk}. The peak
location of the chiral susceptibility defines the chiral crossover temperature.
(Right) Continuum extrapolation of the chiral crossover temperature for (2+1)-flavor
physical QCD, yielding $T_c=(154\pm9)$~MeV \cite{Bazavov:2011nk}.}
\label{fig:Tc1}
\end{figure}

For example, the scaling behavior of the chiral susceptibility for two light, degenerate flavors,
\begin{eqnarray}
\chi_{m,l}(T)&=& {2}
\frac{\partial \langle \bar\psi \psi \rangle_{l}}{\partial m_l} \; ,
\label{chi_tot}
\end{eqnarray}
is related to that of $\chi_h$. 
The renormalization group invariant product of the chiral
susceptibility $\chi_{m,l}$ and the square of the strange quark mass is 
related to the scaling functions by, 
\begin{equation}
\frac{m_s^2\chi_{m,l}}{T^4} = \frac{1}{h_0} h^{1/\delta-1} f_\chi(z) + 
\mathrm{regular\ terms} \;.
\end{equation}
At vanishing quark chemical potential the chiral susceptibility thus diverges at the chiral 
critical temperature $T_c^0$ in the
chiral limit $h\to0$. The peak location of the chiral susceptibility will then be
associated with the peak of the scaling function $f_\chi(z)$, located at $z=z_p$, and
defines the chiral pseudo-critical temperature $T_{c}$ for $h>0$,
\begin{equation}
T_{c} = T_c^0 \left[ 1 + \frac{z_p}{z_0} \left(
\frac{m_l}{m_s}\right)^{1/\beta\delta} \right] +
\mathrm{regular\ terms} \;,
\end{equation}
with $z_0=t_0/h_0^{1/\beta\delta}$. As illustrated in Fig.~\ref{fig:Tc1}, the chiral
crossover temperature for physical QCD with $m_s=m_s^{phys}$ and
$m_l^{phys}=m_s^{phys}/27.5$ was studied in detail \cite{Bazavov:2011nk} by using
such a scaling analysis of the chiral susceptibility, including the influence of
regular terms. By performing scaling fits to the chiral susceptibility, extrapolating
to the physical value of light quark mass and subsequently taking the continuum
limit one obtains a value of 
\begin{equation}
T_{c}=(154\pm9)~{\rm MeV} 
\label{Tc}
\end{equation}
for the chiral crossover temperature in (2+1)-flavor physical QCD 
\cite{Bazavov:2011nk}. Studies with other improved
staggered fermion discretization schemes 
\cite{Aoki:2006br,Aoki:2009sc,Borsanyi:2010bp}, which used other criteria,
not directly related to criticality, to define a crossover temperature, and 
with chiral Domain Wall Fermions \cite{Bhattacharya:2014ara} also 
yielded compatible results for the QCD chiral crossover temperature. 

\subsubsection{Curvature of the pseudo-critical line}
\begin{figure}[t]
\begin{center}
\begin{minipage}[c]{0.49\textwidth}
\includegraphics[width=0.99\textwidth]{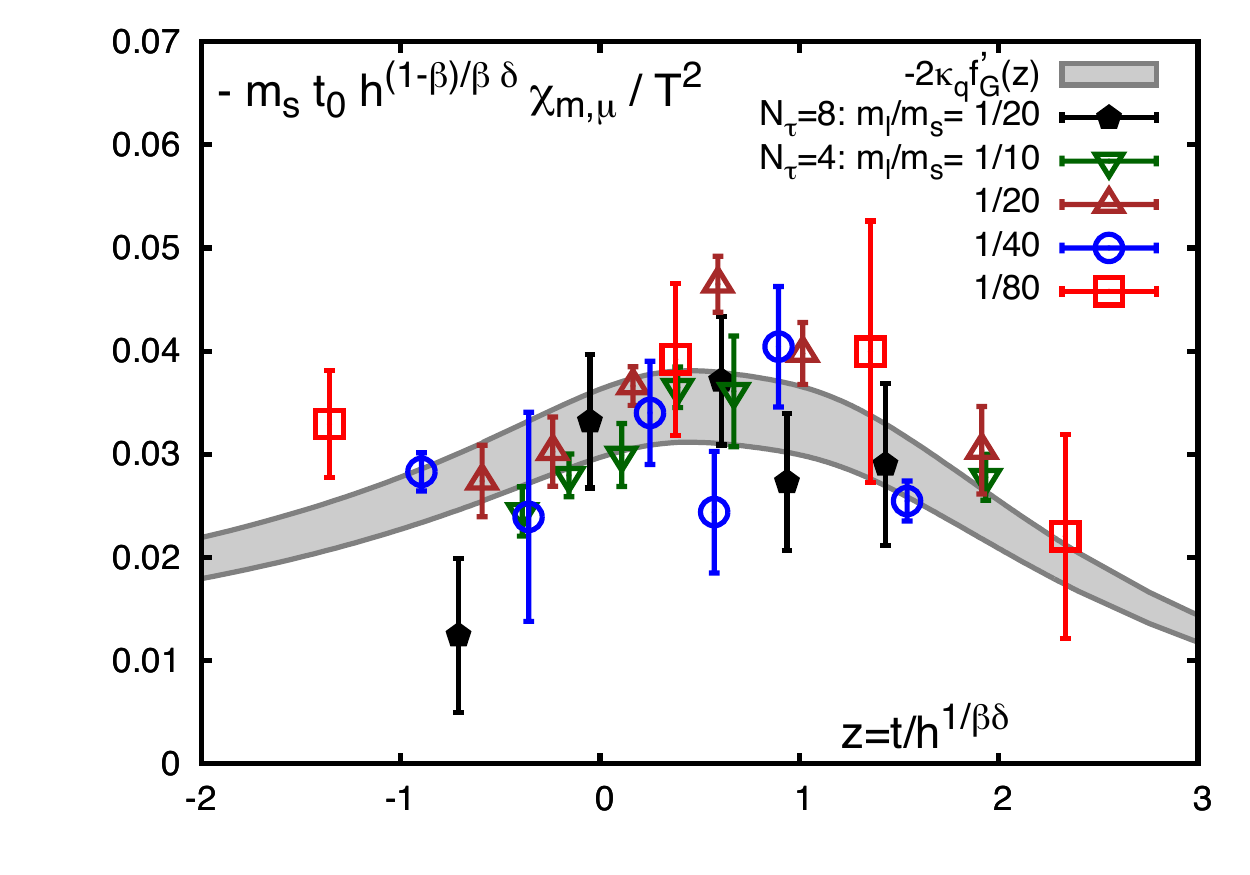}
\end{minipage}
\begin{minipage}[c]{0.50\textwidth}
\includegraphics[width=0.99\textwidth,height=0.20\textheight]{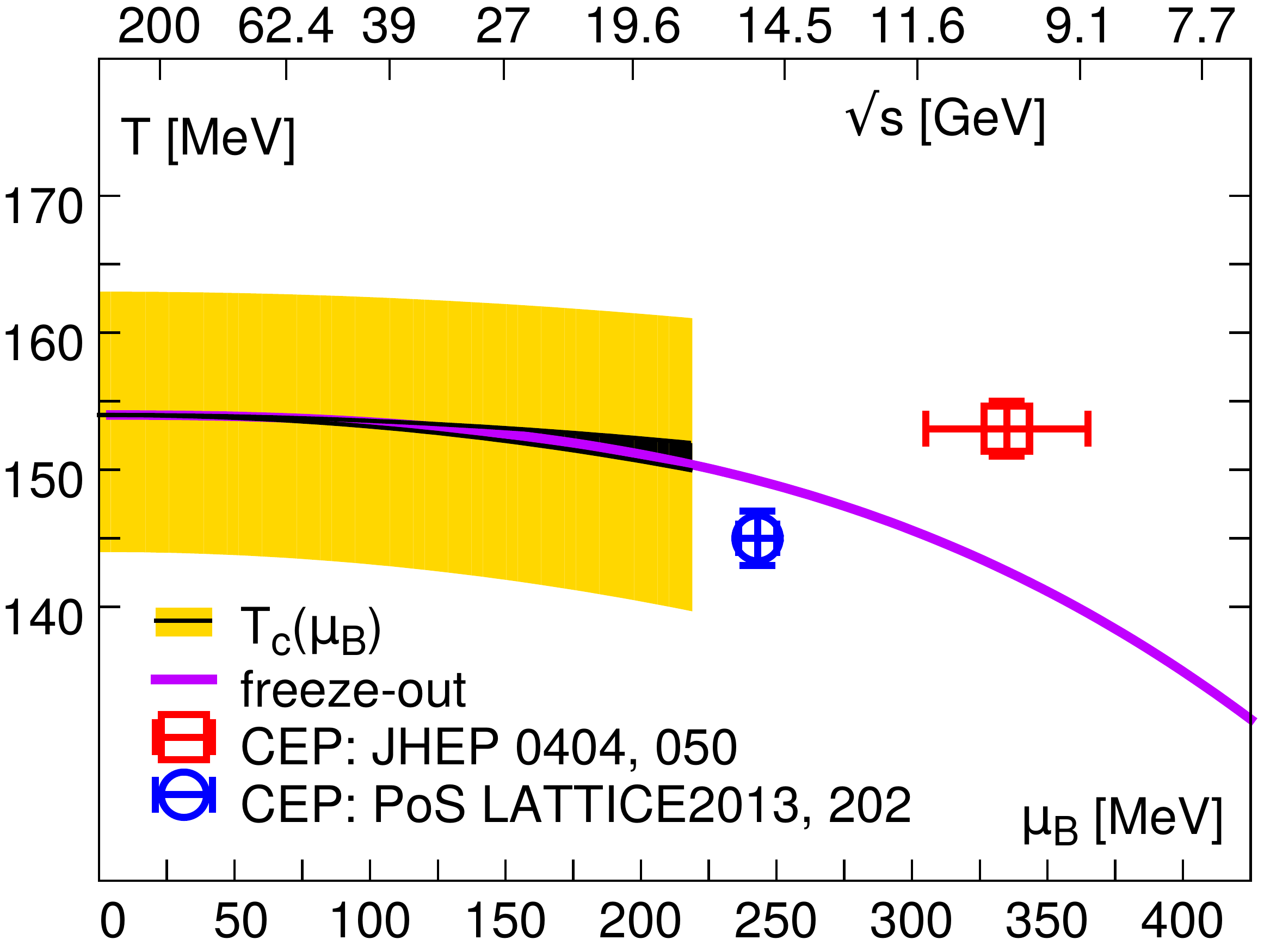}
\end{minipage}
\end{center}
\caption{(Left) $O(N)$ scaling of the susceptibility $\chi_{m,\mu}$ and determination
of the curvature, $\kappa_q$, of the chiral phase boundary in the $T$-$\mu_q$ plane
\cite{Kaczmarek:2011zz}. (Right) Chiral crossover temperature as function of baryon
chemical potential, $T_c(\mu_B)$, compared with the freeze-out temperature 
in heavy-ion
collisions. The parametrization of the freeze-out line is taken from 
Ref.~\citen{Andronic:2011yq}, with the freeze-out temperature in the 
limit of infinite
collision energy adjusted to the crossover temperature at vanishing chemical potential,
$T_{fo}(\sqrt{s}\to\infty)=154$~MeV. The black band for the crossover line,
$T_c(\mu_B)$, reflects the systematic uncertainty arising from a factor of 2
difference in current estimates from calculations with real 
\cite{Kaczmarek:2011zz,Endrodi:2011gv} and
imaginary \cite{Cea:2014xva,Bonati:2014rfa} $\mu_B$ (see text).
Data points are from a reweighting
analysis \cite{Fodor:2004nz} and Taylor expansion \cite{Datta:2014zqa} 
(see text). They also are normalized to $T_c=154$~MeV.}
\label{fig:Tc2}
\end{figure}

The scaling behavior of the mixed susceptibility, $\chi_t$, can be utilized to determine
the chiral phase boundary in the $T$-$\mu$ plane \cite{Kaczmarek:2011zz}. As introduced
in Eq.~\ref{reduced}, to leading order the chemical potential $\mu_q$ does
not contribute to the
explicit chiral symmetry breaking field $h$ and can be treated as part of 
the `thermal' scaling variable $t$. Thus, the derivative of the order 
parameter with respect to the chemical
potential, $\chi_{m,\mu}=\partial^2 M_b/\partial (\mu_q/T)^2$, is equivalent 
to the mixed susceptibility, $\chi_t$. 
The scaling 
behavior of the renormalization group invariant combination of 
$\chi_{m,\mu}$ multiplied with the strange quark mass, which
combines a `thermal' and `field-like' derivative of the free energy, is 
given by 
\begin{equation}
\frac{m_s\chi_{m,\mu}}{T^2} = \frac{2\kappa_q}{t_0} h^{-(1-\beta)/\beta\delta} f_G^\prime(z) + 
\mathrm{regular\ terms} \;.
\end{equation}
In the chiral limit, $h\to0$, this susceptibility diverges at $T_c^0$ for $\mu_q=0$.
For $\mu_q>0$, the chemical potential dependence of the chiral transition temperature
$T_c^0(\mu_q)$ is defined by the peak location, $z=z_t$, of the scaling function
$f_G^\prime(z)$ and in the leading order in $(\mu_q/T)^2$ it is given by 
\begin{equation}
T_c^0(\mu_q) = T_c^0 \left[ 1 - \kappa_q \left(\frac{\mu_q}{T_c^0}\right)^2 + 
\mathcal{O} \left[ \left(\frac{\mu_q}{T_c^0}\right)^4 \right] \right] + 
\mathrm{regular\ terms} \;.
\end{equation}
As illustrated in Fig.~\ref{fig:Tc2}~(left), the curvature of the chiral phase
transition line, $\kappa_q$, can be determined by fitting the scaled susceptibility
$m_s t_0 h^{(1-\beta)/\beta\delta} \chi_{m,\mu}/T^2$ to the universal scaling function
$f_G^\prime(z)$ \cite{Kaczmarek:2011zz}. Such a scaling analysis provides a value of
the curvature $\kappa_q=0.059(6)$ \cite{Kaczmarek:2011zz}. In terms of the baryon
chemical potential $\mu_B=3 \mu_q$, the chiral transition temperature is then given
by
\begin{equation}
T_c^0(\mu_B) = T_c^0 \left[ 1 - 0.0066(7) \left(\frac{\mu_B}{T_c^0}\right)^2 + 
\mathcal{O} \left[ \left(\frac{\mu_B}{T_c^0}\right)^4 \right] \right] \;.
\end{equation}
This result is in very good agreement with the independent determination of the
continuum extrapolated results of the curvature of the chiral phase boundary 
for  (2+1)-flavor of physical QCD \cite{Endrodi:2011gv}, using different
criteria. However, analytic continuation from
purely imaginary chemical potential yields a factor two larger values for the curvature
\cite{Cea:2014xva,Bonati:2014rfa}. Fig.~\ref{fig:Tc2}~(right) shows the proximity of
the QCD phase boundary to the freeze-out line in heavy-ion collisions
\cite{Andronic:2011yq} in the $T$-$\mu_B$ plane.

\subsubsection{The critical end point}

Information on the location of a critical end point (CEP) in the QCD phase 
diagram (Fig.~\ref{fig:phasediagram}) from lattice QCD is ambiguous. 
Although considerable progress in developing algorithms that would
allow direct calculations in lattice QCD at $\mu_B>0$ have been made
recently (see Ref.~\citen{Sexty:2014dxa} for a recent review) 
these techniques are not yet applicable to realistic QCD parameter values.

The first calculations that provided
hints for the existence of a critical point \cite{Fodor:2004nz} used a 
reweighting technique that is applicable only on rather small lattices.
It had been performed on coarse lattices with temporal extent $N_\tau=4$
using only the naive 1-link staggered fermion action. 
Taste violation effects thus
are large and the reweighting is known to fail when used too far away
from the parameters actually used in the calculations. It has been
argued that this `overlap problem' may have led to the spurious 
identification of a signal for a critical point \cite{Ejiri:2005ts}.
Calculations performed with an imaginary chemical potential do not find 
any evidence for the existence of a critical point~\cite{deForcrand:2002ci}.
However, also these calculations have been performed on coarse lattices
and utilized the naive staggered 1-link action. 

The conclusions drawn in Ref.~\citen{deForcrand:2002ci} are based
on an analysis of the $\mu_B$-dependence of the location of the boundary 
line that separates
the first order transition region shown in Fig.~\ref{fig:Columbia} 
at small quark masses from the crossover region. These calculations,
performed  with staggered fermions using an imaginary chemical potential 
\cite{deForcrand:2002ci}, suggest that this boundary
moves to smaller quark masses with increasing chemical potential. This
disfavors the existence of a critical point connected to the Z(2) boundary
line at vanishing chemical potential.
A similar analysis, performed with clover-improved Wilson fermions, comes 
to the opposite conclusion \cite{Jin:2015taa} and thus does favor the 
existence of a critical point. The apparent differences between 
calculations performed with staggered fermions and Wilson fermions
on coarse lattices underscores that better
control over systematic errors is needed before a definite conclusion
on the existence of a critical point can be drawn. 

Analyzing the convergence properties
of the Taylor series for the pressure at non-zero chemical potential
\cite{Gavai:2003mf,Allton:2005gk} (Eq.~\ref{TaylorBQS}) does, in principle,
allow to relate the location of the CEP to the radius of convergence
of this series. Although this approach has been used closer to the continuum 
limit ($N_\tau=8$) \cite{Gavai:2008zr,Datta:2014zqa} than the studies 
discussed above, the current estimates still are based on calculations 
with the naive staggered 1-link action. They still suffer from large
taste symmetry violations. These calculations yield for the location
of the critical point \cite{Datta:2014zqa} 
$(T^E/T_{c},\mu_B^E/T^E)= (0.94(1),1.68(5))$ 
\cite{Gavai:2008zr} which is significantly lower than the reweighting result
$(T^E/T_{c},\mu_B^E/T^E)= (0.99(1),2.2(2))$
\cite{Fodor:2004nz}. We show these two estimates for the location of the CEP
in Fig.~\ref{fig:Tc2}~(right). The systematic errors of these estimates
are at present difficult to estimate. 

Further information on the location of a critical point comes from the 
calculations of (i) the equation
of state at $\mu_B>0$ performed with improved staggered fermions, which
we will discuss in Section~\ref{sec:bulk}, as well as (ii) the analysis of freeze-out
parameters based on Taylor expansions of cumulants of charge fluctuations,
which are also performed with improved staggered actions and which we will
discuss in Section~\ref{sec:fluctuation}, both suggest that a CEP located at $\mu_B/T < 2$ is
unlikely.

\subsection{Deconfining aspect of QCD}
\label{sec:deconf}
\begin{figure}[thpb]
\begin{center}
\begin{minipage}[c]{0.49\textwidth}
\includegraphics[width=0.99\textwidth]{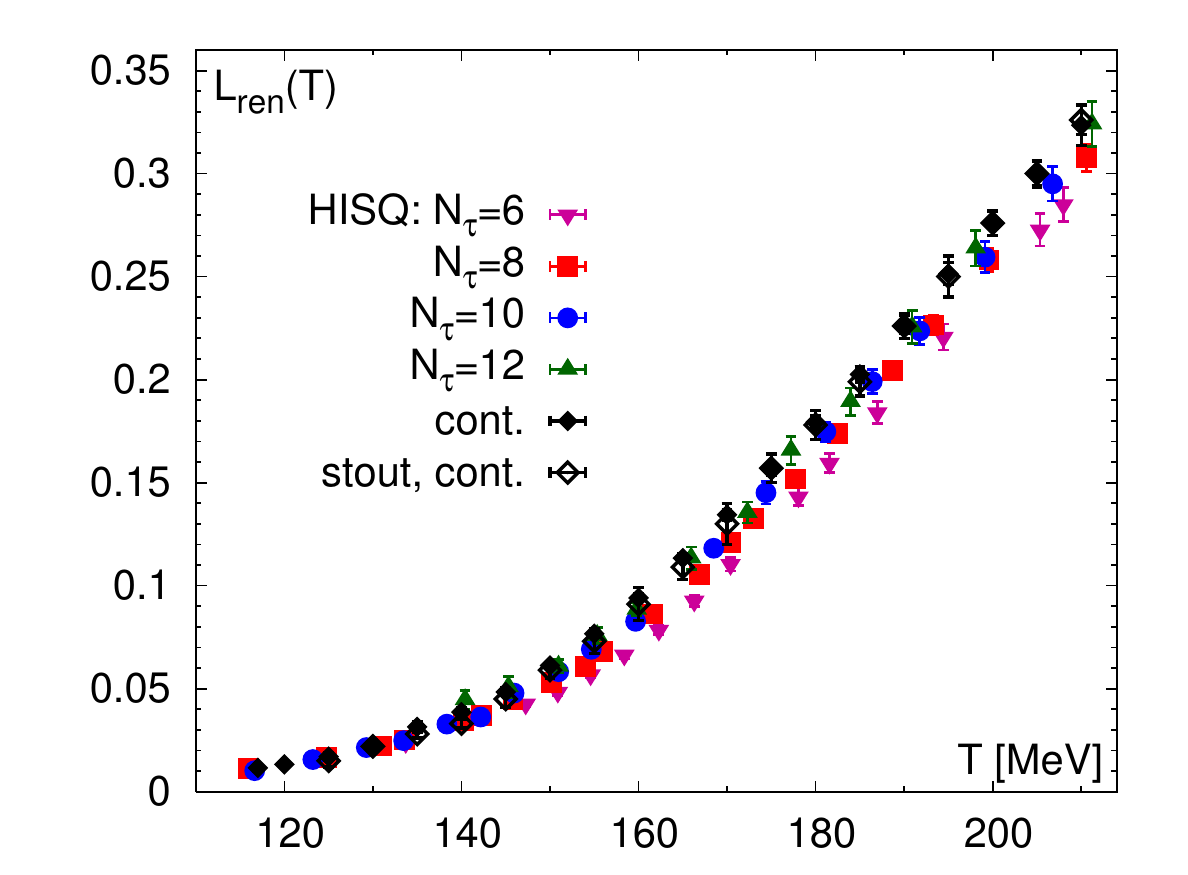}
\end{minipage}
\begin{minipage}[c]{0.49\textwidth}
\includegraphics[width=0.99\textwidth]{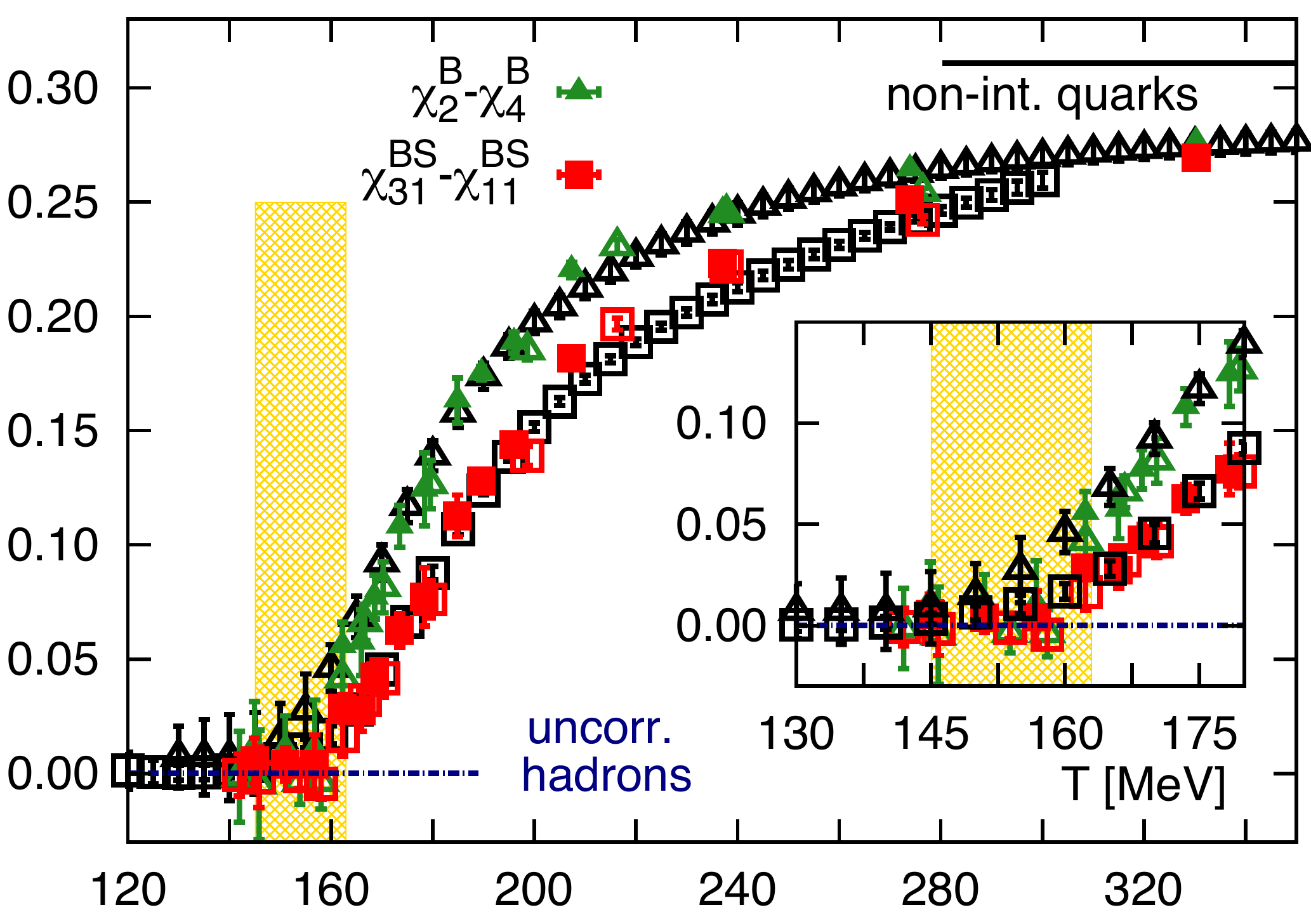}
\end{minipage}
\end{center}
\caption{(Left) The renormalized Polyakov loop in (2+1)-flavor QCD
\cite{Bazavov:2013yv}. (Right) Appearance of the fractionally charged degrees of
freedom in the chiral crossover region $T_c=154(9)$~MeV (shaded region) for the light
as well as the strange quark. The black points show results obtained using the stout
action \cite{Bellwied:2013cta} and the other points have been obtained using 
the HISQ action \cite{Bazavov:2013dta}.} 
\label{fig:deconf1}
\end{figure}

Pure $SU(N_c)$ Yang-Mills theory without quarks possess an exact global $Z(N_c)$
center symmetry, which gets spontaneously broken in the high temperature deconfining
phase \cite{Kuti:1980gh,McLerran:1980pk}. 
The thermal expectation value of the renormalized Polyakov loop
\begin{equation}
L_{\rm{ren}}(T) = e^{-c(g^2)N_\tau} \cdot \frac{1}{VN_c} \sum_{\vec{x}} \left\langle
\rm{Tr} \prod_{x_0=1}^{N_\tau} U_{(x_0,\vec{x}),\hat{0}} \right \rangle \;,
\end{equation}
is not invariant under the $Z(N_c)$ center symmetry. The Polyakov loop can be
interpreted as the free energy difference, $F_\infty(T)$, of a thermal system with
and without an infinitely heavy static quark anti-quark pair separated by infinite
distance, $L_{\rm{ren}}(T)=\exp(-F_\infty(T)/2T)$. The renormalization constant
$c(g^2)$ can be fixed \cite{Kaczmarek:2002mc} by demanding that in the short distance
limit the heavy quark free energy coincides, up to a trivial additive 
constant, with the
Coulombic short distance behavior of the zero temperature heavy quark potential
defined in Eq.~\ref{Cornell}. In the confined phase $F_\infty(T)=\infty$, as a static
quark and anti-quark pair cannot be separated by infinite distance in this phase.
Thus in the $Z(N_c)$ symmetric confined phase $L_{\rm{ren}}(T)=0$. On the other hand,
in the spontaneously $Z(N_c)$ broken deconfined phase a static quark anti-quark pair
can be separated by infinite distance due to the presence of color screening and
$L_{\rm{ren}}(T)\ne0$.  Thus, for a pure $SU(N_c)$ gauge theory, i.e. in the limit
$m_l\to\infty$, the Polyakov loop serves as the order parameter for the
deconfinement transition \cite{McLerran:1981pb}. However, since the mere presence of
quarks explicitly breaks the $Z(N_c)$ center symmetry, the Polyakov loop does not
serve as an order parameter for QCD with realistic light quarks. Furthermore, since
the Polyakov loop is not related to a derivative of the QCD partition function with
respect to the thermal or the symmetry breaking field, its change or fluctuation
across the QCD transition may not capture the true singularities of the QCD
partition function in any limit. Thus, a deconfining temperature
defined from the change of the Polyakov loop may not reflect the 
pseudo-critical properties of QCD with realistic light quark masses. 
As shown in Fig.~\ref{fig:deconf1}
(left), the change of the Polyakov loop within the chiral crossover region,
$T_{c}=154(9)$~MeV, is rather gradual and smooth. 

Deconfinement is associated with the liberation of degrees of freedom, manifested by a
rapid rise in the bulk thermodynamic observables such as the pressure, energy density etc. 
Among such bulk thermodynamic observables fluctuations and correlations
of different conserved quantum numbers, e.g. the baryon number ($B$), electric charge
($Q$) and strangeness ($S$), directly probe the liberation of the quark degrees of
freedom and the consequent appearance of fractionally charged quantum number
carriers. For example, both for an uncorrelated hadron gas 
and in a free quark gas the difference of second to fourth order cumulants of 
baryon number
fluctuations probes the baryon number of degrees of freedom, specifically
$\chi_2^B-\chi_4^B\propto B^2-B^4$. For hadronic degrees of freedom carrying baryon number $B=\pm1$
one thus has $\chi_2^B-\chi_4^B=0$, but for quark-like degrees of freedom with
$B=\pm1/3$ one finds  $\chi_2^B-\chi_4^B\ne0$. Similarly, for the strange 
quark sector higher order baryon-strangeness correlation such as
$\chi_{31}^{BS}-\chi_{11}^{BS} \propto (B^3-B)S$
is non-vanishing if the strangeness carrying degrees of freedom are associated with
quark-like baryon number $B=\pm1/3$ and vanishes when the strangeness is carried by
hadronic degrees of freedom with $B=\pm1$. Thus, such combinations of higher order
fluctuations and correlations of quantum numbers are sensitive probes of appearance
of fractionally charged degrees of freedom in a deconfined medium
\cite{Koch:2005vg,Ejiri:2005wq,Bazavov:2013dta}. The right hand panel of Fig.~\ref{fig:deconf1} shows lattice
QCD results for $\chi_2^B-\chi_4^B$ and $\chi_{31}^{BS}-\chi_{11}^{BS}$ from the
BNL-Bielefeld collaboration \cite{Bazavov:2013dta} as well as the Budapest-Wuppertal
collaboration \cite{Bellwied:2013cta}. Clearly, both quantities start deviating
strongly from zero in the chiral crossover region, $T_c=154(9)$~MeV, 
indicating that
fractionally charged degrees of freedom start to appear at these temperatures 
and onset of deconfinement takes place for the light as well as the strange 
quarks. Note, however, that also these fluctuation
observables are not order parameters in the strict sense.

\subsection{Axial symmetry of QCD at high temperature}

Although, the axial $U_A(1)$ symmetry is not an exact symmetry of QCD, as mentioned
before, the magnitude of its breaking is expected to influence the order of the
chiral phase transition. Hence, knowledge of the temperature dependence of the
$U_A(1)$ breaking is essential for a comprehensive understanding of the QCD chiral
transition.  The axial $U_A(1)$ symmetry of the massless QCD Lagrangian is broken
due to quantum fluctuations, resulting in non-conservation of axial
current~\cite{Adler:1969gk,Bell:1969ts} as well as explicit breaking of the global
$U_A(1)$ symmetry induced by topologically non-trivial gauge field configurations,
such as the instantons \cite{tHooft:1976up}. Due to color screening the instanton
density gets suppressed as the temperature increases \cite{Gross:1980br} and in the
$T\to\infty$ limit the $U_A(1)$ symmetry becomes exact. 

Since $U_A(1)$ is not an exact symmetry of QCD one cannot define an order parameter
associated with its breaking. However, for two massless flavors, the pion ($\pi$) 
and the
iso-vector scalar $\delta$ $(a_1)$ mesons transform into each other under a $U_A(1)$
rotation. Since the presence of an exact $U_A(1)$ will render these meson states
degenerate, the difference of the integrated two-point correlation functions of pion
and $\delta$ meson,
\begin{equation}
\chi_\pi - \chi_\delta = \int d^4x \,\Big( \left\langle \pi^+(x) \pi^-(0) \right\rangle
- \left\langle \delta^+(x) \delta^-(0) \right\rangle \Big) \;,
\label{eq:chi_pi_delta}
\end{equation}
will also vanish in the limit of exact $U_A(1)$ symmetry. Corrections due to small
non-vanishing light quark masses will only contribute at $\mathcal{O}(m_l^2)$. Thus,
this quantity can be taken as a measure of the $U_A(1)$ breaking
\cite{Shuryak:1993ee}. 

\begin{figure}[htb]
\begin{center}
\includegraphics[width=0.65\textwidth]{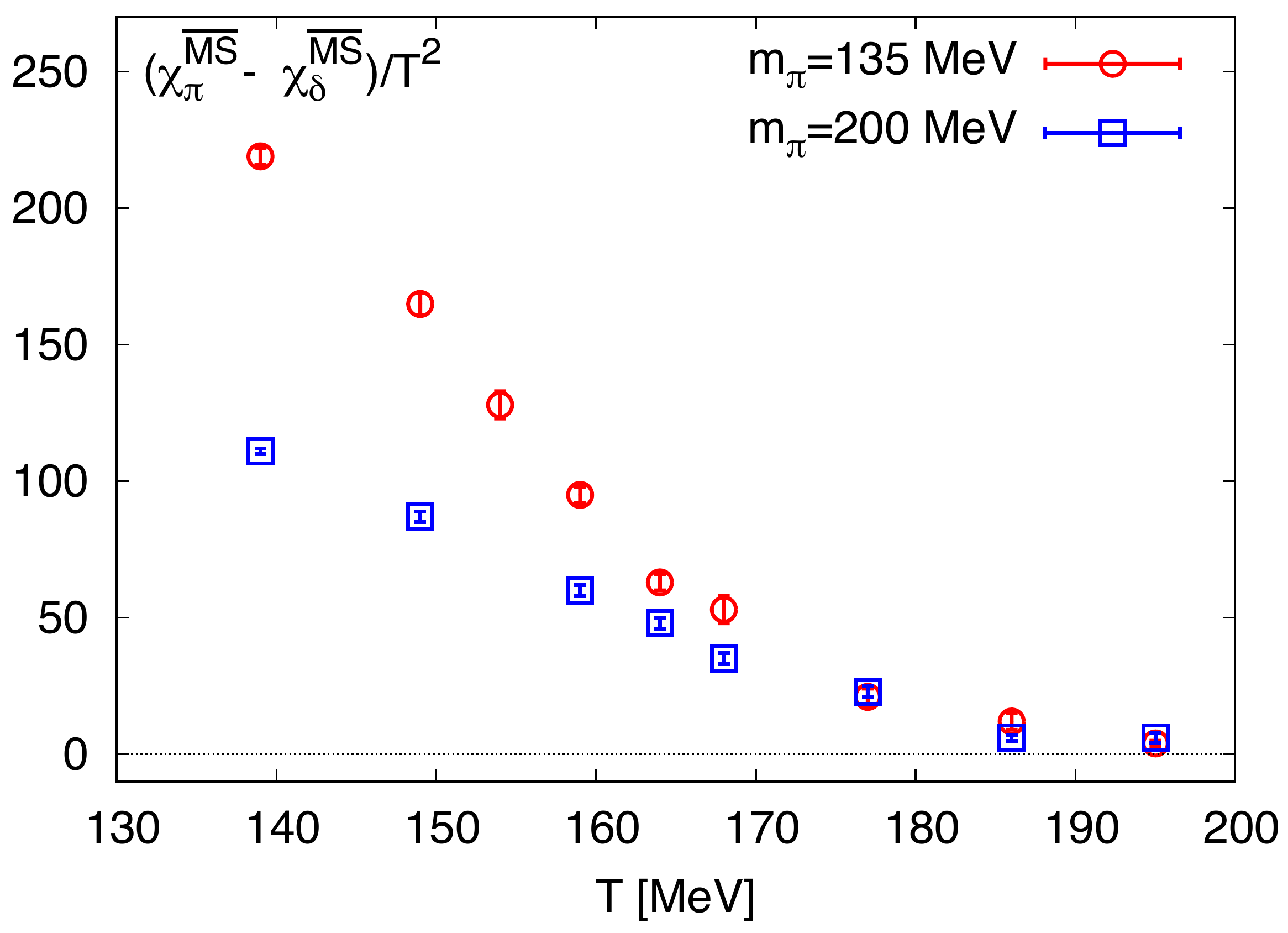}
\end{center}
\caption{The $U_A(1)$ breaking measure $\chi_\pi-\chi_\delta$ renormalized in 
the $\overline{\rm MS}$ scheme as function of temperature. Results are
from a calculation with Domain Wall fermions \cite{Bhattacharya:2014ara}. 
The difference is non-vanishing for $T\ge T_{c}=154(9)$~MeV. It becomes 
independent of quark mass for $T\gtrsim168$~MeV and rises with
decreasing quark mass at lower temperature. In fact, it will diverge in the 
chiral limit for $T<T_c$.}
\label{fig:ua1-1}
\end{figure}

This measure of $U_A(1)$ breaking was studied in detail using improved p4 staggered
fermions \cite{Cheng:2010fe}, and it was found that $\chi_\pi-\chi_\delta$ remains
non-vanishing for $T\lesssim1.2T_c$. However, for staggered fermions the issue
of axial anomaly is quite subtle and the correct anomaly may only emerge in the
continuum limit \cite{Sharpe:2006re,Donald:2011if}. On the other hand, emergence of
the axial anomaly is more straightforward for the chiral DWF formulation
\cite{Furman:1994ky}. For DWF, axial symmetry is broken by the same topologically
non-trivial configurations as in the continuum. Lattice artifacts appear only at
$\mathcal{O}(m_{\rm res}^2)$, due to the explicit chiral symmetry breaking residual
mass \cite{Antonio:2008zz}, $m_{\rm{res}}$, arising from finiteness of the 
fifth dimension ($L_s<\infty$). 
Thus, the DWF
action turns out to be a natural candidate for investigation of the temperature
dependence of $U_A(1)$ breaking in QCD. The temperature dependence of the $U_A(1)$
breaking measure, $\chi_\pi-\chi_\delta$, has been extensively studied using the DWF
formalism for several volumes as well as quark masses
\cite{Bazavov:2012qja,Buchoff:2013nra,Bhattacharya:2014ara}. As depicted in 
Fig.~\ref{fig:ua1-1}, calculations with DWF clearly show that $\chi_\pi-\chi_\delta$ does
not vanish around the chiral crossover temperature $T_{c}$ and remains
non-vanishing for $165~{\rm MeV}\lesssim T \lesssim 195$~MeV, independent of the light quark
masses. These results indicate that $U_A(1)$ symmetry may remain broken at these
temperatures even in the chiral limit.  

\begin{figure}[htb]
\begin{center}
\begin{minipage}[c]{0.6\textwidth}
\includegraphics[width=0.99\textwidth]{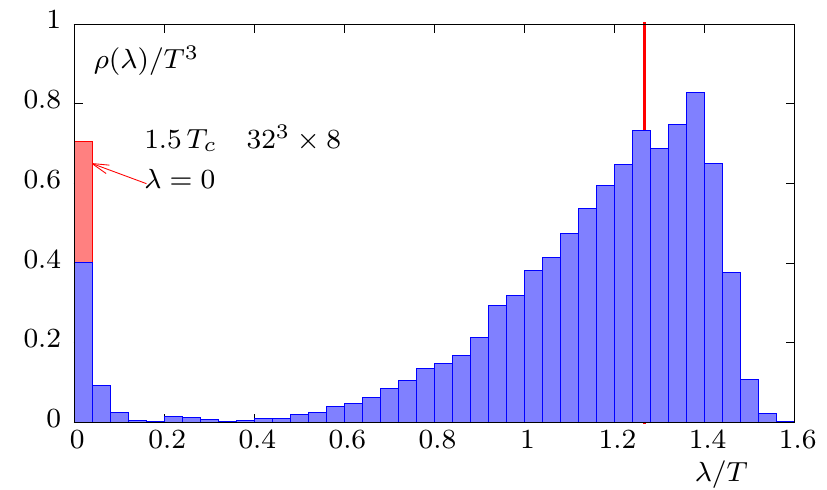}
\end{minipage}
\begin{minipage}[c]{0.47\textwidth}
\includegraphics[width=0.99\textwidth]{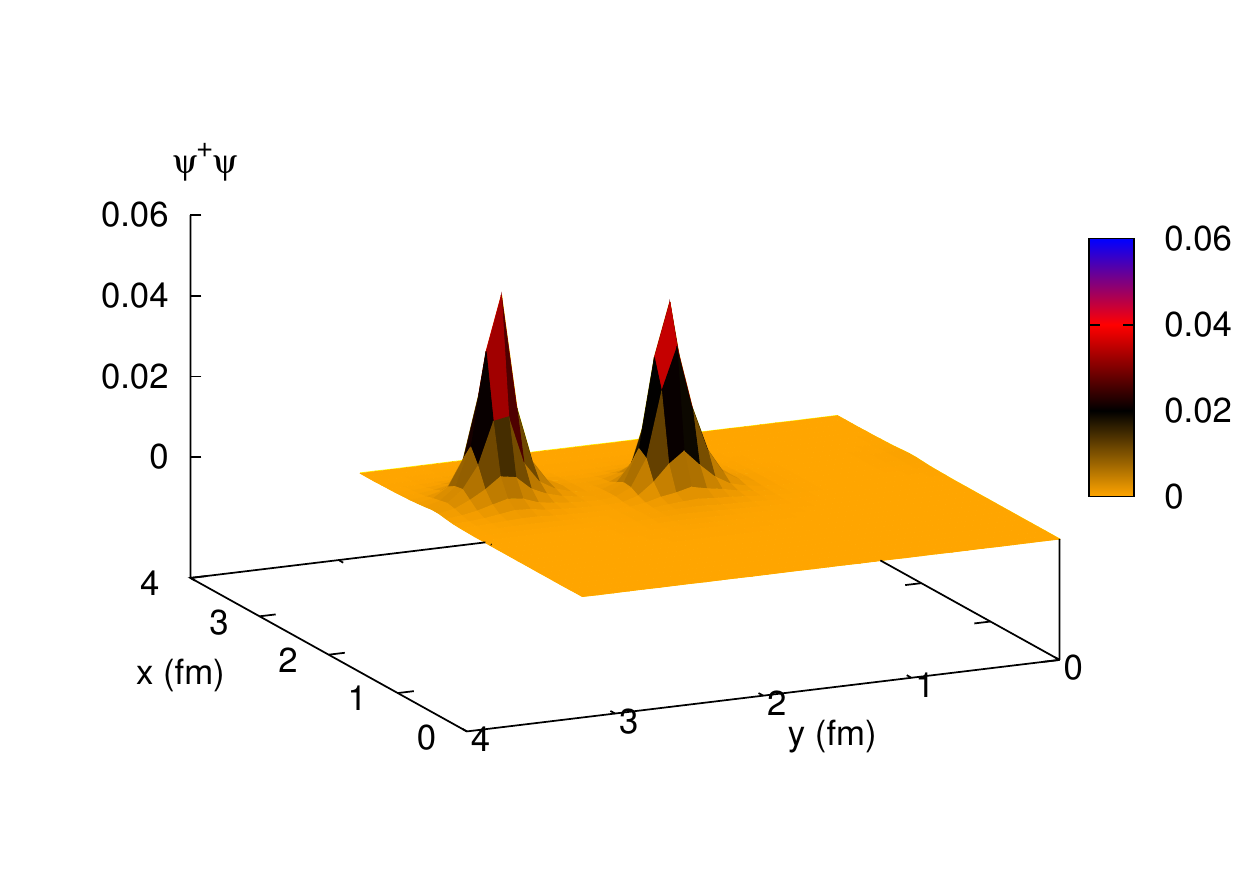}
\end{minipage}
\begin{minipage}[c]{0.47\textwidth}
\includegraphics[width=0.99\textwidth]{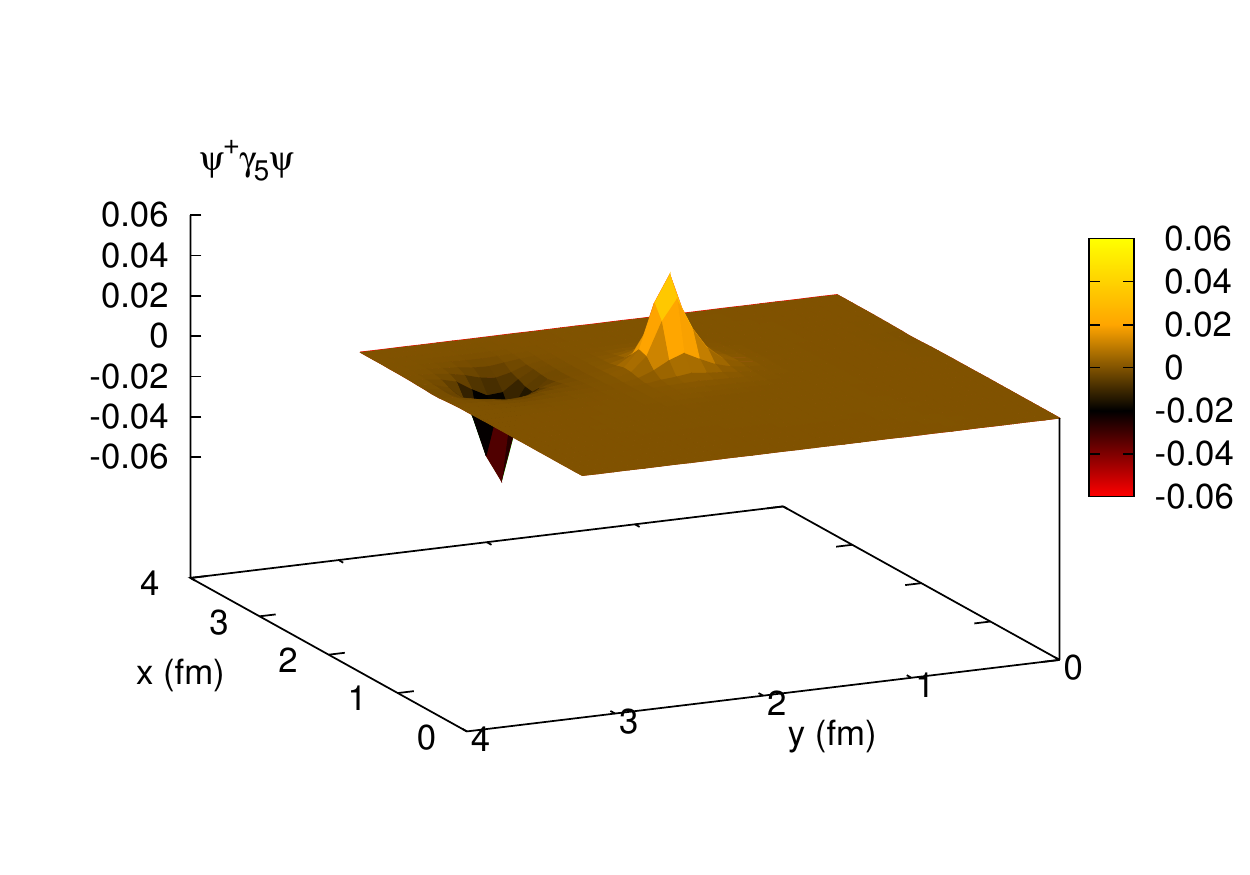}
\end{minipage}
\end{center}
\caption{(Top) Accumulation of near-zero eigenmodes of the Overlap-Dirac 
fermion matrix at $T=1.5T_c$ \cite{Dick:2015twa}. (Bottom) Spatial profile of a typical near-zero 
mode (left) and the spatial
profile of the chirality of the same near-zero mode (right) 
at $T=1.5T_c$ \cite{Dick:2015twa}.} 
\label{fig:ua1-2}
\end{figure}

Since $U_A(1)$ breaking arises due to topology of the gauge fields, it is intimately
related to the infrared modes of the Dirac fermions.  In the limit of infinite
volume, both the chiral order parameter, $\left\langle\bar\psi\psi\right\rangle_l$, and
the $U_A(1)$ breaking measure, $\chi_\pi-\chi_\delta$, can be represented in terms of
the eigenvalue density, $\rho(\lambda)$, of the Dirac fermions
\begin{equation}
\left\langle\bar\psi\psi\right\rangle_l = \int_{0}^{\infty} \hspace{-0.2cm}
d\lambda \,
\frac{2m_l\,\rho(\lambda)}{\lambda^2 + m_l^2} \;, \qquad\mathrm{and}\qquad
\chi_\pi - \chi_\delta = \int_{0}^{\infty} \hspace{-0.2cm} 
d\lambda\, \frac{4m_l^2\,\rho(\lambda)}
{\left( \lambda^2 + m_l^2 \right)^2} \;.
\end{equation}
In the chiral symmetric phase and in the chiral limit
$\left\langle\bar\psi\psi\right\rangle_l$ must vanish, but $\chi_\pi-\chi_\delta$ can
remain non-zero. Identification of the infrared fermionic modes, i.e. the form of
$\rho(\lambda)$, that give rise to such a phenomenon naturally leads to the underlying
non-perturbative mechanism of axial symmetry breaking. Studies with DWF
\cite{Bazavov:2012qja,Buchoff:2013nra} as well as with overlap fermions
\cite{Dick:2015twa}, possessing even better chiral properties and an exact index
theorem, suggest that an accumulation of near-zero eigenmodes of the form
$\rho(\lambda)\sim m_l^2\delta(\lambda)$ may largely account for the observed
$U_A(1)$ breaking in $\chi_\pi-\chi_\delta$ at high temperature. Such accumulation of
the near-zero modes are depicted in Fig.~\ref{fig:ua1-2} (top). More detailed
studies \cite{Dick:2015twa} of the space-time profiles, as illustrated in 
Fig.~\ref{fig:ua1-2} (bottom), localization properties and distributions of
these near-zero modes indicate that their behavior is consistent with underlying
presence of a dilute instanton gas--- a gas of widely separated, weakly interacting,
small instantons and anti-instantons. This suggests that even at temperatures
$T\sim1.5T_c$ weakly interacting instanton anti-instanton pairs are largely
responsible for the $U_A(1)$ breaking.

The analysis of $U_A(1)$ symmetry breaking close to the chiral limit
of 2 or (2+1)-flavor QCD, however, is far from being settled. Detailed
systematic studies of the quark mass and volume dependence as well as the
cut-off dependence are still missing. Moreover, results obtained with
chiral fermions are still controversial. Calculations performed with
dynamical overlap fermions \cite{Cossu:2013uua,Tomiya:2014mma} and
so-called optimal domain wall fermions \cite{Chiu:2013wwa} at present
suggest that $U_A(1)$ does get
restored already at $T_c$, which would still allow for the occurrence
of a first order transition when approaching the chiral limit.

\section{Bulk thermodynamics}
\label{sec:bulk}

\subsection{The QCD equation of state at vanishing chemical potential}

Lattice QCD calculations of 
the equation of state, or more general, of basic bulk thermodynamic
observables like the pressure ($P$), energy density ($\epsilon$)
or entropy density ($s$), have reached now a precision where continuum
extrapolated results for physical light and strange quark masses
can be obtained. This also allows detailed comparisons with perturbative
calculations at high temperature and model calculations at low 
temperature. 

The procedure to calculate bulk thermodynamic observables in lattice
QCD is well established. The starting point for these calculations 
is the evaluation of the trace anomaly, $\Theta^{\mu\mu}(T)\equiv \epsilon -3P$.
The trace anomaly is directly obtained as a temperature derivative of $P/T^4$
which also is related to the QCD partition function, $Z(T,V)$,
\begin{equation}
 \frac{\epsilon -3P}{T^4} = T\frac{{\rm d} P/T^4}{{\rm d} T} \; ,
\label{Theta2}
\end{equation}
with $P/T^4 = \lim_{V\rightarrow \infty} (VT^3)^{-1} \ln Z(T,V)$. 
As the trace anomaly arises
as a derivative of the logarithm of the QCD partition function with
respect to temperature, it can be expressed in terms of expectation
values of rather simple observables, e.g. a combination of the 
gauge action and light and strange quark chiral 
condensates. Nonetheless, the calculations become rather time consuming 
because a subtraction of the zero temperature contributions is
needed to eliminate divergent vacuum contributions\footnote{For more
details we refer, for instance, to Ref.~\citen{Karsch:2001cy}.}.

As can be seen from Eq.~\ref{Theta2} up to an integration
constant at a temperature $T_0$ the pressure can be obtained from the 
trace anomaly,
\begin{equation}
\frac{P(T)}{T^4} - \frac{P(T_0)}{T_0^4} = \int_{T_0}^{T}\ {\rm d}T'
\frac{\epsilon -3P}{T'^5} \; .
\label{pressure}
\end{equation}
This allows to reconstruct the energy density as well as the entropy
density $s/T^3=(\epsilon +P)/T^4$.

\begin{figure}[t]
\includegraphics[scale=0.6]{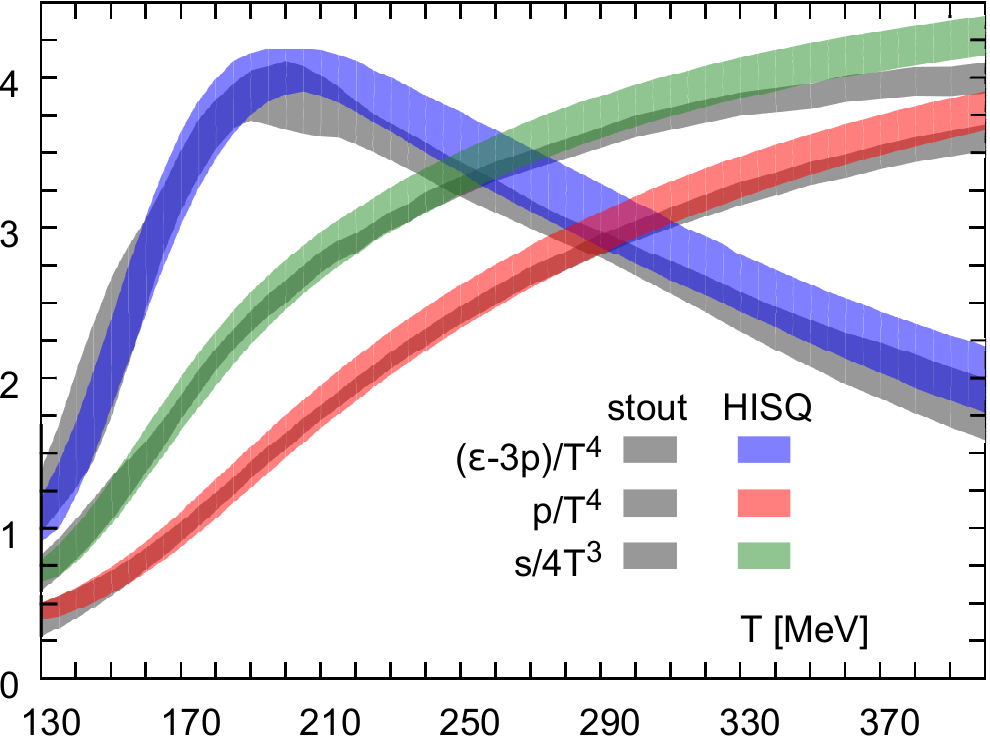}
\includegraphics[scale=0.6]{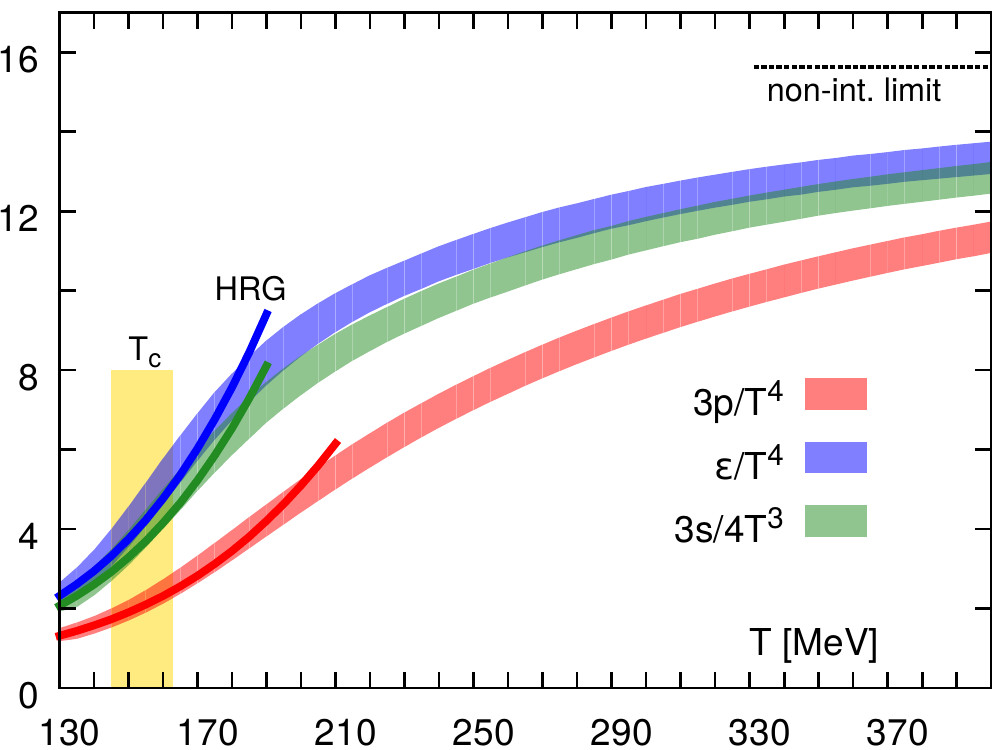}
\caption{(Left) Comparison of the trace anomaly 
$(\epsilon -3P)/T^4$, pressure and entropy density
calculated with the HISQ (colored) \cite{Bazavov:2014pvz} and 
stout (grey) \cite{Borsanyi:2013bia} discretization schemes 
for staggered fermions. (Right) Continuum extrapolated results for 
pressure, energy density and entropy density 
obtained with the HISQ action \cite{Bazavov:2014pvz}. Solid lines on the 
low temperature side correspond to results obtained from hadron resonance 
gas (HRG) model calculations. The dashed line at high temperatures shows the
result for a non-interacting quark-gluon gas.}
\label{fig:eos}
\end{figure}

The determination of thermodynamic quantities in QCD is a parameter free
calculation. All input parameters needed in the calculation, e.g. the
quark masses ($m_u=m_d,\ m_s$) and the relation between the lattice
cut-off, $a$, and the bare gauge coupling, $\beta = 6/g^2$, are determined
through calculations at zero temperature. Likewise, there is only a single
independent thermodynamic observable that is calculated in a lattice QCD
calculation, for instance the trace anomaly, $\Theta^{\mu\mu} (T)$.
All other bulk thermodynamic observables are obtained from 
$\Theta^{\mu\mu} (T)$ through standard thermodynamic relations. 
In Fig.~\ref{fig:eos}~(left) we show recent results for the trace anomaly
of (2+1)-flavor QCD \cite{Borsanyi:2013bia,Bazavov:2014pvz}
obtained with two different discretization schemes by two different groups.
The results are extrapolated to the continuum limit and are obtained 
with a strange quark mass tuned to its physical value and light
quark masses that differ slightly ($m_s/m_l = 27$ \cite{Borsanyi:2013bia}
and 20 \cite{Bazavov:2014pvz}). 
The right hand panel in this figure shows results for the pressure,
energy density and entropy density obtained from the trace anomaly by
using Eqs.~\ref{Theta2} and \ref{pressure}.

Also shown in Fig.~\ref{fig:eos} are results obtained from a 
hadron resonance gas (HRG) model calculation of bulk thermodynamics.
As can be seen this describes the QCD equation of state quite well
also in the transition region, although it may be noted that the HRG 
calculations yield
results for all observables that are at the lower error band of the
current QCD results. It has been speculated that this may indicate
contributions from additional, experimentally not yet observed resonances 
which could contribute to the thermodynamics \cite{Majumder:2010ik}. 
Indeed evidence for the contribution of a large number of strange
baryons has recently been found in lattice QCD calculations of
conserved charge fluctuations \cite{Bazavov:2014xya} (see also the discussion
in Section \ref{sec:fluctuation} and \ref{sec:hadrons}).

\begin{figure}[t]
\begin{center}
\includegraphics[scale=0.6]{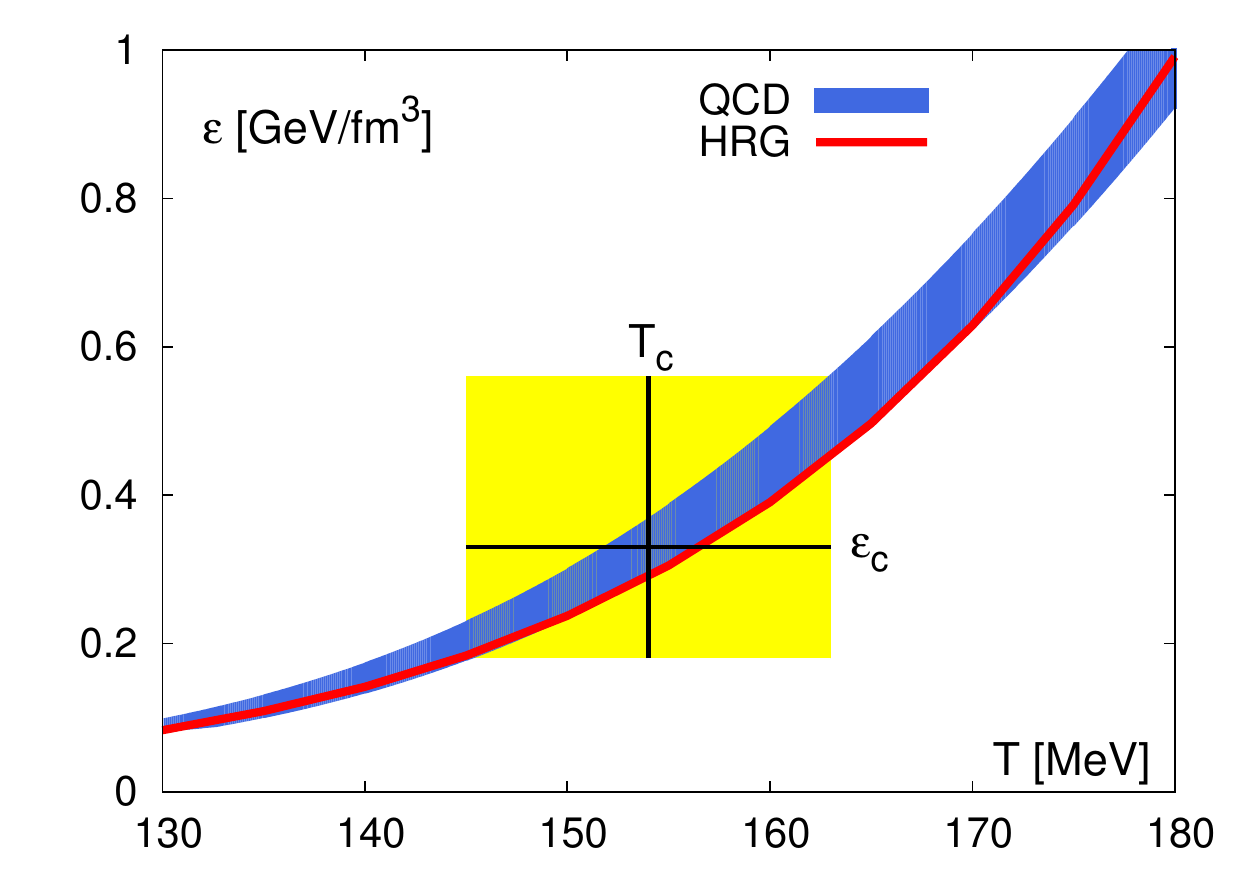}
\end{center}
\caption{The critical energy density $\epsilon_c$ in (2+1)-flavor QCD.
The band gives the continuum extrapolated result for the energy
density taken from Ref.~\citen{Bazavov:2014pvz}. The HRG curve is based
on all resonance with mass less than 2.5~GeV listed by the Particle Data 
Group \cite{Agashe:2014kda}.}
\label{fig:ec}
\end{figure}

In Fig.~\ref{fig:ec} we show the energy density in the low temperature 
region. The box highlights the transition region characterized
by the crossover temperature, $T_{c}=(154\pm 9)$~MeV. 
From this we deduce the energy density in the crossover 
region, $\epsilon_c = (0.34\pm0.16)~{\rm GeV/fm}^3$. This is a rather 
small value for the energy density needed to convert ordinary hadronic
matter into a medium made up from quarks and gluons. It may be compared
to the energy density of ordinary nuclear matter, 
$\epsilon^{\rm nuclear~matter} \simeq 0.15 ~{\rm GeV/fm}^3$ or the energy 
density inside a nucleon, $\epsilon^{\rm nucleon} \simeq 0.45 ~{\rm GeV/fm}^3$,
assuming the radius of nucleon $R_N\simeq 0.8$~fm. In fact, $\epsilon_c$ is close 
to the energy density reached in the dense packing limit of nucleons with 
radius $R_N$.

\begin{figure}[t]
\includegraphics[scale=0.55]{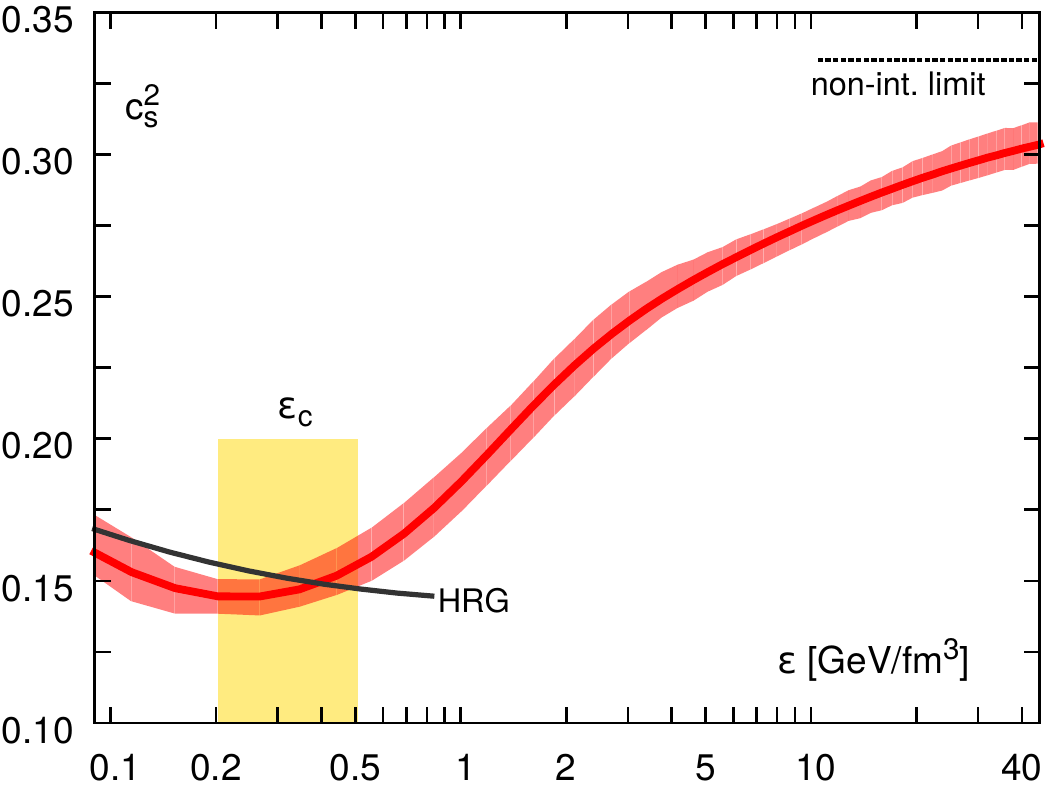}
\includegraphics[scale=0.52]{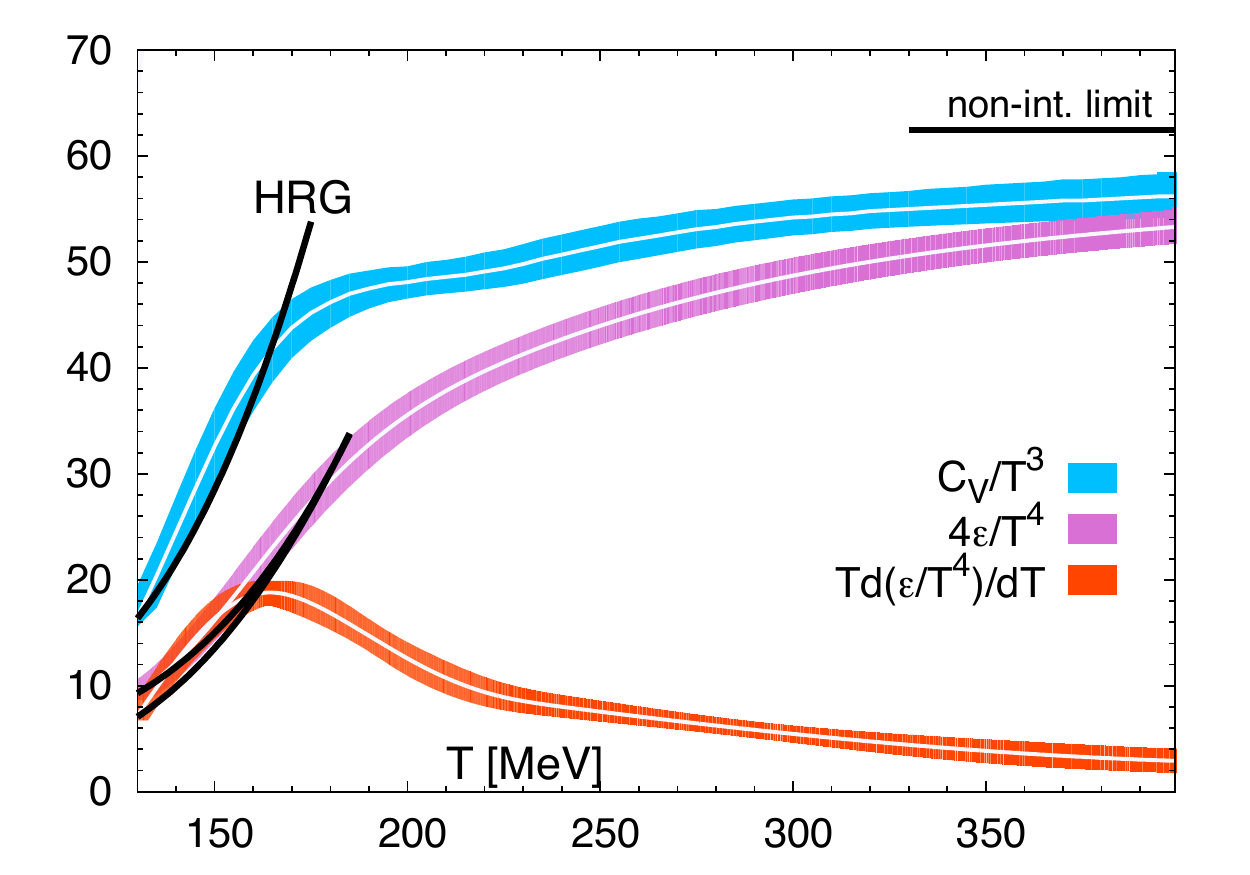}
\caption{The velocity of sound in (2+1)-flavor QCD (left) and the 
specific heat $C_V/T^4$ together with the two
components (see Eq.~\ref{CV}) contributing to it  (right). Solid black 
lines in the low
and high temperature regions show the corresponding hadron resonance gas
(HRG) and non-interacting quark-gluon gas results, respectively.
}
\label{fig:cs2}
\end{figure}

The simple bulk thermodynamic observables like pressure and energy density
give the impression that an HRG model also provides a good description of 
the equation of state for temperatures above the crossover region, i.e.
for $T\gtrsim 165$~MeV. However, as discussed in the previous section
the analysis of conserved charge fluctuations clearly shows that 
at temperatures $T\gtrsim 160$~MeV thermodynamics can no longer be described 
in terms of hadronic degrees of freedom (see for instance 
Fig.~\ref{fig:deconf1}~(right) and the discussion in Section \ref{sec:hadrons}). 
This also becomes evident in second order derivatives of the QCD
partition function with respect to temperature. The speed of sound, 
$c_s^2 =  {\rm dp}/{\rm d\epsilon}$,
is related to the inverse of the specific heat, 
$C_V= {\rm d}\epsilon /{\rm d}T$,
\begin{eqnarray}
c_s^2 &=& \frac{{\rm d} p}{{\rm d} \epsilon}  = 
\frac{{\rm d} p/{\rm d} T}{{\rm d} \epsilon/{\rm d} T} 
= \frac{s}{C_V} \; ,
\label{cs} 
\\
\frac{C_V}{T^3}&=&\left. \frac{\partial \epsilon}{\partial T}\right|_V \equiv \left( 
4 \frac{\epsilon}{T^4} + T 
\left. \frac{{\partial} (\epsilon/T^4)}{\partial T}\right|_V \right)
\, .
\label{CV}
\end{eqnarray}
Both quantities are shown in Fig.~\ref{fig:cs2}.
The specific heat does not develop a pronounced peak in the
transition region as one could have expected from pseudo-critical
behavior of energy density fluctuations close to a critical point. 
This may be understood \cite{Bazavov:2014pvz} from the temperature dependence
of the two terms contributing to $C_V/T^3$. The 
dominant singular contribution arises from the
temperature derivative of $\epsilon/T^4$, which has a peak. This, however,
is overwhelmed by the large energy density contribution
at high temperature which reflects the liberation of many partonic degrees
of freedom.
Furthermore, even in the chiral limit, where QCD is expected to have a 
second order phase transition belonging to the universality class
of 3-d, $O(4)$ symmetric spin models, 
the specific heat will not diverge as the relevant
critical exponent $\alpha\simeq -0.2$ that controls its singular behavior,
$C_V/T^3 \sim (|T-T_c|/T_c)^{-\alpha}+{\rm const.}$, 
is negative for this universality class 
(see Table~\ref{tab:parameter}).
The speed of sound will therefore stay non-zero at $T_c$ also in the chiral 
limit.

\subsection{The QCD equation of state at non-vanishing chemical potential}

In Section \ref{sec:phase} we have discussed lattice QCD results on the dependence
of the QCD crossover temperature on the baryon chemical potential and its
relation to the freeze-out temperatures determined in heavy ion
experiments. These experiments, in particular the beam energy scan program
performed at RHIC, will probe properties of strong-interaction
matter at non-vanishing baryon chemical potential in the temperature
range $0.9 \lesssim T/T_{c} \lesssim 2$ and $0\lesssim \mu_B/T\lesssim 3$, with $T_{c}$ 
denoting the crossover temperature at $\mu_B=0$. For the hydrodynamic
modeling of matter in this $(T,\mu_B)$ regime it thus is of importance
to also know the equation of state at non-vanishing $\mu_B/T$. As direct
numerical calculations at non-zero $\mu_B$ are not yet possible, a viable
approach is to analyze the equation of state using a Taylor
expansion in terms of chemical potentials \cite{Gavai:2001fr, Allton:2002zi}
as given in Eq.~\ref{TaylorBQS}. 
In this way some results for the EoS at non-zero baryon chemical potential 
have already been obtained on coarse lattices 
\cite{Allton:2002zi,Gavai:2003mf,Allton:2003vx,Ejiri:2005uv}.
Continuum extrapolated results for Taylor expansion coefficients of the 
pressure at ${\cal O} (\mu_f^2)$ have been obtained 
\cite{Borsanyi:2011sw,Bazavov:2012jq}. These have been used to
construct the EoS at ${\cal O} (\mu_B^2)$ \cite{Borsanyi:2012cr}  
which implements the strangeness neutrality constraint \cite{Ejiri:2005uv} 
($\langle n_S\rangle =0$) and the electric charge to baryon number relation 
($\langle n_Q\rangle = 0.4 \langle n_B\rangle$) suitable for 
conditions met in heavy ion collisions.

\begin{figure}[hpbt]
\includegraphics[scale=0.5]{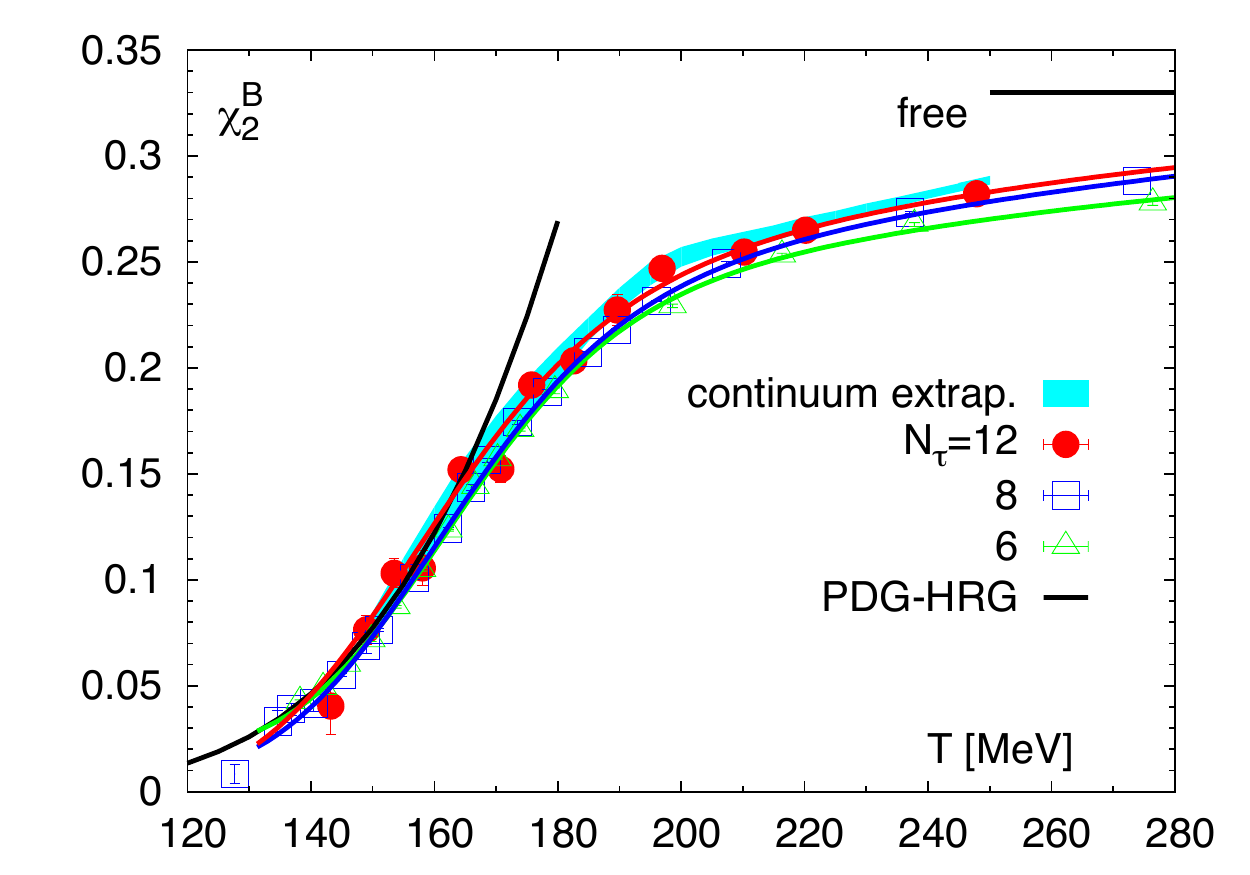}\includegraphics[scale=0.5]{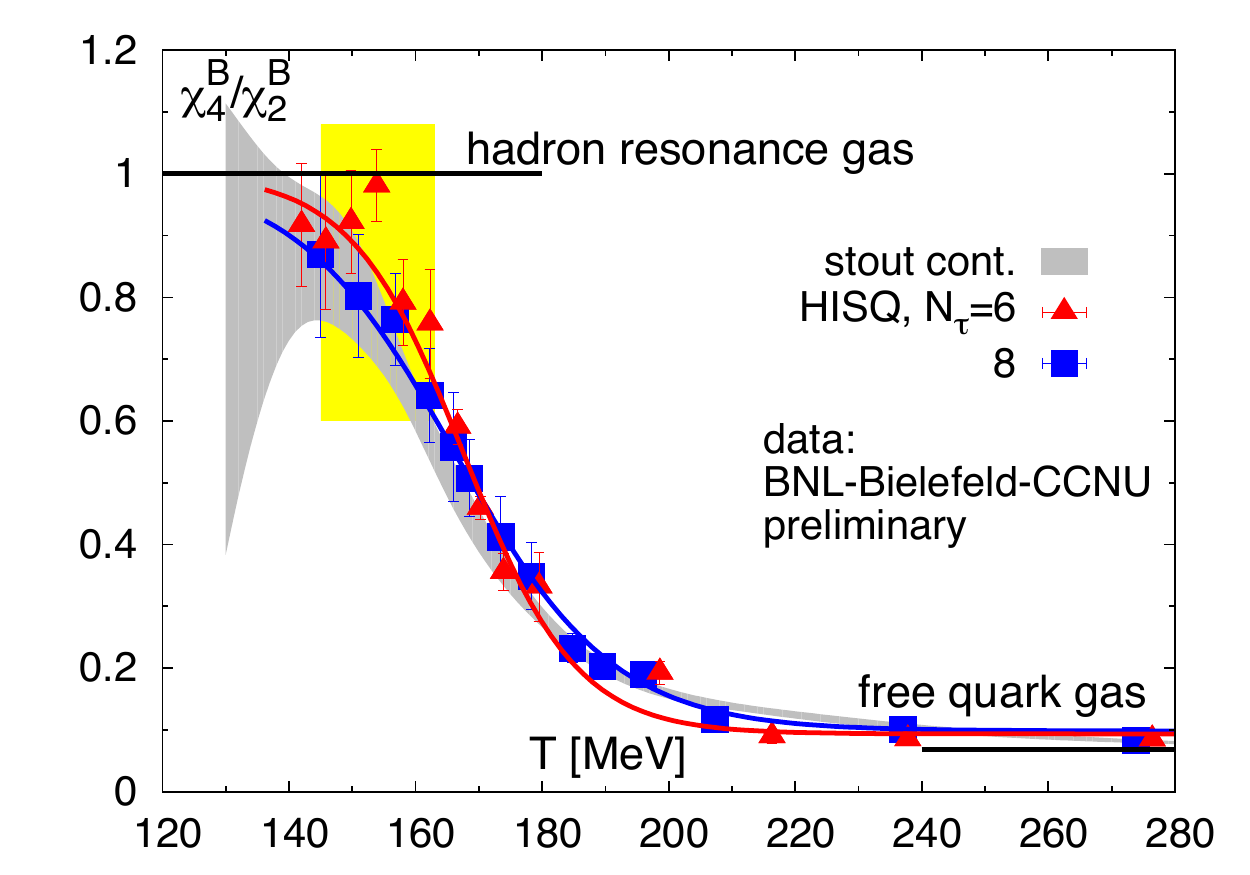}
\caption{Expansion coefficients of the pressure at non-zero baryon
chemical potential. The left hand figure shows the leading order
correction~\cite{Bazavov:2012jq} and the right hand figure shows the relative contribution
of the next to leading order correction. The continuum extrapolated result
obtained with the stout action is taken from Ref.~\citen{Borsanyi:2013hza}.}
\label{fig:Bcumulants}
\end{figure}

We will discuss here only the case $\mu_Q=\mu_S=0$ in some detail in order
to illustrate the relative importance of higher order corrections in 
different temperature and $\mu_B$ regions.
The Taylor series for the pressure is given by,
\begin{equation}
\frac{P(T,\mu_B)-P(T,0)}{T^4} =
\frac{1}{2} \chi_2^B(T) \left(\frac{\mu_B}{T} \right)^2 \left( 1 +
\frac{1}{12}\frac{\chi_4^B(T)}{\chi_2^B(T)}  
\left(\frac{\mu_B}{T} \right)^2 \right) + {\cal O}(\mu_B^6) 
\; .
\label{PTaylor}
\end{equation}
The leading order correction to the pressure thus is proportional
to the quadratic fluctuations of net baryon number. The next to leading
order corrections are proportional to the quartic fluctuations. When written
in the form given in Eq.~\ref{PTaylor} it becomes easy to identify
the relative importance of leading and next to leading order corrections. 
In Fig.~\ref{fig:Bcumulants} we show $\chi_2^B(T)$ (left) and 
$\chi_2^B(T)/\chi_4^B(T)$ (right). With increasing temperature the 
${\cal O}(\mu_B^4)$ correction rapidly looses importance relative to
the leading ${\cal O}(\mu_B^2)$ term.

\begin{figure}[hpbt]
\includegraphics[scale=0.5]{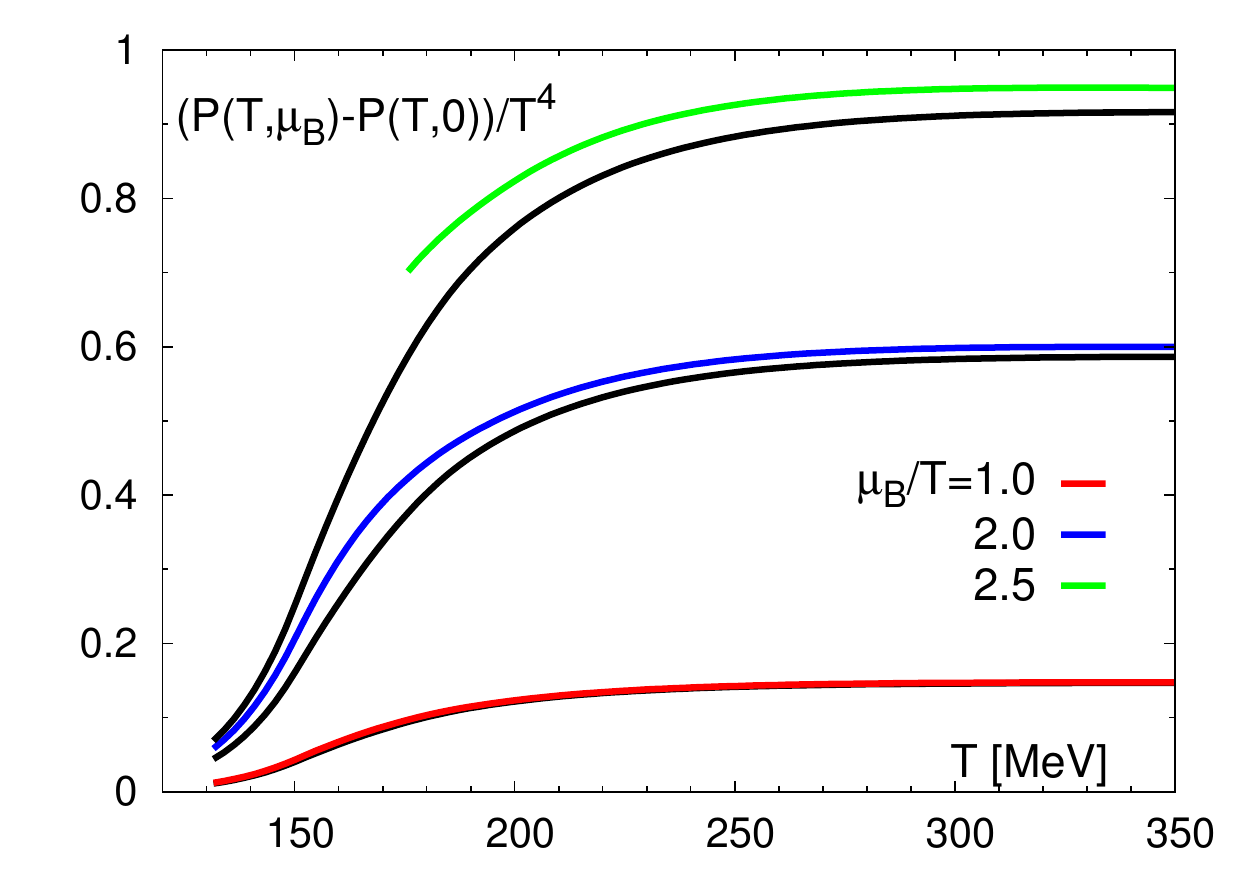}\includegraphics[scale=0.5]{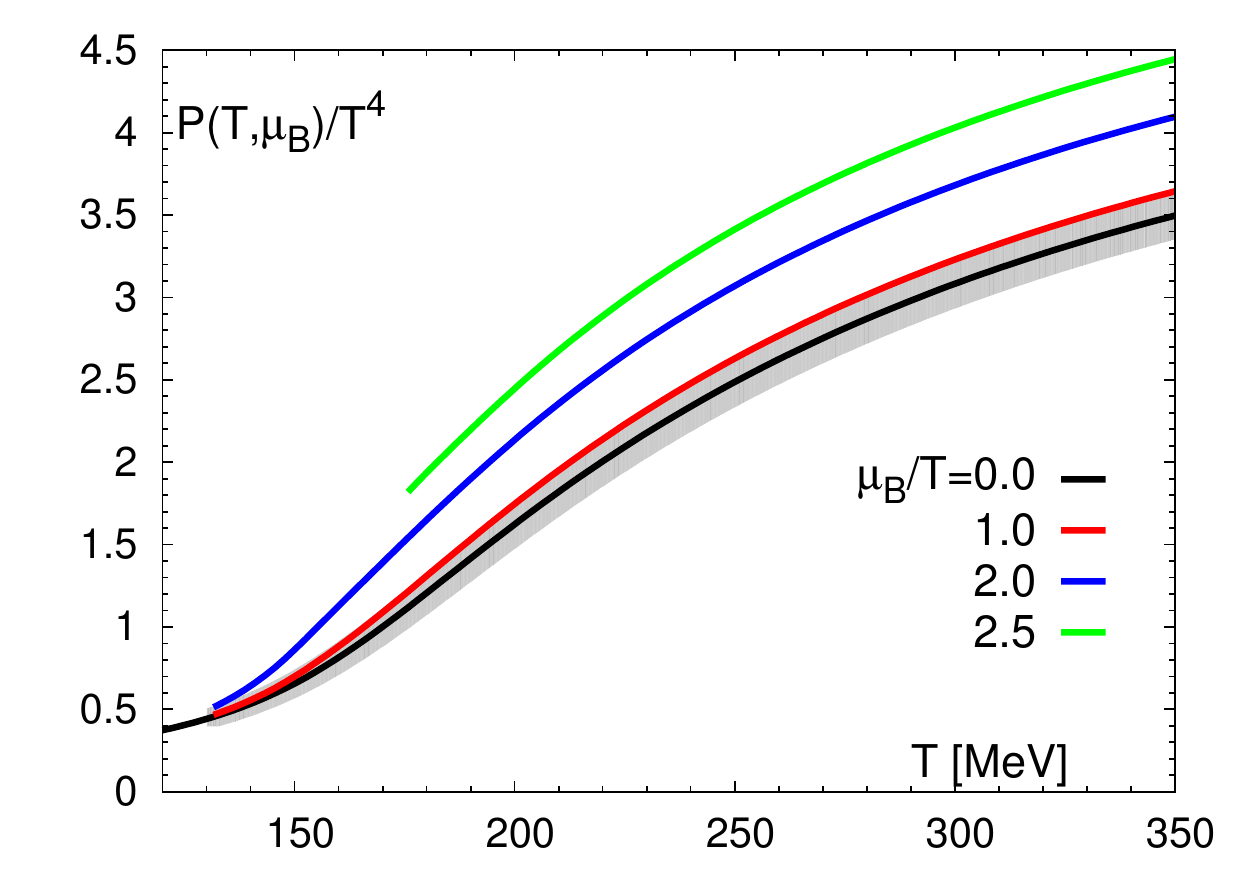}
\caption{(Left) The $\mu_B$-dependent part of the pressure at
${\cal O}((\mu_B/T)^2)$ (black) and ${\cal O}((\mu_B/T)^4)$ (colored) \cite{Hegde:2014wga}.
The latter is shown only in the temperature regime where neglected
corrections at ${\cal O}((\mu_B/T)^6)$  are estimated to contribute
less than 10\%. (Right) When combined with the $\mu_B=0$ contribution for the
pressure shown
in Fig.~\ref{fig:eos} these neglected terms contribute less
than 3\% to the total pressure \cite{Hegde:2014wga}. The grey band gives the uncertainty on
$P(T,0)/T^4$ and the central line in the band is the parametrization
of $P(T,0)/T^4$ given in Ref.~\citen{Bazavov:2014pvz}.
}
\label{fig:Bpressure}
\end{figure}

In Fig.~\ref{fig:Bpressure} we show preliminary results for the $\mu_B$-dependent 
contribution to the total pressure evaluated for different values of
$\mu_B/T$ and taking into account corrections up to 
${\cal O}((\mu_B/T)^4)$~\cite{Hegde:2014wga}. These results suggests that an 
${\cal O}((\mu_B/T)^4)$ Taylor expansion of the pressure (and energy
density) is well controlled for all values of the chemical potential
below $\mu_B/T =2$. This covers a wide range of the QCD phase 
diagram accessible in the beam energy scan (BES) at RHIC, i.e. the region
of beam energies $\sqrt{s}\ge 20$~GeV.

\subsection{Perturbation theory, hadron resonance gas and the
strongly interacting liquid}

In our discussion of bulk thermodynamic observables we have 
compared with HRG model calculations. We have seen that the 
HRG does provide a rather good description of bulk thermodynamics 
below and even in the crossover region. However, the HRG does depend
on the resonance spectrum used in the calculation. This
is of some significance when one looks at more selective observables
like, e.g., fluctuations and correlations of strange baryons (see Sections
\ref{sec:fluctuation} and \ref{sec:hadrons}). 
Spectrum independent HRG results are, for instance, related to
fluctuations of net baryon number. The ratio of fourth to second 
order net-baryon number cumulants shown in Fig.~\ref{fig:Bcumulants}~(right)
is unity in any HRG model irrespective of details of the spectrum.
This figure thus shows that an HRG model
calculation can be expected to provide a good description of QCD 
thermodynamics on a 10\% level only at temperatures less than about 
$T\simeq (140-145)$~MeV. 

\begin{figure}[t]
\begin{center}
\includegraphics[scale=0.55]{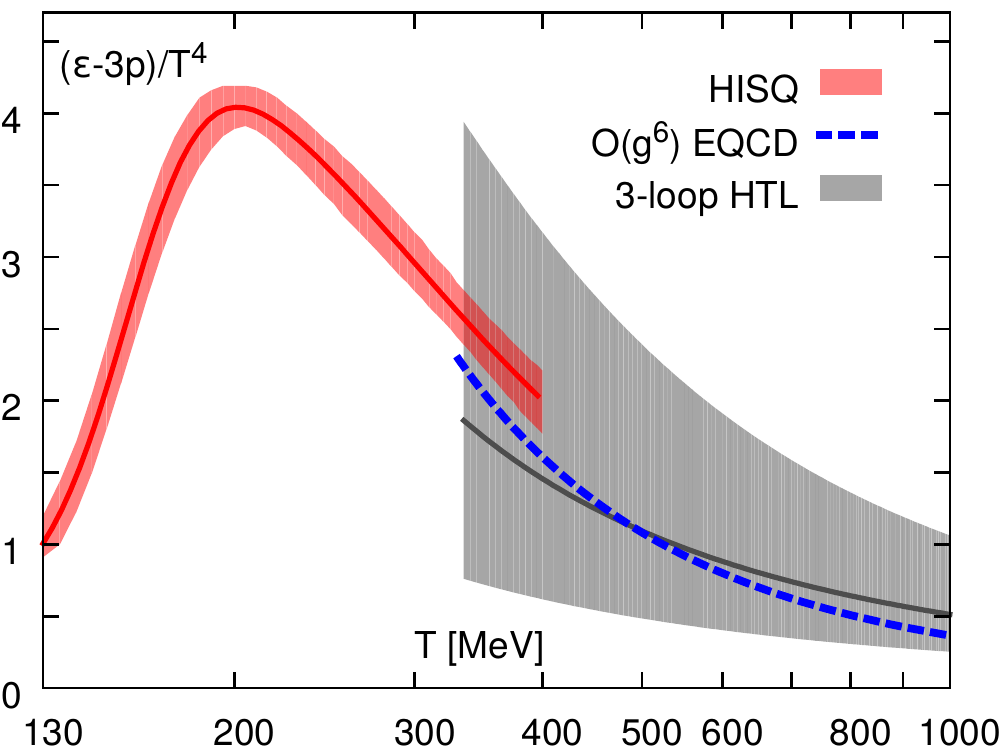}\hspace{0.3cm}\includegraphics[scale=0.55]{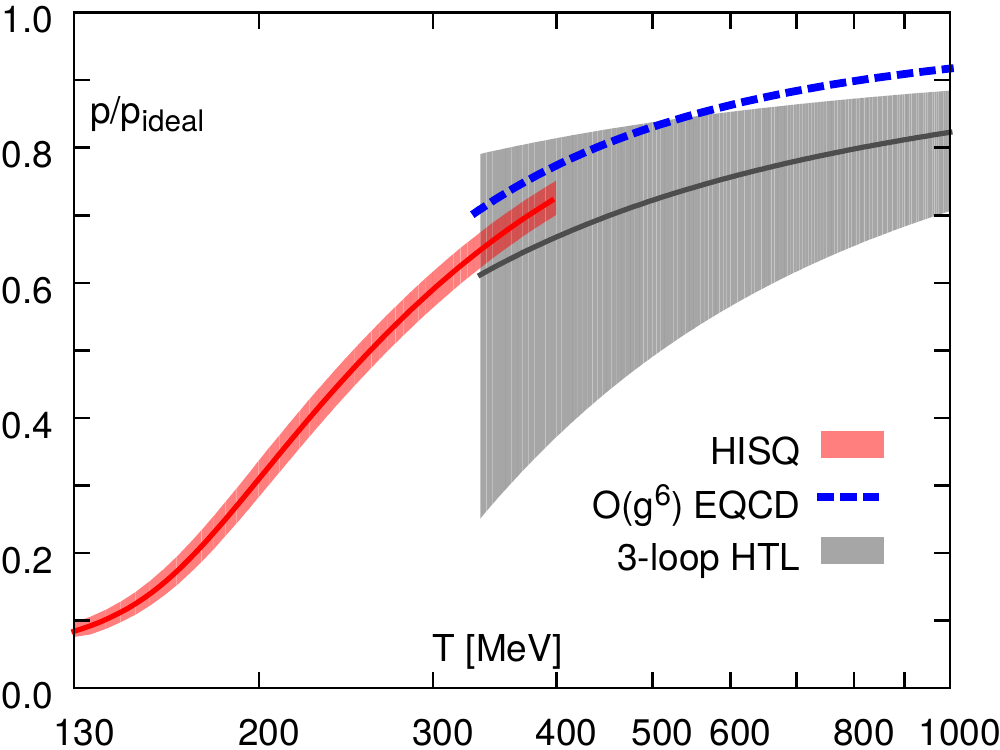}
\end{center}
\caption{Comparison of the (2+1)-flavor calculation \cite{Bazavov:2014pvz} of 
the trace anomaly (left) and pressure (right) with HTL and EQCD (dashed line)
calculations.  The black line corresponds to the HTL calculation 
\cite{Haque:2014rua} with renormalization scale $\mu=2 \pi T$.  
Note that this solid line would move up for the trace anomaly and move down 
for the pressure if the scale $\mu$ in HTL is reduced. }
\label{fig:HTLpressure}
\end{figure}

\begin{figure}[t]
\begin{center}
\includegraphics[width=0.52\textwidth]{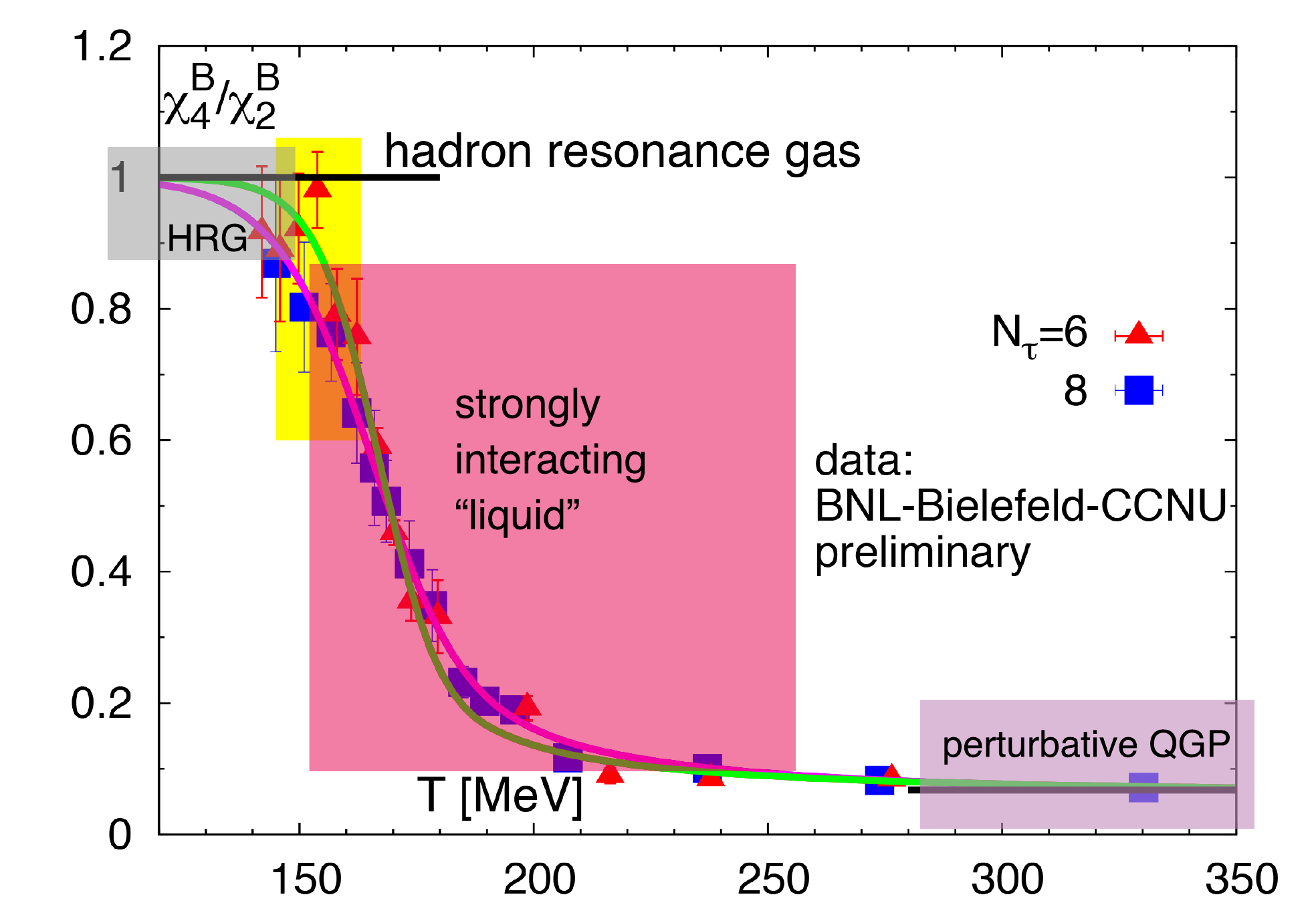}~\includegraphics[width=0.44\textwidth]{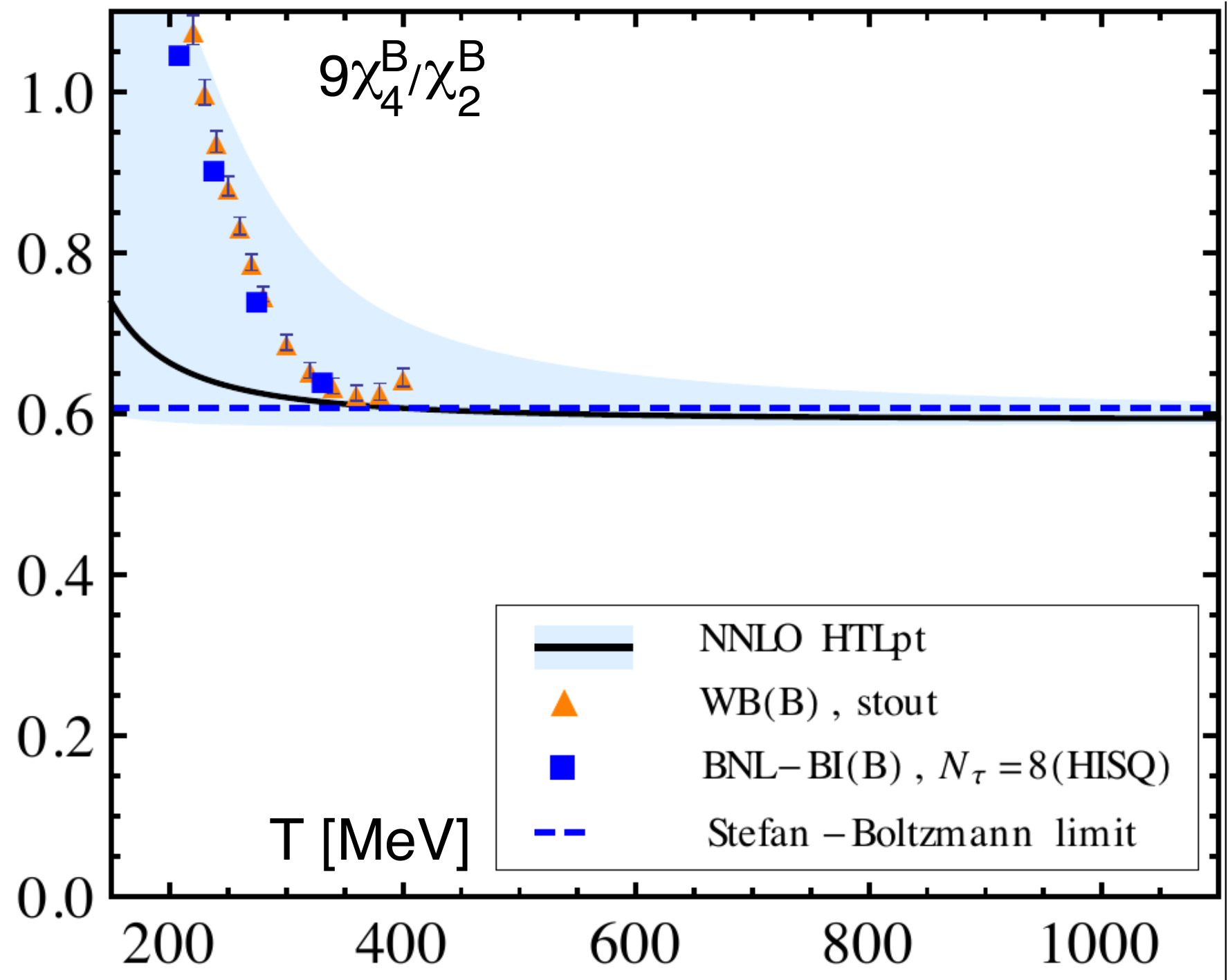}
\end{center}
\caption{The ratio of quartic and quadratic net-baryon number fluctuations
versus temperature. The left hand panel shows temperature ranges in which
HRG and resummed perturbative calculations, respectively, provide good
approximations to lattice QCD results. The right hand panel shows the
result from a HTL-resummed calculations \cite{Haque:2014rua}.}
\label{fig:B4B2comparison}
\end{figure}

At asymptotically high temperatures hard thermal loop, perturbative 
calculations \cite{Haque:2014rua} or dimensionally reduced QCD (EQCD)
\cite{Laine:2006cp} provide a well-established framework for the
analysis of bulk thermodynamics and fluctuations of conserved 
charges. We show in Fig.~\ref{fig:HTLpressure} a comparison 
of the trace anomaly (left) and the pressure (right) calculated in lattice QCD
with HTL and EQCD calculations. These calculations seem to give a good 
description of bulk thermodynamics at temperatures $T\gtrsim 400$~MeV.
However, even in this regime there remain differences between HTL 
and EQCD calculations that are of the order of 10\%. Moreover, within
the HTL calculations there seems to be no unique choice for the
renormalization scale $\mu$, that would allow to match all observables
simultaneously. Similar conclusions can be drawn from the HTL analysis
of net-baryon number fluctuations shown in 
Fig.~\ref{fig:B4B2comparison}~(right).
Although observables that are dominated at high temperature by quark 
rather than gluon contributions seem to approach perturbative 
behavior earlier, it still is evident that 
agreement with lattice QCD results on the 10\% level 
only is possible for $T\gtrsim (250-300)$~MeV. In general the temperature 
range $T_{c} \le T\le 2\ T_{c}$ is highly non-perturbative and obviously not 
accessible to hadronic model calculations. This is highlighted in the
left hand panel of  Fig.~\ref{fig:B4B2comparison}. We will discuss in
the following sections properties of strong-interaction matter in this
temperature range.

\section{Fluctuations of conserved charges}
\label{sec:fluctuation}

Proximity of a second order criticality, such as the $O(4)$ chiral phase transition
or the QCD critical point, is universally manifested through long-range correlations
at all length scales, resulting in increased fluctuations of the order parameter.
These fluctuations can be quantified through the Gaussian (variance) as well as
non-Gaussian (skewness, kurtosis {\sl etc.}) cumulants of the distribution of the
order parameter. Higher order non-Gaussian cumulants become increasingly 
sensitive to proximity of a critical point \cite{Ejiri:2005wq} which is
reflected in their grows with higher powers of the
correlation length \cite{Stephanov:2008qz}. 
Moreover, even qualitative features, such as the sign
change and the associated non-monotonicity, of these non-Gaussian cumulants can
encode the presence of a nearby critical region
\cite{Asakawa:2009aj,Friman:2011pf,Stephanov:2011pb}. Non-Gaussian 
cumulants 
can be accessed in heavy-ion experiments via the event-by-event
fluctuations of various conserved charges and particle multiplicities
\cite{Stephanov:1998dy,Stephanov:1999zu,Jeon:2003gk,Hatta:2003wn,Athanasiou:2010kw}.
In this vein, a major focus of the Beam Energy Scan (BES) program \cite{STAR-wp} at
the RHIC is measurements of the event-by-event fluctuations of particle
multiplicities and conserved charges
\cite{Aggarwal:2010wy,Adamczyk:2013dal,Adamczyk:2014fia,Sahoo:2014bqa,Luo:2015ewa,Mitchell:2012mx}.

Among several conserved charges, the net electric charge is of special 
interest. Cumulants of net electric charge fluctuations can be measured in 
experiments \cite{Adamczyk:2014fia,Mitchell:2012mx} and are 
calculable in lattice QCD \cite{Bazavov:2012jq,Borsanyi:2011sw}. 
Cumulants of net electric charge fluctuations and correlations with other
conserved charges are well defined in lattice QCD at vanishing chemical 
potentials where standard lattice QCD techniques can be used to compute them.  
Furthermore, Taylor expansions in powers of $\mu_B$ around $\mu_B=0$, can be 
employed to obtain generalized  susceptibilities for $\mu_B>0$, 
\begin{equation}
\chi^Q_n \left( T, \mu_B \right) = \sum_{k=0}^\infty \frac{1}{k!} \: 
\chi^{BQ}_{kn}(T) \left( \frac{\mu_B}{T} \right)^k
\;.
\end{equation}
These susceptibilities directly relate to cumulants of 
the net electric charge fluctuations,
\begin{eqnarray}
&& \chi^Q_1 \left( T, \mu_B \right) = \frac{1}{VT^3} \left\langle N_Q \right\rangle
\;, \nonumber \\ 
&& \chi^Q_2 \left( T, \mu_B \right) = \frac{1}{VT^3} 
\left\langle \left( \delta N_Q \right)^2 \right\rangle
\;, \nonumber \\
&& \chi^Q_3 \left( T, \mu_B \right) = \frac{1}{VT^3} 
\left\langle \left( \delta N_Q \right)^3 \right\rangle
\;, \nonumber \\
&& \chi^Q_4 \left( T, \mu_B \right) = \frac{1}{VT^3} \left[
\left\langle \left( \delta N_Q \right)^4 \right\rangle - 3 \left\langle \left( \delta
N_Q \right)^2 \right\rangle^2 \right]
\;,
\end{eqnarray}
where $N_Q$ is the net (positive minus negative) charge and $\delta N_Q = N_Q -
\left\langle N_Q \right\rangle$.  

On the other hand, through the measurements of the event-by-event distributions of
the net electric charge, heavy-ion experiments provide various cumulants, 
mean ($M_Q$), variance ($\sigma_Q$), skewness ($S_Q$), and kurtosis
($\kappa_Q$), of the electric charge fluctuations for given beam energy ($\sqrt{s}$)
\cite{Adamczyk:2014fia}
\begin{eqnarray}
M_Q \left( \sqrt{s} \right) = \left\langle N_Q \right\rangle
\;, &\qquad\qquad&  
\sigma_Q^2 \left( \sqrt{s} \right) = \left\langle \left( \delta N_Q \right)^2 \right\rangle
\;, \nonumber \\ 
S_Q \left( \sqrt{s} \right) = \frac {\left\langle \left( \delta N_Q \right)^3 \right\rangle} {\sigma_Q^3}
\;, &\qquad\qquad& 
\kappa_Q \left( \sqrt{s} \right) = 
\frac {\left\langle \left( \delta N_Q \right)^4 \right\rangle} {\sigma_Q^4} - 3
\;.
\end{eqnarray}
Thus, the charge susceptibilities obtained from lattice QCD calculations and the
cumulants measured in the heavy-ion experiments are directly related to each other
through the  appropriate volume-independent ratios \cite{Bazavov:2012vg}
\begin{subequations}
\begin{eqnarray}
\frac {M_Q \left( \sqrt{s} \right)} {\sigma_Q^2 \left( \sqrt{s} \right)} &=& 
\frac {\chi^Q_1 \left( T, \mu_B \right)} {\chi^Q_2 \left( T, \mu_B \right)} 
\equiv R_{12}^Q \;, \label{eq:R12Q} \\
\frac {S_Q \left( \sqrt{s} \right) \sigma_Q^3 \left( \sqrt{s} \right)}
{M_Q \left( \sqrt{s} \right)} &=& 
\frac {\chi^Q_3 \left( T, \mu_B \right)} {\chi^Q_1 \left( T, \mu_B \right)} 
\equiv R_{31}^Q \;. \label{eq:R31Q}
\end{eqnarray}
\end{subequations}
As in many other cases, one still has to worry about 
directly confronting thermodynamic QCD calculations performed in
equilibrium and in the thermodynamic limit to the rapidly expanding
dense matter created in heavy ion collisions in a small volume.
In the analysis of higher order cumulants additional complications 
may arise from more complicated efficiency corrections and the 
limited experimental acceptance windows
\cite{Kitazawa:2012at,Bzdak:2012ab,Bzdak:2013pha}.  

\begin{figure}[t]
\begin{center}
\begin{minipage}[c]{0.42\textwidth}
\includegraphics[width=0.99\textwidth]{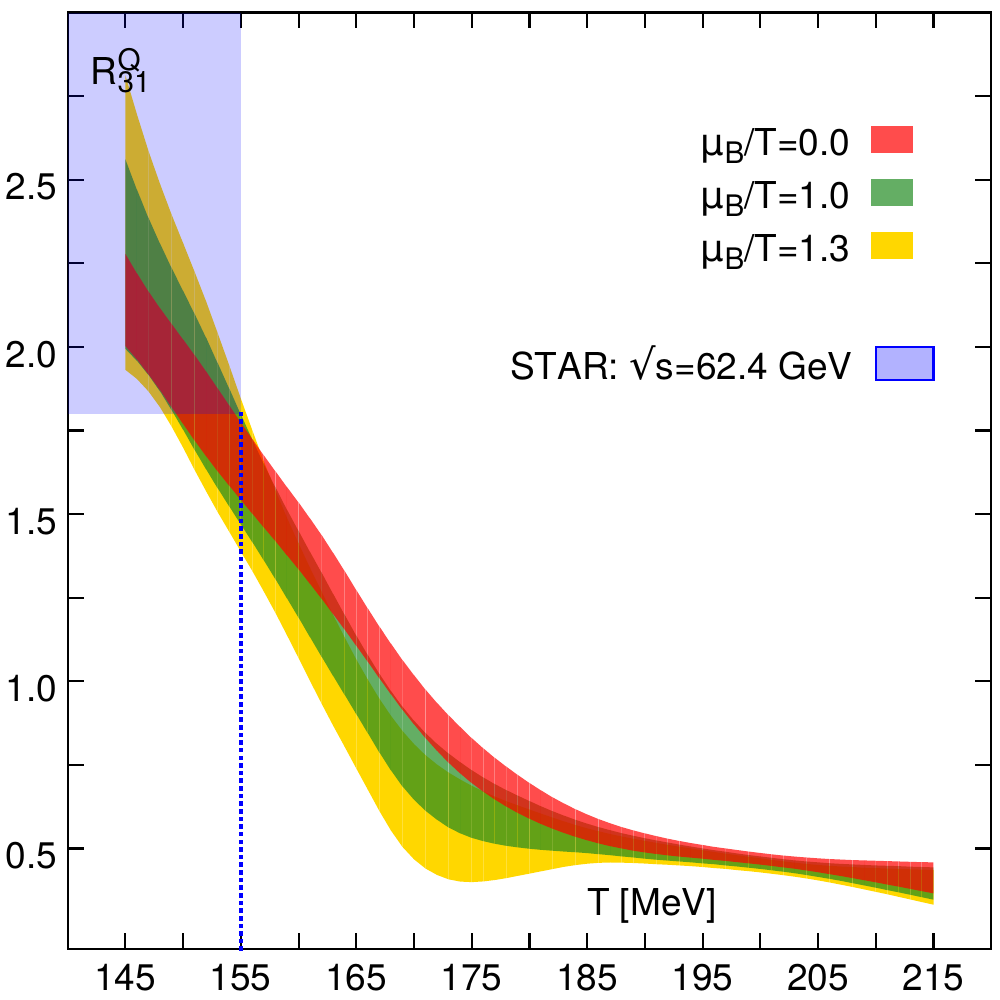}
\end{minipage}
\begin{minipage}[c]{0.56\textwidth}
\includegraphics[width=0.99\textwidth]{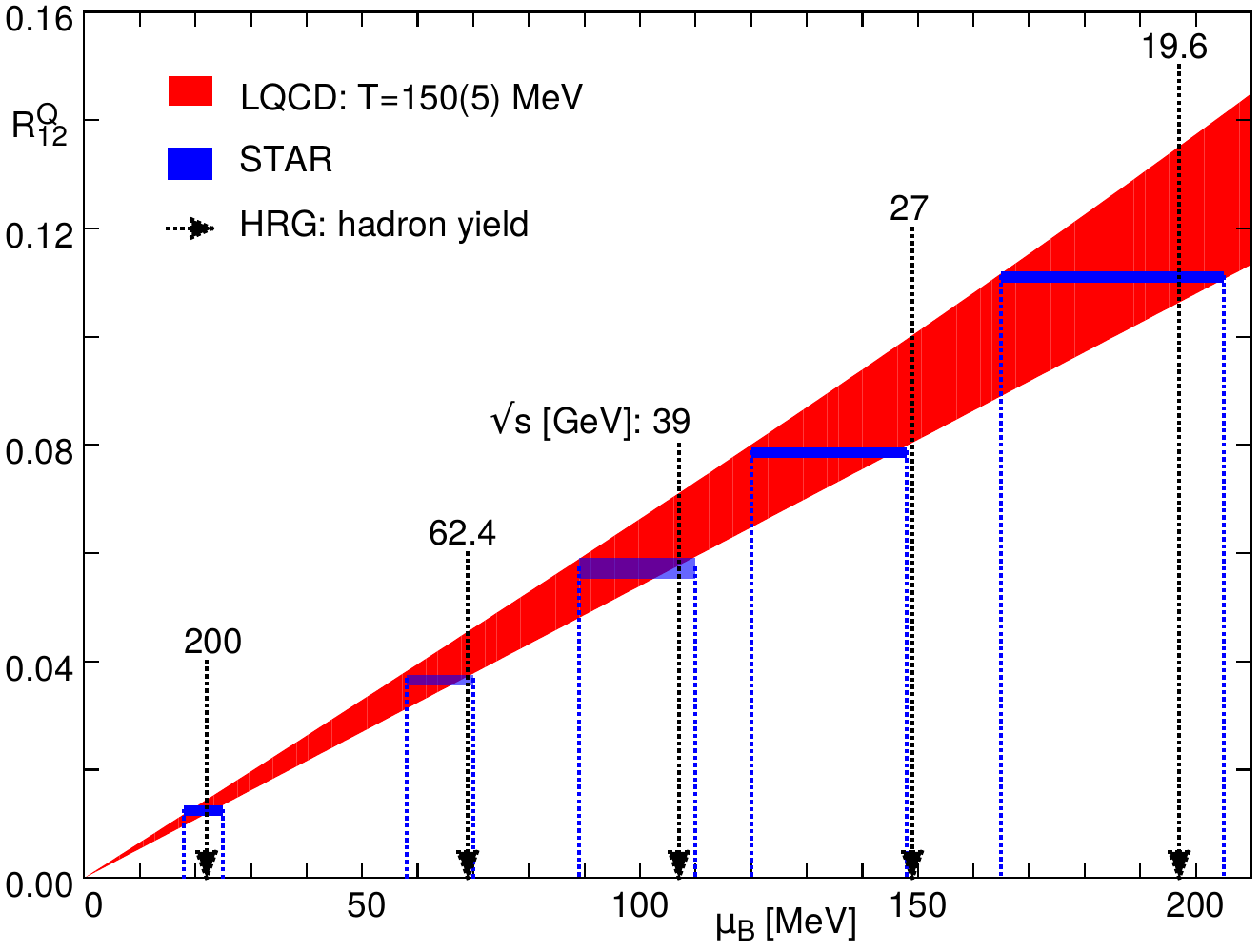}
\end{minipage}
\caption{(Left) A comparison between the lattice QCD results \cite{MukherjeeCPOD14}
for $R_{31}^Q$ and the STAR data \cite{Adamczyk:2014fia} for $(S_Q \sigma_Q^3)/M_Q$
at $\sqrt{s}=62.4$~GeV. The overlap of the experimental results with the lattice QCD
calculations provides an upper bound on the freeze-out temperature $T^f\le155$~MeV.
(Right) Lattice QCD results \cite{MukherjeeCPOD14} for $R_{12}^Q$ as a function of
$\mu_B$ compared with the STAR data \cite{Adamczyk:2014fia} for $M_Q/\sigma_Q^2$ in
the temperature range $T^f=150(5)$~MeV. The overlap regions of the experimentally
measured results with the lattice QCD calculations provide estimates for the
freeze-out chemical potential $\mu_B^f$ for a given $\sqrt{s}$. The arrows indicate
the corresponding values of $\mu_B^f$ obtained from the statistical hadronization
model fits to experimentally measured hadron yields \cite{Andronic:2011yq}.}
\label{fig:fo_lqcd}
\end{center}
\end{figure}

In heavy-ion collision experiments the only tunable parameter is the beam energy,
$\sqrt{s}$. However, to gain access to the information regarding the QCD phase
diagram this only tunable parameter needs to be related to the thermodynamic
variables, e.g. temperature and baryon chemical potential. Traditionally, this
$\sqrt{s}\leftrightarrow(T,\mu_B)$ mapping has been done by relying on the
statistical hadronization model based analysis \cite{Andronic:2011yq}. Recent
advances in heavy-ion experiments as well as in lattice QCD calculations have 
placed us in a unique situation where, for the first time, this mapping can now be
obtained through direct comparisons between the experimental results and rigorous
(lattice) QCD calculations. Recently, it has been shown 
\cite{Karsch:2012wm,Bazavov:2012vg} that by
directly comparing lattice QCD calculations for $R_{31}^Q$ (Eq.~\ref{eq:R31Q}) and
$R_{12}^Q$ (Eq.~\ref{eq:R12Q}) with their corresponding cumulant ratios measured
in heavy-ion experiments it is possible to extract the thermal parameters, namely the
freeze-out temperature, $T^f$, and the freeze-out baryon chemical potential
$\mu_B^f$.  The feasibility of such a procedure has been demonstrated in
Refs.~\citen{Mukherjee:2013lsa,Borsanyi:2013hza,Borsanyi:2014ewa}. 
Fig.~\ref{fig:fo_lqcd}
\cite{MukherjeeCPOD14} illustrates a recent example of such a comparison and
subsequent determination of the freeze-out parameters.

As can be seen from Fig.~\ref{fig:fo_lqcd}, due to large errors on the
experimental results for $(S_Q \sigma_Q^3)/M_Q$, at present, only an upper limit on
the freeze-out temperature can be determined using the method described above. Thus, a
complementary procedure for the determination of $T^f$, relying on a separate observable
that can be extracted both from heavy-ion experiments and lattice QCD calculations,
is certainly welcome. Recently, such a complementary procedure for the determination of
$T^f$ has been proposed in Ref.~\citen{Bazavov:2014xya}. This procedure takes
advantage of the fact that the initially colliding nuclei in heavy-ion collisions are
free of net strangeness. Thus, the conservation of strangeness under strong
interaction ensures that the QGP medium created during the collisions of these
heavy-ions is also strangeness neutral.  

\begin{figure}[t]
\begin{center}
\begin{minipage}[c]{0.49\textwidth}
\includegraphics[width=0.99\textwidth,height=0.25\textheight]{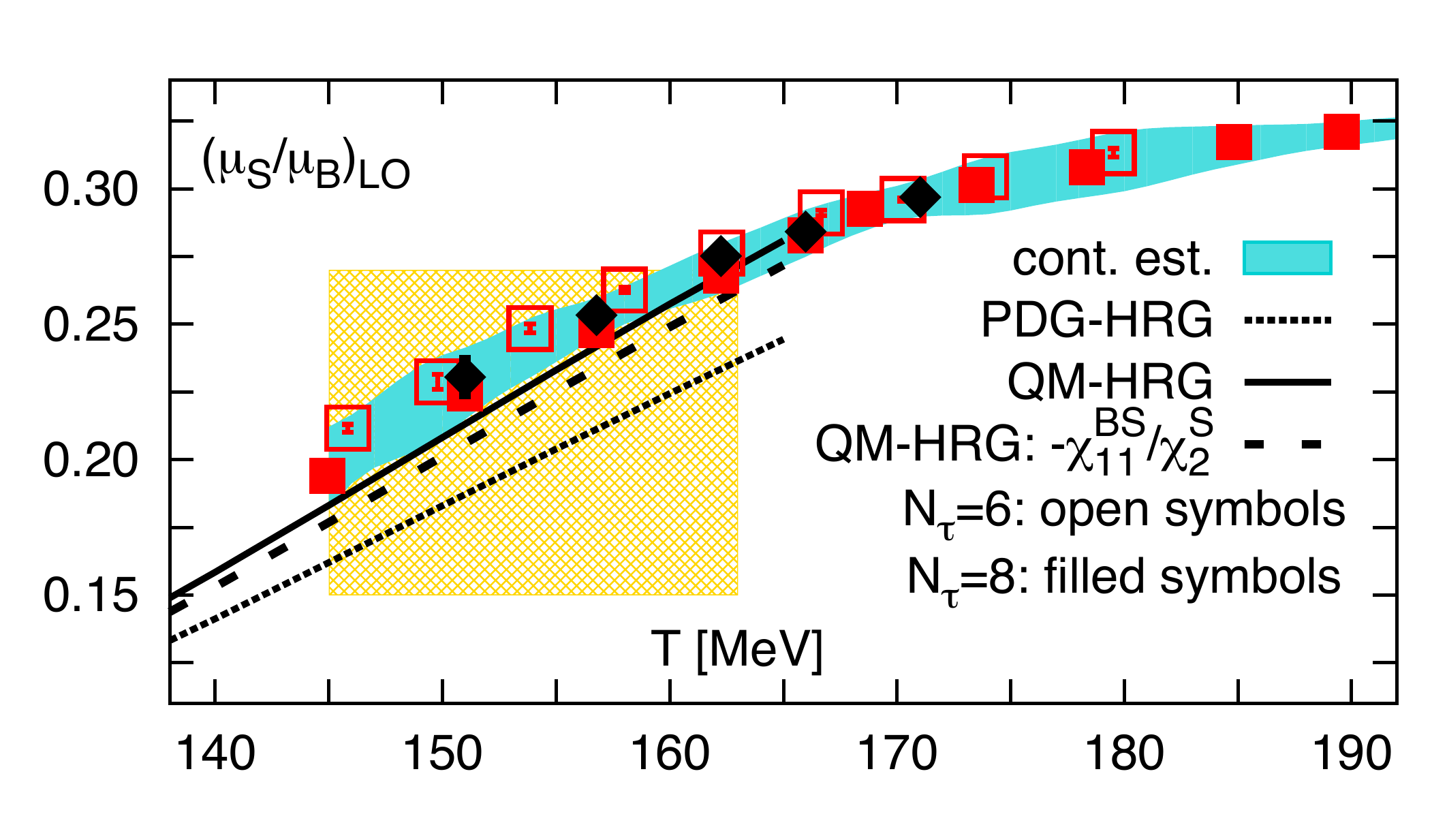}
\end{minipage}
\begin{minipage}[c]{0.49\textwidth}
\includegraphics[width=0.99\textwidth,height=0.25\textheight]{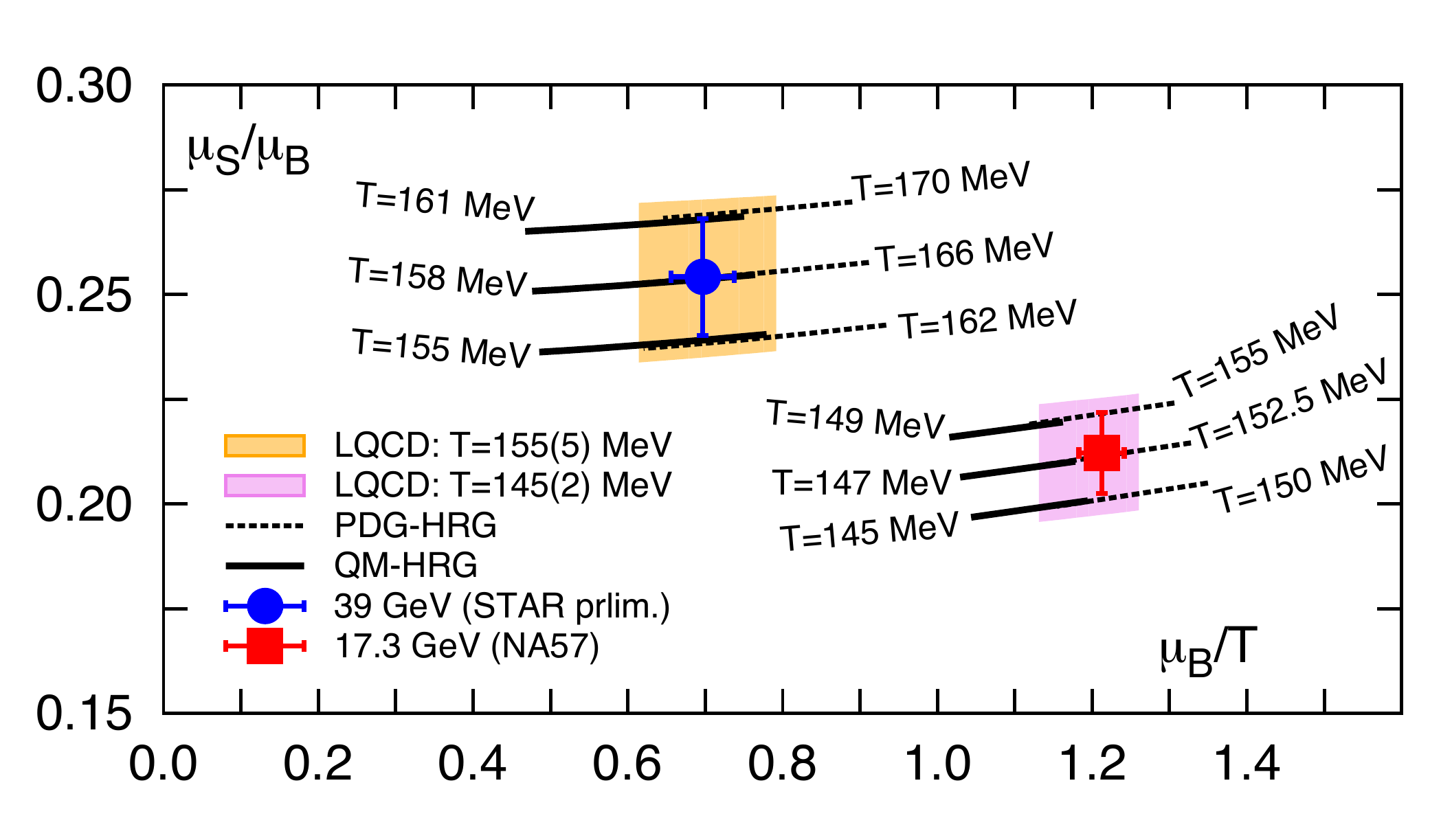}
\end{minipage}
\caption{(Left) Lattice QCD results \cite{Bazavov:2014xya} for $\mu_S/\mu_B$ at the
leading order, i.e. $s_1(T)$ (see Eq.~\ref{eq:muS}). The dotted line
(PDG-HRG) shows the results of a hadron resonance gas model containing only hadrons
listed by the Particle Data Group \cite{Agashe:2014kda}. The solid line (QM-HRG)
depicts the result for a hadron gas when additional, experimentally yet unobserved, quark model
predicted strange hadrons \cite{Capstick:1986bm,Ebert:2009ub} are included. The
shaded region indicates the chiral crossover region $T_c=154(9)$~MeV. (Right) A
comparison between experimentally extracted values of
$(\mu_S^f/\mu_B^f,\mu_B^f/T^f)$ (filled symbols) and lattice QCD results for
$\mu_S/\mu_B$ (shaded bands) \cite{Bazavov:2014xya}. The lattice QCD results are
shown for $\mu_B/T=\mu_B^f/T^f$. The temperature range where lattice QCD results
match with $\mu_S^f/\mu_B^f$ provide the values of $T^f$, i.e. $T^f=155(5)$~MeV
and $145(2)$~MeV for $\sqrt{s}=39$~GeV and $17.3$~GeV, respectively.}
\label{fig:muS_lqcd}
\end{center}
\end{figure}

By Taylor expanding the net strangeness density, $\left\langle n_S
\right\rangle(\mu_B,\mu_S)$, in $\mu_B$ and $\mu_S$ and subsequently 
imposing the strangeness
neutrality condition, $\left\langle n_S \right\rangle (\mu_B,\mu_S)=0$, for a
homogeneous thermal medium the strangeness chemical potential, $\mu_S$, can be
obtained as \cite{Bazavov:2014xya} 
\begin{equation}
\frac{\mu_S}{\mu_B} = s_1(T) + s_3(T) \left(\frac{\mu_B}{T}\right)^2 +
\mathcal{O}\left[\left(\frac{\mu_B}{T}\right)^4 \right] \;.
\label{eq:muS}
\end{equation} 
The coefficients $s_1$, $s_3$, etc. consist of various generalized baryon,
charge and strangeness susceptibilities defined at vanishing chemical potentials and
can be calculated through standard lattice QCD computations at zero chemical
potentials \cite{Bazavov:2014xya}. Fig. \ref{fig:muS_lqcd} (left) shows the leading order contribution to
$\mu_S/\mu_B$, {\sl i.e.} $s_1(T)$. A comparison of the lattice result with the
predictions from the hadron resonance gas model reveal that the inclusion of only
experimentally observed hadrons, as listed by the Particle Data Group 
\cite{Agashe:2014kda},
fails to reproduce the lattice results around the crossover region. Note that,
while $\mu_S/\mu_B$ is unique in QCD, for a hadron gas it depends on the relative
abundances of the open strange baryons and mesons. For fixed $T$ and $\mu_B$, a
strangeness neutral hadron gas having a larger relative abundance of strange baryons
over open strange mesons naturally leads to a larger value of $\mu_S$. Astonishingly,
the inclusion of additional, unobserved strange hadrons predicted within the quark
model \cite{Capstick:1986bm,Ebert:2009ub} provides a much better agreement with
lattice results, hinting that these additional hadrons become thermodynamically
relevant close to the crossover temperature \cite{Bazavov:2014xya}. As will be
discussed in Section~\ref{sec:hadrons} , other lattice thermodynamics studies also indicate that additional,
unobserved charm hadrons do become thermodynamically relevant close to the QCD
crossover \cite{Bazavov:2014yba}. 

On the other hand, the experimentally measured yields of the strangeness $S$
anti-baryon to baryon ratios, $R_H$, at the freeze-out are determined by the 
thermal freeze-out parameters ($T^f,\mu_B^f,\mu_S^f$) \cite{Andronic:2011yq},
\begin{equation}
R_H(\sqrt{s}) = \exp\left[ -\frac{2\mu_B^f}{T^f} 
\left( 1 - \frac{\mu_S^f}{\mu_B^f}|S| \right) \right] \;.
\end{equation}
By fitting the experimentally measured values of $R_\Lambda$, $R_\Xi$ and $R_\Omega$,
corresponding to $|S|=1,2$ and $3$, the values of $\mu_S^f/\mu_B^f$ and
$\mu_B^f/T^f$, as `observed' in a heavy-ion experiment at a given $\sqrt{s}$, can
easily be extracted. Matching these experimentally extracted values of
$\mu_S^f/\mu_B^f$ with the lattice QCD results for $\mu_S/\mu_B$ as a function of
temperature, one can determine the freeze-out temperature $T^f$. 
Fig.~\ref{fig:muS_lqcd} (right) illustrates this procedure. Once again, the 
inclusion of
additional unobserved strange hadrons in the hadron resonance gas model 
(QM-HRG) leads to very similar
values of the freeze-out temperatures as obtained using the lattice data. 
However, including only the hadrons listed by the 
Particle Data Group \cite{Agashe:2014kda} (PDG-HRG)
yields freeze-out temperatures that are $5$-$8$~MeV larger. 

 \section{Transport properties}
\label{sec:transport}

As we have seen in previous sections the analysis of bulk thermodynamics
and charge fluctuations provides plenty of evidence that thermodynamics
of strong-interaction matter above the crossover transition temperature, 
$T_{c}$, and up to temperatures of about $(1.5-2)T_{c}$ is highly 
non-perturbative. In this temperature range the velocity of sound 
is significantly smaller than that in a non-interacting quark gluon gas
suggesting also significant modifications of its transport properties.
One of the striking results of heavy-ion experiments at RHIC and LHC is that 
strong-interaction matter in this temperature range shows features of a 
strongly coupled medium with a small shear viscosity to entropy density ratio
and small diffusion coefficients. This led to the notion of strong-interaction
matter close to the crossover transition being a `nearly perfect liquid'.

Transport properties of strong-interaction matter can be explored in
lattice QCD calculations through the analysis of current-current 
correlation functions, e.g. for some current $J(\tau,\vec{x})$,
\be
G^{\alpha \beta}_J(\tau,\vecp) =\int\mathrm{d}^3x\ {\rm e}^{i\vecp\vecx} 
\langle J^\alpha (0,0) J^{\beta \dagger} (\tau, \vecx)\rangle =
\int_0^{\infty}\frac{\md\omega}{2\pi}\, \rho^{\alpha \beta}_J(\omega,\vecp) \,
K(\omega,\tau),
\label{eq:cor_spf_relation}
\ee
where the integration kernel $K(\omega,\tau)$ is
\be
K(\omega,\tau) = \frac{\cosh(\omega(\tau-1/2T))}{\sinh(\omega/2T)} \ ,
\label{eq:finiteTK}
\ee
and $\alpha, \beta$ are appropriately chosen (scalar, vector, tensor)
indices specifying the current.

In Euclidean space-time these correlation functions
have a spectral representation. The low frequency and low momentum structure
of the spectral functions $\rho^{\alpha \beta}_J(\omega,\vecp)$ then
gives access to transport properties of strong-interaction matter
while at higher frequencies the spectral functions provide information
on in-medium properties of bound states. We will discuss the latter 
in the Section~\ref{sec:hadrons}. 
Extracting the spectral function from a 
Euclidean correlation function is, however, difficult. This is due
to the fact that Euclidean correlation functions are calculated in finite temperature
lattice QCD only at a few discrete Euclidean time separations, e.g. 
$0\le \tau \le N_\tau$~\footnote{Note that Eq.~\ref{eq:finiteTK} is symmetric 
around $\tau=1/2T$ and as a consequence
only half the number of data points of the Euclidean temporal correlation 
function, i.e. $N_\tau/2$, provide independent information to extract the 
spectral function.}. It requires the 
application of inversion methods like the Maximum Entropy Method (MEM)
\cite{Asakawa:2000tr} or a modeling of the structure of the spectral function 
at low frequencies.

According to the Kubo relations, which express the response of a medium
to small perturbations, transport coefficients can be determined from
spectral functions calculated in equilibrium thermodynamics. They are given 
by the slope of the corresponding spectral functions at vanishing energy. 
For instance, using in Eq.~\ref{eq:cor_spf_relation} the electromagnetic 
current $J^{\mu}(t,\vecx)\equiv J_V(t,\vecx) = 
\bar{\psi}(t,\vecx)\gamma^{\mu}\psi(t,\vecx)$
for a fermionic charge carrier (quarks) with charge $e$ (e.g.
$Q_d=-e/3$ for a down quark) one can extract the
electrical conductivity $\sigma$ which quantifies the response of the QGP 
to small perturbations induced by an electromagnetic field, 
\be
\sigma = \frac{C_{em}}{6} \lim_{\omega \rightarrow 0} \lim_{\vecp \rightarrow 0} 
\sum_{i=1}^{3}\frac{\rho^{ii}(\omega,\vecp, T)}{\omega} \; .
\label{eq:conduct}
\ee
Here $C_{em}$ is the sum of the square of the elementary charges of the 
quark flavor $f$, $C_{em}=\sum_f Q_f^2$.
The electrical conductivity and the vector spectral function are 
also related to the emission rate of soft thermal photons, 
\be
\lim_{\omega \rightarrow 0} \omega \frac{{\rm d} R_\gamma}{{\rm d}^3p} =
\frac{3}{2\pi^2}\, \sigma(T) \,T \,\alpha_{em} \ .
\label{softphoton}
\ee
and the contribution of quark-antiquark annihilation to the thermal 
dilepton rate,
\be
\frac{\md W}{\md\omega\,\md^3\vecp} = \frac{5\alpha_{em}^2}{54\pi^3}\frac{1}{\omega^2(e^{\omega/T} -1)}\,\rho_V(\omega,\vecp,T),
\label{dilepton}
\ee
where $\alpha_{em}$ is the electromagnetic fine structure constant.

Using the vector current $J_V(\tau,\vecx)$ for heavy quarks one obtains
from Eq.~\ref{eq:cor_spf_relation} the corresponding spectral function of 
heavy quarkonium in the vector channel, $\rho^{ii}_V$, which allows to 
determine the heavy quark diffusion coefficient $D$,
\be
D = \frac{1}{6\chi_2^{h}}\lim_{\omega\rightarrow0}\lim_{\vecp\rightarrow 0}\sum_{i=1}^{3}\frac{\rho^{ii}_V(\omega,\vecp,T)}{\omega} .
\label{eq:HQ_diffusion_formula}
\ee
Here $\chi_2^{h}$ is the heavy quark number susceptibility which is defined 
through the zeroth component of the temporal meson correlator 
in the vector channel. E.g., for charm quarks this is just the 
net charm number susceptibility introduced in Eq.~\ref{eq:fluct1}.
Similarly, in the light quark sector, the electric charge diffusion 
coefficient $D_Q$ is defined as the ratio of $\sigma$ to electric charge 
susceptibility $\chi_2^Q$ defined in Eq.~\ref{eq:fluct1BQS}, 
i.e. $D_Q =\sigma/\chi_2^Q$.

The heavy quark diffusion coefficient $D$ can be related to the momentum diffusion 
coefficient $\kappa$ and drag coefficient $\eta$ through the Einstein 
relation
\be
D=\frac{2T^2}{\kappa} = \frac{T}{M\eta}\; .
\label{eq:Einstein}
\ee
It is used to describe the Brownian motion of a heavy quark in the 
hot medium and can also be related to the ratio of shear viscosity over 
entropy density $\eta/s$~\cite{Rapp:2008qc}.

The shear ($\eta$) and bulk ($\zeta$) viscosities of strong-interaction 
matter are extracted from correlation functions of the energy-momentum tensor, 
$T^{\mu\nu} = F^{\mu\alpha}F^{\nu\alpha} - \frac{1}{4}\delta_{\mu\nu}F^{\rho\sigma}F^{\rho\sigma}$. I.e. using the current
$J^{\mu\nu} (\tau,\vecx)\equiv T^{\mu\nu}(t,\vecx)$ in Eq.~\ref{eq:cor_spf_relation} one obtains
\bea
\eta &=&\pi \lim_{\omega\rightarrow0} \lim_{\vecp\rightarrow 0}\frac{\rho^{12,12}(\omega,\vecp)}{\omega}\; , \\
\zeta &=&\frac{\pi}{9} \lim_{\omega\rightarrow0} \lim_{\vecp\rightarrow 0} \sum_{k,l}\frac{\rho^{kk,ll}(\omega,\vecp)}{\omega} \; .
\eea

At present only a few calculations of transport coefficients have been performed
in QCD with dynamical quark degrees of freedom. 
Most calculations utilized the quenched approximation at vanishing baryon 
number density \cite{Ding:2014xha}.
In the following subsections we will summarize current results for the
transport coefficients introduced above. For further details on the
calculation of transport coefficients on the lattice and earlier results see, for instance, 
Ref.~\citen{Meyer:2011gj}.

\subsection{Electrical conductivity, charge diffusion and dilepton rates}

Probably the best analyzed spectral function is that of the light quark
vector current which gives access to the electrical 
conductivity, $\sigma (T)$ (Eq.~\ref{eq:conduct}), soft photon emission 
rates (Eq.~\ref{softphoton}) and thermal dilepton rates (Eq.~\ref{dilepton}) 
as well as the electric charge diffusion constant $D_Q = \sigma /\chi_2^Q$.
The vector spectral function $\rho_V(\omega)\equiv \rho_V(\omega,0,T)$ 
has been calculated first in quenched QCD \cite{Karsch:2001uw} using the MEM
approach \cite{Nakahara:1999vy,Asakawa:2000tr} in which the analysis had 
only been constrained at large frequencies through a default model based on 
the free fermion spectral 
function and the largest lattices used only had $N_\tau=16$ points in the
temporal direction. It has, however, been noticed \cite{Aarts:2007wj} that 
this is not sufficient
to get access to transport coefficients, which require the spectral
functions at low frequencies to be linear in $\omega$.
Furthermore, a larger time extent \cite{Ding:2010ga} is needed to control 
the low frequency part of the spectral functions.

\begin{figure}[hbpt]
\begin{center}
    \hspace{-0.3cm}\includegraphics[width=0.52\textwidth]{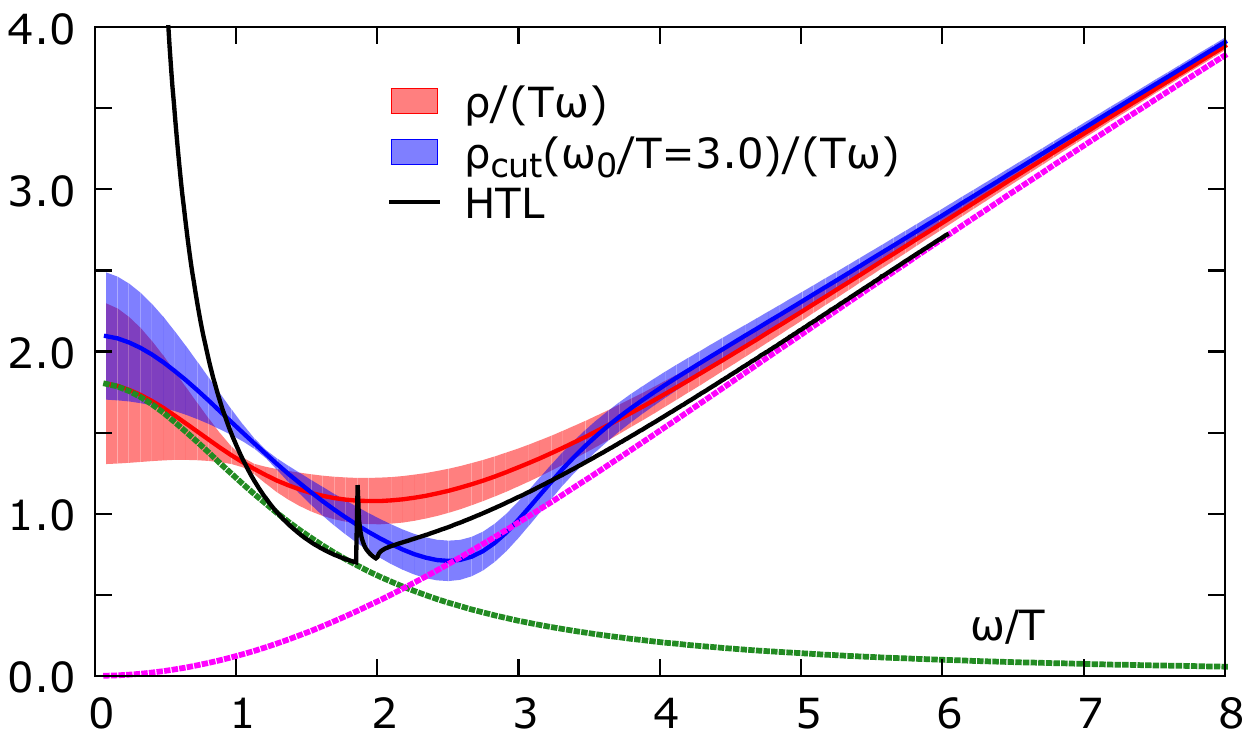}\hspace{-0.1cm}\includegraphics[width=0.49\textwidth]{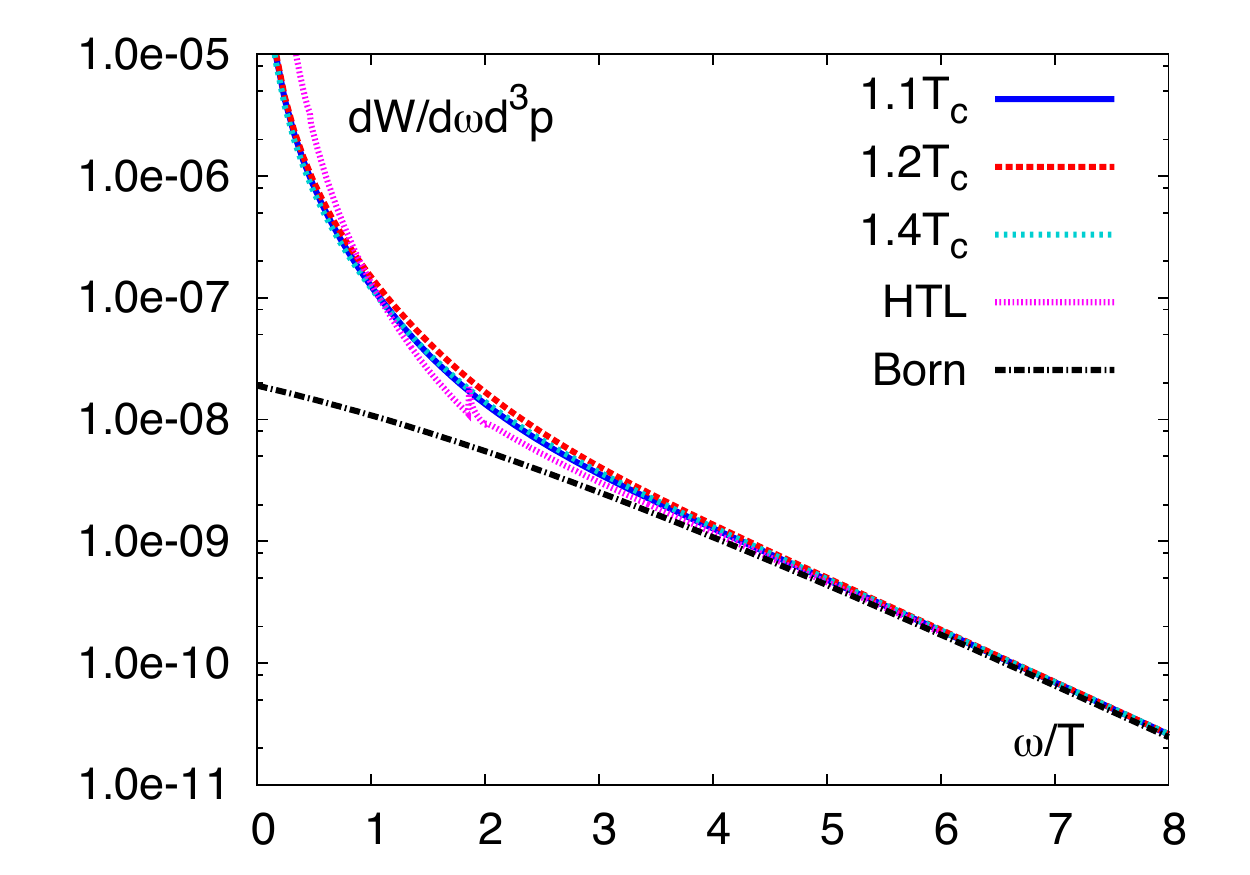}
\caption{(Left) Thermal spectral function at $1.1\ T_c$ calculated in quenched
QCD~\cite{Ding:2014dua}. (Right) Temperature dependence of the quark-antiquark annihilation 
contribution to the thermal dilepton rate above 
$T_c$~\cite{Ding:2014dua}.}
\label{fig:dilepton}
  \end{center}
\end{figure}

First continuum extrapolated results for the vector spectral function were 
obtained in quenched QCD using clover-improved Wilson fermions \cite{Ding:2010ga} 
at $T\simeq 1.4~T_c$. These results have been
extended to three different temperatures 
recently~\cite{Ding:2013qw,Kaczmarek:2013dya,Ding:2014dua}.
In these calculations the thermal spectral functions are obtained by fitting 
the  continuum-extrapolated two-point Euclidean correlation functions to 
an ansatz for the spectral function that is motivated by kinetic theory and 
perturbation theory 
\be
\rho^{model}(\omega) = \chi_q c_{BW} \, \frac{\omega\Gamma}{\omega^2 + 
\Gamma^2/4} + \frac{3}{2\pi}\,(1+k)\,\omega^2\,\tanh\left(\frac{\omega}{4T}\right)\; .
\label{eq:ModelSpf}
\ee
Here $\chi_q$ is similar to $\chi_2^h$ appearing in 
Eq.~\ref{eq:HQ_diffusion_formula} but for light quarks and it is obtained
by summing the
zeroth component of the light temporal vector correlation function over space-time coordinates. The unknown parameters are thus the amplitude $c_{BW}$ and the width $\Gamma$ of the Breit-Wigner function 
as well as the parameter $k$, which parametrizes deviations from the free 
theory behavior at large $\omega$.
The thermal spectral functions obtained in this way and the resulting thermal dilepton rates 
are shown in Fig.~\ref{fig:dilepton}. In the high frequency region, i.e. 
for $\omega/T \gtrsim 7$, all the thermal spectral functions at the three 
temperatures above $T_c$ are well described by leading order perturbation
theory (Born rate) as well as Hard Thermal Loop (HTL) calculations. At 
small frequencies the lattice QCD results are significantly 
enhanced over the Born rate, indicating the presence of a transport peak,
but are smaller than the HTL rates. In fact, the latter rises too rapidly
at low frequency and would give rise to a divergent electrical conductivity.

The finite intercept of $\rho(\omega)/(\omega T)$ at $\omega =0$ seen in 
Fig.~\ref{fig:dilepton} gives the electrical conductivity (see Eq.~\ref{eq:conduct}), 
$\sigma/(C_{\rm em}T)\in [0.21, 0.44]$~\cite{Ding:2014dua}. In these calculations $\sigma/T$ does
not show any significant temperature dependence\footnote{This is 
also reflected in the temperature dependence of the temporal
correlation functions divided by $T^3$ which are temperature 
independent for this temperature range~\cite{Ding:2014dua}.}, 
which may be due to the
use of the quenched approximation. The results are consistent with a 
calculation that used staggered fermions, but no renormalized vector
currents \cite{Aarts:2007wj}, $\sigma/(C_{\rm em}T)\simeq 0.4(1)$ 
for $1.5 \lesssim T/T_c \lesssim 2.25$. These results are, however, 
substantially smaller than earlier calculations performed on coarse lattices 
\cite{Gupta:2003zh}.

\begin{figure}[hbpt]
\begin{center}
\hspace{-0.4cm}\includegraphics[width=0.49\textwidth]{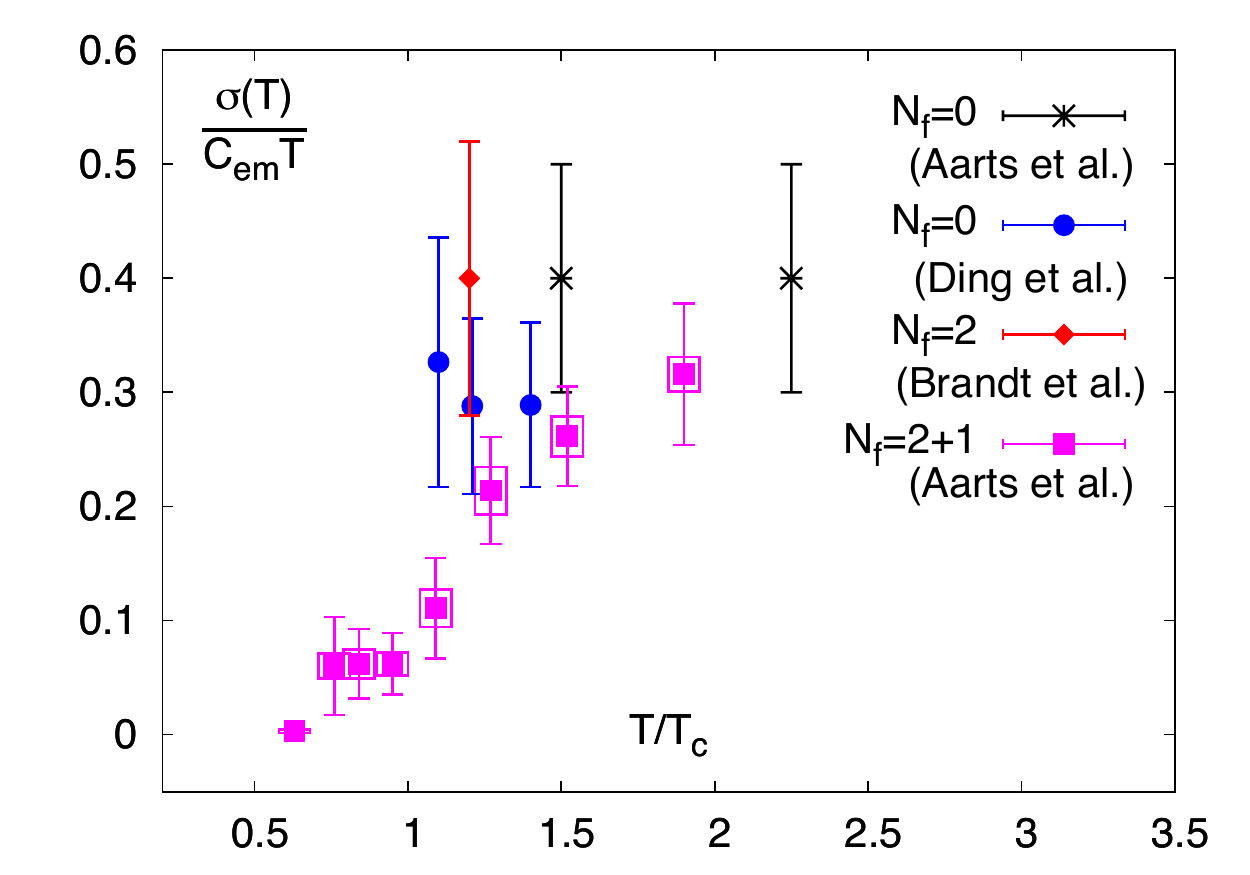}\includegraphics[width=0.53\textwidth]{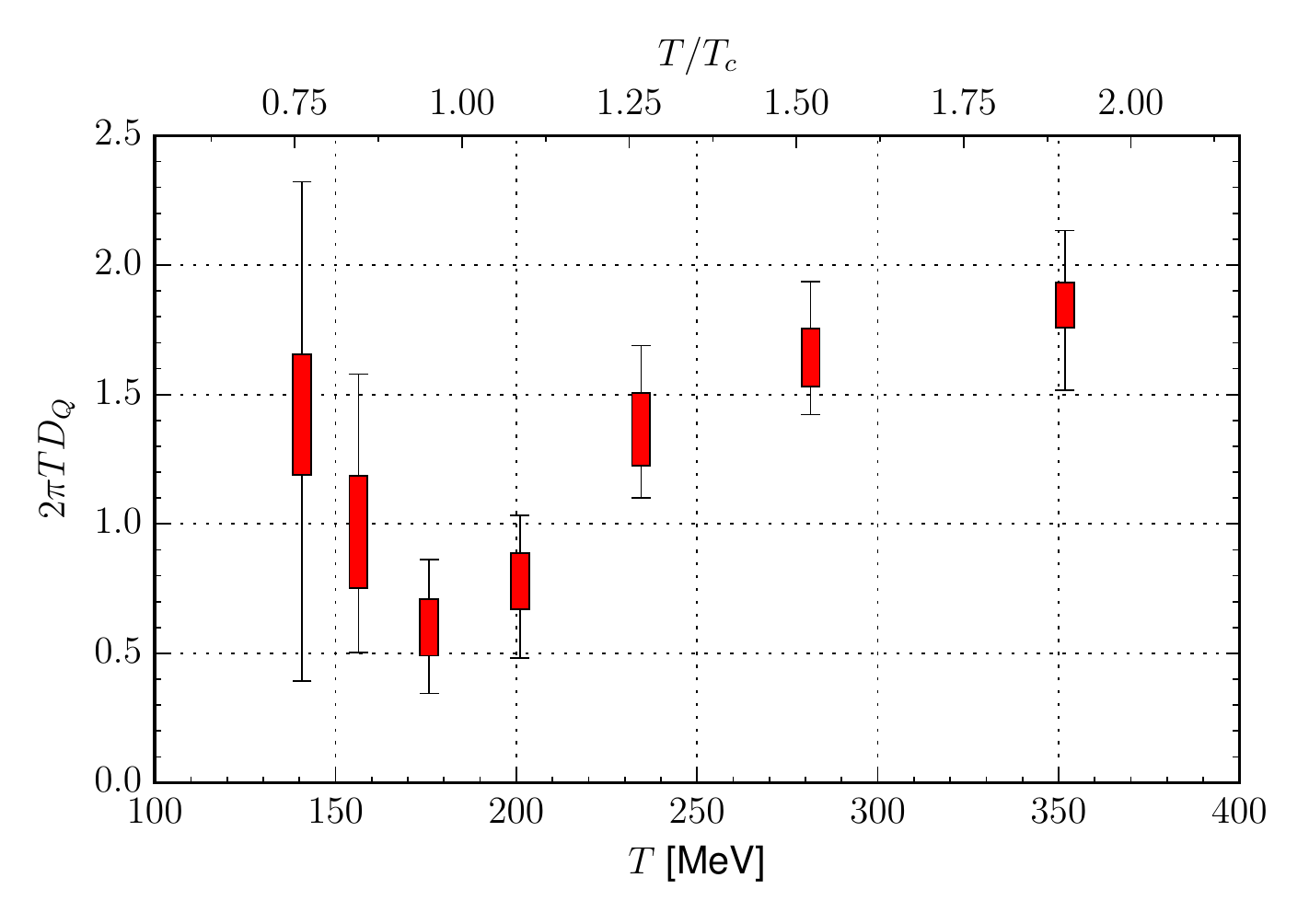}
                \caption{ (Left) Lattice QCD results for the electrical conductivity 
$\sigma/(C_{\rm em}T)$ as function of $T/T_c$ including results obtained in quenched QCD~\cite{Aarts:2007wj,Ding:2010ga,Ding:2014dua}, 2-flavor QCD with $m_\pi\simeq 270$ MeV~\cite{Brandt:2012jc} and 
(2+1)-flavor QCD with $m_\pi\simeq380 $MeV~\cite{Aarts:2014nba}. Note that the values of $T_c$ obtained in these calculations are different, i.e. $T_c^{N_f=0}\simeq 270$ MeV, $T_c^{N_f=2}\simeq 208$ MeV and $T_c^{N_f=2+1}\simeq 185$ MeV. 
(Right) Temperature dependence of charge diffusion coefficient 
$2\pi TD_Q=2\pi T\sigma/\chi_2^Q$. The results are obtained in (2+1)-flavor 
QCD calculations on anisotropic lattices at one value of the spatial lattice 
cut-off, $a\simeq 0.12$~fm and with light quark masses corresponding to 
$m_{\pi}\simeq 380$~MeV~\cite{Aarts:2014nba}.}
\label{fig:sigma_QCD}
  \end{center}
\end{figure}  

Recently the calculation of the electrical conductivity has also been 
performed in QCD with dynamical quarks. The first calculation was done 
in 2-flavor QCD using $\mathcal{O}(a)$ improved Wilson quarks corresponding 
to $m_{\pi}\simeq270~$MeV. The electrical conductivity of the quark-gluon plasma 
at $T\simeq 250$~MeV was found to be similar to the quenched result, 
$\sigma/(C_{\rm em}T)=0.40(12)$~\cite{Brandt:2012jc}.

First results from a calculation in (2+1)-flavor QCD performed on anisotropic 
lattices with quark masses corresponding to a pion mass
$m_\pi\simeq380~$MeV~\cite{Amato:2013naa,Aarts:2014nba} are shown in
Fig.~\ref{fig:sigma_QCD}. The simulation is performed at one value
of the lattice cut-off. Varying the temporal extent of
the lattice, $N_\tau$, from between 48 and 16 a temperature range
from 0.63~$T_c$ to 1.90~$T_c$ is covered.
The left hand panel of this figure shows
temperature dependence of $\sigma/(C_{\rm em}T)$ (square points).
It stays constant within errors in the transition region and increases with 
temperature for $T>T_{c}$. Also shown in the left hand panel of Fig.~\ref{fig:sigma_QCD}
are electrical conductivities obtained in quenched QCD~\cite{Aarts:2007wj,Ding:2010ga,Ding:2014dua} 
and 2-flavor QCD~\cite{Brandt:2012jc}. 
The right hand panel of Fig.~\ref{fig:sigma_QCD} shows the temperature dependence of charge diffusion coefficient $D_Q$ multiplied 
by $2\pi T$. $2\pi T D_Q$ shows a dip near $T_c$ which arises from the rapid exponential drop of the 
electric charge susceptibility at low temperature where the electrical conductivity of a pion gas \cite{FernandezFraile:2005ka} 
remains nonzero and vanishes only like $\sqrt{T}$. 

\subsection{Heavy quark diffusion}
\label{sec:HQ_diffusion}

Contrary to earlier expectations it has been found in heavy ion
collision experiments at RHIC~\cite{Adare:2006nq,Abelev:2006db} and 
LHC~\cite{ALICE:2012ab} that heavy-quark mesons show a substantial
elliptic flow that is comparable to that of light-quark mesons.
Moreover, also somewhat unexpectedly, heavy quarks are found to
lose a significant amount of energy while traversing through the QGP.
The latter has been called the heavy quark energy loss puzzle.
Phenomenological explanations of these phenomena try to model
the heavy quark diffusion in a hot and dense medium. This requires
knowledge about the heavy quark diffusion coefficients 
$D$~\cite{He:2014epa,Rapp:2008qc,Cao:2014fna} which can be determined
in lattice QCD calculations.

\begin{figure}[t]
\begin{center}
  \includegraphics[width=0.45\textwidth]{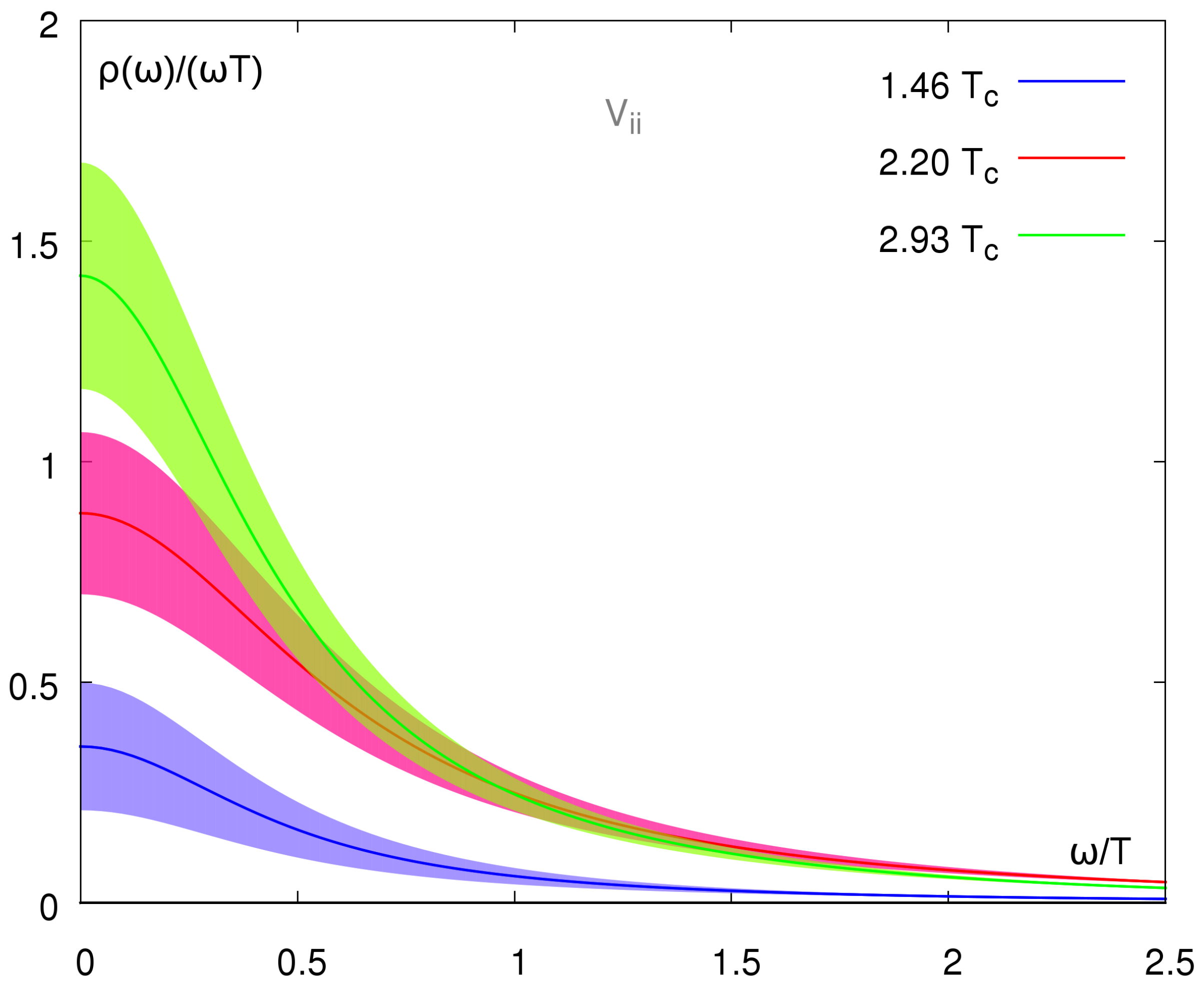}~\includegraphics[width=0.5\textwidth,height=0.35\textwidth]{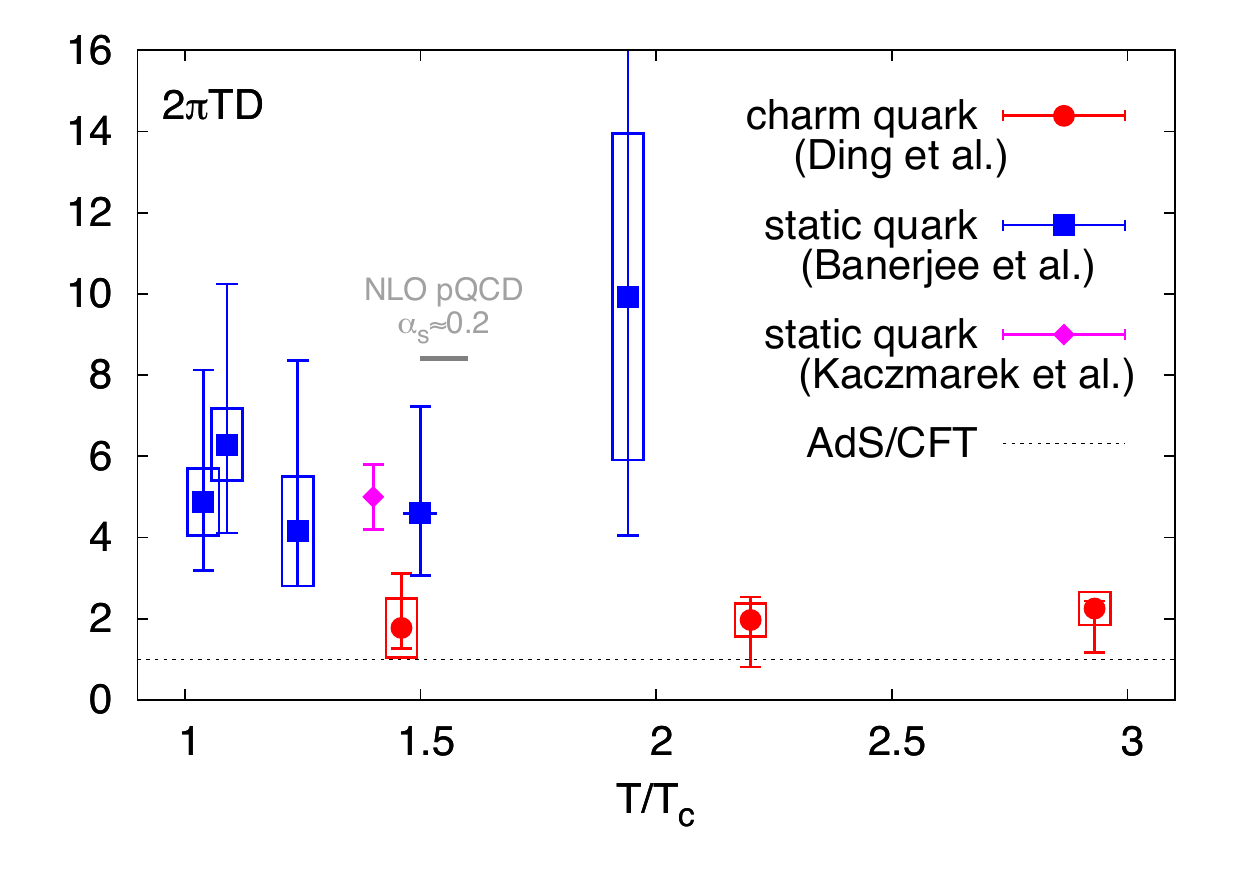}
\caption{(Left) The low frequency transport part of the 
charmonium vector spectral function obtained from a MEM analysis of vector
correlation functions in quenched lattice QCD~\cite{Ding:2011hr}. The central lines are mean 
values while the bands reflect the statistical uncertainties. 
(Right) Temperature dependence of heavy quark diffusion coefficients multiplied by $2\pi T$ in pure gluonic matter.
The charm diffusion coefficients are obtained from the spectral
functions shown in the left plot~\cite{Ding:2011hr}.
Also shown are results for the static quark diffusion coefficient 
obtained by using the lattice discretized versions of HQEFT \cite{Kaczmarek:2014jga,Banerjee:2011ra}. 
The boxes show the statistical error while the error bars reflect the 
systematic uncertainties.
The horizontal dotted line labels the value of $2\pi T D$ in the heavy quark limit from the AdS/CFT correspondence
~\cite{Kovtun:2003wp} and the
short, horizontal solid line indicates the value of $2\pi T D$ from next-to-leading order pQCD calculations at $\alpha_s\simeq0.2$~\cite{CaronHuot:2007gq}.
}.
\label{fig:spf_trans_statistics_errors}
\end{center}
\end{figure}

The charm-quark diffusion coefficient has been calculated 
in quenched QCD at three temperatures in the deconfined phase~\cite{Ding:2011hr}
using lattices with temporal extent $N_\tau=48,32$ and 24 at a fixed value of
the cut-off. Results from this calculation are shown in  
Fig.~\ref{fig:spf_trans_statistics_errors}. The left hand panel  
shows the transport peak in the charmonium vector spectral function at three 
different temperatures above $T_{c}$. These transport peaks are obtained using 
a MEM analysis where an ansatz of Lorentzian form is used as the prior information
for the low frequency part of the spectral function, 
i.e. $\omega\eta/(\omega^2 + \eta^2)$. The intercept at $\omega/T=0$ gives
the charm quark diffusion coefficient (c.f. Eq.~\ref{eq:HQ_diffusion_formula}).
It is found that  $2\pi T D\simeq 2$ and,
with current understanding of uncertainties, this product is 
independent of temperature. These values of $D$
are much smaller than the results from perturbative 
calculations~\cite{CaronHuot:2007gq} and are very close to the value obtained 
from the AdS/CFT correspondence~\cite{Kovtun:2003wp}.
These different results are compared in the right hand panel of 
Fig.~\ref{fig:spf_trans_statistics_errors}. 

A direct determination of the diffusion coefficient for bottom quarks from
bottomonium correlation functions is difficult as it would require quite
small lattice spacings to accommodate the bottomonium states on the lattice.
The method of choice here is to start with the infinite quark mass limit 
where one can use Heavy Quark Effective Theory (HQEFT) to compute the 
diffusion coefficient of a static quark.
In this approach the propagation of a heavy quark carrying color charge and 
its response to a colored Lorentz force can be described using linear 
response theory. This leads to the spectral analysis of a ``color-electric
correlator''~\cite{CaronHuot:2009uh,CasalderreySolana:2006rq}  
\be
G_E(\tau) = -\frac{1}{3} \sum_{i=1}^{3} \frac{ \left \langle  \mathrm{Re} \mathrm{Tr} \left [ U(1/T;\tau) gE_i(\tau,\vecnull) \, U(\tau;0) gE_i(0,\vecnull)\right ] \right\rangle}
{\left \langle\mathrm{Re}\,\mathrm{Tr} \left[ U(1/T;0)\right] \right\rangle},
\label{eq:Corr_HQ}
\ee
where  $U(\tau_2;\tau_1)$ is a Wilson line in the Euclidean time direction 
and  $gE_i(\tau,\vecx)$ denotes the color-electric field at time-space $(\tau,\vecx)$. The momentum diffusion 
coefficient $\kappa$ can then be extracted from the slope of the
corresponding spectral function in the limit of vanishing frequency $\omega$, 
\begin{eqnarray}
\kappa/T^3 = \lim_{\omega\rightarrow 0} \frac{2\, \rho_E(\omega)}{\omega T^2}.
\end{eqnarray}
Here  $\rho_E(\omega)$ is the corresponding spectral function that is related
to $G_E(\tau)$ via Eq.~\ref{eq:cor_spf_relation}.
In the non-relativistic limit, i.e. for a heavy quark mass $M\gg \pi T$, 
the momentum diffusion coefficient $\kappa$ is related to the 
(spatial) heavy quark diffusion coefficient $D$ via 
Eq.~\ref{eq:Einstein}.

\begin{figure}[t]
\begin{center}
  \includegraphics[width=0.49\textwidth]{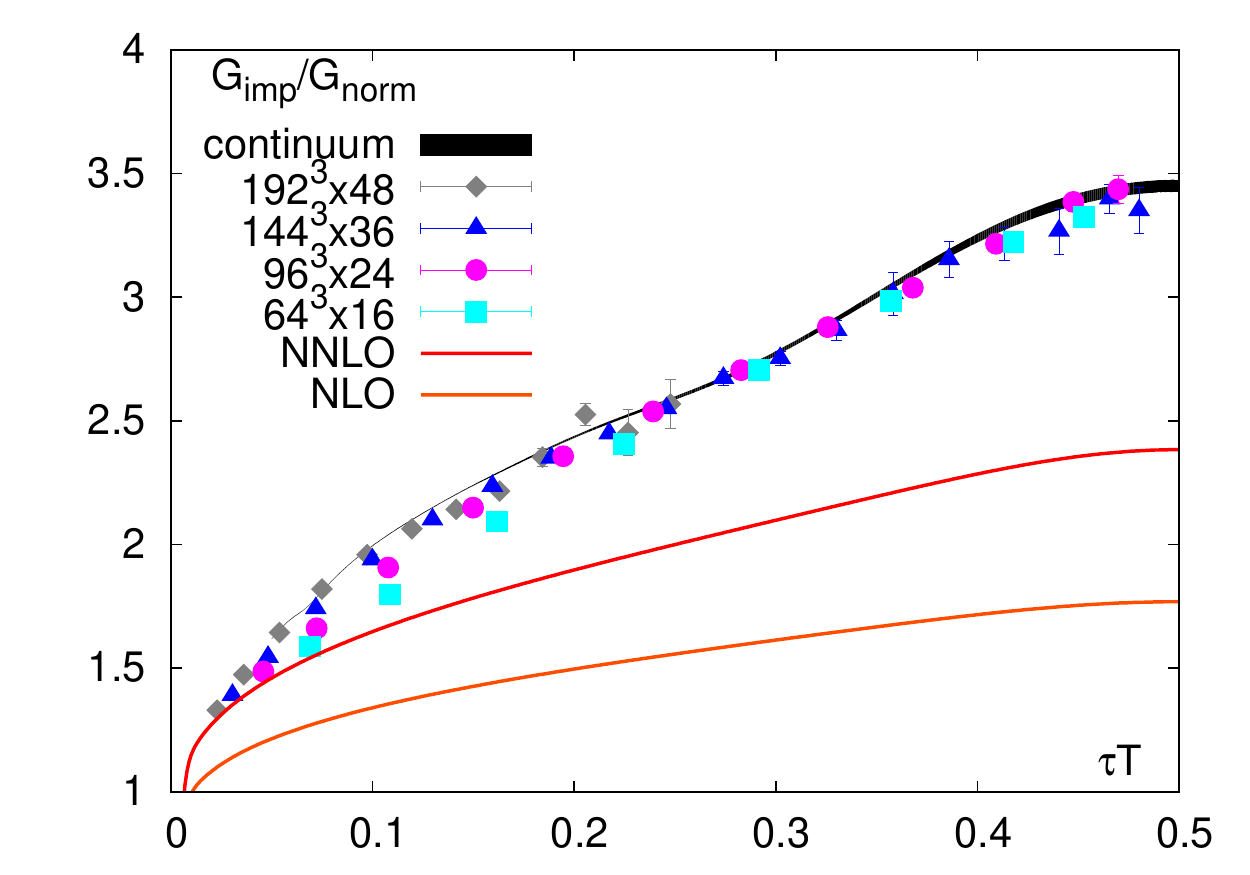}~~      
  \includegraphics[width=0.49\textwidth]{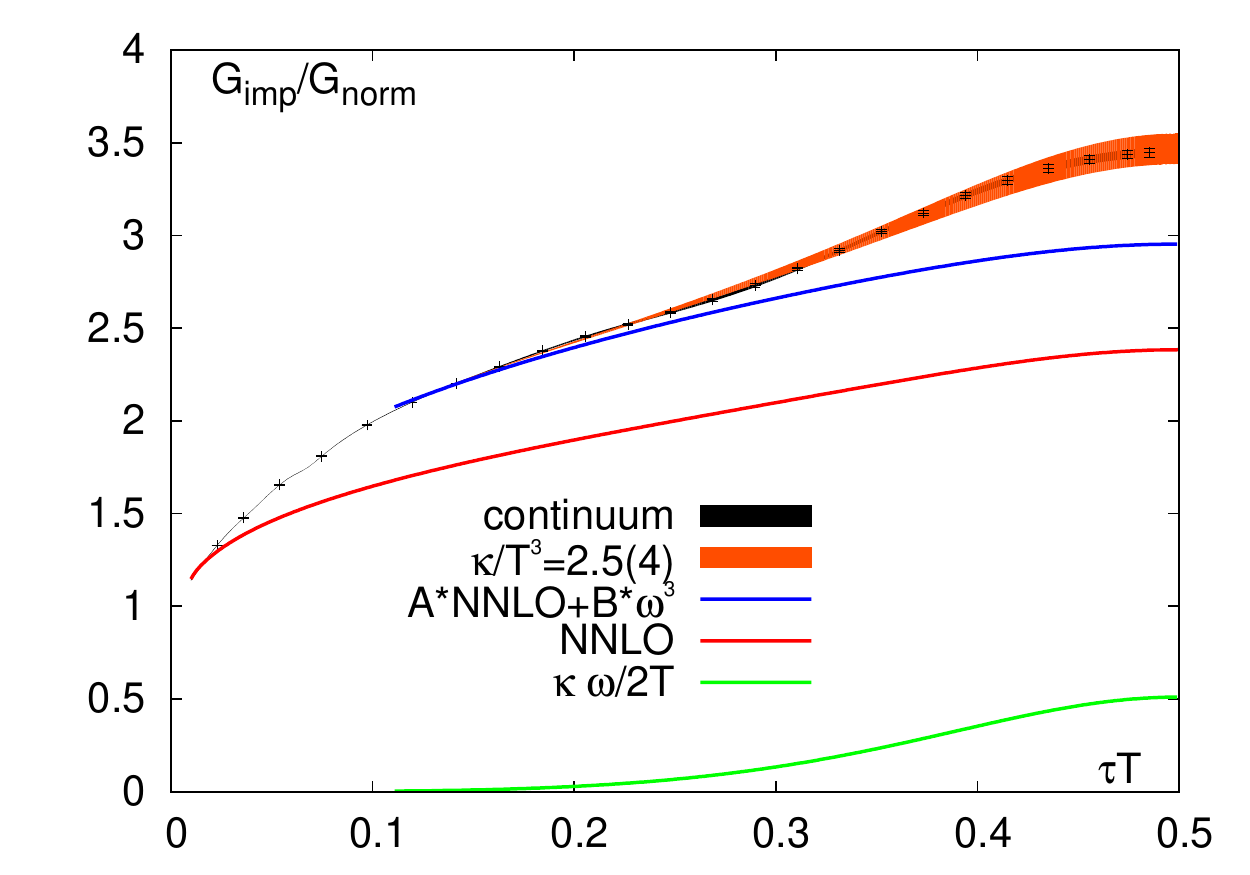}          
  \caption{(Left) Continuum extrapolation of improved color-electric correlator 
$G_E(\tau)\equiv G_{imp}(\tau)$ (see Eq.~\ref{eq:Corr_HQ}) normalized to the 
high temperature, free field theory result
$G_{norm} (\tau)$ and the comparison with NNLO and NLO perturbative QCD
  calculations at 1.4$~T_c$ on quenched lattice QCD \cite{Kaczmarek:2014jga}.
(Right) $\chi^2$ fit to the continuum extrapolated $G_{imp}/G_{norm}$ using an 
ansatz $\rho^{model}_E(\omega)=\mathrm{max}\left\{A\rho_{NNLO}(\omega)+B\omega^3, \frac{\omega\kappa}{2T}\right\}$. Also shown are the separate contributions 
to the correlator arising from the different parts of the fit ansatz for the
spectral function \cite{Kaczmarek:2014jga}.}
\label{fig:transport_HQ}
  \end{center}
\end{figure}  

The color-electric correlator $G_E(\tau)$ is defined in terms of gluonic 
observables only. It thus suffers from a small signal to noise ratio,
a problem similar to what one encounters in the calculation of 
viscosities~\cite{Meyer:2007ic} (see next subsection). In both cases efficient 
noise reduction techniques are needed for a calculation of the gluonic 
correlation functions. In a recent study \cite{Kaczmarek:2014jga}
the correlator $G_E(\tau)$ has been calculated using  multi-level updates
\cite{Luscher:2001up,Meyer:2007ic} as well as the link integration
technique \cite{Parisi:1983hm,DeForcrand:1985dr}. The improved correlator
$G_{imp}(\tau T)$ obtained this way at $T\simeq 1.4\ T_c$ for four 
different values of the lattice cut-off in quenched QCD has been extrapolated
to the continuum limit. Results are shown in the left hand panel of 
Fig.~\ref{fig:transport_HQ}. In order
to extract the momentum diffusion coefficient from this correlator an ansatz
for the spectral function has been used \cite{Kaczmarek:2014jga}. 
The corresponding fit is shown in the right hand panel of 
Fig.~\ref{fig:transport_HQ}. It yields,
\be
\kappa/T^3 = 2.5(4)~~~~\mathrm{at~~} T\sim1.4~T_c.
\ee

This corresponds to $2\pi T D= 5.0(8)$ which is consistent with similar 
lattice QCD studies using HQEFT~\cite{Banerjee:2011ra}. However, it is 
a factor 2 to 3 larger
than the charm quark diffusion coefficient discussed above~\cite{Ding:2011hr}
(see Fig.\ref{fig:spf_trans_statistics_errors}).

The heavy quark diffusion coefficient $D$ has also been calculated in
perturbative QCD in both leading and next-to-leading order as well as from 
the AdS/CFT correspondence. At $\alpha_s\approx 0.2$ 
the leading order pQCD calculation \cite{Moore:2004tg} gives $2\pi TD\approx 71.2$
while $2\pi TD\approx 8.4$ is obtained in next-to-leading order 
calculations~\cite{CaronHuot:2007gq}. In the strong coupling limit 
$2\pi TD=1$ is obtained from the AdS/CFT correspondence~\cite{Kovtun:2003wp}
(see Fig.\ref{fig:spf_trans_statistics_errors}).

\subsection{Shear \& bulk viscosities}

\begin{figure}[hbpt]
\begin{center}
  \includegraphics[width=0.50\textwidth]{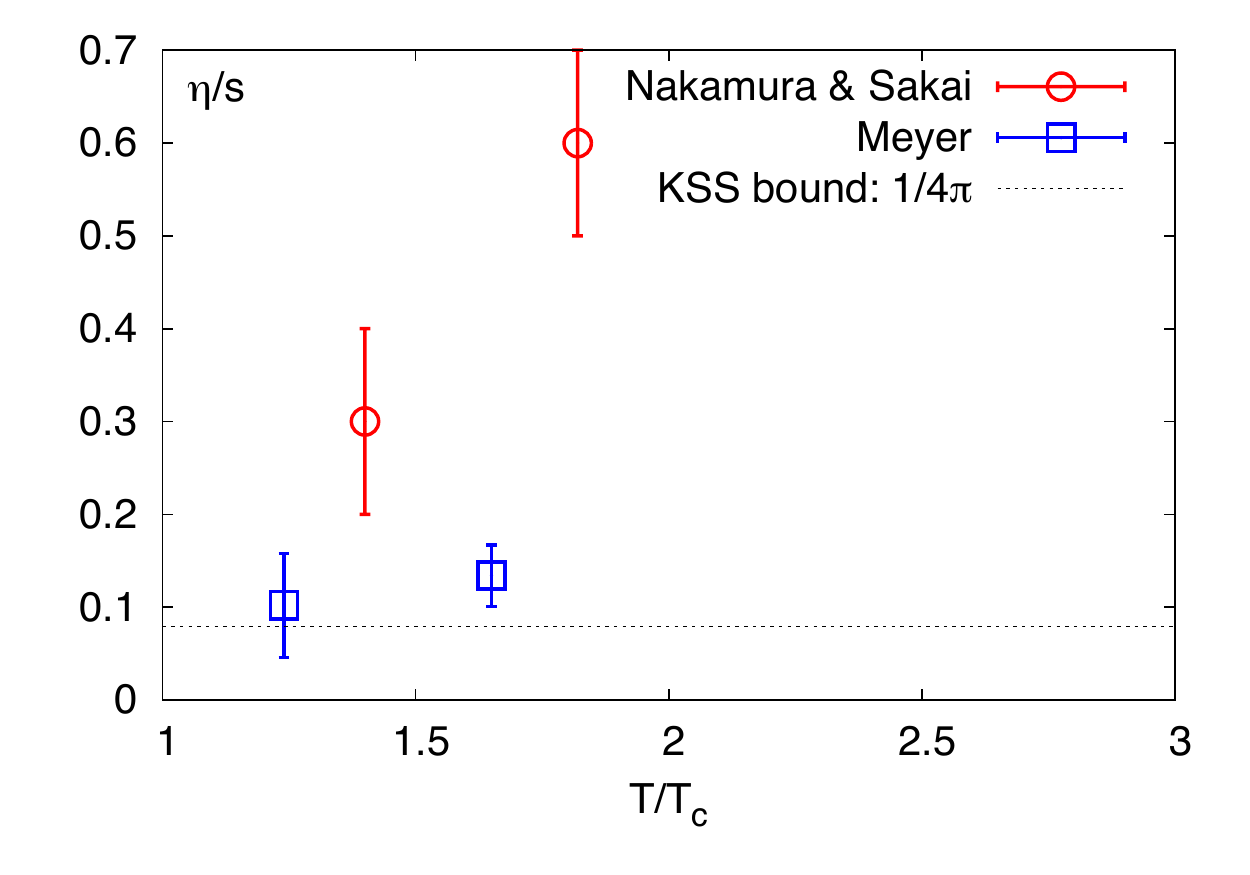}\hspace{-0.2cm}\includegraphics[width=0.50\textwidth]{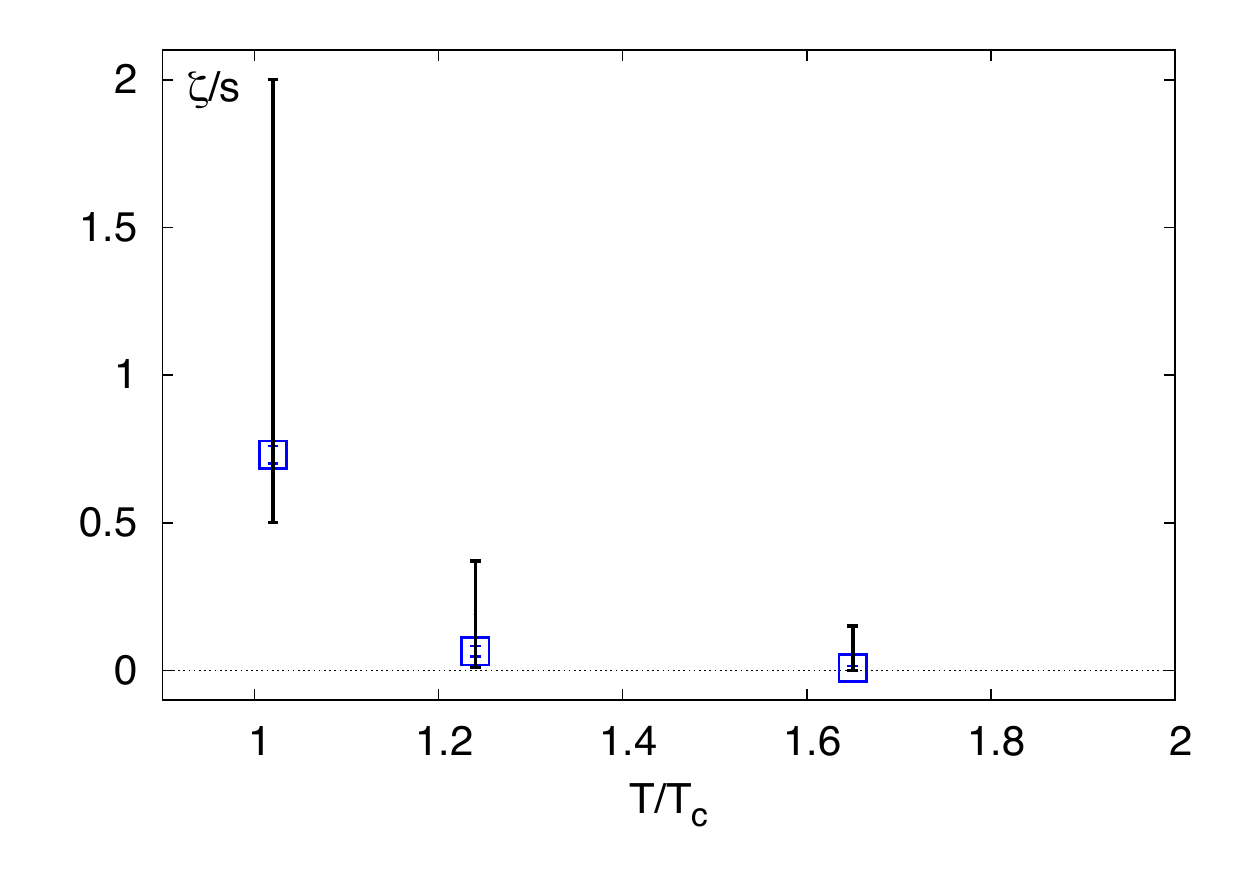}
\caption{(Left) Temperature dependence of the shear viscosity, $\eta/s$, of 
gluonic matter obtained by using noise-reduction techniques 
(squares) \cite{Meyer:2007ic}
and direct high statistical calculations (circles) \cite{Nakamura:2004sy} 
on $N_\tau=8$ lattices. The dotted line corresponds
to the conjectured lower bound of $\eta/s$ from AdS/CFT correspondence~\cite{Kovtun:2003wp}. 
(Right) Temperature dependence of the bulk viscosity, $\zeta/s$, 
of gluonic matter~\cite{Meyer:2007dy}.
The square points include the statistical uncertainties while the solid black 
bars denote the systematic uncertainties.
Results for the same quantity from Ref.~\citen{Nakamura:2004sy} which are not shown are consistent with zero at $T\in[1.4\ T_c,1.8\ T_c]$.
Data are taken from Refs.~\citen{Nakamura:2004sy,Meyer:2007dy,Meyer:2007ic}.}
\label{fig:viscosity}
  \end{center}
\end{figure}

The computation of shear and bulk viscosities has been attempted
already in the 80's \cite{Karsch:1986cq}. However, it turned
out that the signal to noise ratio in calculating Euclidean 
energy-momentum correlation functions is very small and high
statistics or improved algorithms are needed. The latter became
available for quenched QCD calculations with the multi-level algorithms
\cite{Luscher:2001up} which have been used for the calculation of 
shear and bulk viscosities \cite{Meyer:2003hy,Meyer:2007ic}.
Results from such an analysis by Meyer~\cite{Meyer:2007ic} and those with 
direct high statistics calculations by Nakamura \& Sakai~\cite{Nakamura:2004sy} are shown in Fig.~\ref{fig:viscosity}. 
The ratio obtained in the former calculation
are smaller than those obtained in the earlier direct simulations;
$\eta/s = 0.102(56)$ at 1.24~$T_c$ and 0.134(33) at $1.65~T_c$
obtained in quenched QCD on lattices with temporal extent $N_\tau=8$
\cite{Meyer:2007ic}.
This shows that noise reduction
techniques like the  multi-level algorithm are mandatory for a successful 
calculation of viscosities.
Results on the bulk viscosity to entropy ratio $\zeta/s$ 
show that this ratio rapidly becomes small above $T_c$~\cite{Meyer:2007dy}.
At $T\gtrsim 1.2\ T_c$ it is smaller than $\eta/s$ and, in fact, within
errors it is consistent with zero 
\cite{Nakamura:2004sy,Meyer:2007dy,Meyer:2007ic}.

Still the calculations of viscosities are performed on lattices with 
rather small temporal
extent compared to those used in calculations of the electrical conductivity
and diffusion constants. Systematic uncertainties in these calculations 
need to be better controlled in future. Here it will also be helpful to
combine lattice QCD calculations with information from analytic approaches
that put constraints on the structure of spectral functions,  e.g. QCD sum 
rules~\cite{Karsch:2007jc,Kharzeev:2007wb,Romatschke:2009ng}.

\subsection{Transport coefficients of second order hydrodynamics}

Due to the importance of viscous effects in the evolution of hot and
dense matter created in heavy ion collision there is considerable interest
in extending the hydrodynamic modeling beyond leading order gradient
expansions of the energy momentum tensor.
A second order gradient expansion is parameterized
by additional transport coefficients \cite{Romatschke:2009kr},
which may become accessible to lattice
QCD calculations \cite{Moore:2010bu,Moore:2012tc,Denicol:2012cn}.
One of these new transport coefficients\footnote{Although this particular 
second
order coefficient is also called $\kappa$ it should be noted that it is
not related to the momentum diffusion coefficient $\kappa$ introduced in Eq.~\ref{eq:Einstein}}
is $\kappa$, which controls the
momentum dependence of space-like correlations of the energy momentum
tensor. It is defined as the
leading order coefficient of the Taylor expansion of the retarded correlator of energy-momentum tensor in momentum $\vecp$ about zero~\cite{Baier:2007ix,Romatschke:2009ng},
\begin{eqnarray}
G^R(\omega =0,\vecp) &=& \int^{\infty}_{-\infty} \mathrm{d}t \,\mathrm{d}\vecx \ 
{\rm e}^{i \vecp\vecx} \langle [T_{12}(\tau,\vecx), T_{12}(0,0) ] \rangle 
\theta(t) \nonumber \\
&=& G^R(0) + \frac{\kappa}{2}\,|\vecp|^2 + \mathcal{O}(|\vecp|^4) \; .
\end{eqnarray}
Since the difference between the retarded correlator $G^{R}$ and the 
Euclidean correlator at vanishing frequency $\omega=0$ is just a contact term, 
which has been shown to be momentum independent~\cite{Kohno:2011aa},
$\kappa$ can directly be extracted from the corresponding Euclidean correlator,
\be
\kappa = 2  \lim_{|\vecp|\rightarrow 0}\frac{\mathrm{d}G^E(\omega=0,\vecp)}{\mathrm{d} |\vecp|^2} \; .
\ee
This is considerably simpler to compute than first order
coefficients such as the shear or bulk viscosity as it does not require analytic continuation from imaginary time to real time.
The only complication is that large spatial lattice sizes are required to get access 
to small momenta in the lattice QCD calculation.

\begin{figure}[hbpt]
\begin{center}
  \scalebox{0.7}{\input{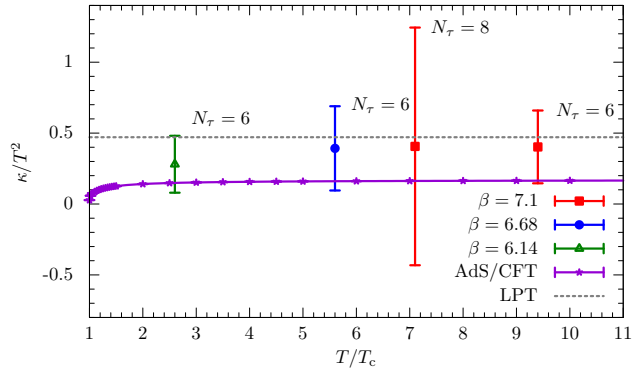}}
            \caption{Transport coefficient of second order hydrodynamics ${\kappa}/{T^2}$ as a function of $T$. The lines denote
          the results from AdS/CFT correspondence \cite{Baier:2007ix} and lattice perturbation theory (LPT)
          \cite{Philipsen:2013nea}, respectively.}
                           \label{fig:kappa}
  \end{center}
\end{figure}

The second order hydro-coefficient $\kappa$ has been computed in quenched QCD 
on lattices with temporal extent $N_\tau=8$ and 6 and a large spatial size 
\cite{Philipsen:2013nea}. Results from this calculation are shown in
Fig.~\ref{fig:kappa}. It yields $\kappa/T^2 = 0.36(15)$. 
Within the current large errors no further temperature dependence is 
observed in the temperature range $2T_c< T < 10T_c$ and results are
compatible with both perturbative QCD~\cite{Moore:2012tc,Romatschke:2009ng} 
and AdS/CFT results~\cite{Romatschke:2009im}.

\section{Open heavy flavors and heavy quarkonia}
\label{sec:hadrons}

Owing to their large masses heavy flavors, such as charm and bottom, are produced
only during the earliest stages of a heavy ion collision and get affected by 
the hot and dense medium during its entire
evolution. They then emerge as hadrons with open heavy flavors or 
as quarkonia. Thus, heavy flavors provide unique identifiable probes against 
which the temperature and coupling of the medium created during heavy-ion 
collisions can be calibrated. 
The melting of charmonium states has long been considered as an important
signature for the formation of quark-gluon plasma~\cite{Matsui:1986dk}. 
Understanding its production rates and abundance can provide detailed 
information on properties of the matter formed in a heavy ion collision. 
We will discuss current results on the temperature dependence of 
quarkonium bound states in subsection ~\ref{sec:HeavyQuarkonia}. 
In addition,
understanding the fate of open flavor bound states also is of importance
for the understanding of the strongly interacting regime in the 
QGP close to the transition temperature. The question whether or not heavy 
flavor mesons can survive above $T_c$ plays a crucial role in the
phenomenological modeling of the heavy quark energy loss 
\cite{Adil:2006ra,Sharma:2009hn,He:2011qa}.
Lattice QCD results on the dissociation of open
charm hadrons will be discussed in the next subsection.

\subsection{Melting and abundances of open charm hadrons}
\label{sec:HQLight}

Similar to the case of strangeness \cite{Bazavov:2013dta}, discussed in 
Section~\ref{sec:deconf}, the deconfinement or melting of open charm 
hadrons can be probed \cite{Bazavov:2014yba} by using fluctuations of 
the charm quantum number ($C$) and its correlations with baryon number 
($B$) fluctuations. To this end one may consider appropriately
chosen ratios of correlations between net-baryon and net-charm number 
fluctuations. One such ratio is, for instance, $\chi^{BC}_{22}/\chi^{BC}_{13}$
\cite{Bazavov:2014yba}. Since at low temperatures, for a gas of uncorrelated 
charmed hadrons, the thermodynamic contributions of the much heavier 
$|C|=2,3$ baryons can be safely neglected, this particular ratio is dominated 
by $|C|=1$ baryons and thus yields $\chi^{BC}_{22}/\chi^{BC}_{13}=|B|=1$. 
On the other hand, at high temperatures, for a free gas of massive charm 
quarks with masses much larger than $T$, the above ratio is\footnote{Here
one makes use of the fact that the Boltzmann gas approximation 
is still a good approximation for a charm quark gas at a few times the 
transition temperature. This is the temperature range of interest to heavy 
ion collision experiments.}
$\chi^{BC}_{22}/\chi^{BC}_{13}=|B|=1/3$.
Thus, this quantity can be viewed as a measure for the baryon number 
associated with the predominant degrees of freedom that are 
carriers of the charm quantum number in a given temperature region. 
Its behavior can indicate the liberation of quark-like degrees of freedom 
with fractional baryon number. 

\begin{figure}[htp]
\begin{center} 
\begin{minipage}[c]{0.5\textwidth}
\includegraphics[width=0.99\textwidth,height=0.25\textheight]{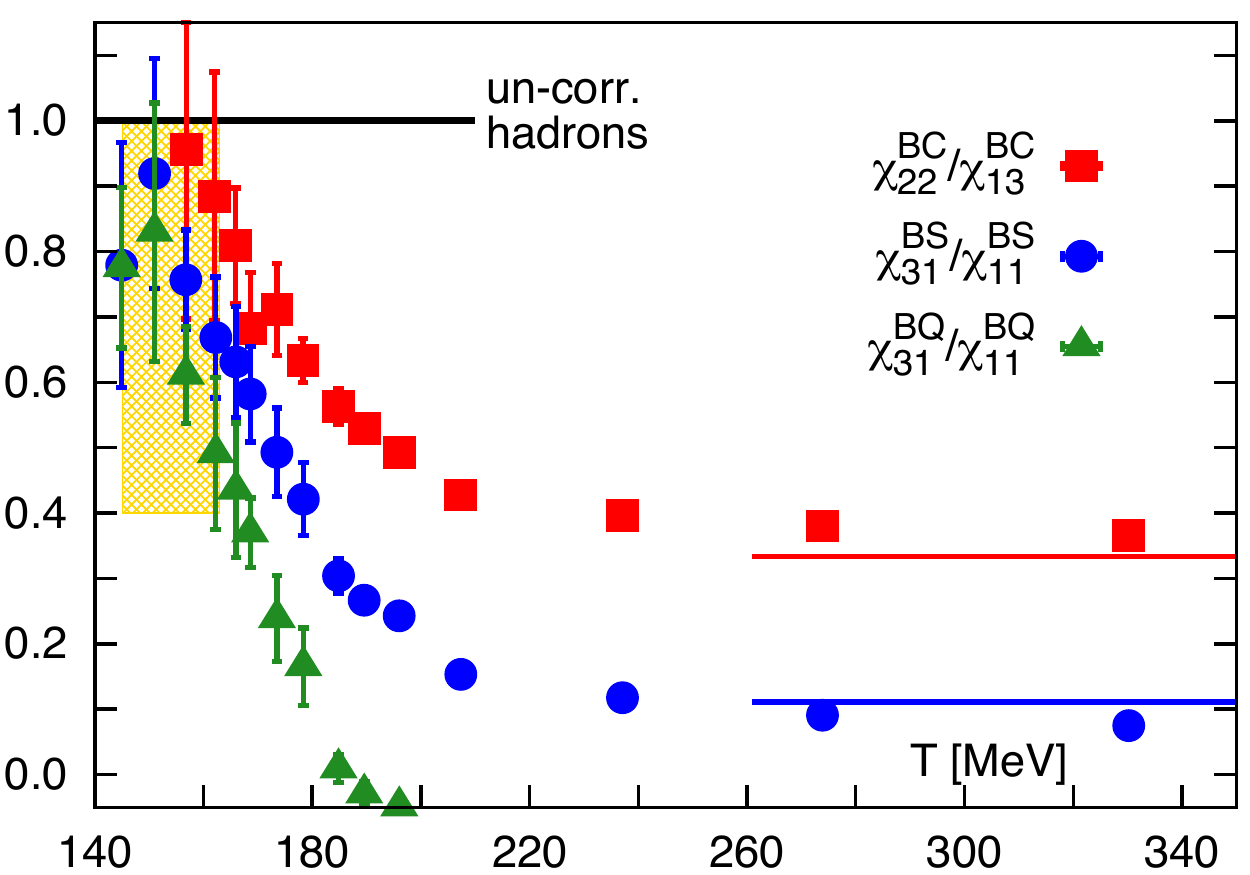}
\end{minipage}
\begin{minipage}[c]{0.4\textwidth}
\includegraphics[width=0.99\textwidth,height=0.25\textheight]{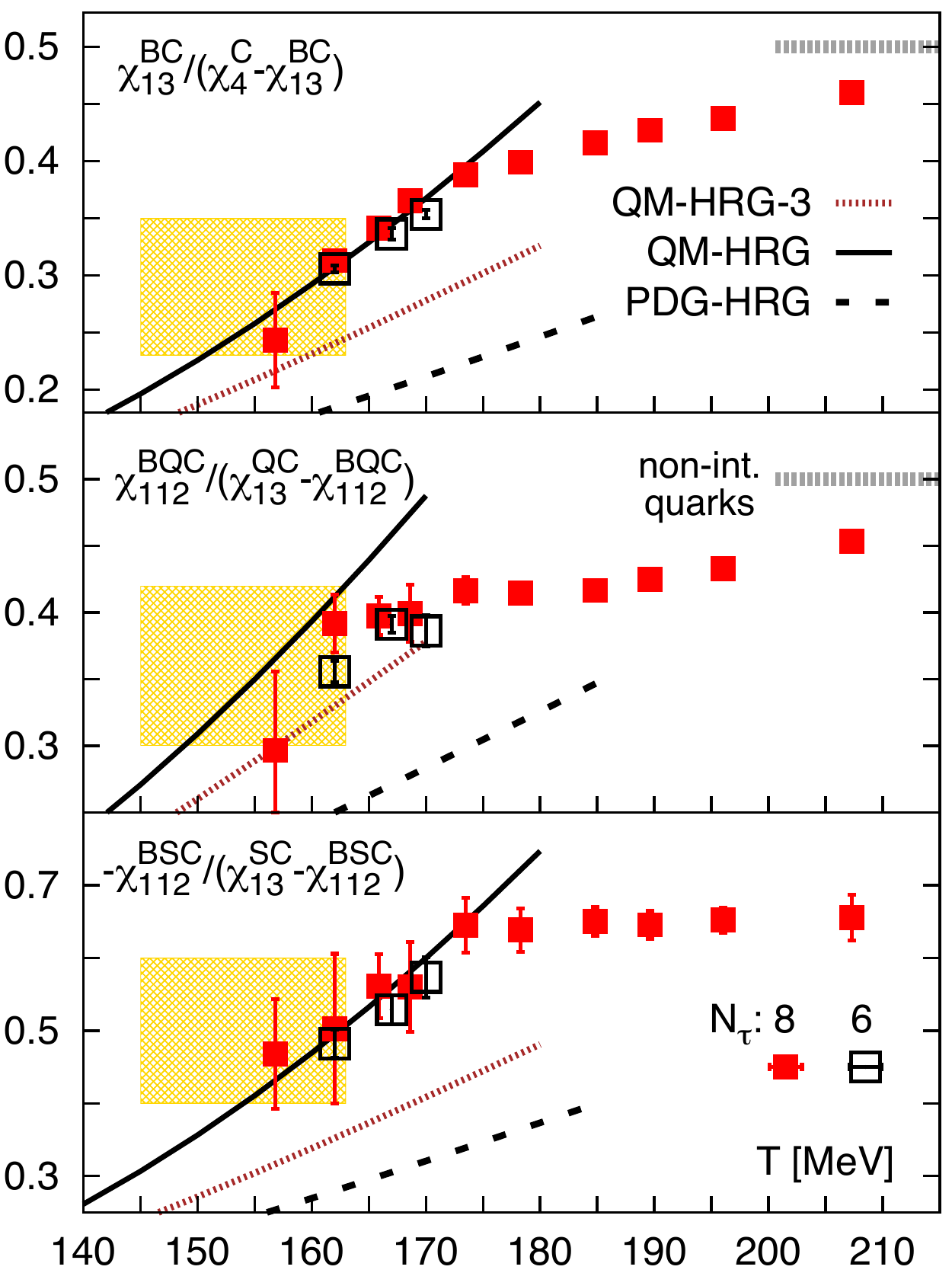}
\end{minipage}
\caption{(Left) Ratios between cumulants of 
correlations of net baryon number with net charm,
net strangeness and net electric charge fluctuations \cite{Bazavov:2014yba}. 
Deviations of these ratios from unity in the crossover region (shaded band) 
suggest that the onset of the melting/deconfinement 
of open charm hadrons as well as open strange hadrons and hadrons with
light up/down quarks 
starts in the same temperature range (see text for details). (Right) Thermodynamic
contributions of all charmed baryons (top), all charged charmed baryons (middle) and
all strange charmed baryons (bottom) relative to that of corresponding charmed mesons
\cite{Bazavov:2014yba}. The dashed lines (PDG-HRG) are predictions for an
uncorrelated hadron gas using only the PDG states \cite{Agashe:2014kda} . The solid
lines (QM-HRG) are similar HRG predictions including also the states predicted by the
quark model \cite{Ebert:2009ua,Ebert:2011kk}. The dotted lines (QM-HRG-3) are the
same QM predictions, but only including states having masses less than 3 GeV.}
\label{fig:deconf}
\end{center}
\end{figure}

Fig.~\ref{fig:deconf} (left)
\cite{Bazavov:2014yba} shows that close to the chiral crossover temperature,
$T_c=154(9)$~MeV, the quantity $\chi^{BC}_{22}/\chi^{BC}_{13}$ starts deviating from unity.
This indicates that beyond this temperature fractionally charged quark-like degrees
of freedom start appearing and open charm hadrons start to melt. The left hand
panel of Fig.~\ref{fig:deconf} also shows a comparison with similar quantities 
involving both baryon-strangeness correlations as well as baryon-electric 
charge correlations. The first of these
quantities is sensitive to the deconfinement of strange quarks and the second 
one is also sensitive to the deconfinement of light up and down quarks. 
It is clear from the figure that the onset of deconfinement for the charm, 
strange as well as the up/down quarks happens in the same chiral crossover 
region. 

The right hand panel of Fig.~\ref{fig:deconf} shows ratios of generalized
susceptibilities that are constructed such that in a gas of 
uncorrelated hadrons the numerator would correspond to the partial pressure
of baryons with quantum numbers selected by the other indices, while 
the denominator would correspond to the partial pressure of corresponding mesons. E.g. 
$\chi_{13}^{BC}$ will give the partial pressure of charmed baryons and
$\chi_4^C-\chi_{13}^{BC}$ filters out the corresponding partial pressure
of charmed mesons. The ratios
 $\chi_{13}^{BC}/(\chi_4^C-\chi_{13}^{BC})$, $\chi_{112}^{BQC}/(\chi_{13}^{QC}-\chi_{112}^{BQC})$ and $-\chi_{112}^{BSC}/(\chi_{13}^{SC}-\chi_{112}^{BSC})$
thus correspond to ratios of partial pressures arising from charmed baryons 
to that from open charm mesons, charged-charmed baryons to open charm charged 
mesons and strange-charmed baryons to strange-charmed mesons, respectively. 
As can be seen in Fig.~\ref{fig:deconf}~(right) results for these ratios at
low temperatures are well described by a hadron gas that uses open charm
resonances obtained in quark model calculations (QM-HRG) while, not
surprisingly, it differs
strongly from a simple hadron resonance gas based only on the very few
experimentally known charmed hadrons (PDG-HRG). This observation is similar
to that made for strange hadron fluctuations in Section~\ref{sec:fluctuation}
(right panel of Fig.~\ref{fig:muS_lqcd}). These observations clearly
provide evidence for contributions from experimentally yet unobserved 
charmed/strange hadrons to charm and strangeness fluctuations in the 
vicinity of the crossover transition temperature. In additions, the
agreement between lattice QCD data and QM-HRG model calculations below
$T_c$ and the onset of deviations just in the transition region also 
suggests that open charm hadrons start to dissociate in the 
crossover region. This conclusion is consistent with that drawn from
the temperature dependence of the ratio in the left hand panel 
of Fig.~\ref{fig:deconf}, which has the advantage of being independent
of details of the charmed hadron spectrum.

\subsection{Heavy Quarkonia}
\label{sec:HeavyQuarkonia}
Inside the QGP the interaction between a heavy quark anti-quark pair gets weakened due to
the screening effects of the intervening deconfined colored medium. This telltale
signature of the presence of a color deconfined medium is expected to be manifested
through melting of heavy quarkonium states \cite{Matsui:1986dk}, i.e. bound states of a 
quark anti-quark pair. Melting of heavy quarkonium states in the QGP is also expected to
follow a distinctive sequential pattern with the smallest, most tightly bound
quarkonium state surviving up to the highest temperature, effectively serving as a
`thermometer' to probe the temperature of the medium created at RHIC (for a recent
review see Ref. \citen{Mocsy:2013syh}). Since quarkonia do not carry flavor quantum
numbers, unlike the open heavy flavor hadrons, their melting cannot be accessed in
lattice QCD calculations through quantum number correlation based studies as discussed in Section~\ref{sec:HQLight}.
Currently, at least three approaches are being actively pursued to study thermal
modifications of heavy quarkonia on the lattice. The first approach is to use
potential models with heavy quark potentials computed on the lattice. The second
approach relies on the extraction of quarkonium spectral functions from Euclidean
temporal correlators. The third approach involves spatial correlation functions of
quarkonia and the study of their in-medium screening properties. Here, we summarize the
recent progresses and developments concerning these three approaches. 
More details about earlier lattice results can also be found in Refs.
\citen{Bazavov:2009us,Brambilla:2010cs,Mocsy:2013syh}. 

\subsubsection{Heavy quark potential}

Properties of heavy quarkonia in the vacuum have been successfully described by the
potential model approach. In this approach, the interaction between a heavy quark
pair forming the quarkonium is described by an instantaneous potential
\cite{Eichten:1978tg,Eichten:1979ms}. Due to its success at zero temperature, the
potential model approach has also been applied at nonzero temperature, by making the
potential between the heavy quarks temperature dependent \cite{Mocsy:2008eg}.  The
temperature dependent potentials used in these calculations are based either on model
calculations or on finite temperature lattice QCD results for the free energy and the
internal energy.  However, justification for the use of these potentials in a 
Schr\"odinger equation is mostly based on phenomenological arguments and  has 
no rigorous connections
to the real-time evolution of heavy quarkonia described by the 
Schr\"odinger equation.

Recently, a non-relativistic effective theory approach at nonzero temperature,
relying on a particular hierarchy of the relevant scales, has been
developed~\cite{Laine:2006ns,Beraudo:2007ky,Brambilla:2008cx}.  Generally by integrating out the hard energy scale, i.e.
the rest mass of the heavy quark $m$, Non-Relativistic QCD (NRQCD) effective theory~\cite{Caswell:1985ui,Bodwin:1994jh} is
obtained, and by further integrating out the soft scale, i.e. the typical momentum
exchange between the bounded quarks of ${\cal O} ( mv)$, so-called
potential non-relativistic QCD (pNRQCD), is obtained~\cite{Brambilla:1999xf}.  
In pNRQCD a heavy quark bound state can be described by a
two-point function satisfying a Schr\"odinger equation~\cite{Laine:2006ns,Beraudo:2007ky,Brambilla:2008cx},
\beq
i\partial_t D^{>}_{NR}(t,\vec{r}) = \left( -\frac{1}{m}\nabla_r^2 + V(t,r) \right) D^{>}_{NR}(t,\vec{r}),
\eeq
where $D^{>}_{NR}(t,\vec{r})$ is the real-time forward heavy quark pair correlation
function in the non-relativistic limit. The potential $V(t,r)$ in such a
Schr\"odinger equation is well defined and turns out to be complex
valued \cite{Laine:2006ns,Beraudo:2007ky,Brambilla:2008cx}. Its real part reflects
color Debye screening effects while the imaginary part is related to
Landau-damping, i.e.  the scattering of quarks with the constituents of the 
medium and the absorption of gluons from the medium via singlet-octet 
transition \cite{Brambilla:2008cx}.  It has been shown in
Ref.~\citen{Rothkopf:2011db} that in the limit $m\to\infty$ the leading part
of the potential $V(t,r)$ can be obtained as 
\beq
V(r) = \lim_{t\rightarrow\infty} \frac{i\partial_t W(t,r)}{W(t,r)} \;,
\label{eq:ComplexV}
\eeq 
where $W(t,r)$ is the real-time thermal Wilson loop. Through a Fourier transformation
one obtains,
\beq
W(t,r)=\int^{+\infty}_{-\infty} \mathrm{d}\omega \, \rho(\omega,r)\ 
{\rm e}^{-i\omega t}
\;,
\eeq
where $\rho(\omega,r)$ is the spectral function of the real-time Wilson loop.  The
above equation can be analytically continued to the Euclidean time, giving
\beq
W(\tau,r)=\int^{+\infty}_{-\infty} \mathrm{d}\omega 
\,\rho(\omega,r) \ {\rm e}^{-\omega \tau}\; .
\label{eq:WilsonSpf}
\eeq
In Eq.~\ref{eq:WilsonSpf} the Wilson loop, $W(\tau,r)$ is the usual 
Euclidean-time Wilson loop which can be obtained from lattice QCD 
calculations.  
As if there exists a well defined lowest lying peak structure in the spectral function $\rho(\omega,r)$ it would
dominate the dynamics in the Wilson loop in the late time limit thus giving to the complete
information of the heavy quark potential as shown in Eq.~\ref{eq:ComplexV}. However, as argued in
Ref.~\citen{Burnier:2012az} the late time dynamics is not completely separated from the early time non-potential
physics and to take these effects into account a skewed Lorentzian function including the early time contribution
for $\rho(\omega,r)$ is required to fit the Wilson loop.
The position and width of this skewed Lorentzian peak are connected to the real and the imaginary parts
of the potential, respectively. Thus, by extracting the spectral function,
$\rho(\omega,r)$, from the Euclidean-time Wilson loop\footnote{As illustrated 
in Ref.~\citen{Burnier:2013fca} in order to avoid the cusp divergence and 
the large noise to signal ratio in computations of Wilson loops, 
some other observables, e.g. Wilson line correlation functions defined in the 
Coulomb gauge, can be a substitution and are actually used in lattice 
calculations.}, $W(\tau,r)$, obtained from
lattice QCD calculations and by subsequently fitting this $\rho(\omega,r)$ with a
skewed Lorentzian ansatz the real and the imaginary part of the heavy quark potential,
$V(r)$, can be obtained. 

\begin{figure}[t]
\begin{center} 
\includegraphics[width=0.44\textwidth]{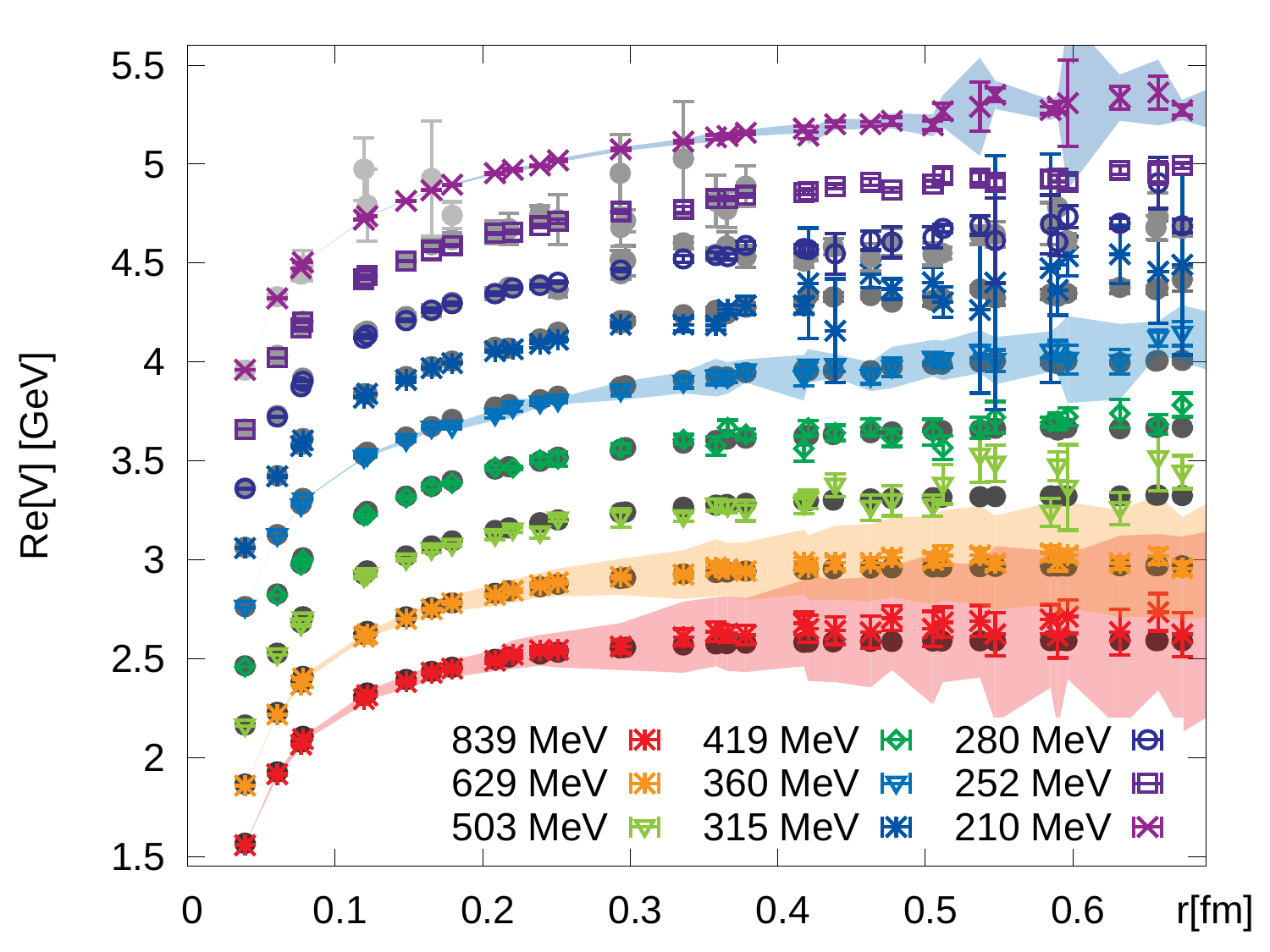}
\includegraphics[width=0.44\textwidth]{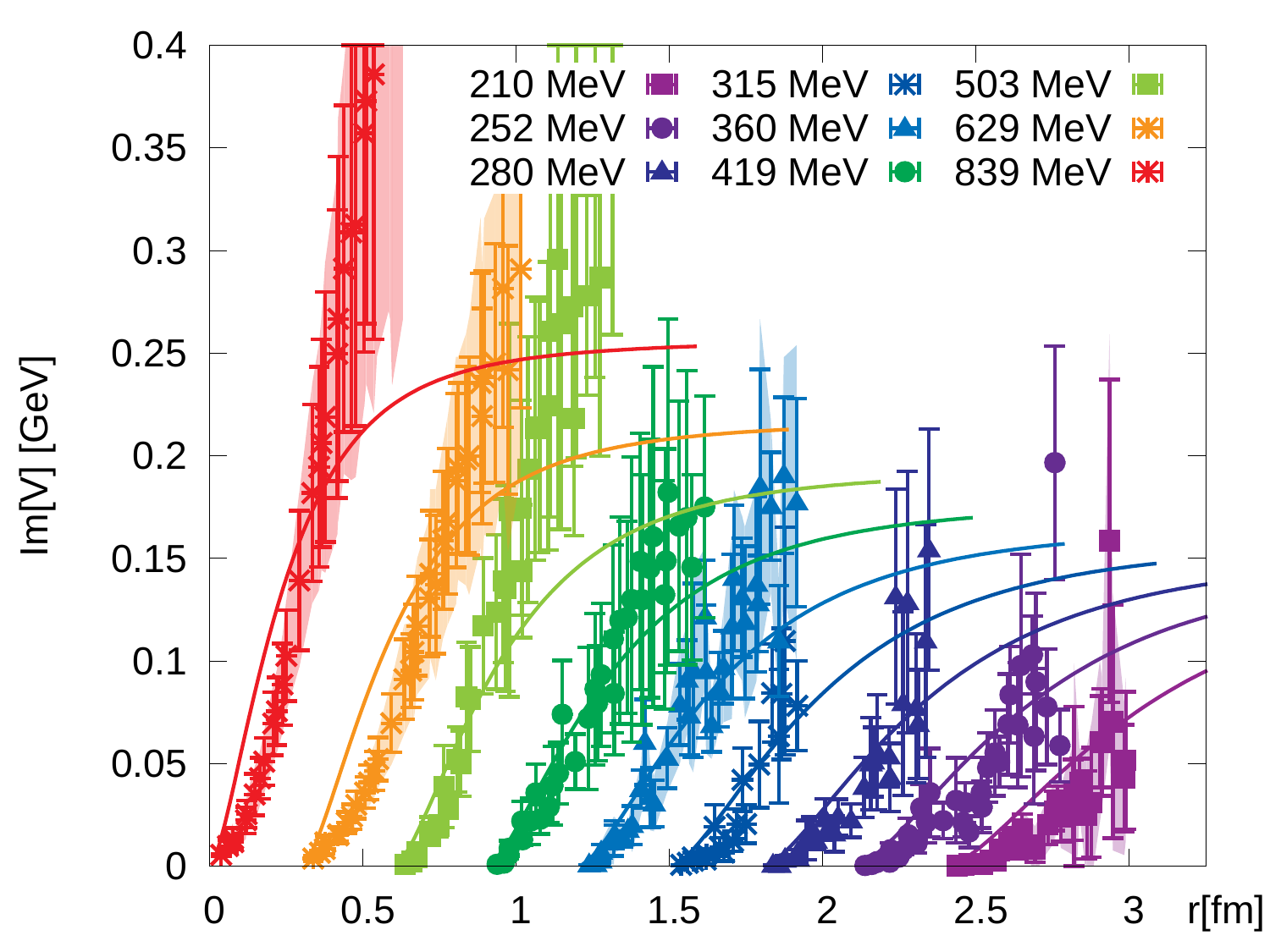}
\caption{Real (left) and imaginary (right) parts of the heavy quark potential 
calculated in 
quenched lattice QCD \cite{Burnier:2014ssa}. Values are shifted for better visibility.
Gray circles denote color singlet free energies and solid lines are leading order
hard thermal loop results.}
\label{fig:HQV_quenched}
\end{center}
\end{figure}

This procedure has been tested using the Wilson loop calculated using hard thermal
loop perturbation theory \cite{Burnier:2013fca} as well as for the pure gauge theory~\cite{Burnier:2014ssa}. Fig.~\ref{fig:HQV_quenched} shows the real (left) and
imaginary (right) parts of the heavy quark potential in the pure gluonic medium
\cite{Burnier:2014ssa}. The real part of the heavy quark potential turns out to be
close to the heavy quark singlet free energy, defined through the spatial correlation
function of the Polyakov loop and its conjugate \cite{Nadkarni:1986as}, at all
temperatures. At high temperatures the imaginary part of the heavy quark potential is
close to the leading order hard thermal loop result, but at low temperatures it 
lies below hard thermal loop result.

Using the same procedure the real part of the heavy quark potential has also
been extracted in (2+1)-flavor QCD \cite{Burnier:2014ssa} on lattices with
temporal extent of only $N_\tau=12$. 
Of course, on such small lattices it is hardly possible to extract reliable 
information on the imaginary part of the heavy quark potential. Nonetheless,
similar to the quenched limit result the real part
of the heavy quark potential is found to be close to the singlet free 
energies. However, string breaking is not observed at $T<T_c\approx174$ MeV up 
to $\sim1.2$ fm. This could be due to the large pion mass 
($m_\pi\approx300$ MeV) employed in this exploratory QCD calculation.

The heavy quark potential has also been extracted from the Wilson line 
correlation function calculated on $48^3\times12$ lattices at $T = 250$~MeV 
and $305$~MeV using a fit ansatz motivated by hard thermal loop 
calculations~\cite{Bazavov:2014kva}. These calculations used
HISQ action with quark masses that would correspond to $m_\pi\approx160$~MeV 
in the continuum limit. In this calculation
the real part of the potential was found to be equal or larger than the 
singlet free energies at these two temperatures 
and it always was smaller than the zero temperature potential. 
Moreover, the imaginary part is found to be of similar size as predicted by
leading order hard thermal loop perturbation theory \cite{Bazavov:2014kva}.

\subsubsection{Spectral functions of quarkonia}

The spectral function of quarkonium can be extracted from lattice QCD
calculations of the two-point correlation functions of the heavy quark
currents in the Euclidean time
\beq
G(\tau,T) = \int d^3\vec{x} \left\langle 
\Big( \bar{q}(\tau,\vec{x}) \,\Gamma \,q(\tau,\vec{x}) \Big)
\Big(\bar{q}(0,\vec{0})\, \Gamma \,q(0,\vec{0}) \Big)^\dagger \right\rangle
\;,
\eeq
where $q$ represents the heavy quark field. $\Gamma$ are Dirac matrices 
defining the spin structure of quarkonium, in particular
$\Gamma=1,\gamma_5,\gamma_\mu,\gamma_5\gamma_\mu$ corresponds to the scalar,
pseudo-scalar, vector and axial-vector quarkonium states, respectively. Signatures of
medium modification and melting of quarkonium are reflected in the structure of
the  spectral density $\rho(\omega,T)$. 
It can be obtained from the quarkonium correlator in the Euclidean-time
\beq
G(\tau,T) = \int_0^\infty \frac{d\omega}{2\pi} \rho(\omega,T)
\frac{\cosh\left(\omega\left(\tau-1/2T\right)\right)}{\sinh\left(\omega/2T\right)}
\;.
\label{eq:temporal}
\eeq
While $\rho(\omega,T)$ is a continuous function of the frequency $\omega$ over an
infinite range,  $G(\tau,T)$ is calculable on the lattice only at a finite
number of discrete points along the temporal direction due to the finiteness 
of the lattice. 
As has been discussed in Section~\ref{sec:transport} 
inverting Eq.~\ref{eq:temporal} to get spectral function $\rho(\omega,T)$
from the correlation function $G(\tau,T)$ is a typical ill-posed problem 
and a suitable inversion method is needed to solve the problem.
The commonly used inversion methods for the extractions of quarkonium spectral
functions is the Maximum Entropy Method (MEM) \cite{Asakawa:2000tr}. Variants 
of this have recently been suggested which include a modified MEM with an 
extended search space \cite{Rothkopf:2011ef,Rothkopf:2012vv} and a Bayesian 
method which is analogous to MEM but uses a different prior distribution 
\cite{Burnier:2013nla}.

\begin{figure}[t]
\begin{center}
\includegraphics[width=0.45\textwidth]{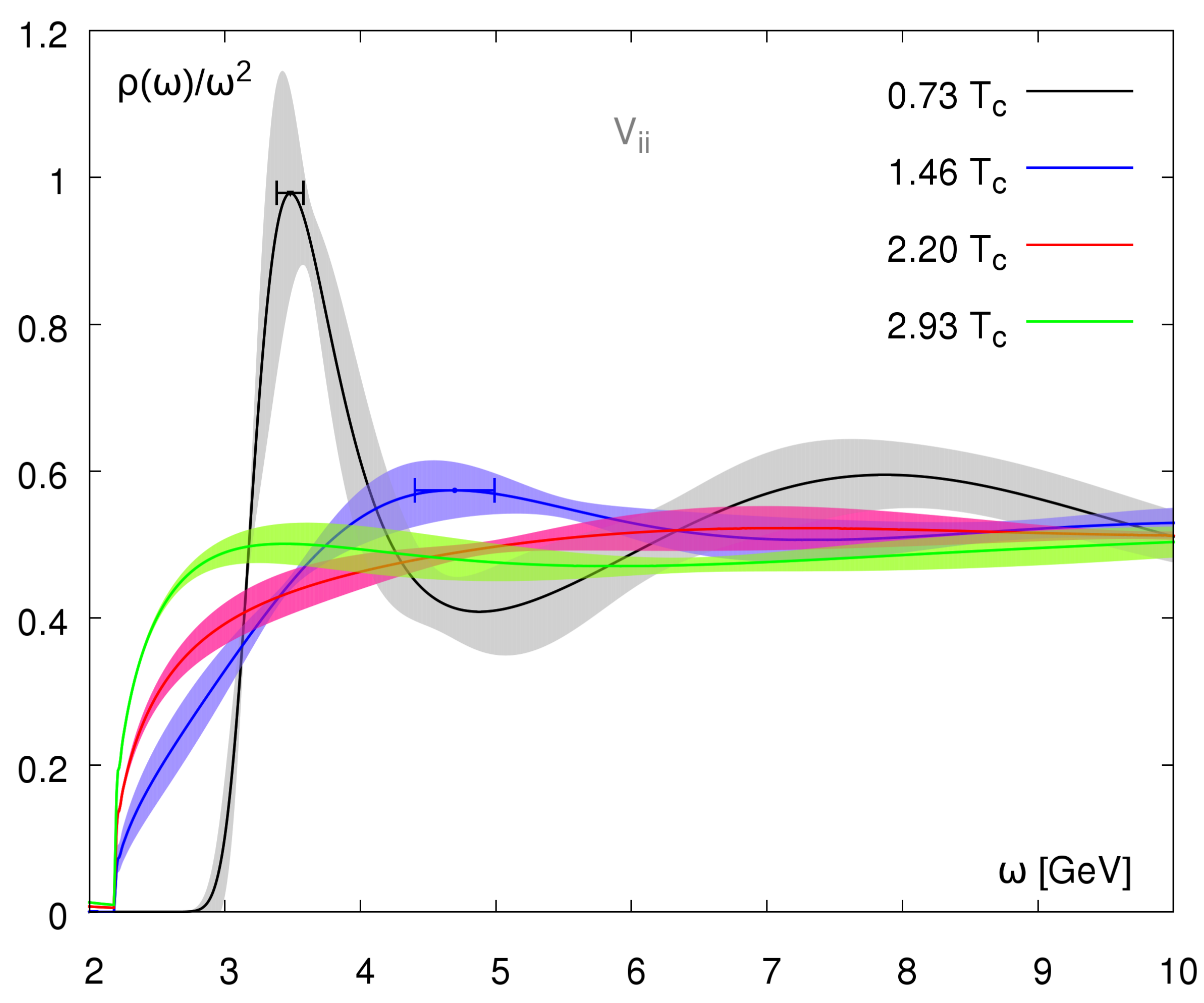}
\includegraphics[width=0.45\textwidth]{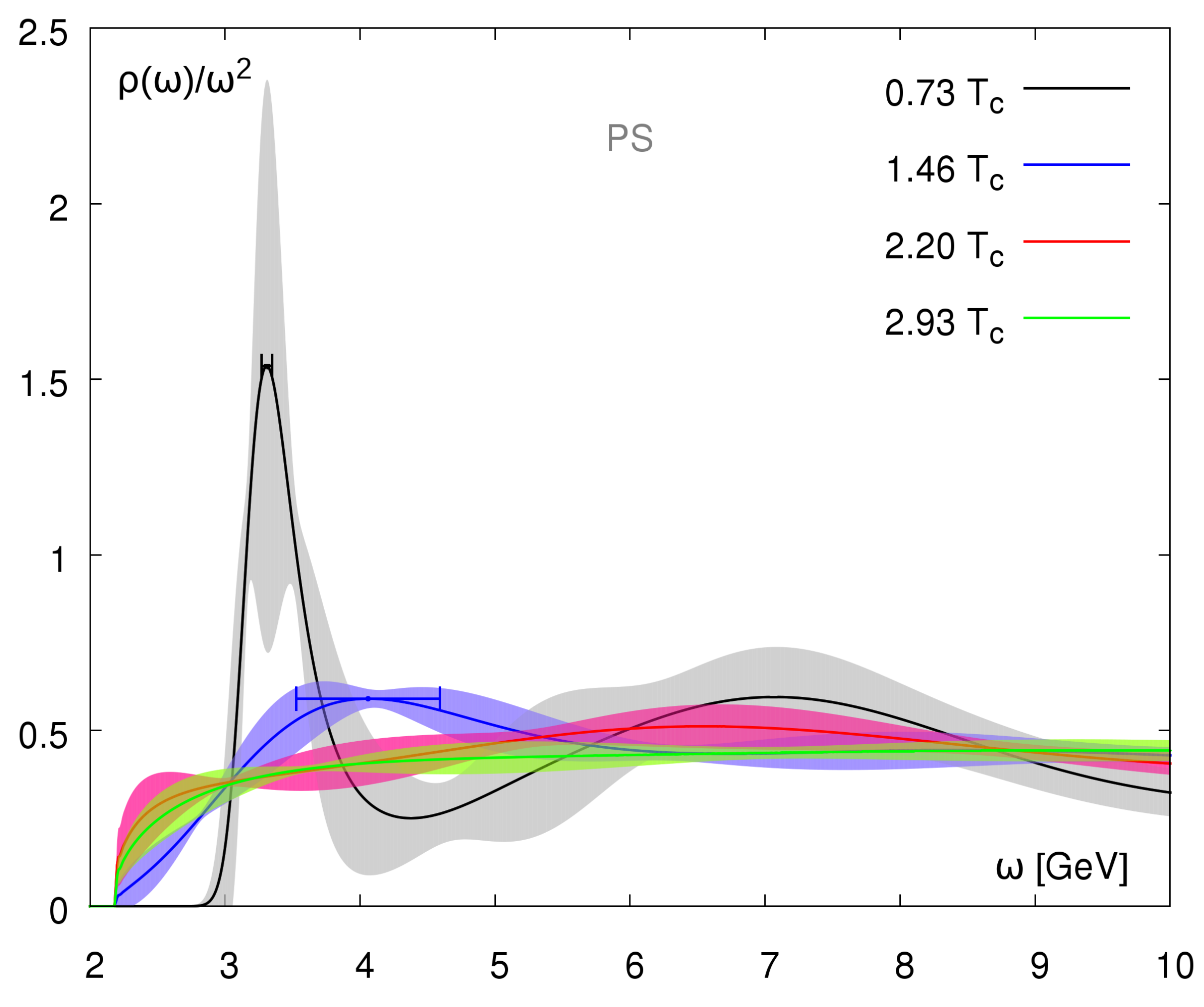}
\caption{Temperature dependence of $J/\Psi$ (left) and $\eta_c$ (right) 
spectral functions obtained from quenched lattice QCD calculations 
\cite{Ding:2012sp}.}
\label{fig:charmonium1}
\end{center}
\end{figure}  

The charmonium spectral functions at finite temperature have been studied extensively
for quenched lattice QCD\cite{Asakawa:2003re,Datta:2003ww,Jakovac:2006sf,Iida:2006mv,Ohno:2011zc}. Recently
quenched lattice QCD calculations \cite{Ding:2012sp} have been performed using
lattices with large temporal extents, leading to quite reliable extractions of
the charmonium spectral functions. Fig.~\ref{fig:charmonium1} illustrates the spectral
functions of $J/\Psi$ (left) and $\eta_c$ (right) for various temperatures.  
These studies suggest that all charmonium states are dissociated for 
$T\gtrsim1.5T_c$ in a
gluonic plasma \cite{Ding:2012sp}. Investigations on charmonium spectral functions in
lattice QCD calculations with dynamical quarks lead to consistent conclusions
\cite{Aarts:2007pk,Borsanyi:2014vka}.

\begin{figure}[t]
\begin{center}
~~\includegraphics[width=0.75\textwidth]{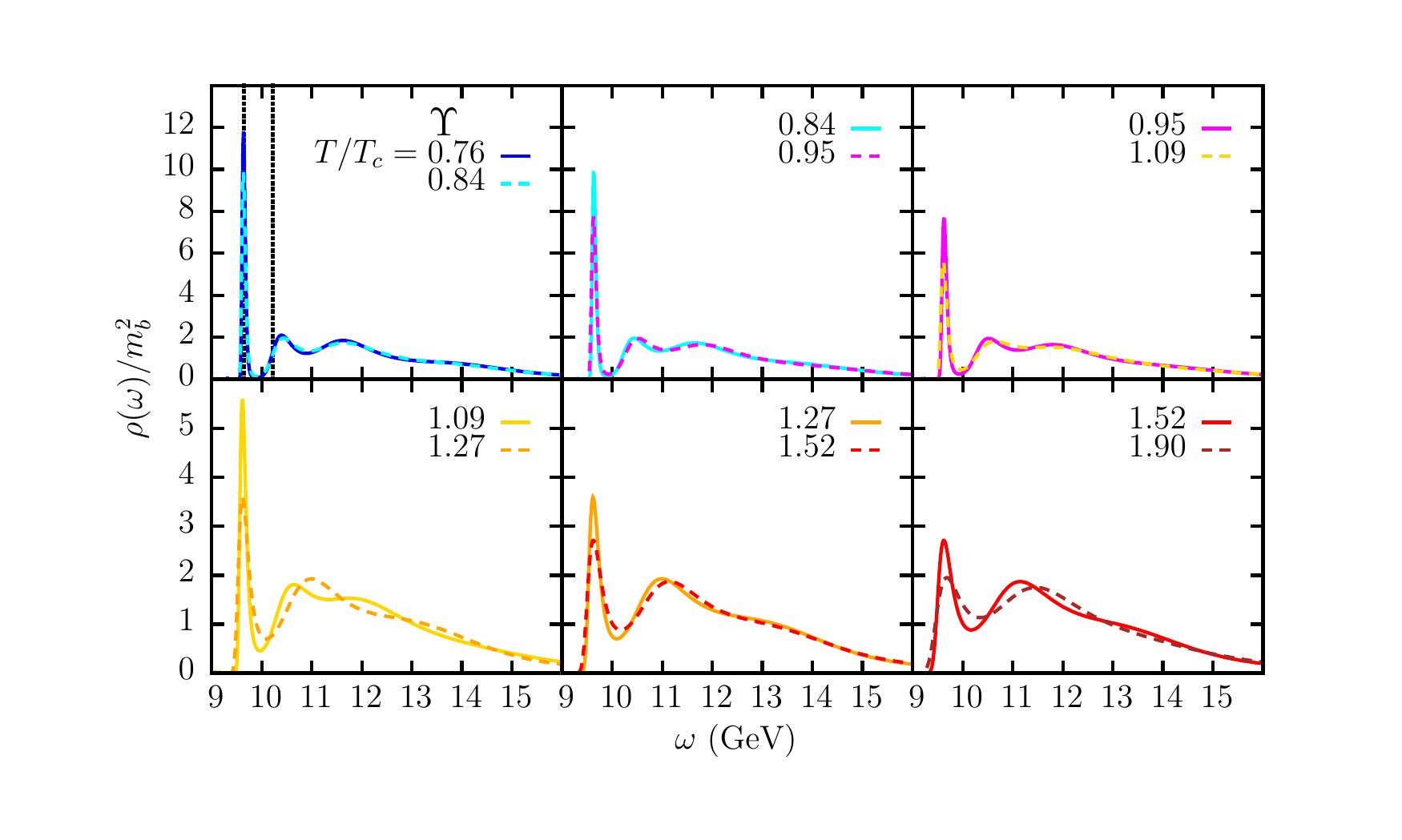}\vspace{-0.05cm}
\includegraphics[width=0.774\textwidth]{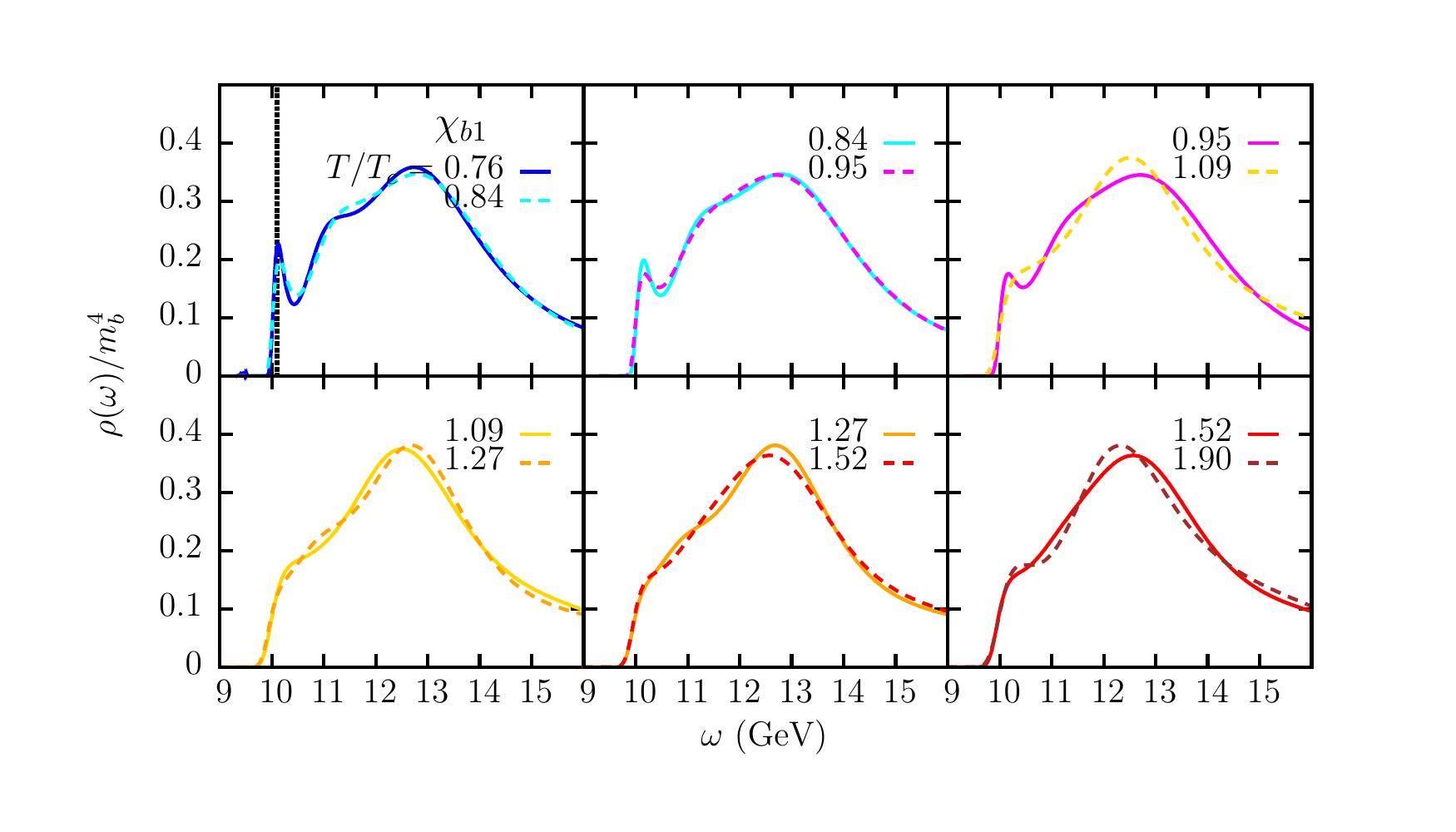}
\caption{Temperature dependence of spectral functions for the $\Upsilon$ 
(top)  and
$\chi_{b1}$ states (bottom). Results are obtained by using a 
lattice discretized form of NRQCD \cite{Aarts:2014cda}.}
\label{fig:NRQCD1}
\end{center}
\end{figure}

Lattice QCD calculations for bottom quarks are technically challenging since 
present computational limitations do not permit calculations with sufficiently 
fine lattice spacings that are much smaller than the inverse of the heavy 
bottom quark mass. Large discretization errors arise from the large
heavy-quark bare mass in units of the lattice spacing, $m_ba$. 
Nonetheless, although it is computationally very demanding relativistic 
bottomonium correlations functions have been computed recently in quenched 
QCD using isotropic lattices with $N_\sigma$=192 and large temporal extent \cite{Ohno:2014uga},
$N_\tau =96$ and $48$ at the same value of the lattice spacing,
$a\simeq0.01$~fm, i.e. $m_ba\simeq 0.2$.
The temporal lattice sizes correspond to temperatures
$T=0.7\ T_c$ and $1.4\ T_c$. In this calculation it has been found
that at $1.4~T_c$ only the bottomonium correlation function in the vector channel show
significantly smaller thermal modifications than the charmonium correlation function in the 
same channel \cite{Ohno:2014uga}.

To circumvent the problem of large cut-off effects in calculations with
bottom quarks without performing calculations at extremely small lattice
spacings, 
one needs to adapt a lattice discretization scheme for  
heavy quarks that is capable to describe heavy quark physics
in some chosen kinematic regime also at moderately small values of the cut-off. 
One such approach is Non-Relativistic QCD
(NRQCD) \cite{Thacker:1990bm,Lepage:1992tx} where the heavy quark mass is 
assumed to be much larger than the inverse lattice spacing, but the 
momentum dependence of the heavy quark energy is included in the 
non-relativistic limit. NRQCD
does not possess a proper continuum limit because the radiative corrections
to coefficients of the NRQCD Lagrangian diverge as $m_ba\to0$. However,
since the non-relativistic quark field is not compactified along the 
Euclidean-time direction, the relation between Euclidean-time quarkonium 
correlation functions and the corresponding spectral functions becomes 
simpler,
\be
G(\tau,T) = \int_0^{\infty}\frac{\md\omega}{2\pi}\, \rho(\omega ,T)\
{\rm e}^{-\omega \tau}  \;,
\label{cor_spf_relation_nrqcd}
\ee
giving rise to a temperature independent integration kernel, 
$\exp(-\omega\tau)$. Moreover, the transport peak is also absent in the 
NRQCD spectral function, which makes its determination much easier.    

Treating the bottom quark within the NRQCD framework the fate of bottomonium states
has been studied on anisotropic lattices using $N_f=2$ and $2+1$ of
improved Wilson quarks with $m_\pi\approx~400$ MeV and $m_\pi\approx~500$ MeV,
respectively
\cite{Aarts:2010ek,Aarts:2013kaa,Aarts:2012ka,Aarts:2011sm,Aarts:2014cda}.  
Fig.~\ref{fig:NRQCD1} shows the temperature dependence of the $\Upsilon$ (top) and
$\chi_{b1}$ (bottom) spectral functions extracted using the MEM \cite{Aarts:2014cda}.
This indicates that the S-wave ground state $\Upsilon$ survives up to $1.9\ T_c$ while
the P-wave ground state $\chi_{b1}$ melts just above $T_c$
\cite{Aarts:2014cda}. However, a recent lattice QCD study \cite{Kim:2014iga} that
uses a different Bayesian method suggested 
in Ref.~\citen{Burnier:2013nla} for the extraction of the spectral functions finds that the P-wave ground state $\chi_{b1}$ may 
survive up to $1.6\ T_c$. This analysis, however, has been performed on 
lattices with a rather small temporal extents, $N_\tau=12$.

\subsubsection{Spatial correlation functions of quarkonia}

Similarly as the temporal correlation functions spatial quarkonium correlation functions are defined as the sum of heavy quark current-current correlation functions instead 
over a `funny' space of $\tau$, $x$ and $y$
\beq
G(z,T) =  \int_0^{1/T} \mathrm{d}\tau \int \mathrm{d}x\, \mathrm{d}y \left\langle 
\Big( \bar{q}(\tau,\vec{x}) \,\Gamma \,q(\tau,\vec{x}) \Big)
\Big( \bar{q}(0,\vec{0}) \,\Gamma \,q(0,\vec{0}) \Big)^\dagger \right\rangle
\;,
\eeq
are also related to the quarkonium spectral functions as follows
\be
G(z,T) = \int_{-\infty}^{\infty} \frac{\mathrm{d}p_z}{2\pi}\, {\rm e}^{ip_z z} \int_{0}^{\infty} \frac{\mathrm{d}\omega}{2\pi}\, \frac{\rho(\omega,p_z, T)}{\omega}.
\label{eq:spatial}
\ee
The spatial correlation functions, however, are sensitive also to the 
momentum dependence of spectral functions which is reflected in an additional
integration over the momentum $p_z$ along the $z$-direction.
Despite of its more complicated connection to the spectral function compared 
to that of temporal correlators (see Eq. \ref{eq:temporal}), the spatial 
correlation function has some distinct advantages.  Since the spatial 
separation is not limited by the inverse
temperature, the spatial correlation function can be studied at separations larger
than $1/T$ and thus is more sensitive to in-medium modifications of quarkonia. 
Furthermore, in contrast to
the temporal correlation functions, the spatial correlation functions can be directly
compared to the corresponding vacuum correlation function to quantify modifications
of the in-medium spectral function as
the entire temperature dependence of spatial correlation functions 
emerges from the temperature dependence of the spectral function (Eq.~\ref{eq:spatial}), while in
the case of temporal correlation functions (Eq.~\ref{eq:temporal}) 
this temperature dependence is folded with the 
temperature dependence of the kernel,
$\cosh\left(\omega\left(\tau-1/2T\right)\right)/\sinh\left(\omega/2T\right)$. 

While the relation between spectral functions and spatial meson correlators
is more involved, in some limiting cases it becomes more intuitive. At large
distances the spatial correlation functions decay exponentially, $G(z,T) \sim
\exp(-M(T)z)$, characterized by the screening mass $M$. At low 
temperatures this exponential drop of a mesonic correlator projecting on
a given quantum number channel $\Gamma$ is controlled by the bound state 
with smallest mass in this channel. The  
spectral function then is dominated by this state,
\be
\rho(\omega,p_z,T)\sim\delta(\omega^2-p_z^2-m_0^2),
\ee
where $M$ becomes equal to the pole mass $m_0$ of the meson. At high temperatures, when the mesonic
excitations have completely dissolved and become uncorrelated, the spatial 
meson correlation functions describe the propagation of a free quark 
anti-quark pair and the screening mass is given by~\cite{Florkowski:1993bq}
\be
M_{\rm free}=\sqrt{m_{q_1}^2+(\pi T)^2} + \sqrt{m_{q_2}^2+(\pi T)^2},
\ee
where $m_{q_1}$ and $m_{q_2}$ are the masses of the quark
and anti-quark that form the meson. This form of the screening mass in the
non-interacting limit is a consequence of the fermionic (anti-periodic) boundary
condition along the Euclidean-time direction, which leads to the appearance of a smallest
non-zero Matsubara frequency, $\pi T$, in the quark and anti-quark propagators.  As
the bosonic meson state is dissolved in the non-interacting limit the screening mass
reflects two independently propagating fermionic degrees of freedom. Thus, the
transition between these two limiting values for the screening mass can be used as an
indicator for the thermal modification and eventual dissociation of charmonia and open
charm mesons.

\begin{figure}[t]
\begin{center}
\includegraphics[width=0.47\textwidth]{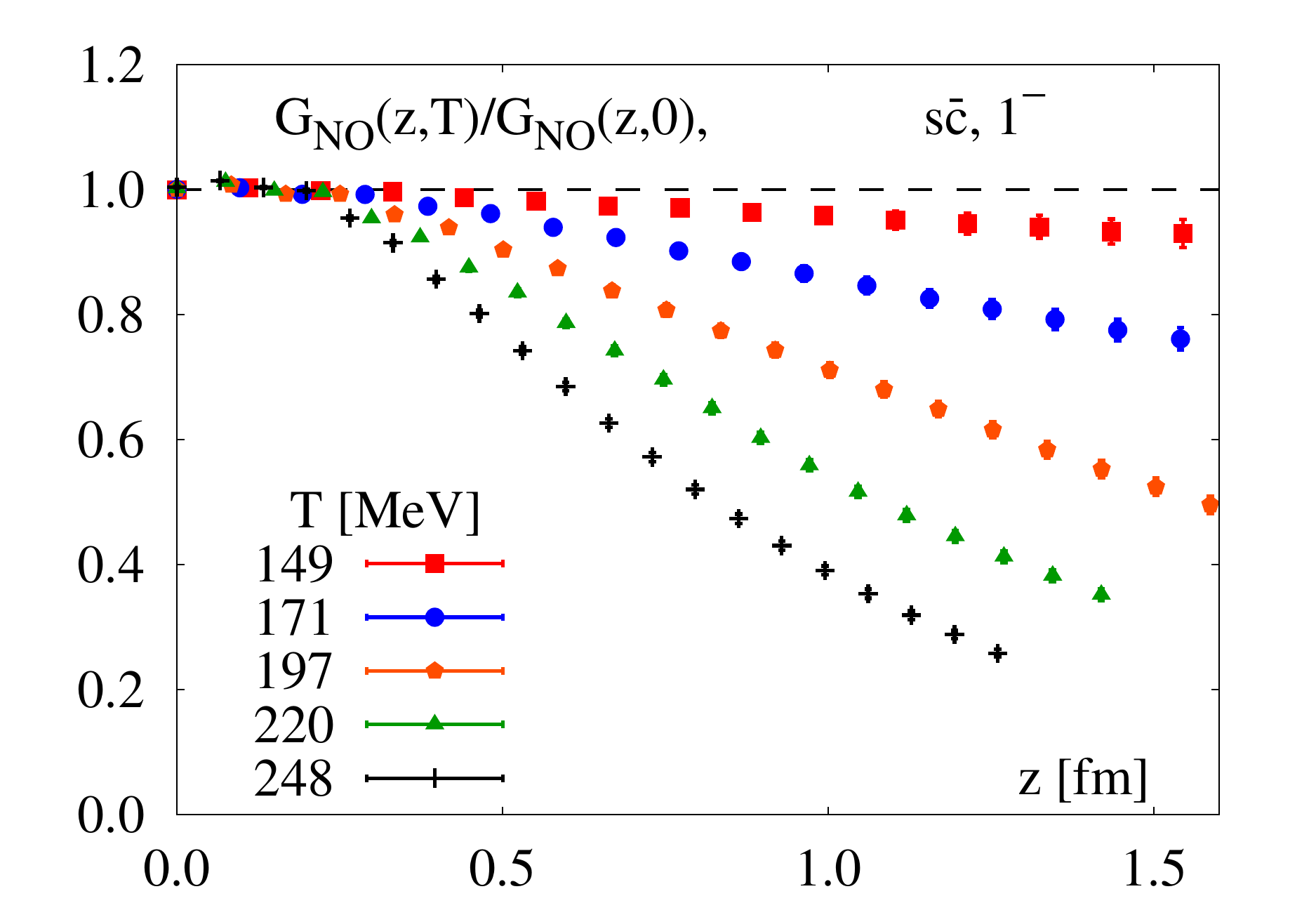}
\includegraphics[width=0.47\textwidth]{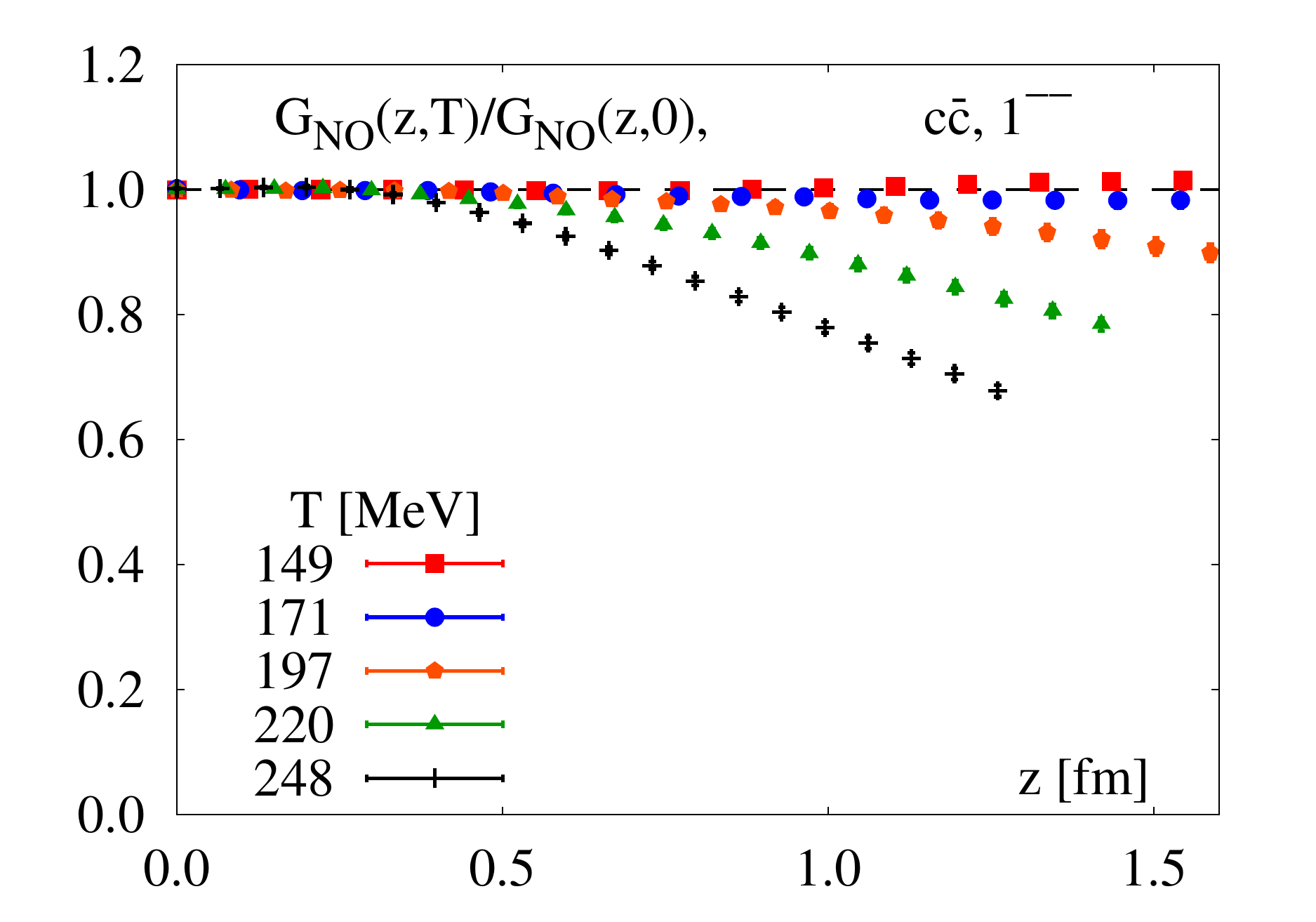} \\
\includegraphics[width=0.47\textwidth]{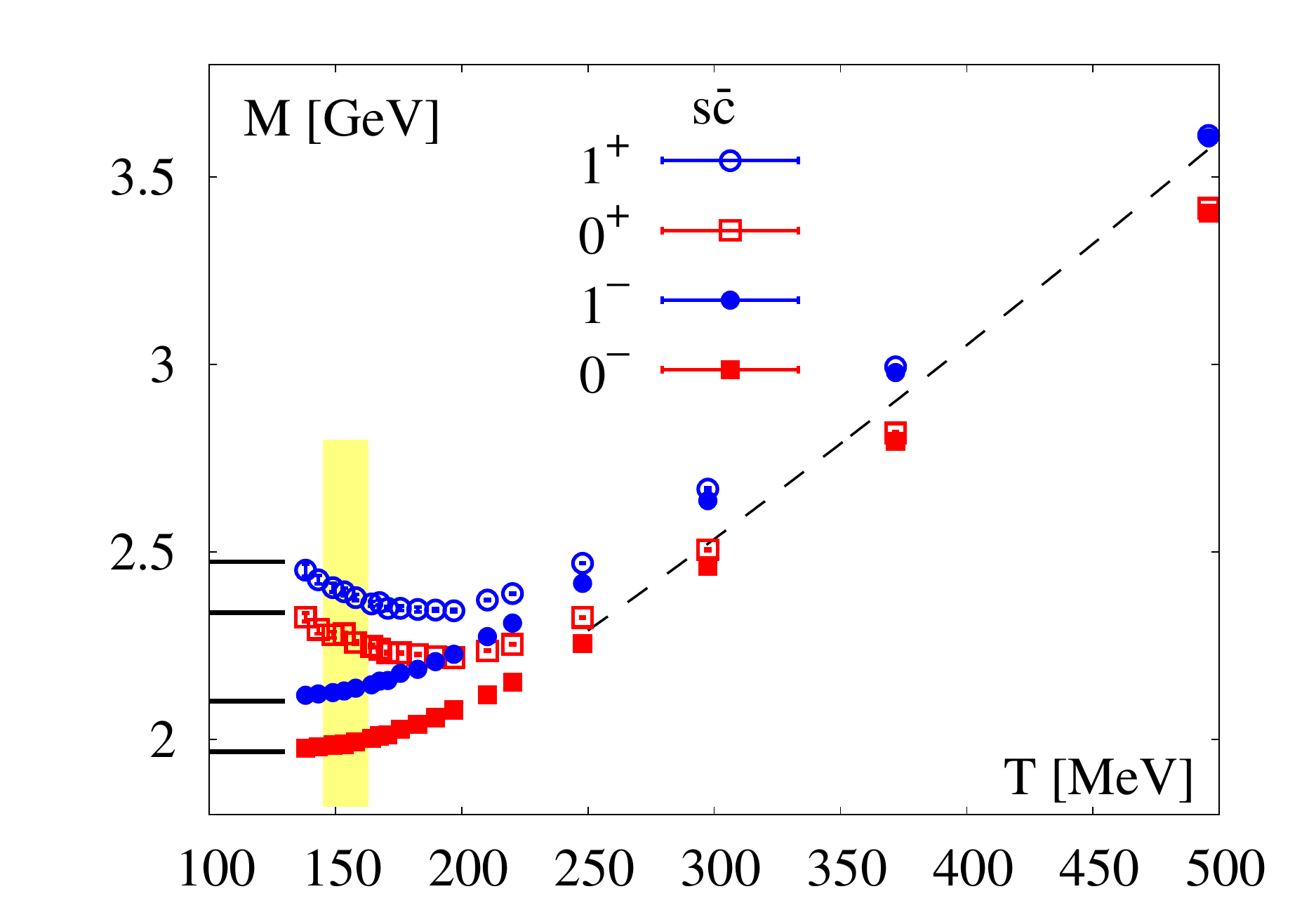} 
\includegraphics[width=0.47\textwidth]{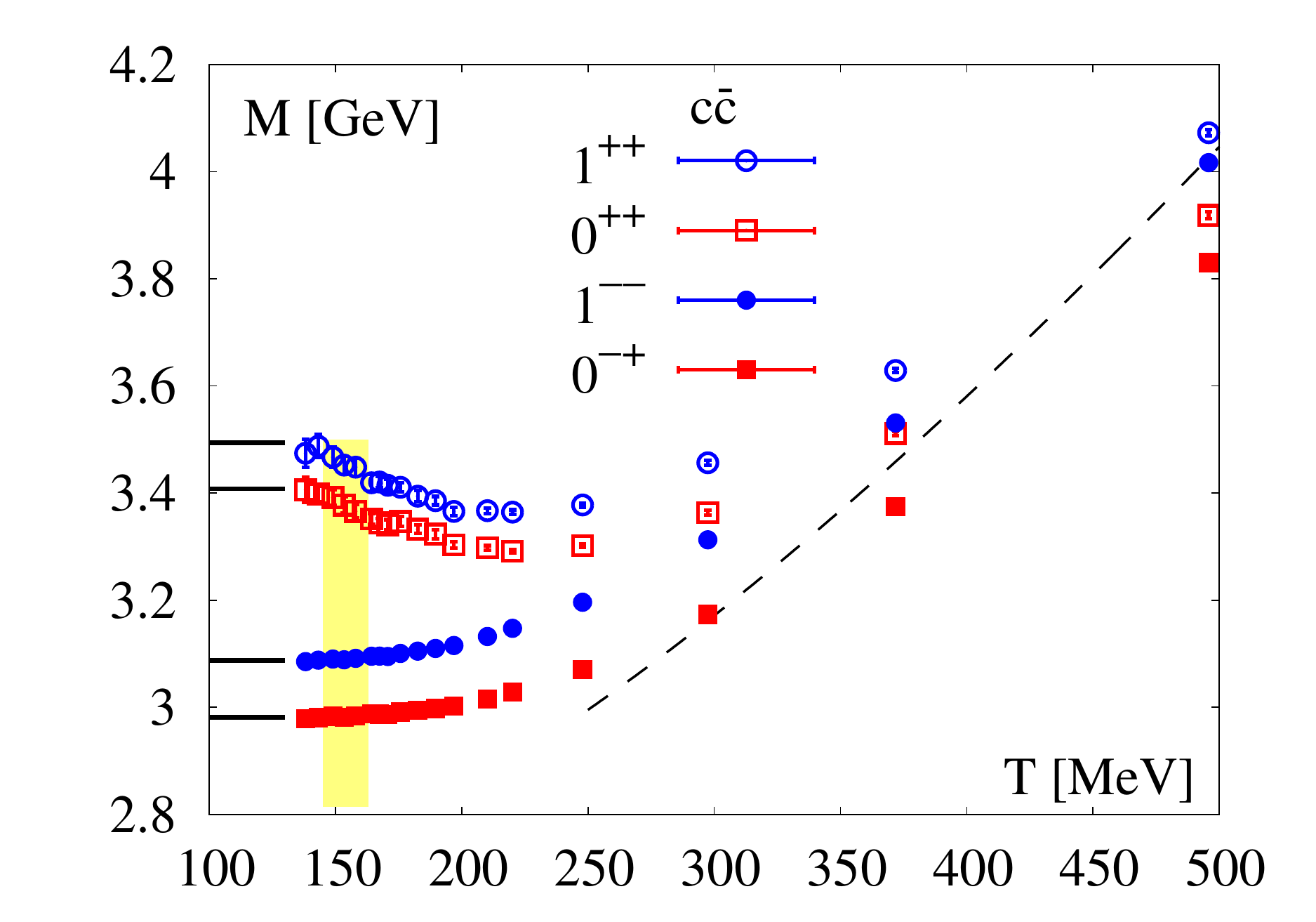}
\caption{(Top) Temperature dependence of ratios of spatial correlation 
functions to those at zero temperature
for $D_s^\ast$ (top left) and $J/\Psi$ (top right)\cite{Bazavov:2014cta}. 
(Bottom) Temperature dependence of screening masses of ground state 
$s\bar{c}$ mesons (left) $D_{s1}^\ast$ ($1^{+}$),
$D_{s0}^\ast$ ($0^{+}$), $D_{s}^\ast$ ($1^{-}$) and $D_{s}$ ($0^{-}$) and
ground state charmonia (right)
$\chi_{c_1}$ ($1^{++}$), $\chi_{c_0}$ ($0^{++}$), $J/\Psi$ ($1^{--}$) and $\eta_c$ ($0^{-+}$)~\cite{Bazavov:2014cta}. The horizontal solid
lines denote the corresponding zero-temperature meson masses and the dashed line
depicts the non-interacting theory limit of a freely propagating  
quark anti-quark pair. The yellow band denotes the chiral crossover temperature region, i.e. $T_c=154\pm9$ MeV.}
\label{fig:ratn}
\end{center}
\end{figure} 

A first analysis of spatial correlation functions of charmonia was
performed using staggered fermions with the p4 action \cite{Karsch:2012na}. 
Recently, a detailed comparative study of
the spatial correlation functions of charmonia and open charm mesons was performed
using the HISQ action \cite{Bazavov:2014cta}. Fig.~\ref{fig:ratn} shows ratios of
the spatial correlation functions for the $D_s^*$ (top left) and the $J/\Psi$ (top right) at
different temperatures to the corresponding zero temperature results. As discussed
before, such a ratio can directly probe the thermal modifications of the spectral
functions. If there is no change in the meson spectral functions, these ratios will
be equal to unity and deviations from unity will indicate in-medium modification of the
meson spectral functions at non-zero temperature. As can be seen, even for $T\sim
T_c\approx154$ MeV the ratio for the open charm sector (Fig.~\ref{fig:ratn}~(top left)) shows 
significant deviation from unity. This
is completely consistent with the previously discussed quantum number correlation
based studies in Section~\ref{sec:HQLight} which concluded that open charm 
mesons start to melt in the
vicinity of $T_c$~\cite{Bazavov:2014yba}. On the other hand, for the charmonia 
the ratio of non-zero to zero temperature correlation functions remains 
very close to unity for $T\lesssim170$~MeV and significant 
deviations, comparable to the open charm sector at $T\sim T_c$, only show up 
for $T\gtrsim200$~MeV (Fig.~\ref{fig:ratn}~(top right)). This indicates
that up to $T\simeq 1.1\ T_c$ $J/\Psi$ exists almost like a vacuum state and
significant thermal modifications appears only for $T\gtrsim1.3T_c$. 
The same physics is also reflected in temperature dependence of the screening 
masses of open charm mesons and charmonia as shown in Fig.~\ref{fig:ratn}
(bottom). The screening masses of all open charm mesons (bottom left)
start to deviate from unity significantly already in the chiral crossover 
region. The $J/\Psi$ and $\eta_c$ screening masses (bottom right), 
however, remain to be almost identical to their vacuum pole masses up to 
$T\simeq 1.1\ T_c$. The screening masses for $\chi_{c_0}$
and $\chi_{c_1}$ deviate from their vacuum masses even below $T_c$ indicating
significant thermal modifications of these states. At higher temperatures, 
i.e. $T\gtrsim2\ T_c$ the screening masses in all channels increase linearly 
with temperature. This rise agrees well with that of a freely propagating quark anti-quark pair.

\section{QCD in external magnetic fields}
\label{sec:externalB}

In recent past it has been realized that large magnetic fields created during
the early stages of heavy-ion collisions may give rise to fascinating 
observable effects induced through
the coupling between the magnetic field and the chiral anomaly
\cite{Kharzeev:2007jp,Kharzeev:2012ph}. This observation also motivated a 
plethora of
activities including lattice QCD studies of hot-dense strong-interaction matter under
the influence of external magnetic fields. This section presents a very brief summary
of these lattice QCD studies. More comprehensive reviews on this topic can be found
in Refs. \citen{DElia:2012tr,Yamamoto:2012bi,Endrodi:2014vza,DElia:2015rwa}.

An external magnetic field along the $z$-direction, $\vec{B}=B\hat{z}$, can be
induced by choosing electro-magnetic gauge fields: $A_{\hat y}=Bx$ and $A_{\hat
x}=A_{\hat z}=A_{\hat\tau}=0$.  On the lattice this can be implemented simply by
multiplying the $SU(3)$ gauge field variables $U_{n,\hat{\mu}}$ with the 
corresponding $U(1)$ phase factors:
$u_{n,{\hat y}}=e^{ia^2qBn_x}$ and $u_{n,{\hat x}}=u_{n,{\hat
z}}=u_{n,{\hat\tau}}=1$. Here, $q$ is the (electric) charge of a quark and
$n=(n_x,n_y,n_z,n_\tau)$ denotes a lattice site, with $n_{x,y,z}=1\;\dots\;N_\sigma$ and
$n_\tau=1\;\dots\;N_\tau$. This choice ensures that the magnetic flux, $a^2B$, is
constant through all the plaquettes in the $x$-$y$ plane, except at the boundary
$(N_\sigma,n_y,n_z,n_\tau)$ owing to the periodic boundary condition on the gauge fields
along the spatial directions. Thus, to preserve the smoothness of the external
magnetic field across the boundary and the gauge invariance of the fermion action
the $U(1)$ factor on the boundary links must be modified: 
$u_{(N_\sigma,n_y,n_z,n_\tau),{\hat x}}=e^{-ia^2qN_\sigma Bn_y}$. 
By making such a choice one pays the price that the
magnetic flux becomes quantized, $a^2B=2\pi n_B/(qN_\sigma^2)$ with
$n_B=0\;\dots\;N_\sigma^2$. 
Thus, at a given temperature $T$ and on a finite volume there is a
minimal value for the constant, external magnetic field $B=2\pi
T^2N_\tau^2/(qN_\sigma^2)$. 
On the other hand, for given temporal extent, $N_\tau$, of the
lattice the maximal value of the magnetic field achieved in this case 
is $B=2\pi T^2 N_\tau^2/q$. 
Most importantly, since the external magnetic field is induced by
multiplying the $SU(3)$ gauge field variables with purely imaginary 
phase factors the fermion determinant remains real and positive, 
allowing direct Monte-Carlo sampling.  

\begin{figure}[t]
\begin{center}
\hspace{-0.2cm}\includegraphics[width=0.48\textwidth]{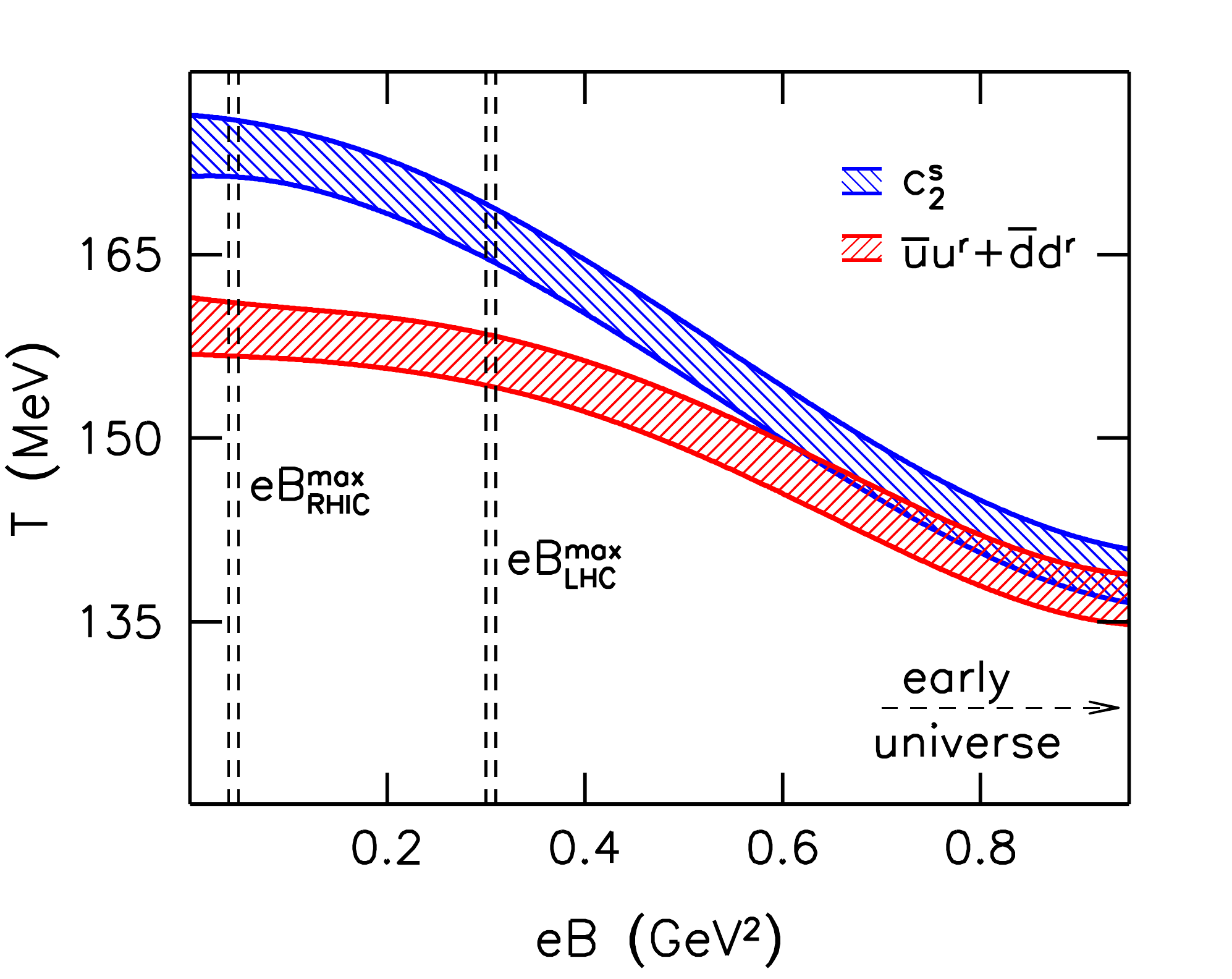}\hspace{-0.2cm}\includegraphics[width=0.51\textwidth]{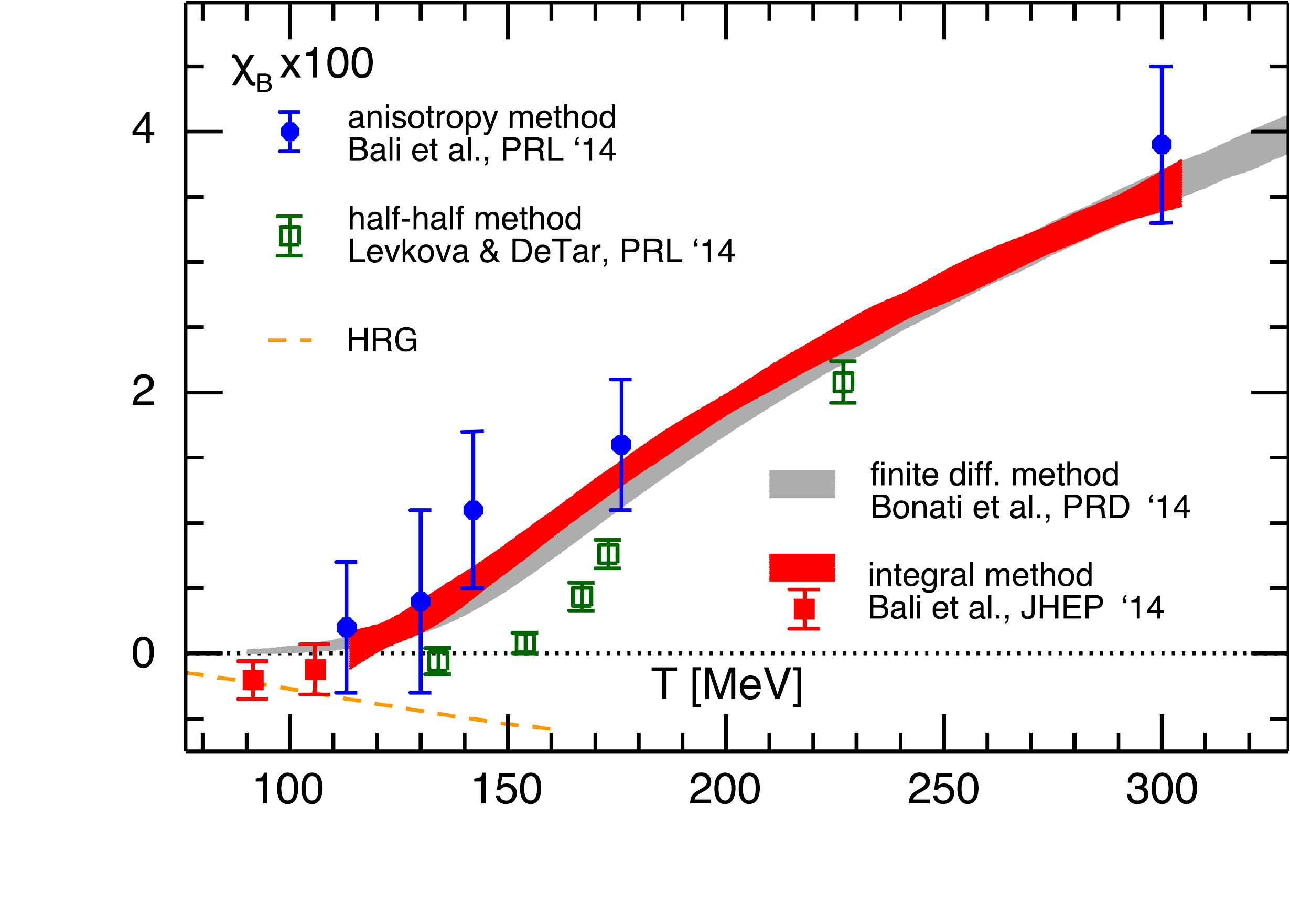}
\caption{(Left) The QCD crossover temperature as a function of the magnetic field
obtained from the inflection point of the quark condensate (red band) and  strange
quark susceptibility (blue band) \cite{Bali:2011qj}. (Right) The magnetic
susceptibility $\chi_B$, i.e. the correction to the pressure proportional to 
quadratic power
of the magnetic field, as a function of temperature \cite{Bali:2014kia}.}
\label{fig:extB}
\end{center}
\end{figure}

The QCD crossover transition in an external magnetic field has been 
extensively
studied using lattice QCD. In contrast to the earlier effective model based
predictions and the results of first lattice QCD calculations
\cite{DElia:2010nq,Ilgenfritz:2012fw} using unimproved staggered fermions with heavy
pion masses, the QCD crossover temperature $T_c$ turns out to decease with increasing
magnetic field \cite{Bali:2011qj,Bornyakov:2013eya,Ilgenfritz:2013ara}. 
Fig.~\ref{fig:extB} (left) illustrates the decrease of $T_c$ as function of 
the magnetic field \cite{Bali:2011qj}. The reduction of $T_c$ follows from 
the reduction of the
chiral condensate with increasing $B$, a phenomenon known as the inverse magnetic
catalysis, for the relevant intermediate temperature range. However, for low as well
as high enough temperatures the chiral condensate shows the expected magnetic
catalysis, i.e. an increase with increasing $B$
\cite{DElia:2011zu,Bali:2012zg,Ilgenfritz:2013ara}.

The response of the hot strong-interaction matter to an external magnetic field has also
been investigated through lattice QCD calculations of the magnetic susceptibility.
The magnetic susceptibility is the second derivative of the QCD partition function
with respect to the external magnetic field $B$ and, thus, gives the order $B^2$
corrections to the pressure. Since the standard approach leads to discrete, quantized
values for $B$ on the lattice, unlike for the susceptibilities of the chemical
potentials, taking direct derivatives of the QCD partition function with respect to
$B$ is not straightforward. Lattice calculations of the magnetic susceptibility have
been performed using three different approaches: (i) by computing the free energy
difference through an integration over the varying magnetic field
\cite{Bonati:2013lca,Bonati:2013vba,Bonati:2013vba}; (ii) through computations of the
magnetization via pressure anisotropies parallel and orthogonal to the direction
of the magnetic field \cite{Bali:2013esa,Bali:2013owa}; (iii) by adopting a
non-uniform external field such that the net magnetic flux across the $x$-$y$ plane of
the whole lattice is zero, resulting in a non-quantized magnetic field and allowing
for a direct computation of the derivatives of the partition function with respect to
$B$ \cite{Levkova:2013qda}. As shown in Fig.~\ref{fig:extB} (right), all these
different approaches produce more or less consistent results. The QGP phase turns out
to be paramagnetic. However, for low enough temperatures QCD is weakly diamagnetic,
in accordance with the hadron resonance gas model \cite{Endrodi:2013cs}. These results also
suggests that for $T\lesssim300$~MeV the order $B^2$ magnetic contribution to the
QCD pressure is at best a few percent for a magnetic field of the order of $10^{15}$~T,
which is relevant for heavy-ion collision experiments \cite{Kharzeev:2007jp}.

Lattice QCD studies have also been carried out to investigate the chiral magnetic
effect \cite{Kharzeev:2007jp}, namely the phenomenon of (electric) charge separation
along the direction of the magnetic field in presence of chiral anomaly induced
topological fluctuations. Various different avenues have been pursued by studying:
(i) charge fluctuations on given topological background \cite{Abramczyk:2009gb}; (ii)
enhanced fluctuations of the electric current along the magnetic field
\cite{Buividovich:2009wi}, (iii) dependence of the electric current on a given axial
chemical potential \cite{Buividovich:2009wi}, (iv) correlations of the electric
polarization and the topological charge \cite{Bali:2014vja}. In all cases, the chiral
magnetic effect has been observed, however, in a suppressed magnitude compared 
to the model based expectations. Obviously, in lattice QCD studies of the
chiral magnetic effect, which is closely related to the chiral 
anomaly of QCD, it would be of great interest to have lattice QCD calculations
perfomed with a chiral fermion formulation. Fully dynamical calculations
with DWF or overlap fermions at present do not yet exist.

\section{Summary}
\label{sec:summary}

In this review we have discussed selected topics in finite temperature
QCD that have a direct link to the ongoing experimental study of
the phase diagram of strong-interaction matter and the properties
of matter formed in heavy ion collisions. We naturally focused on
topics to which lattice QCD calculations can contribute and have lead 
or may lead to definite, quantitative answers in the future. 

Lattice QCD calculations at finite temperature started more than
30 years ago with simulations of pure $SU(N)$ gauge theories on
computers that delivered less than one Mega-Flop/s peak-performance. 
Nowadays these calculations are performed with dynamical quarks
and the correct mass spectrum of QCD on computers that reach a
peak-performance of more than ten Peta-Flop/s. The field thus 
could utilize computing resources with a peak-performance that 
increased by more than ten orders of magnitude, or doubled the 
speed almost every year. Although we know quite well from the
physics of the hadron gas that an exponential rise cannot continue
for ever, we still do not seem to have reached the critical point
and we may also in the future expect to gain further inside into 
the physics of strong-interaction matter through numerical 
lattice QCD calculations. 

During the last years lattice QCD calculations have accomplished two  
long-standing goals: (i) the pseudo-critical temperature of strong-interaction 
matter with physical light and strange quark masses has been determined
at vanishing net-baryon number, $T_c=(154\pm 9)$~MeV, and (ii)
the equation of state has been calculated in a wide temperature range.
It now is frequently used in the hydrodynamic modeling of heavy 
ion collisions. The extension of these results to non-zero 
chemical potential is well controlled at least in leading order
Taylor expansion of the pressure in terms of
baryon, electric charge and strangeness chemical potentials
and we will soon have results for bulk thermodynamics in a range
of baryon chemical potentials $\mu_B/T\lsim 3$, which will cover
most of the parameter space that can be explored in heavy ion
experiments at RHIC and the LHC. These Taylor expansions also provide
specific information on fluctuations of and correlations among
conserved charges. Lattice QCD calculations of these observables have 
reached a stage where guidance can be provided to the experimental
search for the critical end point and, with some care, a comparison
between experimental results for measured proton number, strangeness and, 
in particular, electric charge fluctuations on the one hand and lattice
QCD calculations on the other hand starts to become possible.

A major focus in the ongoing experimental heavy ion program is to explore 
in detail the properties of matter formed in heavy ion collisions.
The analysis of transport coefficients, thermal masses and screening
lengths does play an important role in this effort. Lattice QCD
calculations of these quantities are possible, although in many
cases they are complicated as an analytic continuation from Euclidean time to Minkowski
time is needed. For this reason most of the existing calculations
have been performed in the quenched approximation where fermionic 
degrees of freedom are taken into account only as valence quarks and
the influence of virtual quark loops on the gluonic background is 
neglected. Still these calculations prepare the ground for future
fully dynamical calculations of transport properties of strong-interaction
matter, for which in some cases exploratory
work has already started.

There remain many obvious, open questions in the study of the
phase diagram of strong-interaction matter and the properties of the
strongly interacting medium created in heavy ion collisions which 
can be addressed in future lattice QCD calculations and which will profit 
from future increasing computing resources as well as algorithmic advances.
Calculations at non-zero baryon number density and the search for 
the critical point, the calculation of transport properties with
fully dynamical, light quarks as well as calculations of thermodynamic
quantities using chiral fermion discretization schemes are among the
most demanding problems which hopefully can be addressed in the coming
years with improved techniques and resources.

\section*{Acknowledgements}
This review would not have been possible without the input from many
of our colleagues, in particular, Prasad Hegde, Olaf Kaczmarek, 
Christian Schmidt and Mathias Wagner, with whom we could collaborate 
over many years in the HotQCD and Bielefeld-BNL-CCNU collaborations. 
This work has been supported in part through contract 
DE-SC0012704 with the U.S. Department of Energy, the BMBF under grant 
05P12PBCTA and the GSI BILAER grant.

\bibliographystyle{JHEP}
\bibliography{qgp5_refs}

\providecommand{\href}[2]{#2}\begingroup\raggedright\begin{thebibliography}{100}

\bibitem{Hagedorn:1965st}
R.~Hagedorn, {\it {Statistical thermodynamics of strong interactions at
  high-energies}},  {\em Nuovo Cim.Suppl.} {\bf 3} (1965) 147--186.

\bibitem{Baym:1979}
G.~Baym, {\it {Confinement of quarks in nuclear matter}},  {\em Physica A} {\bf
  96} (1979) 131.

\bibitem{Celik:1980td}
T.~Celik, F.~Karsch, and H.~Satz, {\it {A percolation approach to strongly
  interacting matter}},  {\em Phys.Lett.} {\bf B97} (1980) 128--130.

\bibitem{Redlich:2015bga}
K.~Redlich and H.~Satz, {\it {The Legacy of Rolf Hagedorn: Statistical
  Bootstrap and Ultimate Temperature}},
  \href{http://arxiv.org/abs/1501.0752}{{\tt arXiv:1501.0752}}.

\bibitem{Politzer:1973fx}
H.~D. Politzer, {\it {Reliable Perturbative Results for Strong Interactions?}},
   {\em Phys.Rev.Lett.} {\bf 30} (1973) 1346--1349.

\bibitem{Gross:1973id}
D.~J. Gross and F.~Wilczek, {\it {Ultraviolet Behavior of Nonabelian Gauge
  Theories}},  {\em Phys.Rev.Lett.} {\bf 30} (1973) 1343--1346.

\bibitem{Cabibbo:1975ig}
N.~Cabibbo and G.~Parisi, {\it {Exponential Hadronic Spectrum and Quark
  Liberation}},  {\em Phys.Lett.} {\bf B59} (1975) 67--69.

\bibitem{Collins:1974ky}
J.~C. Collins and M.~Perry, {\it {Superdense Matter: Neutrons Or Asymptotically
  Free Quarks?}},  {\em Phys.Rev.Lett.} {\bf 34} (1975) 1353.

\bibitem{Shuryak:1980tp}
E.~V. Shuryak, {\it {Quantum Chromodynamics and the Theory of Superdense
  Matter}},  {\em Phys.Rept.} {\bf 61} (1980) 71--158.

\bibitem{Pisarski:1983ms}
R.~D. Pisarski and F.~Wilczek, {\it {Remarks on the Chiral Phase Transition in
  Chromodynamics}},  {\em Phys.Rev.} {\bf D29} (1984) 338--341.

\bibitem{Rajagopal:2000wf}
K.~Rajagopal and F.~Wilczek, {\it {The Condensed matter physics of QCD}},
  \href{http://arxiv.org/abs/hep-ph/0011333}{{\tt hep-ph/0011333}}.

\bibitem{Fukushima:2010bq}
K.~Fukushima and T.~Hatsuda, {\it {The phase diagram of dense QCD}},  {\em
  Rept.Prog.Phys.} {\bf 74} (2011) 014001,
  [\href{http://arxiv.org/abs/1005.4814}{{\tt arXiv:1005.4814}}].

\bibitem{Wilson:1974sk}
K.~G. Wilson, {\it {Confinement of Quarks}},  {\em Phys.Rev.} {\bf D10} (1974)
  2445--2459.

\bibitem{Creutz:1980zw}
M.~Creutz, {\it {Monte Carlo Study of Quantized SU(2) Gauge Theory}},  {\em
  Phys.Rev.} {\bf D21} (1980) 2308--2315.

\bibitem{McLerran:1980pk}
L.~D. McLerran and B.~Svetitsky, {\it {A Monte Carlo Study of SU(2) Yang-Mills
  Theory at Finite Temperature}},  {\em Phys.Lett.} {\bf B98} (1981) 195.

\bibitem{Kuti:1980gh}
J.~Kuti, J.~Polonyi, and K.~Szlachanyi, {\it {Monte Carlo Study of SU(2) Gauge
  Theory at Finite Temperature}},  {\em Phys.Lett.} {\bf B98} (1981) 199.

\bibitem{Engels:1980ty}
J.~Engels, F.~Karsch, H.~Satz, and I.~Montvay, {\it {High Temperature SU(2)
  Gluon Matter on the Lattice}},  {\em Phys.Lett.} {\bf B101} (1981) 89.

\bibitem{Matsui:1986dk}
T.~Matsui and H.~Satz, {\it {$J/\psi$ Suppression by Quark-Gluon Plasma
  Formation}},  {\em Phys.Lett.} {\bf B178} (1986) 416.

\bibitem{Rapp:1999ej}
R.~Rapp and J.~Wambach, {\it {Chiral symmetry restoration and dileptons in
  relativistic heavy ion collisions}},  {\em Adv.Nucl.Phys.} {\bf 25} (2000) 1,
  [\href{http://arxiv.org/abs/hep-ph/9909229}{{\tt hep-ph/9909229}}].

\bibitem{Kharzeev:2012ph}
D.~E. Kharzeev, K.~Landsteiner, A.~Schmitt, and H.-U. Yee, {\it {'Strongly
  interacting matter in magnetic fields': an overview}},  {\em Lect.Notes
  Phys.} {\bf 871} (2013) 1--11, [\href{http://arxiv.org/abs/1211.6245}{{\tt
  arXiv:1211.6245}}].

\bibitem{Kharzeev:2007jp}
D.~E. Kharzeev, L.~D. McLerran, and H.~J. Warringa, {\it {The Effects of
  topological charge change in heavy ion collisions: 'Event by event P and CP
  violation'}},  {\em Nucl.Phys.} {\bf A803} (2008) 227--253,
  [\href{http://arxiv.org/abs/0711.0950}{{\tt arXiv:0711.0950}}].

\bibitem{Linde:1980ts}
A.~D. Linde, {\it {Infrared Problem in Thermodynamics of the Yang-Mills Gas}},
  {\em Phys.Lett.} {\bf B96} (1980) 289.

\bibitem{Braaten:1989mz}
E.~Braaten and R.~D. Pisarski, {\it {Soft Amplitudes in Hot Gauge Theories: A
  General Analysis}},  {\em Nucl.Phys.} {\bf B337} (1990) 569.

\bibitem{Haque:2014rua}
N.~Haque, A.~Bandyopadhyay, J.~O. Andersen, M.~G. Mustafa, M.~Strickland,
  et~al., {\it {Three-loop HTLpt thermodynamics at finite temperature and
  chemical potential}},  {\em JHEP} {\bf 1405} (2014) 027,
  [\href{http://arxiv.org/abs/1402.6907}{{\tt arXiv:1402.6907}}].

\bibitem{Braaten:1995cm}
E.~Braaten and A.~Nieto, {\it {Effective field theory approach to high
  temperature thermodynamics}},  {\em Phys.Rev.} {\bf D51} (1995) 6990--7006,
  [\href{http://arxiv.org/abs/hep-ph/9501375}{{\tt hep-ph/9501375}}].

\bibitem{Hietanen:2008tv}
A.~Hietanen, K.~Kajantie, M.~Laine, K.~Rummukainen, and Y.~Schroder, {\it
  {Three-dimensional physics and the pressure of hot QCD}},  {\em Phys.Rev.}
  {\bf D79} (2009) 045018, [\href{http://arxiv.org/abs/0811.4664}{{\tt
  arXiv:0811.4664}}].

\bibitem{Montvay94}
I.~Montvay and G.~Muenster, {\em {Quantum Fields on a Lattice }}.
\newblock Cambridge Monographs on Mathematical Physics, 2010.

\bibitem{Rothe05}
H.~J. Rothe, {\em Lattice Gauge Theories: An Introduction (Fourth Edition)}.
\newblock World Scientific Lecture Notes in Physics, 2012.

\bibitem{Degrand06}
T.~Degrand and C.~Detar, {\em Lattice Methods for Quantum Chromodynamics}.
\newblock World Scientific, 2006.

\bibitem{Gattringer10}
C.~Gattringer and C.~Lang, {\em Quantum Chromodynamics on the Lattice-An
  Introductory Presentation}.
\newblock Springer, 2010.

\bibitem{Agashe:2014kda}
{\bf Particle Data Group} Collaboration, K.~Olive et~al., {\it {Review of
  Particle Physics}},  {\em Chin.Phys.} {\bf C38} (2014) 090001.

\bibitem{Hasenfratz:1983ba}
P.~Hasenfratz and F.~Karsch, {\it {Chemical Potential on the Lattice}},  {\em
  Phys.Lett.} {\bf B125} (1983) 308.

\bibitem{Gavai:1985ie}
R.~Gavai, {\it {Chemical Potential on the Lattice Revisited}},  {\em Phys.Rev.}
  {\bf D32} (1985) 519.

\bibitem{Aarts:2011ax}
G.~Aarts, F.~A. James, E.~Seiler, and I.-O. Stamatescu, {\it {Complex Langevin:
  Etiology and Diagnostics of its Main Problem}},  {\em Eur.Phys.J.} {\bf C71}
  (2011) 1756, [\href{http://arxiv.org/abs/1101.3270}{{\tt arXiv:1101.3270}}].

\bibitem{Aarts:2013uxa}
G.~Aarts, L.~Bongiovanni, E.~Seiler, D.~Sexty, and I.-O. Stamatescu, {\it
  {Controlling complex Langevin dynamics at finite density}},  {\em
  Eur.Phys.J.} {\bf A49} (2013) 89, [\href{http://arxiv.org/abs/1303.6425}{{\tt
  arXiv:1303.6425}}].

\bibitem{Sexty:2014dxa}
D.~Sexty, {\it {New algorithms for finite density QCD}},  {\em PoS} {\bf
  Lattice2015} (2014) 016, [\href{http://arxiv.org/abs/1410.8813}{{\tt
  arXiv:1410.8813}}].

\bibitem{Cristoforetti:2012su}
{\bf AuroraScience} Collaboration, M.~Cristoforetti, F.~Di~Renzo, and
  L.~Scorzato, {\it {New approach to the sign problem in quantum field
  theories: High density QCD on a Lefschetz thimble}},  {\em Phys.Rev.} {\bf
  D86} (2012) 074506, [\href{http://arxiv.org/abs/1205.3996}{{\tt
  arXiv:1205.3996}}].

\bibitem{Fujii:2013sra}
H.~Fujii, D.~Honda, M.~Kato, Y.~Kikukawa, S.~Komatsu, et~al., {\it {Hybrid
  Monte Carlo on Lefschetz thimbles - A study of the residual sign problem}},
  {\em JHEP} {\bf 1310} (2013) 147, [\href{http://arxiv.org/abs/1309.4371}{{\tt
  arXiv:1309.4371}}].

\bibitem{Cristoforetti:2014gsa}
M.~Cristoforetti, F.~Di~Renzo, G.~Eruzzi, A.~Mukherjee, C.~Schmidt, et~al.,
  {\it {An efficient method to compute the residual phase on a Lefschetz
  thimble}},  {\em Phys.Rev.} {\bf D89} (2014) 114505,
  [\href{http://arxiv.org/abs/1403.5637}{{\tt arXiv:1403.5637}}].

\bibitem{Gavai:2003mf}
R.~V. Gavai and S.~Gupta, {\it {Pressure and nonlinear susceptibilities in QCD
  at finite chemical potentials}},  {\em Phys.Rev.} {\bf D68} (2003) 034506,
  [\href{http://arxiv.org/abs/hep-lat/0303013}{{\tt hep-lat/0303013}}].

\bibitem{Allton:2003vx}
C.~Allton, S.~Ejiri, S.~Hands, O.~Kaczmarek, F.~Karsch, et~al., {\it {The
  Equation of state for two flavor QCD at nonzero chemical potential}},  {\em
  Phys.Rev.} {\bf D68} (2003) 014507,
  [\href{http://arxiv.org/abs/hep-lat/0305007}{{\tt hep-lat/0305007}}].

\bibitem{Symanzik:1983dc}
K.~Symanzik, {\it {Continuum Limit and Improved Action in Lattice Theories. 1.
  Principles and phi**4 Theory}},  {\em Nucl.Phys.} {\bf B226} (1983) 187.

\bibitem{Symanzik:1983gh}
K.~Symanzik, {\it {Continuum Limit and Improved Action in Lattice Theories. 2.
  O(N) Nonlinear Sigma Model in Perturbation Theory}},  {\em Nucl.Phys.} {\bf
  B226} (1983) 205.

\bibitem{Nielsen:1981hk}
H.~B. Nielsen and M.~Ninomiya, {\it {No Go Theorem for Regularizing Chiral
  Fermions}},  {\em Phys.Lett.} {\bf B105} (1981) 219.

\bibitem{Kogut:1974ag}
J.~B. Kogut and L.~Susskind, {\it {Hamiltonian Formulation of Wilson's Lattice
  Gauge Theories}},  {\em Phys.Rev.} {\bf D11} (1975) 395--408.

\bibitem{Kaplan:1992bt}
D.~B. Kaplan, {\it {A Method for simulating chiral fermions on the lattice}},
  {\em Phys.Lett.} {\bf B288} (1992) 342--347,
  [\href{http://arxiv.org/abs/hep-lat/9206013}{{\tt hep-lat/9206013}}].

\bibitem{Shamir:1993zy}
Y.~Shamir, {\it {Chiral fermions from lattice boundaries}},  {\em Nucl.Phys.}
  {\bf B406} (1993) 90--106, [\href{http://arxiv.org/abs/hep-lat/9303005}{{\tt
  hep-lat/9303005}}].

\bibitem{Shamir:1992im}
Y.~Shamir, {\it {The Euclidean spectrum of Kaplan's lattice chiral fermions}},
  {\em Phys.Lett.} {\bf B305} (1993) 357--365,
  [\href{http://arxiv.org/abs/hep-lat/9212010}{{\tt hep-lat/9212010}}].

\bibitem{Furman:1994ky}
V.~Furman and Y.~Shamir, {\it {Axial symmetries in lattice QCD with Kaplan
  fermions}},  {\em Nucl.Phys.} {\bf B439} (1995) 54--78,
  [\href{http://arxiv.org/abs/hep-lat/9405004}{{\tt hep-lat/9405004}}].

\bibitem{Neuberger:1998wv}
H.~Neuberger, {\it {More about exactly massless quarks on the lattice}},  {\em
  Phys.Lett.} {\bf B427} (1998) 353--355,
  [\href{http://arxiv.org/abs/hep-lat/9801031}{{\tt hep-lat/9801031}}].

\bibitem{Neuberger:1997fp}
H.~Neuberger, {\it {Exactly massless quarks on the lattice}},  {\em Phys.Lett.}
  {\bf B417} (1998) 141--144, [\href{http://arxiv.org/abs/hep-lat/9707022}{{\tt
  hep-lat/9707022}}].

\bibitem{Sheikholeslami:1985ij}
B.~Sheikholeslami and R.~Wohlert, {\it {Improved Continuum Limit Lattice Action
  for QCD with Wilson Fermions}},  {\em Nucl.Phys.} {\bf B259} (1985) 572.

\bibitem{PhysRevD.75.054502}
E.~Follana, Q.~Mason, C.~Davies, K.~Hornbostel, G.~P. Lepage, J.~Shigemitsu,
  H.~Trottier, and K.~Wong, {\it Highly improved staggered quarks on the
  lattice with applications to charm physics},  {\em Phys. Rev. D} {\bf 75}
  (2007) 054502.

\bibitem{Morningstar:2003gk}
C.~Morningstar and M.~J. Peardon, {\it {Analytic smearing of SU(3) link
  variables in lattice QCD}},  {\em Phys.Rev.} {\bf D69} (2004) 054501,
  [\href{http://arxiv.org/abs/hep-lat/0311018}{{\tt hep-lat/0311018}}].

\bibitem{Heller:1999xz}
U.~M. Heller, F.~Karsch, and B.~Sturm, {\it {Improved staggered fermion actions
  for QCD thermodynamics}},  {\em Phys.Rev.} {\bf D60} (1999) 114502,
  [\href{http://arxiv.org/abs/hep-lat/9901010}{{\tt hep-lat/9901010}}].

\bibitem{Antonio:2008zz}
{\bf RBC, UKQCD} Collaboration, D.~J. Antonio et~al., {\it {Localization and
  chiral symmetry in three flavor domain wall QCD}},  {\em Phys.Rev.} {\bf D77}
  (2008) 014509, [\href{http://arxiv.org/abs/0705.2340}{{\tt
  arXiv:0705.2340}}].

\bibitem{Renfrew:2009wu}
D.~Renfrew, T.~Blum, N.~Christ, R.~Mawhinney, and P.~Vranas, {\it {Controlling
  Residual Chiral Symmetry Breaking in Domain Wall Fermion Simulations}},  {\em
  PoS} {\bf LATTICE2008} (2008) 048,
  [\href{http://arxiv.org/abs/0902.2587}{{\tt arXiv:0902.2587}}].

\bibitem{Ginsparg:1981bj}
P.~H. Ginsparg and K.~G. Wilson, {\it {A Remnant of Chiral Symmetry on the
  Lattice}},  {\em Phys.Rev.} {\bf D25} (1982) 2649.

\bibitem{Buchoff:2013nra}
M.~I. Buchoff, M.~Cheng, N.~H. Christ, H.~T. Ding, C.~Jung, et~al., {\it {QCD
  chiral transition, U(1)A symmetry and the dirac spectrum using domain wall
  fermions}},  {\em Phys.Rev.} {\bf D89} (2014) 054514,
  [\href{http://arxiv.org/abs/1309.4149}{{\tt arXiv:1309.4149}}].

\bibitem{Cossu:2013uua}
G.~Cossu, S.~Aoki, H.~Fukaya, S.~Hashimoto, T.~Kaneko, et~al., {\it {Finite
  temperature study of the axial U(1) symmetry on the lattice with overlap
  fermion formulation}},  {\em Phys.Rev.} {\bf D87} (2013) 114514,
  [\href{http://arxiv.org/abs/1304.6145}{{\tt arXiv:1304.6145}}].

\bibitem{Bhattacharya:2014ara}
T.~Bhattacharya, M.~I. Buchoff, N.~H. Christ, H.-T. Ding, R.~Gupta, et~al.,
  {\it {QCD Phase Transition with Chiral Quarks and Physical Quark Masses}},
  {\em Phys.Rev.Lett.} {\bf 113} (2014) 082001,
  [\href{http://arxiv.org/abs/1402.5175}{{\tt arXiv:1402.5175}}].

\bibitem{Dick:2015twa}
V.~Dick, F.~Karsch, E.~Laermann, S.~Mukherjee, and S.~Sharma, {\it {Microscopic
  Origin of \boldmath{$U_A(1)$} Symmetry Violation in the High Temperature
  Phase of QCD}},  \href{http://arxiv.org/abs/1502.0619}{{\tt
  arXiv:1502.0619}}.

\bibitem{Saito:2011fs}
{\bf WHOT-QCD} Collaboration, H.~Saito et~al., {\it {Phase structure of finite
  temperature QCD in the heavy quark region}},  {\em Phys.Rev.} {\bf D84}
  (2011) 054502, [\href{http://arxiv.org/abs/1106.0974}{{\tt
  arXiv:1106.0974}}].

\bibitem{Saito:2013vja}
H.~Saito, S.~Ejiri, S.~Aoki, K.~Kanaya, Y.~Nakagawa, et~al., {\it {Histograms
  in heavy-quark QCD at finite temperature and density}},  {\em Phys.Rev.} {\bf
  D89} (2014) 034507, [\href{http://arxiv.org/abs/1309.2445}{{\tt
  arXiv:1309.2445}}].

\bibitem{Karsch:2001nf}
F.~Karsch, E.~Laermann, and C.~Schmidt, {\it {The Chiral critical point in
  three-flavor QCD}},  {\em Phys.Lett.} {\bf B520} (2001) 41--49,
  [\href{http://arxiv.org/abs/hep-lat/0107020}{{\tt hep-lat/0107020}}].

\bibitem{deForcrand:2006pv}
P.~de~Forcrand and O.~Philipsen, {\it {The Chiral critical line of N(f) = 2+1
  QCD at zero and non-zero baryon density}},  {\em JHEP} {\bf 0701} (2007) 077,
  [\href{http://arxiv.org/abs/hep-lat/0607017}{{\tt hep-lat/0607017}}].

\bibitem{Endrodi:2007gc}
G.~Endrodi, Z.~Fodor, S.~Katz, and K.~Szabo, {\it {The Nature of the finite
  temperature QCD transition as a function of the quark masses}},  {\em PoS}
  {\bf LAT2007} (2007) 182, [\href{http://arxiv.org/abs/0710.0998}{{\tt
  arXiv:0710.0998}}].

\bibitem{Ding:2011du}
H.-T. Ding, A.~Bazavov, P.~Hegde, F.~Karsch, S.~Mukherjee, et~al., {\it
  {Exploring phase diagram of $N_f=3$ QCD at $\mu=0$ with HISQ fermions}},
  {\em PoS} {\bf LATTICE2011} (2011) 191,
  [\href{http://arxiv.org/abs/1111.0185}{{\tt arXiv:1111.0185}}].

\bibitem{Butti:2003nu}
A.~Butti, A.~Pelissetto, and E.~Vicari, {\it {On the nature of the finite
  temperature transition in QCD}},  {\em JHEP} {\bf 0308} (2003) 029,
  [\href{http://arxiv.org/abs/hep-ph/0307036}{{\tt hep-ph/0307036}}].

\bibitem{Pelissetto:2013hqa}
A.~Pelissetto and E.~Vicari, {\it {Relevance of the axial anomaly at the
  finite-temperature chiral transition in QCD}},  {\em Phys.Rev.} {\bf D88}
  (2013) 105018, [\href{http://arxiv.org/abs/1309.5446}{{\tt
  arXiv:1309.5446}}].

\bibitem{Grahl:2013pba}
M.~Grahl and D.~H. Rischke, {\it {Functional renormalization group study of the
  two-flavor linear sigma model in the presence of the axial anomaly}},  {\em
  Phys.Rev.} {\bf D88} (2013) 056014,
  [\href{http://arxiv.org/abs/1307.2184}{{\tt arXiv:1307.2184}}].

\bibitem{Aoki:2006we}
Y.~Aoki, G.~Endrodi, Z.~Fodor, S.~Katz, and K.~Szabo, {\it {The Order of the
  quantum chromodynamics transition predicted by the standard model of particle
  physics}},  {\em Nature} {\bf 443} (2006) 675--678,
  [\href{http://arxiv.org/abs/hep-lat/0611014}{{\tt hep-lat/0611014}}].

\bibitem{Riedel72}
E.~K. Riedel and F.~J. Wegner, {\it Tricritical exponents and scaling fields},
  {\em Phys. Rev. Lett.} {\bf 29} (1972) 349--352.

\bibitem{Engels:2001bq}
J.~Engels, S.~Holtmann, T.~Mendes, and T.~Schulze, {\it {Finite size scaling
  functions for 3-d O(4) and O(2) spin models and QCD}},  {\em Phys.Lett.} {\bf
  B514} (2001) 299--308, [\href{http://arxiv.org/abs/hep-lat/0105028}{{\tt
  hep-lat/0105028}}].

\bibitem{Engels:2003nq}
J.~Engels, L.~Fromme, and M.~Seniuch, {\it {Correlation lengths and scaling
  functions in the three-dimensional O(4) model}},  {\em Nucl.Phys.} {\bf B675}
  (2003) 533--554, [\href{http://arxiv.org/abs/hep-lat/0307032}{{\tt
  hep-lat/0307032}}].

\bibitem{Ejiri:2009ac}
S.~Ejiri, F.~Karsch, E.~Laermann, C.~Miao, S.~Mukherjee, et~al., {\it {On the
  magnetic equation of state in (2+1)-flavor QCD}},  {\em Phys.Rev.} {\bf D80}
  (2009) 094505, [\href{http://arxiv.org/abs/0909.5122}{{\tt
  arXiv:0909.5122}}].

\bibitem{Bonati:2014kpa}
C.~Bonati, P.~de~Forcrand, M.~D'Elia, O.~Philipsen, and F.~Sanfilippo, {\it
  {Chiral phase transition in two-flavor QCD from an imaginary chemical
  potential}},  {\em Phys.Rev.} {\bf D90} (2014) 074030,
  [\href{http://arxiv.org/abs/1408.5086}{{\tt arXiv:1408.5086}}].

\bibitem{Bazavov:2011nk}
A.~Bazavov, T.~Bhattacharya, M.~Cheng, C.~DeTar, H.~Ding, et~al., {\it {The
  chiral and deconfinement aspects of the QCD transition}},  {\em Phys.Rev.}
  {\bf D85} (2012) 054503, [\href{http://arxiv.org/abs/1111.1710}{{\tt
  arXiv:1111.1710}}].

\bibitem{Aoki:2006br}
Y.~Aoki, Z.~Fodor, S.~Katz, and K.~Szabo, {\it {The QCD transition temperature:
  Results with physical masses in the continuum limit}},  {\em Phys.Lett.} {\bf
  B643} (2006) 46--54, [\href{http://arxiv.org/abs/hep-lat/0609068}{{\tt
  hep-lat/0609068}}].

\bibitem{Aoki:2009sc}
Y.~Aoki, S.~Borsanyi, S.~Durr, Z.~Fodor, S.~D. Katz, et~al., {\it {The QCD
  transition temperature: results with physical masses in the continuum limit
  II.}},  {\em JHEP} {\bf 0906} (2009) 088,
  [\href{http://arxiv.org/abs/0903.4155}{{\tt arXiv:0903.4155}}].

\bibitem{Borsanyi:2010bp}
{\bf Wuppertal-Budapest} Collaboration, S.~Borsanyi et~al., {\it {Is there
  still any Tc mystery in lattice QCD? Results with physical masses in the
  continuum limit III}},  {\em JHEP} {\bf 1009} (2010) 073,
  [\href{http://arxiv.org/abs/1005.3508}{{\tt arXiv:1005.3508}}].

\bibitem{Kaczmarek:2011zz}
O.~Kaczmarek, F.~Karsch, E.~Laermann, C.~Miao, S.~Mukherjee, et~al., {\it
  {Phase boundary for the chiral transition in (2+1) -flavor QCD at small
  values of the chemical potential}},  {\em Phys.Rev.} {\bf D83} (2011) 014504,
  [\href{http://arxiv.org/abs/1011.3130}{{\tt arXiv:1011.3130}}].

\bibitem{Endrodi:2011gv}
G.~Endrodi, Z.~Fodor, S.~Katz, and K.~Szabo, {\it {The QCD phase diagram at
  nonzero quark density}},  {\em JHEP} {\bf 1104} (2011) 001,
  [\href{http://arxiv.org/abs/1102.1356}{{\tt arXiv:1102.1356}}].

\bibitem{Cea:2014xva}
P.~Cea, L.~Cosmai, and A.~Papa, {\it {Critical line of 2+1 flavor QCD}},  {\em
  Phys.Rev.} {\bf D89} (2014) 074512,
  [\href{http://arxiv.org/abs/1403.0821}{{\tt arXiv:1403.0821}}].

\bibitem{Bonati:2014rfa}
C.~Bonati, M.~D'Elia, M.~Mariti, M.~Mesiti, F.~Negro, et~al., {\it {Curvature
  of the chiral pseudocritical line in QCD}},  {\em Phys.Rev.} {\bf D90} (2014)
  114025, [\href{http://arxiv.org/abs/1410.5758}{{\tt arXiv:1410.5758}}].

\bibitem{Andronic:2011yq}
A.~Andronic, P.~Braun-Munzinger, K.~Redlich, and J.~Stachel, {\it {The thermal
  model on the verge of the ultimate test: particle production in Pb-Pb
  collisions at the LHC}},  {\em J.Phys.} {\bf G38} (2011) 124081,
  [\href{http://arxiv.org/abs/1106.6321}{{\tt arXiv:1106.6321}}].

\bibitem{Fodor:2004nz}
Z.~Fodor and S.~Katz, {\it {Critical point of QCD at finite T and mu, lattice
  results for physical quark masses}},  {\em JHEP} {\bf 0404} (2004) 050,
  [\href{http://arxiv.org/abs/hep-lat/0402006}{{\tt hep-lat/0402006}}].

\bibitem{Datta:2014zqa}
S.~Datta, R.~V. Gavai, and S.~Gupta, {\it {QCD at finite chemical potential
  with $N_t = 8$}},  {\em PoS} {\bf LATTICE2013} (2014) 202.

\bibitem{Ejiri:2005ts}
S.~Ejiri, {\it {Lee-Yang zero analysis for the study of QCD phase structure}},
  {\em Phys.Rev.} {\bf D73} (2006) 054502,
  [\href{http://arxiv.org/abs/hep-lat/0506023}{{\tt hep-lat/0506023}}].

\bibitem{deForcrand:2002ci}
P.~de~Forcrand and O.~Philipsen, {\it {The QCD phase diagram for small
  densities from imaginary chemical potential}},  {\em Nucl.Phys.} {\bf B642}
  (2002) 290--306, [\href{http://arxiv.org/abs/hep-lat/0205016}{{\tt
  hep-lat/0205016}}].

\bibitem{Jin:2015taa}
X.-Y. Jin, Y.~Kuramashi, Y.~Nakamura, S.~Takeda, and A.~Ukawa, {\it {Curvature
  of the critical line on the plane of quark chemical potential and pseudo
  scalar meson mass for three-flavor QCD}},
  \href{http://arxiv.org/abs/1504.0011}{{\tt arXiv:1504.0011}}.

\bibitem{Allton:2005gk}
C.~Allton, M.~Doring, S.~Ejiri, S.~Hands, O.~Kaczmarek, et~al., {\it
  {Thermodynamics of two flavor QCD to sixth order in quark chemical
  potential}},  {\em Phys.Rev.} {\bf D71} (2005) 054508,
  [\href{http://arxiv.org/abs/hep-lat/0501030}{{\tt hep-lat/0501030}}].

\bibitem{Gavai:2008zr}
R.~Gavai and S.~Gupta, {\it {QCD at finite chemical potential with six time
  slices}},  {\em Phys.Rev.} {\bf D78} (2008) 114503,
  [\href{http://arxiv.org/abs/0806.2233}{{\tt arXiv:0806.2233}}].

\bibitem{Kaczmarek:2002mc}
O.~Kaczmarek, F.~Karsch, P.~Petreczky, and F.~Zantow, {\it {Heavy quark
  anti-quark free energy and the renormalized Polyakov loop}},  {\em
  Phys.Lett.} {\bf B543} (2002) 41--47,
  [\href{http://arxiv.org/abs/hep-lat/0207002}{{\tt hep-lat/0207002}}].

\bibitem{McLerran:1981pb}
L.~D. McLerran and B.~Svetitsky, {\it {Quark Liberation at High Temperature: A
  Monte Carlo Study of SU(2) Gauge Theory}},  {\em Phys.Rev.} {\bf D24} (1981)
  450.

\bibitem{Bazavov:2013yv}
A.~Bazavov and P.~Petreczky, {\it {Polyakov loop in 2+1 flavor QCD}},  {\em
  Phys.Rev.} {\bf D87} (2013) 094505,
  [\href{http://arxiv.org/abs/1301.3943}{{\tt arXiv:1301.3943}}].

\bibitem{Bellwied:2013cta}
R.~Bellwied, S.~Borsanyi, Z.~Fodor, S.~D. Katz, and C.~Ratti, {\it {Is there a
  flavor hierarchy in the deconfinement transition of QCD?}},  {\em
  Phys.Rev.Lett.} {\bf 111} (2013) 202302,
  [\href{http://arxiv.org/abs/1305.6297}{{\tt arXiv:1305.6297}}].

\bibitem{Bazavov:2013dta}
A.~Bazavov, H.~T. Ding, P.~Hegde, O.~Kaczmarek, F.~Karsch, et~al., {\it
  {Strangeness at high temperatures: from hadrons to quarks}},  {\em
  Phys.Rev.Lett.} {\bf 111} (2013) 082301,
  [\href{http://arxiv.org/abs/1304.7220}{{\tt arXiv:1304.7220}}].

\bibitem{Koch:2005vg}
V.~Koch, A.~Majumder, and J.~Randrup, {\it {Baryon-strangeness correlations: A
  Diagnostic of strongly interacting matter}},  {\em Phys.Rev.Lett.} {\bf 95}
  (2005) 182301, [\href{http://arxiv.org/abs/nucl-th/0505052}{{\tt
  nucl-th/0505052}}].

\bibitem{Ejiri:2005wq}
S.~Ejiri, F.~Karsch, and K.~Redlich, {\it {Hadronic fluctuations at the QCD
  phase transition}},  {\em Phys.Lett.} {\bf B633} (2006) 275--282,
  [\href{http://arxiv.org/abs/hep-ph/0509051}{{\tt hep-ph/0509051}}].

\bibitem{Adler:1969gk}
S.~L. Adler, {\it {Axial vector vertex in spinor electrodynamics}},  {\em
  Phys.Rev.} {\bf 177} (1969) 2426--2438.

\bibitem{Bell:1969ts}
J.~Bell and R.~Jackiw, {\it {A PCAC puzzle: pi0 -gt; gamma gamma in the sigma
  model}},  {\em Nuovo Cim.} {\bf A60} (1969) 47--61.

\bibitem{tHooft:1976up}
G.~'t~Hooft, {\it {Symmetry Breaking Through Bell-Jackiw Anomalies}},  {\em
  Phys.Rev.Lett.} {\bf 37} (1976) 8--11.

\bibitem{Gross:1980br}
D.~J. Gross, R.~D. Pisarski, and L.~G. Yaffe, {\it {QCD and Instantons at
  Finite Temperature}},  {\em Rev.Mod.Phys.} {\bf 53} (1981) 43.

\bibitem{Shuryak:1993ee}
E.~V. Shuryak, {\it {Which chiral symmetry is restored in hot QCD?}},  {\em
  Comments Nucl.Part.Phys.} {\bf 21} (1994) 235--248,
  [\href{http://arxiv.org/abs/hep-ph/9310253}{{\tt hep-ph/9310253}}].

\bibitem{Cheng:2010fe}
M.~Cheng, S.~Datta, A.~Francis, J.~van~der Heide, C.~Jung, et~al., {\it {Meson
  screening masses from lattice QCD with two light and the strange quark}},
  {\em Eur.Phys.J.} {\bf C71} (2011) 1564,
  [\href{http://arxiv.org/abs/1010.1216}{{\tt arXiv:1010.1216}}].

\bibitem{Sharpe:2006re}
S.~R. Sharpe, {\it {Rooted staggered fermions: Good, bad or ugly?}},  {\em PoS}
  {\bf LAT2006} (2006) 022, [\href{http://arxiv.org/abs/hep-lat/0610094}{{\tt
  hep-lat/0610094}}].

\bibitem{Donald:2011if}
G.~C. Donald, C.~T. Davies, E.~Follana, and A.~S. Kronfeld, {\it {Staggered
  fermions, zero modes, and flavor-singlet mesons}},  {\em Phys.Rev.} {\bf D84}
  (2011) 054504, [\href{http://arxiv.org/abs/1106.2412}{{\tt
  arXiv:1106.2412}}].

\bibitem{Bazavov:2012qja}
{\bf HotQCD} Collaboration, A.~Bazavov et~al., {\it {The chiral transition and
  $U(1)_A$ symmetry restoration from lattice QCD using Domain Wall Fermions}},
  {\em Phys.Rev.} {\bf D86} (2012) 094503,
  [\href{http://arxiv.org/abs/1205.3535}{{\tt arXiv:1205.3535}}].

\bibitem{Tomiya:2014mma}
A.~Tomiya, G.~Cossu, H.~Fukaya, S.~Hashimoto, and J.~Noaki, {\it {Effects of
  near-zero Dirac eigenmodes on axial U(1) symmetry at finite temperature}},
  \href{http://arxiv.org/abs/1412.7306}{{\tt arXiv:1412.7306}}.

\bibitem{Chiu:2013wwa}
{\bf TWQCD} Collaboration, T.-W. Chiu, W.-P. Chen, Y.-C. Chen, H.-Y. Chou, and
  T.-H. Hsieh, {\it {Chiral symmetry and axial U(1) symmetry in finite
  temperature QCD with domain-wall fermion}},  {\em PoS} {\bf LATTICE2013}
  (2014) 165, [\href{http://arxiv.org/abs/1311.6220}{{\tt arXiv:1311.6220}}].

\bibitem{Karsch:2001cy}
F.~Karsch, {\it {Lattice QCD at high temperature and density}},  {\em
  Lect.Notes Phys.} {\bf 583} (2002) 209--249,
  [\href{http://arxiv.org/abs/hep-lat/0106019}{{\tt hep-lat/0106019}}].

\bibitem{Borsanyi:2013bia}
S.~Borsanyi, Z.~Fodor, C.~Hoelbling, S.~D. Katz, S.~Krieg, et~al., {\it {Full
  result for the QCD equation of state with 2+1 flavors}},  {\em Phys.Lett.}
  {\bf B730} (2014) 99--104, [\href{http://arxiv.org/abs/1309.5258}{{\tt
  arXiv:1309.5258}}].

\bibitem{Bazavov:2014pvz}
{\bf HotQCD} Collaboration, A.~Bazavov et~al., {\it {Equation of state in ( 2+1
  )-flavor QCD}},  {\em Phys.Rev.} {\bf D90} (2014) 094503,
  [\href{http://arxiv.org/abs/1407.6387}{{\tt arXiv:1407.6387}}].

\bibitem{Majumder:2010ik}
A.~Majumder and B.~Muller, {\it {Hadron Mass Spectrum from Lattice QCD}},  {\em
  Phys.Rev.Lett.} {\bf 105} (2010) 252002,
  [\href{http://arxiv.org/abs/1008.1747}{{\tt arXiv:1008.1747}}].

\bibitem{Bazavov:2014xya}
A.~Bazavov, H.~T. Ding, P.~Hegde, O.~Kaczmarek, F.~Karsch, et~al., {\it
  {Additional Strange Hadrons from QCD Thermodynamics and Strangeness Freezeout
  in Heavy Ion Collisions}},  {\em Phys.Rev.Lett.} {\bf 113} (2014) 072001,
  [\href{http://arxiv.org/abs/1404.6511}{{\tt arXiv:1404.6511}}].

\bibitem{Gavai:2001fr}
R.~V. Gavai and S.~Gupta, {\it {Quark number susceptibilities, strangeness and
  dynamical confinement}},  {\em Phys.Rev.} {\bf D64} (2001) 074506,
  [\href{http://arxiv.org/abs/hep-lat/0103013}{{\tt hep-lat/0103013}}].

\bibitem{Allton:2002zi}
C.~Allton, S.~Ejiri, S.~Hands, O.~Kaczmarek, F.~Karsch, et~al., {\it {The QCD
  thermal phase transition in the presence of a small chemical potential}},
  {\em Phys.Rev.} {\bf D66} (2002) 074507,
  [\href{http://arxiv.org/abs/hep-lat/0204010}{{\tt hep-lat/0204010}}].

\bibitem{Ejiri:2005uv}
S.~Ejiri, F.~Karsch, E.~Laermann, and C.~Schmidt, {\it {The Isentropic equation
  of state of 2-flavor QCD}},  {\em Phys.Rev.} {\bf D73} (2006) 054506,
  [\href{http://arxiv.org/abs/hep-lat/0512040}{{\tt hep-lat/0512040}}].

\bibitem{Borsanyi:2011sw}
S.~Borsanyi, Z.~Fodor, S.~D. Katz, S.~Krieg, C.~Ratti, et~al., {\it
  {Fluctuations of conserved charges at finite temperature from lattice QCD}},
  {\em JHEP} {\bf 1201} (2012) 138, [\href{http://arxiv.org/abs/1112.4416}{{\tt
  arXiv:1112.4416}}].

\bibitem{Bazavov:2012jq}
{\bf HotQCD} Collaboration, A.~Bazavov et~al., {\it {Fluctuations and
  Correlations of net baryon number, electric charge, and strangeness: A
  comparison of lattice QCD results with the hadron resonance gas model}},
  {\em Phys.Rev.} {\bf D86} (2012) 034509,
  [\href{http://arxiv.org/abs/1203.0784}{{\tt arXiv:1203.0784}}].

\bibitem{Borsanyi:2012cr}
S.~Borsanyi, G.~Endrodi, Z.~Fodor, S.~Katz, S.~Krieg, et~al., {\it {QCD
  equation of state at nonzero chemical potential: continuum results with
  physical quark masses at order $mu^2$}},  {\em JHEP} {\bf 1208} (2012) 053,
  [\href{http://arxiv.org/abs/1204.6710}{{\tt arXiv:1204.6710}}].

\bibitem{Borsanyi:2013hza}
S.~Borsanyi, Z.~Fodor, S.~Katz, S.~Krieg, C.~Ratti, et~al., {\it {Freeze-out
  parameters: lattice meets experiment}},  {\em Phys.Rev.Lett.} {\bf 111}
  (2013) 062005, [\href{http://arxiv.org/abs/1305.5161}{{\tt
  arXiv:1305.5161}}].

\bibitem{Hegde:2014wga}
{\bf for the BNL-Bielefeld-CCNU} Collaboration, P.~Hegde, {\it {The QCD
  equation of state to $\mathcal{O}(\mu_B^4)$}},
  \href{http://arxiv.org/abs/1412.6727}{{\tt arXiv:1412.6727}}.

\bibitem{Laine:2006cp}
M.~Laine and Y.~Schroder, {\it {Quark mass thresholds in QCD thermodynamics}},
  {\em Phys.Rev.} {\bf D73} (2006) 085009,
  [\href{http://arxiv.org/abs/hep-ph/0603048}{{\tt hep-ph/0603048}}].

\bibitem{Stephanov:2008qz}
M.~Stephanov, {\it {Non-Gaussian fluctuations near the QCD critical point}},
  {\em Phys.Rev.Lett.} {\bf 102} (2009) 032301,
  [\href{http://arxiv.org/abs/0809.3450}{{\tt arXiv:0809.3450}}].

\bibitem{Asakawa:2009aj}
M.~Asakawa, S.~Ejiri, and M.~Kitazawa, {\it {Third moments of conserved charges
  as probes of QCD phase structure}},  {\em Phys.Rev.Lett.} {\bf 103} (2009)
  262301, [\href{http://arxiv.org/abs/0904.2089}{{\tt arXiv:0904.2089}}].

\bibitem{Friman:2011pf}
B.~Friman, F.~Karsch, K.~Redlich, and V.~Skokov, {\it {Fluctuations as probe of
  the QCD phase transition and freeze-out in heavy ion collisions at LHC and
  RHIC}},  {\em Eur.Phys.J.} {\bf C71} (2011) 1694,
  [\href{http://arxiv.org/abs/1103.3511}{{\tt arXiv:1103.3511}}].

\bibitem{Stephanov:2011pb}
M.~Stephanov, {\it {On the sign of kurtosis near the QCD critical point}},
  {\em Phys.Rev.Lett.} {\bf 107} (2011) 052301,
  [\href{http://arxiv.org/abs/1104.1627}{{\tt arXiv:1104.1627}}].

\bibitem{Stephanov:1998dy}
M.~A. Stephanov, K.~Rajagopal, and E.~V. Shuryak, {\it {Signatures of the
  tricritical point in QCD}},  {\em Phys.Rev.Lett.} {\bf 81} (1998) 4816--4819,
  [\href{http://arxiv.org/abs/hep-ph/9806219}{{\tt hep-ph/9806219}}].

\bibitem{Stephanov:1999zu}
M.~A. Stephanov, K.~Rajagopal, and E.~V. Shuryak, {\it {Event-by-event
  fluctuations in heavy ion collisions and the QCD critical point}},  {\em
  Phys.Rev.} {\bf D60} (1999) 114028,
  [\href{http://arxiv.org/abs/hep-ph/9903292}{{\tt hep-ph/9903292}}].

\bibitem{Jeon:2003gk}
S.~Jeon and V.~Koch, {\it {Event by event fluctuations}},
  \href{http://arxiv.org/abs/hep-ph/0304012}{{\tt hep-ph/0304012}}.

\bibitem{Hatta:2003wn}
Y.~Hatta and M.~Stephanov, {\it {Proton number fluctuation as a signal of the
  QCD critical endpoint}},  {\em Phys.Rev.Lett.} {\bf 91} (2003) 102003,
  [\href{http://arxiv.org/abs/hep-ph/0302002}{{\tt hep-ph/0302002}}].

\bibitem{Athanasiou:2010kw}
C.~Athanasiou, K.~Rajagopal, and M.~Stephanov, {\it {Using Higher Moments of
  Fluctuations and their Ratios in the Search for the QCD Critical Point}},
  {\em Phys.Rev.} {\bf D82} (2010) 074008,
  [\href{http://arxiv.org/abs/1006.4636}{{\tt arXiv:1006.4636}}].

\bibitem{STAR-wp}
{\it Studying the phase diagram of qcd matter at rhic},  2014.
\newblock
  \url{https://drupal.star.bnl.gov/STAR/files/BES_WPII_ver6.9_Cover.pdf}.

\bibitem{Aggarwal:2010wy}
{\bf STAR} Collaboration, M.~Aggarwal et~al., {\it {Higher Moments of
  Net-proton Multiplicity Distributions at RHIC}},  {\em Phys.Rev.Lett.} {\bf
  105} (2010) 022302, [\href{http://arxiv.org/abs/1004.4959}{{\tt
  arXiv:1004.4959}}].

\bibitem{Adamczyk:2013dal}
{\bf STAR} Collaboration, L.~Adamczyk et~al., {\it {Energy Dependence of
  Moments of Net-proton Multiplicity Distributions at RHIC}},  {\em
  Phys.Rev.Lett.} {\bf 112} (2014) 032302,
  [\href{http://arxiv.org/abs/1309.5681}{{\tt arXiv:1309.5681}}].

\bibitem{Adamczyk:2014fia}
{\bf STAR} Collaboration, L.~Adamczyk et~al., {\it {Beam energy dependence of
  moments of the net-charge multiplicity distributions in Au+Au collisions at
  RHIC}},  {\em Phys.Rev.Lett.} {\bf 113} (2014) 092301,
  [\href{http://arxiv.org/abs/1402.1558}{{\tt arXiv:1402.1558}}].

\bibitem{Sahoo:2014bqa}
{\bf STAR} Collaboration, N.~R. Sahoo, {\it {Recent results on event-by-event
  fluctuations from the RHIC Beam Energy Scan program in the STAR experiment}},
   {\em J.Phys.Conf.Ser.} {\bf 535} (2014) 012007,
  [\href{http://arxiv.org/abs/1407.1554}{{\tt arXiv:1407.1554}}].

\bibitem{Luo:2015ewa}
{\bf STAR} Collaboration, X.~Luo, {\it {Energy Dependence of Moments of
  Net-Proton and Net-Charge Multiplicity Distributions at STAR}},  {\em PoS}
  {\bf CPOD2014} (2014) 019, [\href{http://arxiv.org/abs/1503.0255}{{\tt
  arXiv:1503.0255}}].

\bibitem{Mitchell:2012mx}
{\bf PHENIX} Collaboration, J.~T. Mitchell, {\it {The RHIC Beam Energy Scan
  Program: Results from the PHENIX Experiment}},  {\em Nucl.Phys.} {\bf
  A904-905} (2013) 903c--906c, [\href{http://arxiv.org/abs/1211.6139}{{\tt
  arXiv:1211.6139}}].

\bibitem{Bazavov:2012vg}
A.~Bazavov, H.~Ding, P.~Hegde, O.~Kaczmarek, F.~Karsch, et~al., {\it
  {Freeze-out Conditions in Heavy Ion Collisions from QCD Thermodynamics}},
  {\em Phys.Rev.Lett.} {\bf 109} (2012) 192302,
  [\href{http://arxiv.org/abs/1208.1220}{{\tt arXiv:1208.1220}}].

\bibitem{Kitazawa:2012at}
M.~Kitazawa and M.~Asakawa, {\it {Relation between baryon number fluctuations
  and experimentally observed proton number fluctuations in relativistic heavy
  ion collisions}},  {\em Phys.Rev.} {\bf C86} (2012) 024904,
  [\href{http://arxiv.org/abs/1205.3292}{{\tt arXiv:1205.3292}}].

\bibitem{Bzdak:2012ab}
A.~Bzdak and V.~Koch, {\it {Acceptance corrections to net baryon and net charge
  cumulants}},  {\em Phys.Rev.} {\bf C86} (2012) 044904,
  [\href{http://arxiv.org/abs/1206.4286}{{\tt arXiv:1206.4286}}].

\bibitem{Bzdak:2013pha}
A.~Bzdak and V.~Koch, {\it {Local Efficiency Corrections to Higher Order
  Cumulants}},  {\em Phys.Rev.} {\bf C91} (2015) 027901,
  [\href{http://arxiv.org/abs/1312.4574}{{\tt arXiv:1312.4574}}].

\bibitem{Karsch:2012wm}
F.~Karsch, {\it {Determination of Freeze-out Conditions from Lattice QCD
  Calculations}},  {\em Central Eur.J.Phys.} {\bf 10} (2012) 1234--1237,
  [\href{http://arxiv.org/abs/1202.4173}{{\tt arXiv:1202.4173}}].

\bibitem{Mukherjee:2013lsa}
S.~Mukherjee and M.~Wagner, {\it {Deconfinement of strangeness and freeze-out
  from charge fluctuations}},  {\em PoS} {\bf CPOD2013} (2013) 039,
  [\href{http://arxiv.org/abs/1307.6255}{{\tt arXiv:1307.6255}}].

\bibitem{Borsanyi:2014ewa}
S.~Borsanyi, Z.~Fodor, S.~Katz, S.~Krieg, C.~Ratti, et~al., {\it {Freeze-out
  parameters from electric charge and baryon number fluctuations: is there
  consistency?}},  {\em Phys.Rev.Lett.} {\bf 113} (2014) 052301,
  [\href{http://arxiv.org/abs/1403.4576}{{\tt arXiv:1403.4576}}].

\bibitem{MukherjeeCPOD14}
S.~Mukherjee, {\it "freeze-out condition from lattice qcd and the role of
  additional strange hadrons"},  {\em PoS} {\bf CPOD2014} (2014) 005.

\bibitem{Capstick:1986bm}
S.~Capstick and N.~Isgur, {\it {Baryons in a Relativized Quark Model with
  Chromodynamics}},  {\em Phys.Rev.} {\bf D34} (1986) 2809.

\bibitem{Ebert:2009ub}
D.~Ebert, R.~Faustov, and V.~Galkin, {\it {Mass spectra and Regge trajectories
  of light mesons in the relativistic quark model}},  {\em Phys.Rev.} {\bf D79}
  (2009) 114029, [\href{http://arxiv.org/abs/0903.5183}{{\tt
  arXiv:0903.5183}}].

\bibitem{Bazavov:2014yba}
A.~Bazavov, H.-T. Ding, P.~Hegde, O.~Kaczmarek, F.~Karsch, et~al., {\it {The
  melting and abundance of open charm hadrons}},  {\em Phys.Lett.} {\bf B737}
  (2014) 210--215, [\href{http://arxiv.org/abs/1404.4043}{{\tt
  arXiv:1404.4043}}].

\bibitem{Asakawa:2000tr}
M.~Asakawa, T.~Hatsuda, and Y.~Nakahara, {\it {Maximum entropy analysis of the
  spectral functions in lattice QCD}},  {\em Prog.Part.Nucl.Phys.} {\bf 46}
  (2001) 459--508, [\href{http://arxiv.org/abs/hep-lat/0011040}{{\tt
  hep-lat/0011040}}].

\bibitem{Rapp:2008qc}
R.~Rapp and H.~van Hees, {\it {Heavy Quark Diffusion as a Probe of the
  Quark-Gluon Plasma}},  \href{http://arxiv.org/abs/0803.0901}{{\tt
  arXiv:0803.0901}}.

\bibitem{Ding:2014xha}
H.-T. Ding, {\it {Hard and thermal probes of QGP from the perspective of
  Lattice QCD}},  \href{http://arxiv.org/abs/1404.5134}{{\tt arXiv:1404.5134}}.

\bibitem{Meyer:2011gj}
H.~B. Meyer, {\it {Transport Properties of the Quark-Gluon Plasma: A Lattice
  QCD Perspective}},  {\em Eur.Phys.J.} {\bf A47} (2011) 86,
  [\href{http://arxiv.org/abs/1104.3708}{{\tt arXiv:1104.3708}}].

\bibitem{Karsch:2001uw}
F.~Karsch, E.~Laermann, P.~Petreczky, S.~Stickan, and I.~Wetzorke, {\it {A
  Lattice calculation of thermal dilepton rates}},  {\em Phys.Lett.} {\bf B530}
  (2002) 147--152, [\href{http://arxiv.org/abs/hep-lat/0110208}{{\tt
  hep-lat/0110208}}].

\bibitem{Nakahara:1999vy}
Y.~Nakahara, M.~Asakawa, and T.~Hatsuda, {\it {Hadronic spectral functions in
  lattice QCD}},  {\em Phys.Rev.} {\bf D60} (1999) 091503,
  [\href{http://arxiv.org/abs/hep-lat/9905034}{{\tt hep-lat/9905034}}].

\bibitem{Aarts:2007wj}
G.~Aarts, C.~Allton, J.~Foley, S.~Hands, and S.~Kim, {\it {Spectral functions
  at small energies and the electrical conductivity in hot, quenched lattice
  QCD}},  {\em Phys.Rev.Lett.} {\bf 99} (2007) 022002,
  [\href{http://arxiv.org/abs/hep-lat/0703008}{{\tt hep-lat/0703008}}].

\bibitem{Ding:2010ga}
H.-T. Ding, A.~Francis, O.~Kaczmarek, F.~Karsch, E.~Laermann, et~al., {\it
  {Thermal dilepton rate and electrical conductivity: An analysis of vector
  current correlation functions in quenched lattice QCD}},  {\em Phys.Rev.}
  {\bf D83} (2011) 034504, [\href{http://arxiv.org/abs/1012.4963}{{\tt
  arXiv:1012.4963}}].

\bibitem{Ding:2013qw}
O.~Kaczmarek, E.~Laermann, M.~M{\"u}ller, F.~Karsch, H.~Ding, et~al., {\it
  {Thermal dilepton rates from quenched lattice QCD}},  {\em PoS} {\bf
  ConfinementX} (2012) 185, [\href{http://arxiv.org/abs/1301.7436}{{\tt
  arXiv:1301.7436}}].

\bibitem{Kaczmarek:2013dya}
O.~Kaczmarek and M.~M{\"u}ller, {\it {Temperature dependence of electrical
  conductivity and dilepton rates from hot quenched lattice QCD}},  {\em PoS}
  {\bf LATTICE2013} (2014) 175, [\href{http://arxiv.org/abs/1312.5609}{{\tt
  arXiv:1312.5609}}].

\bibitem{Ding:2014dua}
H.-T. Ding, O.~Kaczmarek, and F.~Meyer, {\it {Vector spectral functions and
  transport properties in quenched QCD}},
  \href{http://arxiv.org/abs/1412.5869}{{\tt arXiv:1412.5869}}.

\bibitem{Gupta:2003zh}
S.~Gupta, {\it {The Electrical conductivity and soft photon emissivity of the
  QCD plasma}},  {\em Phys.Lett.} {\bf B597} (2004) 57--62,
  [\href{http://arxiv.org/abs/hep-lat/0301006}{{\tt hep-lat/0301006}}].

\bibitem{Brandt:2012jc}
B.~B. Brandt, A.~Francis, H.~B. Meyer, and H.~Wittig, {\it {Thermal Correlators
  in the $\rho$ channel of two-flavor QCD}},  {\em JHEP} {\bf 1303} (2013) 100,
  [\href{http://arxiv.org/abs/1212.4200}{{\tt arXiv:1212.4200}}].

\bibitem{Aarts:2014nba}
G.~Aarts, C.~Allton, A.~Amato, P.~Giudice, S.~Hands, et~al., {\it {Electrical
  conductivity and charge diffusion in thermal QCD from the lattice}},
  \href{http://arxiv.org/abs/1412.6411}{{\tt arXiv:1412.6411}}.

\bibitem{Amato:2013naa}
A.~Amato, G.~Aarts, C.~Allton, P.~Giudice, S.~Hands, et~al., {\it {Electrical
  conductivity of the quark-gluon plasma across the deconfinement transition}},
   {\em Phys.Rev.Lett.} {\bf 111} (2013) 172001,
  [\href{http://arxiv.org/abs/1307.6763}{{\tt arXiv:1307.6763}}].

\bibitem{FernandezFraile:2005ka}
D.~Fernandez-Fraile and A.~Gomez~Nicola, {\it {The Electrical conductivity of a
  pion gas}},  {\em Phys.Rev.} {\bf D73} (2006) 045025,
  [\href{http://arxiv.org/abs/hep-ph/0512283}{{\tt hep-ph/0512283}}].

\bibitem{Adare:2006nq}
{\bf PHENIX} Collaboration, A.~Adare et~al., {\it {Energy Loss and Flow of
  Heavy Quarks in Au+Au Collisions at s(NN)**(1/2) = 200-GeV}},  {\em
  Phys.Rev.Lett.} {\bf 98} (2007) 172301,
  [\href{http://arxiv.org/abs/nucl-ex/0611018}{{\tt nucl-ex/0611018}}].

\bibitem{Abelev:2006db}
{\bf STAR} Collaboration, B.~Abelev et~al., {\it {Erratum: Transverse momentum
  and centrality dependence of high-$p_T$ non-photonic electron suppression in
  Au+Au collisions at $\sqrt{s_{NN}} = 200$\,GeV}},  {\em Phys.Rev.Lett.} {\bf
  98} (2007) 192301, [\href{http://arxiv.org/abs/nucl-ex/0607012}{{\tt
  nucl-ex/0607012}}].

\bibitem{ALICE:2012ab}
{\bf ALICE} Collaboration, B.~Abelev et~al., {\it {Suppression of high
  transverse momentum D mesons in central Pb-Pb collisions at
  $\sqrt{s_{NN}}=2.76$ TeV}},  {\em JHEP} {\bf 1209} (2012) 112,
  [\href{http://arxiv.org/abs/1203.2160}{{\tt arXiv:1203.2160}}].

\bibitem{He:2014epa}
M.~He, R.~J. Fries, and R.~Rapp, {\it {Modifications of Heavy-Flavor Spectra in
  $\sqrt{s_{\rm NN}}=62.4~{\rm GeV}$ Au-Au Collisions}},
  \href{http://arxiv.org/abs/1409.4539}{{\tt arXiv:1409.4539}}.

\bibitem{Cao:2014fna}
S.~Cao, Y.~Huang, G.-Y. Qin, and S.~A. Bass, {\it {The Influence of Initial
  State Fluctuations on Heavy Quark Energy Loss in Relativistic Heavy-ion
  Collisions}},  \href{http://arxiv.org/abs/1404.3139}{{\tt arXiv:1404.3139}}.

\bibitem{Ding:2011hr}
H.~Ding, A.~Francis, O.~Kaczmarek, F.~Karsch, H.~Satz, et~al., {\it {Heavy
  Quark diffusion from lattice QCD spectral functions}},  {\em J.Phys.} {\bf
  G38} (2011) 124070, [\href{http://arxiv.org/abs/1107.0311}{{\tt
  arXiv:1107.0311}}].

\bibitem{CaronHuot:2007gq}
S.~Caron-Huot and G.~D. Moore, {\it {Heavy quark diffusion in perturbative QCD
  at next-to-leading order}},  {\em Phys.Rev.Lett.} {\bf 100} (2008) 052301,
  [\href{http://arxiv.org/abs/0708.4232}{{\tt arXiv:0708.4232}}].

\bibitem{Kovtun:2003wp}
P.~Kovtun, D.~T. Son, and A.~O. Starinets, {\it {Holography and hydrodynamics:
  Diffusion on stretched horizons}},  {\em JHEP} {\bf 0310} (2003) 064,
  [\href{http://arxiv.org/abs/hep-th/0309213}{{\tt hep-th/0309213}}].

\bibitem{Kaczmarek:2014jga}
O.~Kaczmarek, {\it {Continuum estimate of the heavy quark momentum diffusion
  coefficient $\kappa$}},  \href{http://arxiv.org/abs/1409.3724}{{\tt
  arXiv:1409.3724}}.

\bibitem{Banerjee:2011ra}
D.~Banerjee, S.~Datta, R.~Gavai, and P.~Majumdar, {\it {Heavy Quark Momentum
  Diffusion Coefficient from Lattice QCD}},  {\em Phys.Rev.} {\bf D85} (2012)
  014510, [\href{http://arxiv.org/abs/1109.5738}{{\tt arXiv:1109.5738}}].

\bibitem{CaronHuot:2009uh}
S.~Caron-Huot, M.~Laine, and G.~D. Moore, {\it {A Way to estimate the heavy
  quark thermalization rate from the lattice}},  {\em JHEP} {\bf 0904} (2009)
  053, [\href{http://arxiv.org/abs/0901.1195}{{\tt arXiv:0901.1195}}].

\bibitem{CasalderreySolana:2006rq}
J.~Casalderrey-Solana and D.~Teaney, {\it {Heavy quark diffusion in strongly
  coupled N=4 Yang-Mills}},  {\em Phys.Rev.} {\bf D74} (2006) 085012,
  [\href{http://arxiv.org/abs/hep-ph/0605199}{{\tt hep-ph/0605199}}].

\bibitem{Meyer:2007ic}
H.~B. Meyer, {\it {A Calculation of the shear viscosity in SU(3)
  gluodynamics}},  {\em Phys.Rev.} {\bf D76} (2007) 101701,
  [\href{http://arxiv.org/abs/0704.1801}{{\tt arXiv:0704.1801}}].

\bibitem{Luscher:2001up}
M.~Luscher and P.~Weisz, {\it {Locality and exponential error reduction in
  numerical lattice gauge theory}},  {\em JHEP} {\bf 0109} (2001) 010,
  [\href{http://arxiv.org/abs/hep-lat/0108014}{{\tt hep-lat/0108014}}].

\bibitem{Parisi:1983hm}
G.~Parisi, R.~Petronzio, and F.~Rapuano, {\it {A Measurement of the String
  Tension Near the Continuum Limit}},  {\em Phys.Lett.} {\bf B128} (1983) 418.

\bibitem{DeForcrand:1985dr}
P.~De~Forcrand and C.~Roiesnel, {\it {REFINED METHODS FOR MEASURING LARGE
  DISTANCE CORRELATIONS}},  {\em Phys.Lett.} {\bf B151} (1985) 77--80.

\bibitem{Moore:2004tg}
G.~D. Moore and D.~Teaney, {\it {How much do heavy quarks thermalize in a heavy
  ion collision?}},  {\em Phys.Rev.} {\bf C71} (2005) 064904,
  [\href{http://arxiv.org/abs/hep-ph/0412346}{{\tt hep-ph/0412346}}].

\bibitem{Karsch:1986cq}
F.~Karsch and H.~Wyld, {\it {Thermal Green's Functions and Transport
  Coefficients on the Lattice}},  {\em Phys.Rev.} {\bf D35} (1987) 2518.

\bibitem{Meyer:2003hy}
H.~B. Meyer, {\it {The Yang-Mills spectrum from a two level algorithm}},  {\em
  JHEP} {\bf 0401} (2004) 030,
  [\href{http://arxiv.org/abs/hep-lat/0312034}{{\tt hep-lat/0312034}}].

\bibitem{Nakamura:2004sy}
A.~Nakamura and S.~Sakai, {\it {Transport coefficients of gluon plasma}},  {\em
  Phys.Rev.Lett.} {\bf 94} (2005) 072305,
  [\href{http://arxiv.org/abs/hep-lat/0406009}{{\tt hep-lat/0406009}}].

\bibitem{Meyer:2007dy}
H.~B. Meyer, {\it {A Calculation of the bulk viscosity in SU(3) gluodynamics}},
   {\em Phys.Rev.Lett.} {\bf 100} (2008) 162001,
  [\href{http://arxiv.org/abs/0710.3717}{{\tt arXiv:0710.3717}}].

\bibitem{Karsch:2007jc}
F.~Karsch, D.~Kharzeev, and K.~Tuchin, {\it {Universal properties of bulk
  viscosity near the QCD phase transition}},  {\em Phys.Lett.} {\bf B663}
  (2008) 217--221, [\href{http://arxiv.org/abs/0711.0914}{{\tt
  arXiv:0711.0914}}].

\bibitem{Kharzeev:2007wb}
D.~Kharzeev and K.~Tuchin, {\it {Bulk viscosity of QCD matter near the critical
  temperature}},  {\em JHEP} {\bf 0809} (2008) 093,
  [\href{http://arxiv.org/abs/0705.4280}{{\tt arXiv:0705.4280}}].

\bibitem{Romatschke:2009ng}
P.~Romatschke and D.~T. Son, {\it {Spectral sum rules for the quark-gluon
  plasma}},  {\em Phys.Rev.} {\bf D80} (2009) 065021,
  [\href{http://arxiv.org/abs/0903.3946}{{\tt arXiv:0903.3946}}].

\bibitem{Romatschke:2009kr}
P.~Romatschke, {\it {Relativistic Viscous Fluid Dynamics and Non-Equilibrium
  Entropy}},  {\em Class.Quant.Grav.} {\bf 27} (2010) 025006,
  [\href{http://arxiv.org/abs/0906.4787}{{\tt arXiv:0906.4787}}].

\bibitem{Moore:2010bu}
G.~D. Moore and K.~A. Sohrabi, {\it {Kubo Formulae for Second-Order
  Hydrodynamic Coefficients}},  {\em Phys.Rev.Lett.} {\bf 106} (2011) 122302,
  [\href{http://arxiv.org/abs/1007.5333}{{\tt arXiv:1007.5333}}].

\bibitem{Moore:2012tc}
G.~D. Moore and K.~A. Sohrabi, {\it {Thermodynamical second-order hydrodynamic
  coefficients}},  {\em JHEP} {\bf 1211} (2012) 148,
  [\href{http://arxiv.org/abs/1210.3340}{{\tt arXiv:1210.3340}}].

\bibitem{Denicol:2012cn}
G.~Denicol, H.~Niemi, E.~Molnar, and D.~Rischke, {\it {Derivation of transient
  relativistic fluid dynamics from the Boltzmann equation}},  {\em Phys.Rev.}
  {\bf D85} (2012) 114047, [\href{http://arxiv.org/abs/1202.4551}{{\tt
  arXiv:1202.4551}}].

\bibitem{Baier:2007ix}
R.~Baier, P.~Romatschke, D.~T. Son, A.~O. Starinets, and M.~A. Stephanov, {\it
  {Relativistic viscous hydrodynamics, conformal invariance, and holography}},
  {\em JHEP} {\bf 0804} (2008) 100, [\href{http://arxiv.org/abs/0712.2451}{{\tt
  arXiv:0712.2451}}].

\bibitem{Kohno:2011aa}
Y.~Kohno, M.~Asakawa, and M.~Kitazawa, {\it {Shear viscosity to relaxation time
  ratio in SU(3) lattice gauge theory}},  {\em Phys.Rev.} {\bf D89} (2014)
  054508, [\href{http://arxiv.org/abs/1112.1508}{{\tt arXiv:1112.1508}}].

\bibitem{Philipsen:2013nea}
O.~Philipsen and C.~Sch{\"a}fer, {\it {The second order hydrodynamic transport
  coefficient $\kappa$ for the gluon plasma from the lattice}},  {\em JHEP}
  {\bf 1402} (2014) 003, [\href{http://arxiv.org/abs/1311.6618}{{\tt
  arXiv:1311.6618}}].

\bibitem{Romatschke:2009im}
P.~Romatschke, {\it {New Developments in Relativistic Viscous Hydrodynamics}},
  {\em Int.J.Mod.Phys.} {\bf E19} (2010) 1--53,
  [\href{http://arxiv.org/abs/0902.3663}{{\tt arXiv:0902.3663}}].

\bibitem{Adil:2006ra}
A.~Adil and I.~Vitev, {\it {Collisional dissociation of heavy mesons in dense
  QCD matter}},  {\em Phys.Lett.} {\bf B649} (2007) 139--146,
  [\href{http://arxiv.org/abs/hep-ph/0611109}{{\tt hep-ph/0611109}}].

\bibitem{Sharma:2009hn}
R.~Sharma, I.~Vitev, and B.-W. Zhang, {\it {Light-cone wave function approach
  to open heavy flavor dynamics in QCD matter}},  {\em Phys.Rev.} {\bf C80}
  (2009) 054902, [\href{http://arxiv.org/abs/0904.0032}{{\tt
  arXiv:0904.0032}}].

\bibitem{He:2011qa}
M.~He, R.~J. Fries, and R.~Rapp, {\it {Heavy-Quark Diffusion and Hadronization
  in Quark-Gluon Plasma}},  {\em Phys.Rev.} {\bf C86} (2012) 014903,
  [\href{http://arxiv.org/abs/1106.6006}{{\tt arXiv:1106.6006}}].

\bibitem{Ebert:2009ua}
D.~Ebert, R.~Faustov, and V.~Galkin, {\it {Heavy-light meson spectroscopy and
  Regge trajectories in the relativistic quark model}},  {\em Eur.Phys.J.} {\bf
  C66} (2010) 197--206, [\href{http://arxiv.org/abs/0910.5612}{{\tt
  arXiv:0910.5612}}].

\bibitem{Ebert:2011kk}
D.~Ebert, R.~Faustov, and V.~Galkin, {\it {Spectroscopy and Regge trajectories
  of heavy baryons in the relativistic quark-diquark picture}},  {\em
  Phys.Rev.} {\bf D84} (2011) 014025,
  [\href{http://arxiv.org/abs/1105.0583}{{\tt arXiv:1105.0583}}].

\bibitem{Mocsy:2013syh}
A.~Mocsy, P.~Petreczky, and M.~Strickland, {\it {Quarkonia in the Quark Gluon
  Plasma}},  {\em Int.J.Mod.Phys.} {\bf A28} (2013) 1340012,
  [\href{http://arxiv.org/abs/1302.2180}{{\tt arXiv:1302.2180}}].

\bibitem{Bazavov:2009us}
A.~Bazavov, P.~Petreczky, and A.~Velytsky, {\it {Quarkonium at Finite
  Temperature}},  \href{http://arxiv.org/abs/0904.1748}{{\tt arXiv:0904.1748}}.

\bibitem{Brambilla:2010cs}
N.~Brambilla, S.~Eidelman, B.~Heltsley, R.~Vogt, G.~Bodwin, et~al., {\it {Heavy
  quarkonium: progress, puzzles, and opportunities}},  {\em Eur.Phys.J.} {\bf
  C71} (2011) 1534, [\href{http://arxiv.org/abs/1010.5827}{{\tt
  arXiv:1010.5827}}].

\bibitem{Eichten:1978tg}
E.~Eichten, K.~Gottfried, T.~Kinoshita, K.~Lane, and T.-M. Yan, {\it
  {Charmonium: The Model}},  {\em Phys.Rev.} {\bf D17} (1978) 3090.

\bibitem{Eichten:1979ms}
E.~Eichten, K.~Gottfried, T.~Kinoshita, K.~Lane, and T.-M. Yan, {\it
  {Charmonium: Comparison with Experiment}},  {\em Phys.Rev.} {\bf D21} (1980)
  203.

\bibitem{Mocsy:2008eg}
A.~Mocsy, {\it {Potential Models for Quarkonia}},  {\em Eur.Phys.J.} {\bf C61}
  (2009) 705--710, [\href{http://arxiv.org/abs/0811.0337}{{\tt
  arXiv:0811.0337}}].

\bibitem{Laine:2006ns}
M.~Laine, O.~Philipsen, P.~Romatschke, and M.~Tassler, {\it {Real-time static
  potential in hot QCD}},  {\em JHEP} {\bf 0703} (2007) 054,
  [\href{http://arxiv.org/abs/hep-ph/0611300}{{\tt hep-ph/0611300}}].

\bibitem{Beraudo:2007ky}
A.~Beraudo, J.-P. Blaizot, and C.~Ratti, {\it {Real and imaginary-time Q anti-Q
  correlators in a thermal medium}},  {\em Nucl.Phys.} {\bf A806} (2008)
  312--338, [\href{http://arxiv.org/abs/0712.4394}{{\tt arXiv:0712.4394}}].

\bibitem{Brambilla:2008cx}
N.~Brambilla, J.~Ghiglieri, A.~Vairo, and P.~Petreczky, {\it {Static
  quark-antiquark pairs at finite temperature}},  {\em Phys.Rev.} {\bf D78}
  (2008) 014017, [\href{http://arxiv.org/abs/0804.0993}{{\tt
  arXiv:0804.0993}}].

\bibitem{Caswell:1985ui}
W.~Caswell and G.~Lepage, {\it {Effective Lagrangians for Bound State Problems
  in QED, QCD, and Other Field Theories}},  {\em Phys.Lett.} {\bf B167} (1986)
  437.

\bibitem{Bodwin:1994jh}
G.~T. Bodwin, E.~Braaten, and G.~P. Lepage, {\it {Rigorous QCD analysis of
  inclusive annihilation and production of heavy quarkonium}},  {\em Phys.Rev.}
  {\bf D51} (1995) 1125--1171, [\href{http://arxiv.org/abs/hep-ph/9407339}{{\tt
  hep-ph/9407339}}].

\bibitem{Brambilla:1999xf}
N.~Brambilla, A.~Pineda, J.~Soto, and A.~Vairo, {\it {Potential NRQCD: An
  Effective theory for heavy quarkonium}},  {\em Nucl.Phys.} {\bf B566} (2000)
  275, [\href{http://arxiv.org/abs/hep-ph/9907240}{{\tt hep-ph/9907240}}].

\bibitem{Rothkopf:2011db}
A.~Rothkopf, T.~Hatsuda, and S.~Sasaki, {\it {Complex Heavy-Quark Potential at
  Finite Temperature from Lattice QCD}},  {\em Phys.Rev.Lett.} {\bf 108} (2012)
  162001, [\href{http://arxiv.org/abs/1108.1579}{{\tt arXiv:1108.1579}}].

\bibitem{Burnier:2012az}
Y.~Burnier and A.~Rothkopf, {\it {Disentangling the timescales behind the
  non-perturbative heavy quark potential}},  {\em Phys.Rev.} {\bf D86} (2012)
  051503, [\href{http://arxiv.org/abs/1208.1899}{{\tt arXiv:1208.1899}}].

\bibitem{Burnier:2013fca}
Y.~Burnier and A.~Rothkopf, {\it {A hard thermal loop benchmark for the
  extraction of the nonperturbative $Q\bar{Q}$ potential}},  {\em Phys.Rev.}
  {\bf D87} (2013) 114019, [\href{http://arxiv.org/abs/1304.4154}{{\tt
  arXiv:1304.4154}}].

\bibitem{Burnier:2014ssa}
Y.~Burnier, O.~Kaczmarek, and A.~Rothkopf, {\it {Static quark-antiquark
  potential in the quark-gluon plasma from lattice QCD}},
  \href{http://arxiv.org/abs/1410.2546}{{\tt arXiv:1410.2546}}.

\bibitem{Nadkarni:1986as}
S.~Nadkarni, {\it {Nonabelian Debye Screening. 2. The Singlet Potential}},
  {\em Phys.Rev.} {\bf D34} (1986) 3904.

\bibitem{Bazavov:2014kva}
A.~Bazavov, Y.~Burnier, and P.~Petreczky, {\it {Lattice calculation of the
  heavy quark potential at non-zero temperature}},
  \href{http://arxiv.org/abs/1404.4267}{{\tt arXiv:1404.4267}}.

\bibitem{Rothkopf:2011ef}
A.~Rothkopf, {\it {Improved Maximum Entropy Analysis with an Extended Search
  Space}},  {\em J.Comput.Phys.} {\bf 238} (2013) 106--114,
  [\href{http://arxiv.org/abs/1110.6285}{{\tt arXiv:1110.6285}}].

\bibitem{Rothkopf:2012vv}
A.~Rothkopf, {\it {Improved Maximum Entropy Method with an Extended Search
  Space}},  {\em PoS} {\bf LATTICE2012} (2012) 100,
  [\href{http://arxiv.org/abs/1208.5162}{{\tt arXiv:1208.5162}}].

\bibitem{Burnier:2013nla}
Y.~Burnier and A.~Rothkopf, {\it {Bayesian Approach to Spectral Function
  Reconstruction for Euclidean Quantum Field Theories}},  {\em Phys.Rev.Lett.}
  {\bf 111} (2013) 182003, [\href{http://arxiv.org/abs/1307.6106}{{\tt
  arXiv:1307.6106}}].

\bibitem{Ding:2012sp}
H.~Ding, A.~Francis, O.~Kaczmarek, F.~Karsch, H.~Satz, et~al., {\it {Charmonium
  properties in hot quenched lattice QCD}},  {\em Phys.Rev.} {\bf D86} (2012)
  014509, [\href{http://arxiv.org/abs/1204.4945}{{\tt arXiv:1204.4945}}].

\bibitem{Asakawa:2003re}
M.~Asakawa and T.~Hatsuda, {\it {J / psi and eta(c) in the deconfined plasma
  from lattice QCD}},  {\em Phys.Rev.Lett.} {\bf 92} (2004) 012001,
  [\href{http://arxiv.org/abs/hep-lat/0308034}{{\tt hep-lat/0308034}}].

\bibitem{Datta:2003ww}
S.~Datta, F.~Karsch, P.~Petreczky, and I.~Wetzorke, {\it {Behavior of
  charmonium systems after deconfinement}},  {\em Phys.Rev.} {\bf D69} (2004)
  094507, [\href{http://arxiv.org/abs/hep-lat/0312037}{{\tt hep-lat/0312037}}].

\bibitem{Jakovac:2006sf}
A.~Jakovac, P.~Petreczky, K.~Petrov, and A.~Velytsky, {\it {Quarkonium
  correlators and spectral functions at zero and finite temperature}},  {\em
  Phys.Rev.} {\bf D75} (2007) 014506,
  [\href{http://arxiv.org/abs/hep-lat/0611017}{{\tt hep-lat/0611017}}].

\bibitem{Iida:2006mv}
H.~Iida, T.~Doi, N.~Ishii, H.~Suganuma, and K.~Tsumura, {\it {Charmonium
  properties in deconfinement phase in anisotropic lattice QCD}},  {\em
  Phys.Rev.} {\bf D74} (2006) 074502,
  [\href{http://arxiv.org/abs/hep-lat/0602008}{{\tt hep-lat/0602008}}].

\bibitem{Ohno:2011zc}
{\bf WHOT-QCD} Collaboration, H.~Ohno et~al., {\it {Charmonium spectral
  functions with the variational method in zero and finite temperature lattice
  QCD}},  {\em Phys.Rev.} {\bf D84} (2011) 094504,
  [\href{http://arxiv.org/abs/1104.3384}{{\tt arXiv:1104.3384}}].

\bibitem{Aarts:2007pk}
G.~Aarts, C.~Allton, M.~B. Oktay, M.~Peardon, and J.-I. Skullerud, {\it
  {Charmonium at high temperature in two-flavor QCD}},  {\em Phys.Rev.} {\bf
  D76} (2007) 094513, [\href{http://arxiv.org/abs/0705.2198}{{\tt
  arXiv:0705.2198}}].

\bibitem{Borsanyi:2014vka}
S.~Borsanyi, S.~D{\"u}rr, Z.~Fodor, C.~Hoelbling, S.~D. Katz, et~al., {\it
  {Charmonium spectral functions from 2+1 flavour lattice QCD}},  {\em JHEP}
  {\bf 1404} (2014) 132, [\href{http://arxiv.org/abs/1401.5940}{{\tt
  arXiv:1401.5940}}].

\bibitem{Aarts:2014cda}
G.~Aarts, C.~Allton, T.~Harris, S.~Kim, M.~P. Lombardo, et~al., {\it {The
  bottomonium spectrum at finite temperature from N$_{f}$ = 2 + 1 lattice
  QCD}},  {\em JHEP} {\bf 1407} (2014) 097,
  [\href{http://arxiv.org/abs/1402.6210}{{\tt arXiv:1402.6210}}].

\bibitem{Ohno:2014uga}
H.~Ohno, H.~T. Ding, and O.~Kaczmarek, {\it {Quark mass dependence of
  quarkonium properties at finite temperature}},
  \href{http://arxiv.org/abs/1412.6594}{{\tt arXiv:1412.6594}}.

\bibitem{Thacker:1990bm}
B.~Thacker and G.~P. Lepage, {\it {Heavy quark bound states in lattice QCD}},
  {\em Phys.Rev.} {\bf D43} (1991) 196--208.

\bibitem{Lepage:1992tx}
G.~P. Lepage, L.~Magnea, C.~Nakhleh, U.~Magnea, and K.~Hornbostel, {\it
  {Improved nonrelativistic QCD for heavy quark physics}},  {\em Phys.Rev.}
  {\bf D46} (1992) 4052--4067,
  [\href{http://arxiv.org/abs/hep-lat/9205007}{{\tt hep-lat/9205007}}].

\bibitem{Aarts:2010ek}
G.~Aarts, S.~Kim, M.~Lombardo, M.~Oktay, S.~Ryan, et~al., {\it {Bottomonium
  above deconfinement in lattice nonrelativistic QCD}},  {\em Phys.Rev.Lett.}
  {\bf 106} (2011) 061602, [\href{http://arxiv.org/abs/1010.3725}{{\tt
  arXiv:1010.3725}}].

\bibitem{Aarts:2013kaa}
G.~Aarts, C.~Allton, S.~Kim, M.~Lombardo, S.~Ryan, et~al., {\it {Melting of P
  wave bottomonium states in the quark-gluon plasma from lattice NRQCD}},  {\em
  JHEP} {\bf 1312} (2013) 064, [\href{http://arxiv.org/abs/1310.5467}{{\tt
  arXiv:1310.5467}}].

\bibitem{Aarts:2012ka}
G.~Aarts, C.~Allton, S.~Kim, M.~P. Lombardo, M.~B. Oktay, et~al., {\it {S wave
  bottomonium states moving in a quark-gluon plasma from lattice NRQCD}},  {\em
  JHEP} {\bf 1303} (2013) 084, [\href{http://arxiv.org/abs/1210.2903}{{\tt
  arXiv:1210.2903}}].

\bibitem{Aarts:2011sm}
G.~Aarts, C.~Allton, S.~Kim, M.~Lombardo, M.~Oktay, et~al., {\it {What happens
  to the $\Upsilon$ and $\eta_b$ in the quark-gluon plasma? Bottomonium
  spectral functions from lattice QCD}},  {\em JHEP} {\bf 1111} (2011) 103,
  [\href{http://arxiv.org/abs/1109.4496}{{\tt arXiv:1109.4496}}].

\bibitem{Kim:2014iga}
S.~Kim, P.~Petreczky, and A.~Rothkopf, {\it {Lattice NRQCD study of S- and
  P-wave bottomonium states in a thermal medium with $N_f=2+1$ light flavors}},
   \href{http://arxiv.org/abs/1409.3630}{{\tt arXiv:1409.3630}}.

\bibitem{Florkowski:1993bq}
W.~Florkowski and B.~L. Friman, {\it {Spatial dependence of the finite
  temperature meson correlation function}},  {\em Z.Phys.} {\bf A347} (1994)
  271--276.

\bibitem{Bazavov:2014cta}
A.~Bazavov, F.~Karsch, Y.~Maezawa, S.~Mukherjee, and P.~Petreczky, {\it
  {In-medium modifications of open and hidden strange-charm mesons from spatial
  correlation functions}},  \href{http://arxiv.org/abs/1411.3018}{{\tt
  arXiv:1411.3018}}.

\bibitem{Karsch:2012na}
F.~Karsch, E.~Laermann, S.~Mukherjee, and P.~Petreczky, {\it {Signatures of
  charmonium modification in spatial correlation functions}},  {\em Phys.Rev.}
  {\bf D85} (2012) 114501, [\href{http://arxiv.org/abs/1203.3770}{{\tt
  arXiv:1203.3770}}].

\bibitem{DElia:2012tr}
M.~D'Elia, {\it {Lattice QCD Simulations in External Background Fields}},  {\em
  Lect.Notes Phys.} (2013) [\href{http://arxiv.org/abs/1209.0374}{{\tt
  arXiv:1209.0374}}].

\bibitem{Yamamoto:2012bi}
A.~Yamamoto, {\it {Chiral Magnetic Effect on the Lattice}},  {\em Lect.Notes
  Phys.} {\bf 871} (2013) 387--397, [\href{http://arxiv.org/abs/1207.0375}{{\tt
  arXiv:1207.0375}}].

\bibitem{Endrodi:2014vza}
G.~Endrodi, {\it {QCD in magnetic fields: from Hofstadter's butterfly to the
  phase diagram}},  \href{http://arxiv.org/abs/1410.8028}{{\tt
  arXiv:1410.8028}}.

\bibitem{DElia:2015rwa}
M.~D'Elia, {\it {Lattice QCD with purely imaginary sources at zero and non-zero
  temperature}},  \href{http://arxiv.org/abs/1502.0604}{{\tt arXiv:1502.0604}}.

\bibitem{DElia:2010nq}
M.~D'Elia, S.~Mukherjee, and F.~Sanfilippo, {\it {QCD Phase Transition in a
  Strong Magnetic Background}},  {\em Phys.Rev.} {\bf D82} (2010) 051501,
  [\href{http://arxiv.org/abs/1005.5365}{{\tt arXiv:1005.5365}}].

\bibitem{Ilgenfritz:2012fw}
E.-M. Ilgenfritz, M.~Kalinowski, M.~Muller-Preussker, B.~Petersson, and
  A.~Schreiber, {\it {Two-color QCD with staggered fermions at finite
  temperature under the influence of a magnetic field}},  {\em Phys.Rev.} {\bf
  D85} (2012) 114504, [\href{http://arxiv.org/abs/1203.3360}{{\tt
  arXiv:1203.3360}}].

\bibitem{Bali:2011qj}
G.~Bali, F.~Bruckmann, G.~Endrodi, Z.~Fodor, S.~Katz, et~al., {\it {The QCD
  phase diagram for external magnetic fields}},  {\em JHEP} {\bf 1202} (2012)
  044, [\href{http://arxiv.org/abs/1111.4956}{{\tt arXiv:1111.4956}}].

\bibitem{Bornyakov:2013eya}
V.~Bornyakov, P.~Buividovich, N.~Cundy, O.~Kochetkov, and A.~Sch{\"a}fer, {\it
  {Deconfinement transition in two-flavor lattice QCD with dynamical overlap
  fermions in an external magnetic field}},  {\em Phys.Rev.} {\bf D90} (2014)
  034501, [\href{http://arxiv.org/abs/1312.5628}{{\tt arXiv:1312.5628}}].

\bibitem{Ilgenfritz:2013ara}
E.~M. Ilgenfritz, M.~Muller-Preussker, B.~Petersson, and A.~Schreiber, {\it
  {Magnetic catalysis (and inverse catalysis) at finite temperature in
  two-color lattice QCD}},  {\em Phys.Rev.} {\bf D89} (2014) 054512,
  [\href{http://arxiv.org/abs/1310.7876}{{\tt arXiv:1310.7876}}].

\bibitem{DElia:2011zu}
M.~D'Elia and F.~Negro, {\it {Chiral Properties of Strong Interactions in a
  Magnetic Background}},  {\em Phys.Rev.} {\bf D83} (2011) 114028,
  [\href{http://arxiv.org/abs/1103.2080}{{\tt arXiv:1103.2080}}].

\bibitem{Bali:2012zg}
G.~Bali, F.~Bruckmann, G.~Endrodi, Z.~Fodor, S.~Katz, et~al., {\it {QCD quark
  condensate in external magnetic fields}},  {\em Phys.Rev.} {\bf D86} (2012)
  071502, [\href{http://arxiv.org/abs/1206.4205}{{\tt arXiv:1206.4205}}].

\bibitem{Bali:2014kia}
G.~Bali, F.~Bruckmann, G.~Endr{\"o}di, S.~Katz, and A.~Sch{\"a}fer, {\it {The
  QCD equation of state in background magnetic fields}},  {\em JHEP} {\bf 1408}
  (2014) 177, [\href{http://arxiv.org/abs/1406.0269}{{\tt arXiv:1406.0269}}].

\bibitem{Bonati:2013lca}
C.~Bonati, M.~D'Elia, M.~Mariti, F.~Negro, and F.~Sanfilippo, {\it {Magnetic
  Susceptibility of Strongly Interacting Matter across the Deconfinement
  Transition}},  {\em Phys.Rev.Lett.} {\bf 111} (2013) 182001,
  [\href{http://arxiv.org/abs/1307.8063}{{\tt arXiv:1307.8063}}].

\bibitem{Bonati:2013vba}
C.~Bonati, M.~D'Elia, M.~Mariti, F.~Negro, and F.~Sanfilippo, {\it {Magnetic
  susceptibility and equation of state of $N_f=2+1$ QCD with physical quark
  masses}},  {\em Phys.Rev.} {\bf D89} (2014) 054506,
  [\href{http://arxiv.org/abs/1310.8656}{{\tt arXiv:1310.8656}}].

\bibitem{Bali:2013esa}
G.~Bali, F.~Bruckmann, G.~Endrodi, F.~Gruber, and A.~Schaefer, {\it {Magnetic
  field-induced gluonic (inverse) catalysis and pressure (an)isotropy in QCD}},
   {\em JHEP} {\bf 1304} (2013) 130,
  [\href{http://arxiv.org/abs/1303.1328}{{\tt arXiv:1303.1328}}].

\bibitem{Bali:2013owa}
G.~Bali, F.~Bruckmann, G.~Endrodi, and A.~Schafer, {\it {Paramagnetic squeezing
  of QCD matter}},  {\em Phys.Rev.Lett.} {\bf 112} (2014) 042301,
  [\href{http://arxiv.org/abs/1311.2559}{{\tt arXiv:1311.2559}}].

\bibitem{Levkova:2013qda}
L.~Levkova and C.~DeTar, {\it {Quark-gluon plasma in an external magnetic
  field}},  {\em Phys.Rev.Lett.} {\bf 112} (2014) 012002,
  [\href{http://arxiv.org/abs/1309.1142}{{\tt arXiv:1309.1142}}].

\bibitem{Endrodi:2013cs}
G.~Endr{\"o}di, {\it {QCD equation of state at nonzero magnetic fields in the
  Hadron Resonance Gas model}},  {\em JHEP} {\bf 1304} (2013) 023,
  [\href{http://arxiv.org/abs/1301.1307}{{\tt arXiv:1301.1307}}].

\bibitem{Abramczyk:2009gb}
M.~Abramczyk, T.~Blum, G.~Petropoulos, and R.~Zhou, {\it {Chiral magnetic
  effect in 2+1 flavor QCD+QED}},  {\em PoS} {\bf LAT2009} (2009) 181,
  [\href{http://arxiv.org/abs/0911.1348}{{\tt arXiv:0911.1348}}].

\bibitem{Buividovich:2009wi}
P.~Buividovich, M.~Chernodub, E.~Luschevskaya, and M.~Polikarpov, {\it
  {Numerical evidence of chiral magnetic effect in lattice gauge theory}},
  {\em Phys.Rev.} {\bf D80} (2009) 054503,
  [\href{http://arxiv.org/abs/0907.0494}{{\tt arXiv:0907.0494}}].

\bibitem{Bali:2014vja}
G.~Bali, F.~Bruckmann, G.~Endr{\"o}di, Z.~Fodor, S.~Katz, et~al., {\it {Local
  CP-violation and electric charge separation by magnetic fields from lattice
  QCD}},  {\em JHEP} {\bf 1404} (2014) 129,
  [\href{http://arxiv.org/abs/1401.4141}{{\tt arXiv:1401.4141}}].

\end{thebibliography}\endgroup

\end{document}